\documentclass[a4paper,12pt]{book}

\usepackage{amsmath}
\usepackage{amsthm}
\usepackage{amsfonts}
\usepackage{amssymb}
\usepackage[T1]{fontenc}
\usepackage[latin1]{inputenc}
\usepackage[english,british,german]{babel}
\usepackage[matrix,arrow,graph]{xy}
\usepackage{epsfig}
\usepackage{graphics}
\usepackage{graphicx,indentfirst}
\numberwithin{equation}{section}
\usepackage{psfrag}
\usepackage{ccaption}

\usepackage{amssymb}
\usepackage{enumerate}
\usepackage{epsfig}
\usepackage{array}
\usepackage{bbm}
\usepackage{color}
\usepackage{a4wide}
\usepackage{cite}
\usepackage{makeidx}
\usepackage{lscape}
\usepackage{longtable}
\usepackage{hhline}
\usepackage{fancyhdr}
\usepackage{caption}

\usepackage{titlesec}

\titleformat{\chapter}[display]
{\normalfont\huge\bfseries}{\chaptertitlename\ \thechapter}{20pt}{\Huge}
\titlespacing*{\chapter} {00pt}{0pt}{15pt}

\makeindex

\title{\Huge \bfseries Branes in Supergroups}

\author{\phantom{AA}\\
\phantom{AA}\\
\phantom{AA}\\ \vspace{.2cm}
\Large \bfseries Dissertation\\ \vspace{.2cm}
\Large \bfseries zur Erlangung des Doktorgrades\\ \vspace{.2cm}
\Large \bfseries des Departments Physik\\ \vspace{.2cm}
	\Large \bfseries der Universit\"at Hamburg\\\\\\\\\\\\\\
	\vspace{.2cm}  vorgelegt von\\ \vspace{.25cm}
	{\Large Thomas Creutzig}\\
	\vspace{.2cm} aus Hannover\\\\\\\\}
\date{Hamburg\\ 2009}

\def \Fig#1#2#3 {
\begin{figure}
\centering
\epsfxsize=#2cm \epsfbox{#1.eps}
\caption{#3}
\label{#1}
\end{figure}
}

\newcount\figno
\def\fig#1#2#3{
\par\begingroup\parindent=0pt\leftskip=1cm\rightskip=1cm\parindent=0pt
\baselineskip=15pt
\global\advance\figno by 1
\epsfxsize=#3
\centerline{\epsfbox{#2}}
\vskip 12pt
{\bf \small Figure \the\figno:} {\small #1}\par
\endgroup\par
}
\def\figlabel#1{\xdef#1{\the\figno
\mbox{ }}}
\def\encadremath#1{\vbox{\hrule\hbox{\vrule\kern8pt\vbox{\kern8pt
\hbox{$\displaystyle #1$}\kern8pt}
\kern8pt\vrule}\hrule}}

\def\L2{{\it Fun\/}\bigl(\text{\GL}\bigr)}

\newcommand{\chim}{\chi_-}
\newcommand{\chip}{\chi_+}
\newcommand{\chipm}{\chi_\pm}
\newcommand{\Pm}{\psi^-}
\newcommand{\Pp}{\psi^+}
\newcommand{\Ppm}{\psi^\pm}
\newcommand{\Pmp}{\psi^\mp}
\newcommand{\Om}{\Omega}
\newcommand{\bm}{b_-}
\newcommand{\bp}{b_+}
\newcommand{\cm}{c_-}
\newcommand{\cp}{c_+}
\newcommand{\cpm}{c_\pm}
\newcommand{\e}{\text{exp}}
\newcommand{\p}{\partial_p}
\newcommand{\n}{\partial_n}
\newcommand{\del}{\partial}
\newcommand{\dm}{d_-}
\newcommand{\dpp}{d_+}
\newcommand{\dpm}{d_\pm}
\newcommand{\ep}{\epsilon}
\newcommand{\etam}{\eta_-}
\newcommand{\etap}{\eta_+}
\newcommand{\etapm}{\eta_\pm}
\newcommand{\zm}{\zeta_-}
\newcommand{\zp}{\zeta_+}
\newcommand{\zpm}{\zeta_\pm}
\newcommand{\Ad}{\text{Ad}}
\newcommand{\ad}{\text{ad}}
\newcommand{\id}{\text{id}}
\def\agl{{\rm $\widehat{\text{gl}}$(1$|$1)}}
\def\gl{{\rm gl(1$|$1)}}
\def\GL{{\rm GL(1$|$1)}}
\newcommand{\rrangle}{{\rangle\!\rangle}}
\newcommand{\llangle}{{\langle\!\langle}}
\newcommand{\g}{\mathfrak{g}}
\newcommand{\Z}{\mathbb{Z}}
\newcommand{\R}{\mathbb{R}}
\newcommand{\K}{\mathcal{K}}
\newcommand{\C}{\mathbb{C}}
\newcommand{\T}{\mathcal{T}}
\newcommand{\str}{\text{str}}
\newcommand{\tr}{\text{tr}}
\newcommand{\sdet}{\text{sdet}}
\newcommand{\sdim}{\text{sdim}}
\newcommand{\sign}{\text{sign}}
\newcommand{\<}{\langle}
\renewcommand{\>}{\rangle}
\newcommand{\gzero}{\mathfrak{g}_{\bar{0}}}
\newcommand{\gone}{\mathfrak{g}_{\bar{1}}}

\newcommand{\m}{\mathfrak{M}}
\newcommand{\nn}{\mathfrak{n}}

\newcommand{\phibdy}{\phi^{\text{bdy}}}
\newcommand{\phibulk}{\phi^{\text{bulk}}}

\renewcommand{\k}{\mathfrak{k}}
\newcommand{\s}{\mathfrak{g}}

\newcommand{\QZ}{Z}
\newcommand{\xib}{\xi^b} 
\newcommand{\xic}{\xi^c} 

\renewcommand{\e}{\varepsilon}

\providecommand{\abs}[1]{\left\lvert#1\right\rvert}

\newcommand{\al}{\alpha} 
\newcommand{\Ga}{\Gamma} 
 \newcommand{\de}{\delta}
 \newcommand{\si}{\sigma}
\newcommand{\la}{\lambda} 

 \newcommand{\De}{\Delta}

\newcommand{\G}{\mathsf{G}}
\newcommand{\HH}{\mathsf{H}}

\newcommand{\half}{\mbox{$\frac{1}{2}$}}

\newcommand{\parij}{{|ij|}}
\newcommand{\parik}{{|ik|}}

\newcommand{\parjk}{{|jk|}}

\newcommand{\ap}{\mathfrak{a_+}}
\newcommand{\am}{\mathfrak{a_-}}
\newcommand{\apm}{\mathfrak{a_\pm}}
\newcommand{\azero}{\mathfrak{a_0}}

\newcommand{\h}{\mathfrak{h}}

\providecommand{\abs}[1]{\left\lvert#1\right\rvert}

\newcommand{\expect}[1]{\langle #1\rangle}

\newcommand{\normord}[1]{{} : #1 : {}}

\newcommand{\delbar}{\bar{\partial}}
\newcommand{\zbar}{\bar{z}}

\renewcommand{\e}{\varepsilon}

\def\agl{{\rm $\widehat{\text{gl}}$(1$|$1)}}
\def\gl{{\rm gl(1$|$1)}}
\def\GL{{\rm GL(1$|$1)}}

\newtheorem{thm}{Theorem}[section]
\newtheorem{prp}[thm]{Proposition}

\newtheorem{definition}[thm]{Definition}
\newtheorem{example}[thm]{Example}

\begin{document}\selectlanguage{british}

\maketitle

\phantom{23}
\thispagestyle{empty}
\vspace{15cm}
\vfill
\begin{tabular}{ll}
Gutachter der Dissertation: 	& Prof.~Dr.~V.~Schomerus \\
						& Prof.~Dr.~K.~Fredenhagen\\
Gutachter der Disputation:	& Prof.~Dr.~V.~Schomerus \\
						& Prof.~Dr.~J.~Louis \\
Datum der Disputation:		& 19.05.2009 \\
Vorsitzende des Pr\"ufungsausschusses:	& Prof.~Dr.~C.~Hagner \\
Vorsitzender des Promotionsausschusses:	& Prof.~Dr.~R.~Klanner  \\
Dekan der Fakult\"at MIN:	& Prof.~Dr.~H.~Graener
\end{tabular}
\newpage

\thispagestyle{empty}

\phantom{23}
\vspace{.1cm}
\begin{center}
{\large \bfseries Abstract}
\end{center}
In this thesis we initiate a systematic study of branes in Wess-Zumino-Novikov-Witten models with Lie supergroup target space. We start by showing that a branes' worldvolume is a twisted superconjugacy class and construct the action of the boundary WZNW model.
Then we consider symplectic fermions and give a complete description of boundary states including twisted sectors. Further we show that the \GL\ WZNW model is equivalent to symplectic fermions plus two scalars. We then consider the \GL\ boundary theory. Twisted and untwisted Cardy boundary states are constructed explicitly and their amplitudes are computed.
In the twisted case we find a perturbative formulation of the model. For this purpose the introduction of an additional fermionic boundary degree of freedom is necessary. We compute all bulk one-point functions, bulk-boundary two-point functions and boundary three-point functions. Logarithmic singularities appear in bulk-boundary as well as pure boundary correlation functions.

Finally we turn to world-sheet and target space supersymmetric models. There is $N=2$ superconformal symmetry in many supercosets and also in certain supergroups. In the supergroup case we find some branes that preserve the topological A-twist and some that preserve the B-twist. 

\vspace{2.3cm}
\vfill
\begin{center}
{\large \bfseries Zusammenfassung}
\end{center}
In dieser Arbeit beginnen wir mit einer systematischen Untersuchung von Branen in Wess-Zumino-Novikov-Witten Modellen mit Lie Supergruppen Zielraum. 
Zuerst zeigen wir, dass das Weltvolumen einer Bran eine getwistete Superkonjugationsklasse ist. Dann konstruieren wir die Wirkung des Rand WZNW Models.
Danach betrachten wir symplektische Fermionen und geben eine komplette Beschreibung von Randzust\"anden einschliesslich getwisteter Sektoren.
Weiterhin zeigen wir, dass das \GL\ WZNW Model \"aquivalent ist zu symplektischen Fermionen plus zwei skalaren Feldern. Danach betrachten wir die \GL\ Randtheorie. 
Getwistete und nicht getwistete Cardy Randzust\"ande sind explizit konstruiert und Amplituden berechnet.
In der getwisteten Randtheorie finden wir eine perturbative Beschreibung des Models.
Daf\"ur ist die Einf\"uhrung eines zus\"atzlichen fermionischen Randfreiheitsgrades notwendig.
Wir berechnen alle Bulk Ein-Punkt Funktionen, Bulk-Rand Zwei-Punkt Funtionen und Rand Drei-Punkt Funktionen. Logarithmische Singularit\"aten treten sowohl in den Bulk-Rand Korrelationsfuntionen auf wie auch in den reinen Randkorrelatoren.

Letzendlich betrachten wir Modelle, deren Zielraum wie auch deren Weltfl\"ache supersymmetrisch ist.
Es gibt $N=2$ superkonforme Symmetrie in vielen Supercosets aber auch in einigen Supergruppen.
Im Supergruppenfall finden wir Branen die den topologischen A-twist erhalten und welche die den B-twist erhalten. 
\vfill
\newpage
\thispagestyle{empty}

\frontmatter

\tableofcontents

\mainmatter

 
\chapter{Introduction}

Conformal field theory (CFT) with supersymmetric target space received increased interest with the discovery of dualities between gauge theories and models of gravity. 
These correspondences are highly valuable since in the strongly coupled regime of one model, where it is almost inaccessible, the dual description is weakly coupled and thus well treatable. The first example of such a duality is due to Juan Maldacena \cite{Maldacena:1997re}. His conjecture is that type IIB string theory compactified on $AdS_5\times S^5$ is exactly equivalent to four-dimensional $N=4$ super Yang-Mills theory. 
The group of global symmetries of these two models is the Lie supergroup PSU(2,2$|$4). In the limit of large rank of the gauge group the dual string theory is described by a conformal field theory with target space being a coset of the supergroup PSU(2,2$|$4). The bosonic subspace of this coset is the ten-dimensional space $AdS_5\times S^5$.

Since Maldacena's discovery many more dualities where conjectured, studied and tested. One of them involves string theory on $AdS_3\times S^3$ whose global symmetry is the Lie supergroup PSU(1,1$|$2) \cite{Maldacena:1997re}. The corresponding sigma model is the principal chiral model of this supergroup. It is an exactly marginal perturbation of the Wess-Zumino-Novikov-Witten (WZNW) model on PSU(1,1$|$2).

Lie supergroup WZNW models are an interesting class of theories in its own right. They describe conformal field theories with target space supersymmetry. Their symmetry algebra consists of two copies of an affine Lie superalgebra. This additional infinite dimensional symmetry is a powerfull aide in solving the theory.
Studying these WZNW models gives valuable insights in the representation theory of the affine Lie superalgebra. Moreover supergroup WZNW models provide a class of non-unitary logarithmic CFTs. Here logarithmic means that some correlation functions possess logarithmic singularities. These are due to fields which transform in representations of the Virasoro algebra that are reducible but indecomposable.

Logarithmic CFTs have important applications in many statistical models.
Some examples are critical polymers and percolation \cite{Saleur:1991hk,Parisi:1980in,Watts:1996yh,Mathieu:2007pe}, two-dimensional turbulences \cite{RahimiTabar:1995nc,Flohr:1996ik}, the quantum Hall effect \cite{Zirnbauer:1999ua} and disordered systems \cite{Guruswamy:1999hi,Gurarie:1999yx,Ludwig:2000em}. Furthermore, supersymmetric target spaces play an important role in the description of polymers and percolation. The integer quantum Hall effect is argued to be described by a sigma model on the supermanifold
$U(1,1|2)/\bigl(U(1|1)\times U(1|1)\bigr)$ \cite{Zirnbauer:1999ua,Weidenmuller:1987gi}.
Further the supergroup $GL(N|N)$ appears in the context of disordered Dirac fermions \cite{Guruswamy:1999hi}.

Problems in condensed matter and statistical physics naturally involve boundaries. In such cases boundary CFT becomes relevant (see e.g. \cite{Saleur:1998hq,Saleur:2000gp}). 
Moreover, in the string theory context a boundary CFT corresponds to an open string starting and ending on two branes. 
In addition, boundary CFT displays a rich mathematical structure.
The understanding of the boundary theory is closely connected to modular properties and fusion. Further twisted K-theory appears in the geometric description of branes \cite{Fredenhagen:2000ei}.
These problems are not understood for boundary logarithmic field theory, see however \cite{Kogan:2000fa,Ishimoto:2001jv,Kawai:2001ur,Bredthauer:2002ct,Bredthauer:2002xb,Pearce:2006sz,Gaberdiel:2006pp} for progress in specific models.
WZNW models on Lie supergroups present themselves as an ideal playground to extend many of the beautiful results of unitary rational CFT to logarithmic models.
\smallskip

Inspired by these applications it is an apparent task to systematically study Lie supergroup conformal field theory. 

For every Lie supergroup, as for every Lie group, there exists one conformal field theory, the Wess-Zumino-Novikov-Witten model. But some Lie supergroups possess an even richer structure. If a Lie supergroup is simple and its dual Coxeter number vanishes then there exists a whole family of exactly marginal deformations of the WZNW model \cite{Bershadsky:1999hk}. In view of the AdS/CFT correspondence the supergroup PSU(1,1$|$2) is an interesting example.

First steps in understanding supergroup WZNW models were done by Rozansky and Saleur, who studied the simplest example the GL(1$|$1) WZNW model \cite{Rozansky:1992td,Rozansky:1992zt,Rozansky:1992rx}. Later, the \GL\ model was reconsidered from a more geometric perspective \cite{Schomerus:2005bf}. This geometric approach was then further generalised to the supergroup PSU(1,1$|$2) \cite{Gotz:2006qp} and to a general class of supergroups \cite{Quella:2007hr}.

These considerations were restricted to bulk WZNW models that is, in the string language, to closed strings. There are two natural tasks.
One also should understand the exactly marginal deformations of WZNW models on simple supergroups with vanishing dual Coxeter number.
In \cite{Quella:2007sg} we considered the deformation of boundary spectra in the PSL(2$|$2) sigma model.

The other task is to understand boundary WZNW models on supergroups.
 The aim of this thesis is to initiate a systematic study of such boundary CFTs. We want to understand how to compute boundary correlation functions and boundary spectra. Moreover we want to investigate characteristic features of logarithmic theories as e.g. indecomposability of representations and logarithmic singularities of correlation functions.
\smallskip

Most of this thesis we will restrict to the boundary \GL\ WZNW model, since this is the simplest example that captures prototypical features. The thesis is organised as follows.

In chapter two, we introduce Lie superalgebras. We start with the example of the Lie superalgebra \gl\ and its representations before we turn to the general class of Lie superalgebras. Representations of Lie superalgebras are sometimes reducible but indecomposable. These representations, which we call atypical, are responsible for logarithmic singularities in correlation functions. We give a geometric interpretation of representations and atypicality in terms of superconjugacy classes. The relevance of this geometric description is then given in chapter three. 

In chapter three, we start with some introductory remarks to two-dimensional CFT and to the concept of boundary states. Then we explain conformal and affine Lie superalgebra symmetry of the WZNW models. Further we explain a method to treat the bulk theory. 
Thereafter we begin with the boundary theory. Boundary fields are supported on a subsupermanifold of the supergroup, the branes' worldvolume. We show that this is a twisted superconjugacy class. This insight is then used to find the action of the boundary WZNW model. 

In chapter four, we investigate symplectic fermions. We start by reviewing the bulk model including twisted sectors. In a twisted sector the modes of the fields are non-integer. We then turn to the boundary theory. The symplectic fermions possess an $SL(2)$-family of boundary conditions that preserve conformal symmetry. Boundary states in twisted and untwisted sectors are constructed. Further we compute the spectrum of an open string stretching between two branes and we construct the corresponding boundary theory.

Chapter five is the main part of this thesis, a detailed study of \GL. We start by reconsidering the bulk model and give a new approach via symplectic fermions. 
Precisely, we show that the \GL\ WZNW model is exactly equivalent to a pair of symplectic fermions and a pair of scalars. Twisted fields are important in this correspondence.

Then we turn to the boundary theory. \GL\ possesses two families of branes. The untwisted family consists of point-like branes in the bosonic directions and generically extending into the fermionic directions. The twisted case contains only one volume-filling brane.
We use the symplectic fermion correspondence to construct all Cardy boundary states explicitly. Then we compute spectra of strings stretching between two branes. 
We identify boundary states with representations and find as in the Lie group case that amplitudes are given by fusion. Further we find that in the semiclassical limit a boundary state is a distribution on a superconjugacy class. This superconjugacy class is identified with the same representation as the boundary state.

Then we restrict to the volume-filling brane.
We find a first order formulation, which allows for a perturbative computation of correlation functions. In finding the set-up, it turned out that it is necessary to introduce additional fermionic boundary degrees of freedom. We then compute those correlation functions that determine the theory completely, i.e. we solve this boundary model. We find logarithmic singularities in bulk-boundary as well as boundary-boundary correlators. 

In chapter six, we turn to a different question. We want to investigate CFTs that possess world-sheet supersymmetry in addition to target space supersymmetry. 
We start with a presentation of topological CFT and gauged $N=1$ supersymmetric WZNW models.
In the spirit of Kazama and Suzuki we then find $N=2$ superconformal field theories with supercoset target. But remarkably we also find some with supergroup target. We show that these models possess two families of branes preserving either the topological A-twist or the B-twist.

In the outlook, we state some open questions for future research. 

The appendix contains modular properties of representations of the affine Lie superalgebra
\agl, as well as some integral formulae. In addition, we discuss the $bc$-ghost system of central charge $c=-2$. This model is non-logarithmic in the bulk, but for our choice of boundary conditions the boundary theory is logarithmic.

The original part of this thesis begins with section \ref{section:boundarycft}. Additional results before this section are indicated.

\chapter{Super algebra}

The aim of this thesis is to study sigma models with Lie supergroup as target space. 
As a first step we need to understand some concepts of supergroups and their associated Lie superalgebras. The theory is in many aspects closely related to its bosonic analogue of Lie theory, but with some important new features. 

This chapter is an introduction of the mathematical concepts necessary to study supergroup sigma models.
Since this thesis will be mainly concerned with the sigma model on the Lie supergroup \GL\ we start with its Lie superalgebra as an example. It exhibits most features that are special to Lie superalgebras. 
In the remaining sections of this chapter, we will then introduce some concepts of the general theory of Lie superalgebras, Lie supergroups and their representations.

\section{An example: The Lie superalgebra \gl}

The Lie superalgebra \gl\ has been discussed in detail in \cite{Schomerus:2005bf}.

\gl\ is generated by two bosonic elements $E,N$ and two fermionic elements $\psi^\pm$. $E$ is central and the other three generators obey
\begin{equation}
 [N,\psi^\pm]\ = \ \pm\psi^\pm\qquad\text{and}\qquad\{\psi^+,\psi^-\}\ = \ E\, .
\end{equation}
This superalgebra is solvable and not semi-simple, it thus has two linear independent choices of invariant bilinear form. The relevant one for our purposes will be
\begin{equation}
    \langle\, N\, ,\, E \,\rangle\ = \ \langle\, E\, ,\, N \,\rangle\ = \ 
     \langle\, \psi^+\, ,\, \psi^- \,\rangle\ = \  -\langle\, \psi^-\, ,\, \psi^+ \,\rangle\ = \  1\, .
\end{equation}
The bilinear form is supersymmetric, i.e. symmetric in the bosonic part and antisymmetric in the fermionic part. There exists another important operator, the quadratic Casimir. For the above choice of metric it is
\begin{equation}
  C\ = \ NE+EN+\psi^-\psi^+-\psi^+\psi^-\, .
\end{equation}
It commutes with every element of the Lie superalgebra.

\subsection{Automorphisms}\label{section:automorphismsgl11}

One important ingredient to boundary sigma models are automorphisms of the Lie superalgebra.
Automorphisms are one-to-one maps of the Lie superalgebra to itself that are compatible with the structure of the Lie superalgebra. 
As in the Lie algebra case, one distinguishes inner and outer automorphisms. An inner automorphism is obtained by conjugation with an element of the bosonic subgroup. Since $E$ is central only conjugation by $\exp i\alpha N$ is non-trivial. It acts as follows
\begin{equation}\label{eq:innerautomorphismsgl11}
        \omega_\alpha(E) \ =\ E \qquad,\qquad
        \omega_\alpha(N) \ = \ N \qquad\text{and}\qquad
        \omega_\alpha(\Ppm) \ = \ e^{\pm i\alpha} \Ppm \, .
\end{equation}
The group of outer automorphism is generated by
\begin{equation}\label{eq:twistedautomorphismgl11}
        \Omega(E) \ = \ - E \qquad, \qquad
        \Omega(N) \ = \ - N \qquad\text{and}\qquad
        \Omega(\Ppm) \ = \ \pm \Pmp\, , \
\end{equation}
by
\begin{equation}\label{eq:piautomorphismgl11}
        \Pi(E) \ = \  E \qquad, \qquad
        \Pi(N) \ = \ - N \qquad\text{and}\qquad
        \Pi(\Ppm) \ = \ \Pmp \, 
\end{equation}
and by the family
\begin{equation}\label{eq:transautomorphismgl11}
        \tau_\alpha(E) \ = \  E \qquad, \qquad
        \tau_\alpha(N) \ = \  N+\alpha E \qquad\text{and}\qquad
        \tau_\alpha(\Ppm) \ = \ \Ppm \, . 
\end{equation}
An automorphism is suitable for boundary conformal field theory if it preserves the metric, or equivalently if it leaves the Casimir invariant. 
We compute that the inner automorphisms $\omega_\alpha$ as well as $\Omega$ leave the Casimir invariant,
\begin{equation}
   \omega_\alpha(C) \ = \ \Omega(C) \ = \ C\, .
\end{equation}
On the other hand $\Pi$ and $\tau_\alpha$ act non-trivially on $C$,
\begin{equation}
 \Pi(C)\ =\ -C\qquad\text{and}\qquad\tau_\alpha(C) \ \neq \ C \ \ \text{for}\ \alpha\ \neq \ 0\, .
\end{equation}

\subsection{Representations}

Representations of Lie superalgebras fall into two types, typical irreducible representations and atypical indecomposable representations\footnote{Atypical representations can also be irreducible}. 

In \gl\ all typical representations are two dimensional, we denote them by $\langle e,n\rangle$. They are constructed from a state $|e,n\rangle$ (with $e\neq0$) satisfying
\begin{equation}\label{eq:gltypical}
 \begin{split}
    E|e,n\rangle \ &= \ e|e,n\rangle \ ,\\
    N|e,n\rangle \ &= \ n|e,n\rangle\ ,\\ 
    \psi^+|e,n\rangle \ &= \ 0 \
\end{split}
\end{equation}
and $\psi^-$ acts freely on this state, hence
\begin{equation}\label{eq:gltypical2} 
             \psi^+\psi^-|e,n\rangle \ = \ e|e,n\rangle\, .
\end{equation}
We summarise the representation in  figure~2.1.
\begin{figure}[htb!]`
	\label{fig:typical}
\centering%
\psfrag{n}{$|e, n \rangle$}
\psfrag{n-}{$\psi^-|e,n \rangle$}
\psfrag{p-}{$\psi^-$}
\psfrag{p+}{$e^{-1}\psi^+$}   
\centering
\includegraphics[width=4cm]{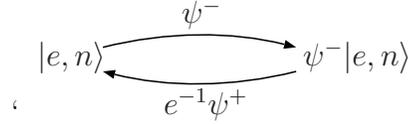}
\caption{\em Typical irreducible representation $\langle e,n\rangle$}
\end{figure}

The action can be stated conveniently in supermatrix form, i.e.  
\begin{alignat}{7}
			E\ &=& \ \begin{pmatrix}e&0\\0&e\end{pmatrix}\qquad , \qquad \qquad  
		N\ &=& \ \begin{pmatrix}n&0\\0&n-1\end{pmatrix}\ ,&& \\ 
		\psi^+ \ &=& \ \begin{pmatrix}0&1\\0&0\end{pmatrix}\qquad \text{and} \qquad  
		\psi^-\ &=& \ \begin{pmatrix}0&0\\e&0\end{pmatrix}\qquad . &&
\end{alignat}
The supertrace in such a matrix representation is a non-degenerate invariant supersymmetric bilinear form 
\begin{equation}
	\str\Bigl( \begin{pmatrix}a&b\\c&d\end{pmatrix}\Bigr) \ = \ a-d \qquad \text{i.e.}\ \str(E\, N)\ = \ \str(\psi^+\,\psi^-)\ = \ e\, .
\end{equation}

For the typical representations we assumed the parameter $e$ to be non-zero. If we set $e=0$, we still obtain a representation of \gl\ but this representation is reducible, i.e. it contains a proper invariant subrepresentation generated by the state $\psi^-|0,n\rangle$. On the other hand this representation is indecomposable, since it does not decompose into a direct some of irreducible representations.
Moreover, this representation is part of a larger representation, the projective cover $\mathcal{P}_n$. 
The projective cover is constructed from a state $|n\rangle$ satisfying
\begin{equation}
     E|n\rangle \ = \ 0 \qquad, \qquad N|n\rangle \ = \ n|n\rangle 
\end{equation}
and $\psi^+,\psi^-$ are acting freely on it. We summarise this in figure~2.2.
\begin{figure}[htb!]
	\label{fig:projectivecover}
\centering%
\psfrag{n}{$|n\rangle$}
\psfrag{n+}{$\psi^+|n\rangle$}
\psfrag{n-}{$\psi^-|n\rangle$}
\psfrag{n+-}{$\pm\psi^+\psi^-|n\rangle$}
\psfrag{p-}{$\psi^-$}
\psfrag{p+}{$\psi^+$}   
\includegraphics[width=6cm]{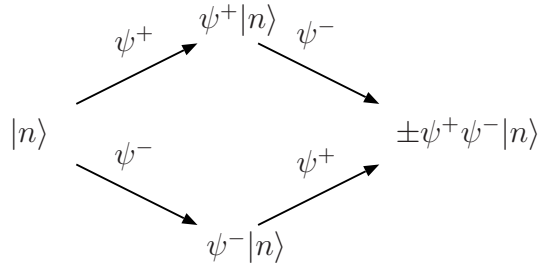}
\caption{{\em Projective cover $\mathcal{P}_n$: $\psi^\pm$ act as indicated. 
There is a 3-dimensional subrepresentation, two 2-dimensional ones and the trivial 1-dimensional subrepresentation.}}
\end{figure}

We observe that this representation contains proper invariant subrepresentations but it is impossible to decompose the representation in a direct sum of irreducible representations.

A generic feature of these indecomposable but reducible representations is that the Casimir is not diagonalisable on this representations, i.e.
\begin{equation}
       C|n\rangle \ = \ 2\psi^-\psi^+|n\rangle \qquad\text{and}
	\qquad C\psi^-|n\rangle\ = \ C\psi^+|n\rangle\ = \ C\psi^-\psi^+|n\rangle\ = \ 0\, . 
\end{equation}

\subsection{Harmonic analysis}

In the Lie algebra case one can obtain the elements of a Lie group by taking exponents of elements of the Lie algebra. The Lie superalgebra case is slightly different. Let $\eta_\pm$ be two Grassmann odd numbers. This means that they satisfy
\begin{equation}
 \eta_\pm\eta_\pm \ = \ 0 \qquad \text{and}\qquad \eta_+\eta_- \ = \ -\eta_-\eta_+\, .
\end{equation}
 Then an element $g$ of the Lie supergroup \GL\ can be written as 
\begin{equation}
 g \ = \ e^{i\eta_-\psi_-}e^{ixE+iyN}e^{i\eta_+\psi^+}\, 
\end{equation}
for some real numbers $x$ and $y$. 
On the other hand given a Lie supergroup one can find a differential operator realisation of the Lie superalgebra in terms of invariant vector fields. They are defined by
\begin{equation}
 R_Xg \ = \ -Xg \qquad\text{and}\qquad L_Xg \ = \ gX\qquad\text{for}\ X \ = \ E,N,\psi^\pm\, .
\end{equation}
Right and left invariant vector fields take the following form
\begin{equation}
R_E\ =\ i\del_x \ , \ \ R_N\ =\ i\del_y+\etam\del_- \ , \ \ R_+\
=\ ie^{-iy}\del_+ -\etam\del_x \ , \ \ R_-\ =\ i\del_-\ ,
\end{equation}
and
\begin{equation}
L_E\ =\ -i\del_x \ , \ \ L_N\ =\ -i\del_y-\etap\del_+ \ , \ \ L_-\
\ =\ -ie^{-iy}\del_- -\etap\del_x \ , \ \ L_+\ =\ -i\del_+\ .
\end{equation}
These differential operators satisfy the relations of the Lie superalgebra \gl, i.e.
\begin{equation}\label{eq:gl11relationvectorfields}
\begin{split}
 R_{[X,Y]}\ &= \ (-1)^{\vert X\vert\vert Y\vert}R_XR_Y-R_YR_X\qquad\text{and}\\
 L_{[X,Y]}\ &= \ (-1)^{\vert X\vert\vert Y\vert}L_XL_Y-L_YL_X\, .
\end{split}
\end{equation}
Note the unusual sign\footnote{Naively one might have expected
\begin{equation}\nonumber
\begin{split}
 R_{[X,Y]}\ = \ R_XR_Y-(-1)^{\vert X\vert\vert Y\vert}R_YR_X\qquad\text{and}\qquad
 L_{[X,Y]}\ = \ L_XL_Y-(-1)^{\vert X\vert\vert Y\vert}L_YL_X\, .
\end{split}
\end{equation}}. In the above formula $|X|$ denotes the parity of $X$, i.e. 
\begin{equation}
\begin{split}
|X|\ = \ \Bigl\{ \begin{array}{cc}
	       \ 0&\qquad X \ \ \text{bosonic}    \\
	       \ 1&\qquad\ \,\, X \ \ \text{fermionic}  \\
                        \end{array} \ .
\end{split}
\end{equation}
The invariant vector fields act on the space of functions of the supergroup $\L2$ spanned by the elements
\begin{equation} \label{eq:basisfunctionsgl}
e_0(e,n)\ =\ e^{iex+iny}\ , \ \ e_\pm(e,n)\ =\ \etapm e_0(e,n)\, \
\ \  e_2(e,n)\ =\ \etam\etap e_0(e,n) \ .
\end{equation}
The invariant Haar measure corresponding to the invariant vector fields is
\begin{equation} d\mu \ = \ e^{-iy}dxdyd\etap d\etam\ \ .
\end{equation}
The decomposition of $\L2$ with respect to both left and right
regular action was analysed in \cite{Schomerus:2005bf}.
In order to illustrate this harmonic analysis, let us review the decomposition of $\L2$ under the left-regular action.
Consider the function $e_0(e,n)$. For $e\neq0$, it satisfies
\begin{equation}\label{eq:gltypicalfunctions}
 \begin{split}
  L_Ee_0(e,n) \ &= \ ee_0(e,n)\, , \\
L_Ne_0(e,n) \ &= \ ne_0(e,n)\, , \\
L_+e_0(e,n) \ &= \ 0\, , \\
L_-e_0(e,n) \ &= \ -iee_+(e,n)\ \text{and} \\
L_+L_-e_0(e,n) \ &= \ -ee_0(e,n). \\
 \end{split}
\end{equation}
If we compare these relations with \eqref{eq:gltypical} and \eqref{eq:gltypical2}, we see that the functions $e_0(e,n)$  and $e_+(e,n)$ span the typical representation $\langle e,n\rangle$. The minus sign in the last equation of \eqref{eq:gltypicalfunctions} is due to the unusual minus sign in \eqref{eq:gl11relationvectorfields}.
Analogously one can see that the functions $e_-(e,n)$ and $e_0(e,n-1)-ee_2(e,n)$ also form the typical representation $\langle e,n\rangle$ (for $e\neq0$).

Let us now consider the case $e=0$. Then the state $e_2(0,n+1)$ satisfies
\begin{equation}
\begin{split}
 L_Ee_2(0,n+1) \ &= \ 0 \qquad\text{and}\\
 L_Ne_2(0,n+1) \ &= \ ne_2(0,n+1)
\end{split}
\end{equation}
and $L_\pm$ act as indicated in figure 2.3. 
\begin{figure}[htb!]
	\label{fig:projectivecover5}
\centering%
\psfrag{n}{$e_2(0,n+1)$}
\psfrag{n+}{$ie_-(0,n+1)$}
\psfrag{n-}{$-ie_+(0,n)$}
\psfrag{n+-}{$\mp e_0(0,n)$}
\psfrag{p-}{$L_-$}
\psfrag{p+}{$L_+$}   
\includegraphics[width=10cm]{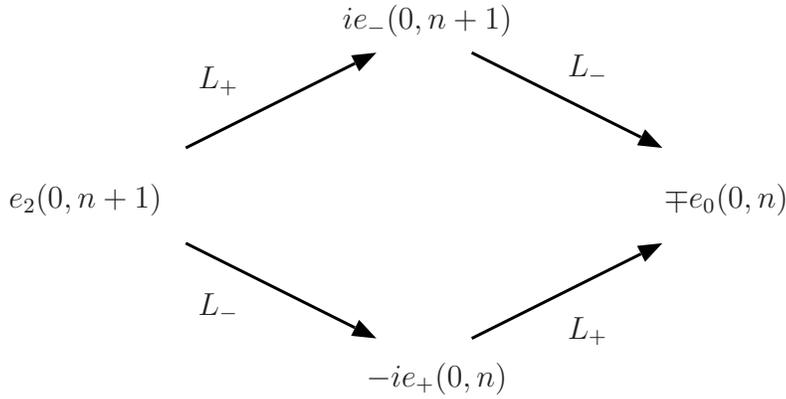}
\caption{{\em Projective cover $\mathcal{P}_n$}}
\end{figure}

We conclude that the functions $e_2(0,n+1)$, $ie_-(0,n+1)$, $e_+(0,n)$ and $e_0(0,n)$ form the projective cover $\mathcal{P}_n$. In summary we have obtained the following result \cite{Schomerus:2005bf}.
\begin{prp}
 The space of functions $\L2$ of the supergroup \GL\ decomposes under the action of the left invariant vector fields as follows
\begin{equation}
 Fun(\text{\GL}) \ = \ \int_{e\neq0}de\,dn\ \bigl( \langle e,n\rangle\oplus\langle e,n\rangle\bigr)
              \,  \oplus\,  \int dn\, \mathcal{P}_n\, .
\end{equation}
\end{prp}

The analysis of the decomposition of $\L2$ under the action of the right invariant vector fields can be performed analogously. 

\subsection{The affine Lie superalgebra \agl}\label{section:affinegl}

The affine Lie superalgebra \agl\ is an infinite dimensional Lie superalgebra with generators $E_n,N_n,\psi^\pm_n$ ($n$ in $\Z$ ) and $K,d$, where $K$ is central and $d$ is a derivation, i.e.
\begin{equation}\label{eq:derivationgl11}
    [d,X_n]\ = \ nX_n\qquad\qquad\text{for} \ X \ \in \ \{E,N,\psi^\pm\}\, .
\end{equation}
The non-vanishing relations of the remaining generators are
\begin{equation}
 \begin{split}
  [E_n,N_m]\ &= \ Kn\delta_{n+m,0}\ , \\
 [N_n,\psi^\pm_m] \ &= \ \psi^\pm_{n+m}\ \text{and}\\
 \{\psi^-_n,\psi^+_m\}\ &=\ E_{n+m}+Kn\delta_{n+m,0}\, .
 \end{split}
\end{equation}
All representations in the WZNW model have fixed $K$-eigenvalue $k$. From now on we restrict to these representations and write the number $k$ instead of the operator $K$.
As in the case of Lie algebras one can construct the generators of another infinite dimensional Lie algebra, the Virasoro algebra, out of the generators of the affine Lie superalgebra. The Virasoro algebra is the symmetry algebra of two dimensional conformal field theory.
The generators are
\begin{eqnarray*}
 L_n & = &  \frac{1}{2k}\  (2 N_n E_0- E_n +
        \Psi^-_n \Psi^+_0 + \Psi^-_0 \Psi^+_n +\frac{1}{k} E_n E_0) \\[2mm]
        & & +  \frac{1}{k} \sum_{m > 0} \, (E_{n-m} N_m +
    N_{n-m} E_m + \Psi^-_{n-m} \Psi^+_{m} - \Psi^+_{n-m} \Psi^-_m
    + \frac{1}{k} E_{n-m} E_m)
\end{eqnarray*} 
and they satisfy the relations of the Virasoro algebra
\begin{equation}
 [L_n,L_m]\ = \ (n-m)L_{n+m}+\frac{c}{12}n(n^2-1)\delta_{n+m,0}
\end{equation}
with central charge $c=0$. This construction is referred to as the Sugawara construction (see e.g. \cite{Kac:1990gs}). The Virasoro zero mode is the affine analogue of the quadratic Casimir. The action of the Virasoro zero mode on the generators of the affine Lie superalgebra coincides with that of the derivation $d$ \eqref{eq:derivationgl11}
\begin{equation}
    [L_0,X_n]\ = \ nX_n\qquad\qquad\text{for} \ X \ \in \ \{E,N,\psi^\pm\}\, .
\end{equation}
In WZNW models one identifies these two operators $d=L_0$.

The zero mode subalgebra is the finite dimensional Lie superalgebra \gl. It is usually called the horizontal subalgebra. Automorphisms that do not leave the horizontal subalgebra invariant are called spectral flow automorphisms. For \agl\ there is a one-parameter family of non-trivial spectral flow automorphisms. They are (for $m\in\Z$)
\begin{equation}
 \gamma_m(E_n)\ = \ E_n+mk\delta_{n,0}\ , \qquad \gamma_m(\psi^\pm_n)\ = \ \psi^\pm_{n\pm m}
\end{equation}
and leave $N_n$ invariant. 
They become relevant in the discussion of representations.

\subsubsection{Representations of \agl}

Representations of the affine Lie superalgebra are constructed as follows. We start with a representation of the horizontal subalgebra, and promote it to a representation of the affine Lie superalgebra by defining all positive mode operators to be annihilation operators and letting the negative mode operators act freely. 
As in the finite dimensional Lie superalgebra case they fall into two classes, typical irreducible and atypical indecomposable but reducible ones.
The typicals are defined on a state $|e,n,k\rangle$, where $n\neq mk$ for any integer $m$. This state satisfies
\begin{equation}\label{eq:affinetypicalgl11}
\begin{split}
    E_0|e,n,k\rangle \ &= \ e|e,n,k\rangle\qquad , \qquad 
    N_0|e,n,k\rangle \ = \ n|e,n,k\rangle\, ,\\
    K|e,n,k\rangle \ &= \ k|e,n,k\rangle\qquad , \qquad 
    \psi^+_0|e,n,k\rangle \ = \ 0\qquad\qquad\text{and}\\    
X_n|0,n\rangle\ &= \ 0 \qquad\qquad \text{for}\ n>0\ \ \text{and}\ \ X\ \in \ \{E,N,\psi^\pm\}\, . 
\end{split}
\end{equation}
The remaining operators act freely on this state. 

Whenever $e=mk$ for some integer $m$ the above construction leads to reducible but indecomposable representations (for a proof see~\cite{Schomerus:2005bf}) and as in the finite dimensional Lie superalgebra case these representations are part of a larger indecomposable representation. All these special atypical representations can be obtained via spectral flow from the projective cover $\widehat{\mathcal{P}}_n^{(0)}$. This representation is already atypical on the level of the horizontal subalgebra. It is defined by a state $|0,n,k\rangle$ satisfying
\begin{equation}\label{eq:affineprojectivegl11}
\begin{split}
    E_0|0,n,k\rangle \ &= \ 0\\
    N_0|0,n,k\rangle \ &= \ n|0,n,k\rangle\\
    K|0,n,k\rangle \ &= \ k|0,n,k\rangle\\
    X_n|0,n,k\rangle\ &= \ 0 \qquad\qquad \text{for}\ n>0\ \ \text{and}\ \ X\ \in \ \{E,N,\psi^\pm\}\, .
\end{split}
\end{equation}
The remaining operators act freely on this state.

The construction is based on a choice of horizontal subalgebra. Any other choice is obtained by application of a spectral flow automorphism. The corresponding representations are called twisted representations. For the projective covers, they can be constructed from a state $|mk,n,k\rangle$, where $m\in\Z$, satisfying
\begin{equation}
\begin{split}
    E_0|mk,n,k\rangle \ &= \ mk|mk,n,k\rangle\\
    N_0|mk,n,k\rangle \ &= \ n|mk,n,k\rangle\\ 
    K|mk,n,k\rangle \ &= \ k|mk,n,k\rangle\\
    E_n|mk,n,k\rangle\ &= \ 0 \qquad\qquad \text{for}\ n>0\\
    N_n|mk,n,k\rangle\ &= \ 0 \qquad\qquad \text{for}\ n>0\\
    \psi^\pm_n|mk,n,k\rangle\ &= \ 0 \qquad\qquad \text{for}\ n>\mp m\, 
\end{split}
\end{equation}
and the remaining modes acting freely on it, it is called $\widehat{\mathcal{P}}_n^{(mk)}$. Since this representation is obtained by applying a spectral flow automorphism to the atypical representation $\widehat{\mathcal{P}}_n^{(0)}$ constructed in \eqref{eq:affineprojectivegl11} it must also be atypical.

It turns out that these are all atypical representations of \agl, i.e. any atypical representation of \agl\ can be obtained from a representation that is already atypical at the level of ground states \cite{Schomerus:2005bf}.

\section{Lie superalgebras}

We turn to the general theory of Lie superalgebras and Lie supergroups.
The theory of Lie superalgebras was developed by Victor Kac \cite{Kac:1977qb,Kac:1977em}, a collection of results is given in \cite{Frappat:1996pb}. As a guideline to Lie supergroups we use the book by Berezin \cite{Berezin:1987wh}. 
Most of this section is contained in \cite{Creutzig:2008ag}.

In the following, the Lie superalgebras will be over the field of real numbers $\R$ or complex numbers $\C$. First we need to define them.
\begin{definition}\label{def:lsa}
Let $\g$ be a $\mathbb{Z}_2$ graded algebra $\g=\gzero\oplus \gone$ with product 
$[\ \ ,\ \ ]:\g\times\g\rightarrow\g$ that respects the grading. The parity of a homogeneous element is denoted by
\begin{equation}
	\begin{split}
|X|\ = \ \Bigl\{ \begin{array}{cc}
	       \ 0&\qquad X \ \ \text{in}\ \ \gzero    \\
	       \ 1&\qquad X \ \ \text{in}\ \ \gone     \\
                        \end{array} \ .
		\end{split}
\end{equation}
Then $\g$ is a {\bf Lie superalgebra} if it satisfies antisupersymmetry and graded Jacobi identity, i.e.
\begin{equation}
	\begin{split}
		0 \ &= \ [X,Y]+(-1)^{|X||Y|}[Y,X]  \qquad\text{and}\\[1mm]
	0\ &= \ (-1)^{|X||Z|}[X,[Y,Z]]+(-1)^{|Y||X|}[Y,[Z,X]]+(-1)^{|Z||Y|}[Z,[X,Y]]\ ,
\end{split}
\end{equation}
for all   $X,Y$ and $Z$ in $\mathfrak{g}$.

Further a bilinear form $B : \mathfrak{g} \times \mathfrak{g} \rightarrow \R (\text{resp.} \C)$
is called a {\bf consistent supersymmetric invariant bilinear form} if
\begin{equation}
	\begin{split}
		B(X,Y) \ &= \ 0 \qquad\forall\, X \in \gzero \land \forall\, Y  \in \gone \\	
        	B(X,Y)-(-1)^{|X||Y|}B(Y,X ) \ &=\ 0 \qquad\forall\, X,Y\in \g \ \text{and}\\ 
	        B([X,Y],Z)-B(X,[Y,Z])\ &= \ 0 \qquad\forall\, X,Y,Z\in \g\ .\\  
\end{split}
\end{equation}	
\end{definition}

A simple Lie superalgebra whose even part is a reductive Lie algebra and which possesses a nonzero supersymmetric invariant bilinear form, is called
a basic Lie superalgebra. They are completely classified \cite{Kac:1977qb,Kac:1977em}. There are the infinite series of unitary superalgebras
$sl(n|m)$ for $m\neq n$, $psl(n|n)$ and the orthosymplectic series $osp(m|2n)$ as well as some exceptional ones. In addition we will also consider Lie superalgebras of type $gl(n|m)$. 
As in the case of Lie algebras it is instructive to keep their fundamental matrix representations in mind.
We provide them for the superalgebras $gl(m|n)$, $sl(m|n)$ and $osp(m|2n)$.
\begin{example}\label{example:glmn}
$gl(n|m)$ is given by
\begin{equation}
	\begin{split}
		\text{gl}(n|m)=\Bigl\{ \left(\begin{array}{cc}A & B \\ C & D\\ \end{array}\right) \Bigr\},
	\end{split}
\end{equation}
where $A$ and $D$ are square matrices of size $n$ and $m$, $B$ is a $n\times m$ matrix and $C$ is a $m\times n$ matrix. The supertrace is a supersymmetric non-degenerate invariant bilinear form and it is defined via 
\begin{equation}
	\begin{split}
		\str  \left(\begin{array}{cc}A & B \\ C & D\\ \end{array}\right) =\tr A -\tr D \ .
	\end{split}
\end{equation}
\end{example}
Then, we have the unitary superalgebra
\begin{example}\label{example:slmn}
 $sl(n|m)$ 
\begin{equation}
	\begin{split}
		\text{sl}(n|m)=\bigl\{ X \ \in \text{gl}(n|m) \ | \ \str X =0\bigr\},
	\end{split}
\end{equation}
for $n\neq m$. If $n=m$ $sl(n|n)$ is not simple, in this case one obtains the projective unitary superalgebra $psl(n|n)$ as the quotient of $sl(n|n)$ by its one dimensional ideal $\mathcal{I}$ generated by the identity matrix $1_{2n}$, i.e. $psl(n|n)=sl(n|n)/\mathcal{I}$.
\end{example} 
Further there is the orthosymplectic series
\begin{example}\label{example:ospmn}
 $osp(m|2n)$
\begin{equation}
	\begin{split}
		\text{osp}(m|2n)=\bigl\{ X \ \in \text{gl}(m|2n) \ | \ X^{st}B_{m,n}+B_{m,n}X=0\bigr\},
	\end{split}
\end{equation}
where the supertranspose is
\begin{equation}\label{eq:supertranspose}
	\begin{split}
		\left(\begin{array}{cc}A & B \\ C & D\\ \end{array}\right)^{st}=\left(\begin{array}{cc}A^t & -C^t \\ B^t & D^t\\\end{array}\right)
	\end{split}
\end{equation}
and 
\begin{equation}
	\begin{split}
		B_{m,n}=\left(\begin{array}{cc}1_m & 0 \\ 0 & J_n\end{array}\right) \ ,\text{where} \ 
			J_n=\left(\begin{array}{cc}0 & 1_n \\ -1_n & 0\end{array}\right).
	\end{split}
\end{equation}
\end{example}

An important ingredient in Lie super theory is the dual Coxeter number.
\begin{definition}\label{definition:coxeter}
 Let $\g$ be a Lie superalgebra, $\{t^a\}$ a basis of $\g$ and ${f^{ab}}_c$ the {\bf structure constants}, i.e. they satisfy
\begin{equation}
 [t^a,t^b] \ = \ {f^{ab}}_ct^c\, ,
\end{equation}
where the summation over the repeated index $c$ is understood.
\newline
Then the {\bf dual Coxeter number} $h^\vee$ is defined via
\begin{equation}
 2h^\vee\ = \  (-1)^n{f^{na}}_m{f^{mb}}_n\kappa_{ab}\, 
\end{equation}
and $\kappa_{ab}$ denotes the supertrace in the adjoint representation, i.e. the Killing form.
\end{definition}

In the case of Lie algebras one usually uses the Killing form as non-degenerate invariant symmetric bilinear form. 
The Lie superalgebras $psl(n|n)$, $gl(n|n)$ and $osp(2n+2|2n)$ are special, because their Killing form and also their dual Coxeter number vanish. Hence in the following we will always use the supertrace $\str$ in the fundamental matrix representation (Example \ref{example:glmn}, \ref{example:slmn} and \ref{example:ospmn}) as supersymmetric non-degenerate invariant bilinear form. 

Classical Lie superalgebras fall into two cases. The fermionic subspace is a representation of the bosonic subalgebra. If this representation is irreducible the Lie superalgebra is of type II and it possesses the following distinguished $\Z$-gradation
\begin{equation}\label{eq:typeIIgradiation}
 \g\ = \ \g_{-2}\oplus\g_{-1}\oplus\g_0\oplus\g_1\oplus\g_2\,.
\end{equation}
Otherwise this representation decomposes into a direct sum of two irreducible representations. In that case the Lie superalgebra is said to be of type I. They possess the following distinguished $\Z$-gradation which will turn out to be very helpful
\begin{equation}\label{eq:typeIgradiation}
 \g\ = \ \g_{-}\oplus\g_0\oplus\g_+\,.
\end{equation}
Here $\g_{\pm}$ are the two irreducible representations forming the fermionic subspace and $\g_0$ is the bosonic Lie subalgebra. 

A Cartan subalgebra of a Lie superalgebra is defined to be a maximal abelian subsuperalgebra.
It turns out that the Cartan subalgebra of the underlying bosonic Lie algebra is also a Cartan subalgebra of the Lie superalgebra.
Fix a Cartan subalgebra $\h$ of $\g$ and denote the dual space by $\h^*$. A non degenerate supersymmetric invariant bilinear form of the classical Lie superalgebras restricts non degenerately to a Cartan subalgebra $\h$ and induces a non degenerate bilinear form on its dual space. We denote it by $(\ \ | \ \ )$. Further a root is defined as follows.
\begin{definition}
	For $\alpha\neq0$ in $\h^*$ one sets
	\begin{equation}
		\g_\alpha \ = \ \{ a \ \in \ \g \ | \ [h,a]\ = \ \alpha(h)a \ \forall \ h \ \in \h \ \}\ .
	\end{equation}
	$\alpha$ is called a {\bf root} if $\g_\alpha\neq0$ and $\g_\alpha$ is called {\bf rootspace}. Further a root is called even if $\g_\alpha\cap\gzero\neq0$ and odd if $\g_\alpha\cap\gone\neq0$. Denote by $\Delta$ the set of roots, by $\Delta_0$ the set of even roots and by $\Delta_1$ the set of odd roots. 
\end{definition}
The Lie superalgebra $\g$ possesses the usual nonunique triangular decomposition
\begin{equation}\label{eq:triangulardecomposition}
	\g\ = \ \nn_- \oplus \h  \oplus \nn_+ \ .\vspace{1mm}
\end{equation}
Here $\nn_\pm$ are isotropic Lie subsuperalgebras. 
One calls a root positive if $\g_\alpha\cap\nn_+\neq0$ and negative if $\g_\alpha\cap\nn_-\neq0$. Let $\rho_0$ be half the sum of even positive roots and $\rho_1$ half the sum of odd positive roots, then the Weyl vector is \vspace{1mm}
\begin{equation}
	\rho\ =\ \rho_0-\rho_1\ .\vspace{1mm}
\end{equation}

\subsection{Lie supergroups}\label{section:liesupergroups}

A Lie supergroup can be obtained from a Lie superalgebra as follows. Let $\{t^a\}$ be a basis of $\gzero$ and $\{s^b\}$ a basis of $\gone$, further let $\Lambda$ be a Grassmann algebra, then the Grassmann envelope $\Lambda(\g)$ of $\g$ consists of formal linear combinations 
\vspace{1mm}
\begin{equation}
	X \ = \ x_{a}t^a +\theta_{b}s^b \vspace{1mm}
\end{equation}
where the $x_a \in\Lambda$ are Grassmann even elements, the $\theta_b\in\Lambda$ are Grassmann odd elements and summation over the indices is implied. Note, that $\Lambda(\g)$ is a Lie algebra. Then following Berezin \cite{Berezin:1987wh}, a supergroup $G$ is the group generated by elements $g$ of the form $g=\exp X$ with $X$ in the Grassmann envelope of $\g$, i.e. the Lie supergroup $G$ of the Lie superalgebra $\g$ is the Lie group of the Lie algebra $\Lambda(\g)$. Further we denote the Lie subgroup of the subalgebra $\gzero$ by $G_0$.


The Lie group $G$ acts on its Lie algebra $\Lambda(\mathfrak{g})$ by conjugation 
\vspace{1mm}
\begin{equation}
\Ad(a):\ \Lambda(\g) \ \rightarrow  \ \Lambda(\g)\ ,\qquad\qquad X\ \mapsto aXa^{-1}\vspace{1mm}
\end{equation}
for $a$ in $G$ and $X$ in $\Lambda(\g)$. Since the invariant bilinear form is the supertrace of a representation it is invariant 
under the adjoint action, i.e. \vspace{1mm}
\begin{equation}\label{eq:innerinvariance}
\str(\Ad(a) X,\Ad(a) Y) \ =\ \str( X, Y)\vspace{1mm}
\end{equation}
for any $X,Y$ in $\Lambda(\g)$ and $a$ in $G$.

Consider a Lie supergroup $G$ with a supersymmetric invariant nonzero bilinear form. We identify the Grassmann envelope of the underlying Lie superalgebra with the tangent space at the
identity, $\Lambda(\mathfrak{g})=\T_e G$. 
On the tangent space $\T_g G$ at $g$ in $G$ we have left and right identification,\vspace{1mm}
\begin{equation}
\begin{split}
		&L_g\ :\ \Lambda(\mathfrak{g})\ \longrightarrow\ \T_gG, \qquad \ \ \ L_g(X)\ =\ gX \ \ \text{and}\\[1mm]
		&R_g\ :\ \Lambda(\mathfrak{g})\ \longrightarrow\ \T_gG, \qquad \ \ \ R_g(X)\ =\ -Xg\ .\\[1mm]
	\end{split}
\end{equation}
The left identification defines a left invariant metric, i.e. $(gX,gY):=\str(X,Y)$. 
This metric is also right invariant, since it is invariant under the adjoint action $\Ad(g^{-1})$.
 
Invariant vector fields will turn out to be a useful tool in the analysis of the conformal field theory on the supergroup, because in the semiclassical limit the invariant vector fields mimic the role of the currents in the full quantum theory.

There is a way of obtaining an explicit differential operator realisation of the Lie superalgebra in terms of invariant vector fields. Let $G$ be a Lie supergroup of type I, then we
parameterise an element according to the distinguished $\Z$-graduation \eqref{eq:typeIgradiation}
\begin{equation}
 g\ = \ e^{\theta^a_-s_a^-}g_0e^{\theta^b_+s_+^b}\, ,
\end{equation}
where the $\theta^a_\pm$ are Grassman odd coordinates, the $s_a^\pm$ generate $\g_{\pm}$  and $g_0$ is an element of the Lie subgroup.
Then we compute recursively
\begin{equation}\label{eq:rightvectorfields}
 \begin{split}
     R_g({s^a_-}) \ &= \ - \del_{\theta^-_a}\\
     R_g(X) \ &= \ R_{g_0}(X) -\theta_a^-R_g({[s^a_-,X]})\\
     R_g({\Ad(g_0)(s^a_+)}) \ &= \ -\del_{\theta_a^+}-\theta_b^-R_g({\{s^b_-,\Ad(g_0)(s^a_+) \}})+\\
&\qquad\qquad\qquad   - \theta_b^-\theta_c^-R_g({[s^b_-,\{s^c_-,\Ad(g_0)(s^a_+)\}]})\, .\\
 \end{split}
\end{equation}
$R_{g_0}(X)$ is the invariant vector field of the Lie subgroup. 
Similar the left-invariant vector fields are
\begin{equation}
 \begin{split}
     L_g({s^a_+}) \ &= \  \del_{\theta^+_a}\\
     L_g(X) \ &= \ L_{g_0}(X) -\theta_a^+L_g({[X,s^a_+]})\\
     L_g({\Ad(g_0^{-1})(s^a_-)}) \ &= \ \del_{\theta_a^-}-\theta_b^-L_g({\{s^b_+,\Ad(g_0^{-1})(s^a_-) \}})+\\
&\qquad\qquad\qquad -  \theta_b^+\theta_c^+L_g({[s^b_+,\{s^c_+,\Ad(g_0^{-1})(s^a_-)\}]})\\
 \end{split}
\end{equation}
Due to the Grassmannian nature of the coordinates one gets some unusual signs.
For example, the invariant vector fields obey the relations of the Lie superalgebra if we take the Lie bracket as
\begin{equation}\label{eq:relationsinvariantvectorfields}
\begin{split}
    R_g({[X,Y]})\ &= \ (-1)^{|X||Y|}R_g(X)R_g(Y)-R_g(Y)R_g(X) \\
    L_g({[X,Y]})\ &= \ (-1)^{|X||Y|}L_g(X)L_g(Y)-L_g(Y)L_g(X) \, .\\
\end{split}
\end{equation}
Later on, we will make analogous observations in the full quantum theory.

The invariant vector fields act on the space of functions of the supergroup. The definition of this space is not obvious and we refer to \cite{Quella:2007hr}. In that article also the invariant measure corresponding to above invariant vector fields is computed. It is
\begin{equation}\label{eq:Haarmeasure}
 d\mu(g) \ = \ d\mu_b(g_0)\det(\Ad(g_0))_-\prod_{a,b} d\theta^a_-d\theta^b_+\, .
\end{equation}
Here $d\mu_b(g_0)$ denotes the invariant measure of the Lie subgroup and $\det(\Ad(g_0))_-$
is the determinant of the adjoint action of $g_0$ on $\g_-$.

\subsection{Automorphisms}
\label{automorphisms}

As in the Lie algebra case, the group of automorphisms consists of inner and outer automorphisms. An inner automorphism of the Grassmann envelope $\Lambda(\g)$ of the Lie superalgebra $\g$ is obtained by conjugating with an element of the corresponding Lie supergroup. Only if this element is in the Lie subgroup the automorphism descends to the Lie superalgebra $\g$. For the study of branes automorphisms that preserve the invariant bilinear form are relevant. We saw that this is true for inner automorphisms \eqref{eq:innerinvariance}.

It remains to find the group of outer automorphisms.
For complex Lie superalgebras it is classified by Vera Serganova \cite{Ser1}.
We start by listing the relevant automorphisms. It is most convenient to state its action in the fundamental matrix representation.
Let $X=\bigl(\substack{A \ B \\ C\ D}\bigr)$ be a supermatrix in $gl(n|m)$, then we have the following list
\begin{equation}\label{eq:minussupertranspose}
 \begin{split}
        (-st)(X)\ &=\ \left(\begin{array}{cc} -A^t & C^t \\ -B^t & -D^t\end{array}\right)\ , \\
        \Pi(X)\ &=\ \left(\begin{array}{cc} D & C \\ B & A\end{array}\right)\  \ \ \ \text{for}\ m\ = \ n\ ,  \\
	\delta_\lambda(X)\ &=\ \left(\begin{array}{cc} A & \lambda B \\ \lambda^{-1}C & D\end{array}\right)\  \ \ \ \text{for}\ m\ = \ n\ \text{and} \ \lambda \ \in\ \C\, 
     \end{split}
\end{equation}
$(-st)$ and $\delta_\lambda$ leave the metric invariant, but $\Pi$ does not.
In addition we introduce the element $J_{m,n}$ in $gl(2m|2n)$, with $\det \ J_{m,n} = -1, J_{m,n}^2=1_{2m+2n}$ and $J_{m,n}B_{2m,n}J_{m,n}=B_{2m,n}$.
\begin{table}[h]\label{table:automorphisms}
\begin{center}
 \begin{tabular}{ | c | c | c | }
    \hline
    $\g$ & Generators of Out $\g\vphantom{\Bigl(\Bigr)}$ & Metric preserving generators \\[1mm] \hline\hline
    sl(n$|$m)\ , $n\neq m$  & $(-st)$ & $(-st)\vphantom{\Bigl(\Bigr)}$ \\  \hline
    psl(n|n)\ , $n\neq 2$ & $(-st)\vphantom{\Bigl(\Bigr)}$, $\Pi$, $\{\delta_\lambda \ |\, \text{for}\ \lambda^n\neq1\}$ & 
     $(-st)$, $\{\delta_\lambda \ |\, \text{for}\ \lambda^n\neq1\}$ \\ \hline
psl(2|2) &  $\Pi$, $SL(2)\vphantom{\Bigl(\Bigr)}$ & 
      $SL(2)$ \\ \hline    
osp(2m+1|2n) & -- & --$\vphantom{\Bigl(\Bigr)}$ \\ \hline
osp(2m|2n) & $\Ad J_{m,n}$ & $\Ad J_{m,n}\vphantom{\Bigl(\Bigr)}$ \\  \hline
\end{tabular}\caption{{\em Outer automorphisms of Lie superalgebras}}
\end{center}
\end{table}

The supergroup psl$(2|2)$ is special, it carries an action of $SL(2)$ induced by a $sl(2)$ bracket of the type
\begin{equation*}
	\ \left[\left(\begin{array}{cc} a & b \\ c & -a\end{array}\right)\ ,\
		\left(\begin{array}{cc} A & B \\ C & D\end{array}\right)\ \right] \ = \ \ 
			\left(\begin{array}{cc} 0 & aB+bJ_1C^tJ_1^{-1} \\ -cJ_1B^tJ_1^{-1}-aC & 0\end{array}\right)\,
\end{equation*}
this defines an automorphism and since it leaves the bosonic subalgebra invariant it preserves the metric. 

We list the group of outer automorphisms and those which preserve the metric in table 2.1.
The group of automorphisms of $gl(n|n)$ coincides with the group of automorphisms of $psl(n|n)$. The only difference is that in $gl(n|n)$ the fermionic dilatation $\delta_\lambda$ is an inner automorphism. 

Note, that the exceptional Lie superalgebras do not admit a metric preserving outer automorphism.

\subsection{Real forms}

Real forms of classical Lie superalgebras are classified in \cite{Serganova:1989fj} and \cite{Parker:1980af}.
As in the case of simple Lie algebras this is done by classifying the involutive semimorphisms of the complex Lie superalgebras. 
A semimorphism $\phi$ of a complex Lie superalgebra $\g$ is a semilinear transformation such that 
\begin{equation}
	[\phi(X),\phi(Y)] \ = \ \phi([X,Y]) \qquad \text{for all} \ X,Y \ \text{in} \ \g \ .
\end{equation}
Then for every involutive semimorphism $\phi$ 
\begin{equation}
	\g^\phi \ = \ \{ X + \phi(X) \ | \ X \ \text{in} \ \g \ \} 
\end{equation}
is a real classical Lie superalgebra and these are all (Theorem 2.5 in \cite{Parker:1980af}). 

Real forms of Lie supergroups correspond to real forms of the underlying Lie algebra, that is the Grassmann envelope $\Lambda(\g)$ of the Lie superalgebra $\g$. There is the following real form that is not induced from a real form of a Lie superalgebra. Define the superstar operation as \cite{Frappat:1996pb}
\begin{equation}
	(c\theta)^{\#} \ = \ \bar{c}\theta^{\#} \ , 
	\qquad \theta^{\#\#}\ = \ -\theta \ , \qquad (\theta_1\theta_2)^{\#} \ = \ \theta_1^{\#}\theta_2^{\#} \ 
\end{equation}
for any Grassmann elements $\theta, \theta_i$ and any complex number $c$. 
Then concatenation of the superstar $\#$ with the automorphism $(-st)$ is a semimorphism of $\Lambda(\g)$ giving rise to a real form of $\Lambda(\g)$. 
\smallskip

Furthermore, an automorphism $\Om$ of the Lie algebra $\Lambda(\g)$ restricts to an automorphism of $\Lambda(\g)^\phi$ if and only if it leaves $\Lambda(\g)^\phi$ invariant that is $\Om$ and $\phi$ commute. 

Let us provide an example of a real form.
\begin{example}
psu(1,1$|$2)

psu(1,1$|$2) is the Lie superalgebra of the Lie supergroup PSU(1,1$|$2), whose bosonic subgroup is $AdS_3\times S^3$. Sigma models on this supermanifold are highly relevant for the string theory in the $AdS_3/CFT_2$ correspondence. 

A good way to describe the real form is in terms of a matrix realisation. Consider
the $4\times 4$ supertraceless supermatrix
\begin{equation}
   	\begin{split}
		X\ = \ \left(\begin{array}{cc}A & B \\ C & D\\ \end{array}\right) \, ,
	\end{split}
\end{equation}
where $A,B,C$ and $D$ are $2\times 2$ matrices. Consider the involutive semimorphism given by
\begin{equation}
  \phi \, :\, X \ \mapsto \ -\eta\bar X^{st}\eta^{-1}\ ,\qquad\qquad\text{where}\ \eta \ = \ \text{diag}(-1,1,1,1)\,
\end{equation}
and the bar denotes complex conjugation.
 If we in addition divide out the ideal generated by the identity matrix we obtain psu(1,1$|$2). Especially the upper diagonal block is a matrix realisation of su(1,1) over $\R$, while the lower diagonal block corresponds to su(2).
\end{example}

\subsection{Representations}
\label{section:finitereps}

In this section all Lie superalgebras are basic simple Lie superalgebras of type I.

The finite dimensional representations of finite dimensional classical Lie superalgebras are described by Kac in \cite{Kac:1977qb} and \cite{Kac:1977hp}.
Gould gives a generalisation to infinite dimensional representations \cite{gould}.

We recall the classification results for irreducible representations of type I Lie superalgebras by Gould \cite{gould}. 
Let $\lambda$ in $\h^*$ be the highest weight of a highest weight representation $V(\lambda)$ and let $Z$ be the centre of the universal enveloping algebra $U(\g)$ of the Lie superalgebra $\g$, then $Z$ takes constant values on $V(\lambda)$. The eigenvalue of $z$ in $Z$ on $V(\lambda)$ is denoted by $\chi_\lambda(z)$, this defines an algebra homomorphism\vspace{1mm}
\begin{equation}
	\chi_\lambda : Z\rightarrow \C \ ,  \qquad z\mapsto \chi_\lambda(z)\vspace{1mm}
\end{equation}
called infinitesimal character. A (not necessarily highest-weight) representation admits an infinitesimal character $\chi_\lambda$ if the elements $z$ in $Z$ take constant values 
$\chi_\lambda(z)$ in the representation. In the case of simple Lie algebras it is well known that every irreducible representation admits an infinitesimal character \cite{humphreys}. The generalisation to type I Lie superalgebras is proved by Gould \cite{gould}:
\begin{thm}
	Every irreducible representation of a Lie superalgebra of type I admits an infinitesimal character $\chi_\lambda$ for some $\lambda$ in $\h^*$. 
\end{thm}
We construct representations explicitly as done by Kac \cite{Kac:1977hp}. 
Recall the triangular decomposition of type I Lie superalgebras $\g=\g_{-}\oplus\g_0\oplus\g_+$ \eqref{eq:typeIgradiation}. Let $V_0$ be a representation of the bosonic subalgebra $\g_0$ of countable dimension then one gets a representation of $\g_0\oplus\g_+$ by promoting the elements in $\g_+$ to annihilation operators $\g_+(V_0)=0$ and the elements in $\g_-$ to creation operators, i.e. we define the Kac module of $V_0$ to be \vspace{1mm}
\begin{equation}
	\K(V_0)\ = \ \text{Ind}_{\g_0\oplus\g_+}^\g(V_0)\ .\vspace{1mm}
\end{equation}
The main results in \cite{gould} are summarised in
\begin{thm} Let $V_0$ be an irreducible representation of $\g_0$ and $\K(V_0)$ the Kac module. Then 
	\begin{itemize}
		\item there exists a maximal proper submodule $M(V_0)$, 
		\item the quotient $\K(V_0)/M(V_0)$ is irreducible and all irreducible representations are of this form. 
	        \item $V_0$ admits an infinitesimal character $\chi_\lambda^0$ and $\K(V_0)$ is irreducible if and 
			only if $(\lambda+\rho|\alpha)\neq0$ for all odd positive roots $\alpha$.
		\end{itemize}
\end{thm}
Denote the collection of representations with infinitesimal character $\chi_\lambda$ by $\m_\lambda$.  
In view of this theorem we call a representation $V_\lambda$ in $\m_\lambda$ atypical if there exists an odd positive root $\alpha$ such that $(\lambda+\rho|\alpha)=0$.

\subsubsection{Geometric interpretation of representations}

This subsection is part of \cite{Creutzig:2008ag}.

The co-adjoint orbit method of Kirilov and Kostant \cite{kirillov} relates co-adjoint orbits of a Lie group to representations of the group. In the case of compact simple Lie groups this correspondence is (\cite{Kirillovmerits} and references therein)\vspace{1mm}
\begin{equation}
	\pi_\lambda \ \longleftrightarrow \ \Omega_{\lambda+\rho}\vspace{1mm}
\end{equation}
where $\pi_\lambda$ is a irreducible highest weight representation of the compact Lie group $G$ with dominant highest weight $\lambda$, $\rho$ the Weyl vector and $\Omega$ the co-adjoint orbit in the dual of the Lie algebra $\g^*$ containing $\lambda+\rho$.  

We seek an analogous description for Lie supergroups encoding information about atypicality.
Let us consider co-adjoint orbits. We fix a Cartan subalgebra $\h$. Since the metric restricts non-degenerately to $\h$ there exists a $h_{\lambda+\rho}$ in $\h$ such that $(\lambda+\rho)(h)=(h_{\lambda+\rho},h)$ for all $h$ in $\h$. We write $\lambda+\rho=(h_{\lambda+\rho},\ \cdot \ )$, then the co-adjoint orbit containing $\lambda+\rho$ is\vspace{1mm}
\begin{equation}
	\Omega_{\lambda+\rho}\ = \ \{\ (gh_{\lambda+\rho}g^{-1}, \ \cdot\ ) \ | \ g \ \text{in} \ G \ \} \ .\vspace{1mm}
\end{equation}
It follows that the orbit extends into the dual space $\g^*_\alpha$ of the root space of the root $\alpha$ if and only if $(\lambda+\rho|\alpha)\neq0$, i.e.
\begin{equation}
 \Omega_{\lambda+\rho}\cap \g^*_\alpha\neq\emptyset\ \ \Longleftrightarrow\ \ (\lambda+\rho|\alpha)\neq0\,.
\end{equation}
In other words, if $(\lambda+\rho|\alpha)=0$ then $\Omega_{\lambda+\rho}\cap\g^*_\alpha=\emptyset$ and we say that the co-adjoint orbit $\Omega_{\lambda+\rho}$ is localised in the direction corresponding to the root $\alpha$.  
This gives us the following relation between representations and co-adjoint orbits. 
\begin{prp}\label{coadjointfinite}
	There is a one-to-one correspondence between collections of representations with infinitesimal character $\chi_\lambda$ and co-adjoint orbits 
	\begin{equation}
		\Omega_{\lambda+\rho} \ \longleftrightarrow \ \m_\lambda \ ,
	\end{equation}
	such that if and only if a representation is atypical the associated co-adjoint orbit is localised in a fermionic direction.
\end{prp}

\section{Affine Lie superalgebras}

This section is part of \cite{Creutzig:2008ag}.
The conformal field theories we are going to consider are Wess-Zumino-Novikov-Witten models on Lie supergroups. These possess an affine Lie superalgebra symmetry. 

References to affine Lie superalgebras are \cite{Kac:1994kn} and \cite{Kac:2000}.
Denote by $\{t^a\}$ a basis of the finite dimensional Lie superalgebra $\g$ with structure constants ${f^{ab}}_c$ and non-degenerate invariant supersymmetric bilinear form $(\ \ , \ \ )$. 
Then the affine Lie superalgebra $\widehat{\g}$ corresponding to $\g$ is the infinite dimensional Lie superalgebra generated by $\{t^a_n, K,d\}$ for $n$ in $\Z$.
$K$ is central and the non-vanishing relations are\vspace{1mm}
\begin{equation}\label{eq:relaffine}
	\begin{split}
[ \, t^a_m , t^b_n \, ]\ &=\  {f^{ab}}_ct^c_{m+n} + m \delta_{m+n} (t^a,t^b) K   \\
[ \, d , t^a_n \, ]\ &=\ n t^a_n   \, .\vspace{1mm}
         \end{split}
\end{equation}
The vector space \vspace{1mm}
\begin{equation}
    \widehat{\h}\ =\ \h \oplus \C K \oplus \C d   \vspace{1mm}
\end{equation}
is a Cartan subalgebra of $\widehat{\g}$. We extend a linear function $\lambda$ on $\h$ to $\widehat{\h}$ by setting $\lambda(K) = \lambda(d) = 0$ and define linear functions $\Lambda_0$ and $\delta$ on $\widehat{\h}$ by\vspace{1mm}
\begin{equation}
	\Lambda_0 (\h \oplus \C d) \ =\  0 \ \ \ \ ,\ \ \ \ \Lambda_0 (K) \ =\  1 \ \ \ \ ,\ \ \ \
	\delta (h \oplus \C K)   \ =\ 0 \ \ \ \ \text{and}\ \ \ \ \delta(d)    \ = \ 1 \, .\vspace{1mm}
\end{equation}
Then $\widehat{\h}^* = \h^* \oplus \C \Lambda_0 \oplus \C \delta$.
We also extend the bilinear form $(\ \ ,\ \ )$ from $\g$ to $\widehat{\g}$ by setting\vspace{1mm}
\begin{equation}
	\begin{split}
		( t^a_m , t^b_n ) \ =\ \delta_{m+n} (t^a,t^b) \qquad &,\qquad
		( t^a_m, K )\ =\ ( t^a_m , d )\ = \ 0\ , \\
		( K , K )\ =\ ( d , d )\ =\ 0 \qquad&\text{and}\qquad(K,d)\ =\ 1\ .\vspace{1mm}
\end{split}
\end{equation}
Further the space of positive roots is\vspace{1mm}
\begin{equation}
	\widehat{\Delta}_+\ = \ \Delta_+\cup\{\alpha+n\delta|n>0\} \ .\vspace{1mm}
\end{equation}
The affine Weyl vector is\vspace{1mm}
\begin{equation}
	\widehat{\rho} \ = \ \rho + h^\vee\Lambda_0 \ ,\vspace{1mm}
\end{equation}
where $h^\vee$ is the dual Coxeter number of $\g$ (definition \ref{definition:coxeter}).

\subsection{Representations}

In this section, we consider representations of affine Lie superalgebras.
Recall the construction of the Virasoro algebra out of the affine Lie superalgebra \agl\ (section \ref{section:affinegl}). This Sugawara construction holds in general as we will demonstrate in the next chapter. The relations of the Virasoro zero mode $L_0$ with elements of the affine Lie superalgebra $\widehat{\g}$ are
\begin{equation}
 [L_0,t^a_n] \ = \ n\, t^a_n\, ,
\end{equation}
i.e. they coincide with the action of the derivation $d$ of the affine Lie superalgebra $\widehat{\g}$ \eqref{eq:relaffine}. In a WZNW model one identifies these two operators $d=L_0$. The highest-weight representations relevant for the WZNW model are constructed as follows. 
\begin{definition}
 Consider a weight $\Lambda=\lambda+k\Lambda_0$ of $\widehat{\h}^*$, then the Verma module $V_+(\Lambda)$ of highest-weight $\Lambda$ is constructed from a state $|\Lambda_+\rangle$ satisfying
\begin{equation}
 \begin{split}
  h|\Lambda_+\rangle \ &= \ \Lambda(h)|\Lambda_+\rangle \qquad\text{for}\ h\,\in\,\h\, ,\\
  K|\Lambda_+\rangle \ &= \ k|\Lambda_+\rangle\, , \\
  t^a_n|\Lambda_+\rangle \ &= \ 0\qquad \qquad\text{for}\ n>0\ \text{and}\\
  t^a_0|\Lambda_+\rangle \ &= \ 0\qquad \qquad\text{for}\ t^a\,\in\,\nn_+\, .\\
 \end{split}
\end{equation}
The Verma module $V_-(\Lambda)$ of lowest-weight $\Lambda$ is constructed analogously from a state $|\Lambda_-\rangle$ satisfying
\begin{equation}
 \begin{split}
  h|\Lambda_-\rangle \ &= \ \Lambda(h)|\Lambda_-\rangle \qquad\text{for}\ h\,\in\,\h\, ,\\
  K|\Lambda_-\rangle \ &= \ k|\Lambda_-\rangle\, , \\
  t^a_n|\Lambda_-\rangle \ &= \ 0\qquad \qquad\text{for}\ n<0\ \text{and}\\
  t^a_0|\Lambda_-\rangle \ &= \ 0\qquad \qquad\text{for}\ t^a\,\in\,\nn_-\, .\\
 \end{split}
\end{equation}
\end{definition}
The highest(lowest)-weight state then has conformal dimension ($L_0$ eigenvalue)\vspace{1mm}
\begin{equation}
	h_\Lambda \ = \ \frac{(\Lambda+2\widehat{\rho}\ |\ \Lambda)}{2(k+h^\vee)}\, .\vspace{1mm}
\end{equation}
A singular vector is a state of the representations that generates a proper subrepresentation.
We call such a Verma module typical if all its singular vectors are inherited from the bosonic subalgebra, otherwise if there are also fermionic singular vectors it is called atypical. 
According to \cite{Quella:2007hr} a necessary condition for atypicality is \vspace{1mm}
\begin{equation}\label{eq:affineatypical1}
	h_{\Lambda-\alpha'} \ = \ h_\Lambda+n\vspace{1mm}
\end{equation}
where $\alpha=\alpha'+n\delta$ for some integer $n$ and an odd root $\alpha'$ of $\g$. If $\alpha$ is a positive odd root the highest-weight representation $V_+(\Lambda)$ can be atypical, and if $\alpha$ is a negative odd root the lowest-weight representation $V_-(\Lambda)$ can be atypical. Equation \eqref{eq:affineatypical1} can be rewritten as \vspace{1mm}
\begin{equation}\label{eq:affineatypical2}
	(\Lambda+\widehat{\rho}\ |\ \alpha) \ = \ 0\ .\vspace{1mm}
\end{equation}
In \cite{Kac:1978ge} it is shown that this is exactly the atypicality condition for basic affine Lie superalgebras of type I.

Atypical representations are closely related to atypical representations of the horizontal subalgebra. We know that $V_\pm(\Lambda)$ is atypical if there is a singular vector on the level of the horizontal subalgebra $\g$. Concatenating the representation with an inner automorphism of $\widehat{\g}$ gives an isomorphic representation that is also atypical. 
The affine Weyl group induces such automorphisms of the affine Lie superalgebra $\widehat{\g}$. 
The affine Weyl group is the automorphism group on the root and coroot systems and hence induces an automorphism on the affine Lie superalgebra since this in return is uniquely defined via its roots, coroots and Cartan subalgebra.  Denote by $M$ the $\Z$ span of the coroots of $\g$ and define the translation $T_\alpha$ as ($\alpha$ in $M$)\vspace{1mm}
\begin{equation}
	T_\alpha(\Lambda) \ = \ \Lambda + \Lambda(K)\alpha-((\Lambda|\alpha)+\frac{1}{2}(\alpha|\alpha)\Lambda(K))\delta \ .\vspace{1mm}
\end{equation}
for $\Lambda$ in $\widehat{\h}^*$.
We denote the group of translations $\{ T_\alpha \ | \ \alpha \ \text{in}\ M \}$ by $T_M$. Then the affine Weyl group is \cite{Kac:1994kn} ($W$ denotes the Weyl group of $\g$)\vspace{1mm}
\begin{equation}
	\widehat{W} \ = \ W \ltimes T_M \ .\vspace{1mm}
\end{equation}
The translation $T_\alpha$ induces an isomorphism $\tilde{T}_\alpha$ on $\widehat{\g}$ which acts explicitly as \vspace{1mm}
\begin{equation}\label{eq:spectralflow}
	\begin{split}
		\tilde{T}_\alpha:\ \ \qquad  h \ &\mapsto\ h+\alpha(h)K \qquad\qquad \text{for} \ h \ \text{in}\ \h \\[1mm]
		                     K \ &\mapsto\ K \\
				     L_0 \ &\mapsto\ L_0 - h_\alpha - \frac{1}{2}(\alpha|\alpha)K \\
				     t^\beta_n \ &\mapsto\ t^\beta_{n-(\alpha|\beta)} \qquad\qquad\text{for}\ t^\beta \ \text{in} \ \g_\beta \ .\\[1mm]
        \end{split}
\end{equation}
These automorphisms are usually called spectral flow automorphisms in the physics literature.
If one knows the characters of the representations of $\widehat{\g}$ then one can identify the representations obtained by an automorphism via (the $h_1,\dots h_r$ form an orthonormal basis of $\h$)\vspace{1mm}
\begin{equation}
	\begin{split}
	\chi_{\rho\circ\tilde{T}_\alpha}(q,z_i) \ &= \ 
	\tr_\rho(q^{\tilde{T}_\alpha(L_0)}\ z_1^{\tilde{T}_\alpha(h_1)}\ \dots\ z_r^{\tilde{T}_\alpha(h_r)}\ (-1)^F) \\[1mm]
	&= \ q^{-\frac{k}{2}(\alpha|\alpha)}\ z_1^{\alpha(h_1)k}\ \dots\ z_r^{\alpha(h_r)k}\ \chi_\rho(q, z_iq^{-\alpha(h_i)}) \ .\\[1mm]
\end{split}
\end{equation}
If every character corresponds uniquely to a representation then this identification is exact. In the cases of \agl\footnote{Even though \gl\ is not classical the above statements hold} \cite{Schomerus:2005bf}, $\widehat{\text{su}}(2|1)$ \cite{Saleur:2006tf} and $\widehat{\text{psu}}(1,1|2)$ \cite{Gotz:2006qp} all atypical representations could be obtained in this way from representations that have a singular vector on the level of the horizontal subalgebra $\g$.

\subsubsection{Geometric interpretation of representations}

We saw that the geometry of co-adjoint orbits provided information whether the associated representations are atypical or not. In a similar manner one can relate superconjugacy classes to representations of the affine Lie superalgebra $\widehat{\g}$. Choose an element $h_{\lambda+\rho}$ of the bosonic subalgebra $\g_0$ and choose a Cartan subalgebra $\h$ containing $h_{\lambda+\rho}$. Then we consider the superconjugacy class containing the point $\exp{\frac{2\pi ih_{\lambda+\rho}}{k+h^\vee}}$,\vspace{1mm}
\begin{equation}
	C_a \ = \ \{\ gag^{-1} \ | \ g \ \text{in} \ G \ \} \ , \qquad a \ = \ \exp{\frac{2\pi ih_{\lambda+\rho}}{k+h^\vee}} \ .\vspace{1mm}
\end{equation}	
The superconjugacy class is localised into a fermionic direction corresponding to an odd root $\alpha$ of $\g$ if and only if 
$\alpha(h_{\lambda+\rho})=n(k+h^\vee)$ for some $n$ in $\Z$. But this is equivalent to \vspace{1mm}
\begin{equation}
	(\lambda+k\Lambda_0+\widehat{\rho}\, |\, \alpha-n\delta) \ = \ (\lambda+\rho\, |\, \alpha)-n(k+h^\vee)(\Lambda_0\, |\, \delta) \ = \ 0 \ .\vspace{1mm}
\end{equation}
Thus we arrive at the affine analogue of proposition \ref{coadjointfinite}.
\begin{prp}\label{proposition:conjugacy}
	There is a one-to-one correspondence between Verma modules $V_\pm(\Lambda)$, $(\Lambda=\lambda+k\Lambda_0)$ and superconjugacy classes 
	\begin{equation}
		C_{\exp{\frac{2\pi ih_{\lambda+\rho}}{k+h^\vee}}} \ \longleftrightarrow \ V_\pm(\Lambda) \ ,
	\end{equation}
	such that a representation is atypical if and only if the associated superconjugacy class is not completely delocalised in the fermionic directions.
\end{prp}
\smallskip 
In the next chapter, we will show that superconjugacy classes have the physical interpretation of a brane. 

In the case of compact simple Lie groups the above correspondence has an interpretation in terms of Cardy boundary states \cite{Cardy:1989ir} (or equivalently branes). 
To each irreducible finite dimensional highest-weight representation of highest-weight $\Lambda=\lambda+k\Lambda_0$ 
there exists a boundary state $B(\Lambda)$ and in the semiclassical limit $k\rightarrow \infty$ 
this state becomes a distribution concentrated on the conjugacy class $C_a$ ($a= \exp{\frac{2\pi ih_{\lambda+\rho}}{k+h^\vee}}$)\cite{Alekseev:1999bs}.

Later we will see that this statement also holds in the example of the Lie supergroup \GL.

\chapter[CFT with Lie supergroup as target space]{Conformal field theory with Lie supergroup as target space}

The aim of this chapter is to introduce bulk and boundary Wess-Zumino-Novikov-Witten models. 
We begin with some general considerations of two-dimensional conformal field theory and its boundary theory.
Then we introduce the relevant models the Wess-Zumino-Novikov-Witten models with Lie supergroup target. 
We show that they are indeed conformal field theories and that they in addition possess an affine Lie superalgebra current symmetry. 
We present the bulk model and explain a perturbative formalism to compute correlation functions. 

Then we start to investigate the boundary theory. First we explain that a boundary Wess-Zumino-Novikov-Witten model is described by boundary conditions for the chiral fields. These conditions describe a subsupermanifold of the supergroup, the branes' worldvolume. We show that these branes are twisted superconjugacy classes. These insights are then used to find the action of the boundary WZNW model.  

\section{Conformal field theory in two dimensions}

The conformal field theories we are considering are two dimensional. As a reference we use \cite{DiFrancesco:1997nk}.

We start with some general considerations about quantum field theory. In general the world-sheet of a two-dimensional field theory is a two-dimensional orientable Riemann surface $\Sigma$. We will mostly restrict to the sphere and to the disc in the boundary case (or equivalently the complex plane and upper half plane) and parameterise it by holomorphic and anti-holomorphic coordinates $z,\bar z$. A field $\Phi(z,\bar z)$ maps the world-sheet to some target space. There exist some distinguished fields, chiral fields, that depend only on one of the two coordinates $z$ or $\bar z$.   
We take a Lagrangian approach, that is a model is described by an action $S$ and fields $\Phi_i(z_i,\bar z_i)$.
Observable quantities are correlation functions.
Let $X=\Phi_1(z_1,\bar z_1)\dots\Phi_n(z_n,\bar z_n)$ be a product of fields and  $\mathcal{D}\Phi$ denote the path integral measure, 
then a correlation function is defined by the path integral
\begin{equation}
   \langle\, X\,\rangle \ = \ \int\mathcal{D}\Phi\ X\ e^{-S}\, .
\end{equation}
For the computation of correlation functions the symmetry of the model is an essential aide.
Let $\delta$ be an infinitesimal symmetry transformation of the theory. Symmetry means that any correlation function is invariant under such a transformation, i.e.
\begin{equation}\label{eq:wardidentity}
    0\ = \ \delta\langle\, X\,\rangle \ = \ \langle\, \delta X\,\rangle -
              \langle\, (\delta S)X\,\rangle \, .
\end{equation}
These constraints to the form of correlation functions are called Ward identities.

The second concept we need is that of an operator product expansion. 
Let $A(z)$ and $B(w)$ be two local chiral fields then their operator product expansion is the Laurent expansion of the product of the two fields at the point $w$. In other words the operator product expansion describes the behaviour of the product of the two fields $A(z)$ and $B(w)$ in a small neighbourhood of $z-w$. The first regular term in this expansion is called the normal ordered product indicated by colons, that is 
\begin{equation}
    :AB:(w) \ = \ \frac{1}{2\pi i }\oint_w\frac{dx}{(x-w)}A(x)B(w)\, .
\end{equation}

Now, let us describe the main ingredients of conformal field theory. The first object is the energy-momentum tensor, it consists of a holomorphic part $T(z)$ and an antiholomorphic one $\bar T(\bar z)$. They satisfy the operator product expansion
\begin{equation}\label{eq:energymomentumtensor}
	\begin{split}
		T(z)T(w)\ &\sim \ \frac{c/2}{(z-w)^4}+\frac{2T(w)}{(z-w)^2}+ \frac{\del T(w)}{(z-w)}\\
		\bar T(\bar z)\bar T(\bar w)\ &\sim \ \frac{c/2}{(\bar z-\bar w)^4}+\frac{2\bar T(\bar w)}{(\bar z-\bar w)^2}+ 
		                    \frac{\bar \del \bar T(\bar w)}{(\bar z-\bar w)}\\
		T(z)\bar T(\bar w)\ &\sim \ 0 \, .\\
	\end{split}
\end{equation}
The number $c$ is called the central charge of the model. 
If we turn to the operator formalism and express the energy-momentum tensor in a Laurent expansion
\begin{equation}
	T(z)\ = \ \sum_{n\,\in\,\Z}L_nz^{-n-2}\qquad,\qquad\bar T(\bar z)\ = \ \sum_{n\,\in\,\Z}\bar L_n\bar z^{-n-2}\, ,
\end{equation}
then the modes of this expansion satisfy the relations of an infinite-dimensional Lie algebra, that is two copies of the Virasoro algebra
\begin{equation}
	\begin{split}
		[L_n,L_m] \ &= \ (n-m)L_{n+m}+\frac{c}{12}n(n^2-1)\delta_{n+m,0}\\
		[\bar L_n,\bar L_m] \ &= \ (n-m)\bar L_{n+m}+\frac{c}{12}n(n^2-1)\delta_{n+m,0}\\
		[L_n,\bar L_m] \ &= \ 0\, .\\
	\end{split}
\end{equation}
This is the infinite dimensional symmetry of a two-dimensional CFT. 

The second ingredient are the primary fields. A field $\phi(z,\bar{z})$ is called a primary field of conformal dimension $(h,\bar h)$ if it satisfies
\begin{equation}
	\begin{split}
		T(z)\phi(w,\bar w) \ &\sim \ \frac{h}{(z-w)^2}+ \frac{\del \phi(w,\bar w)}{(z-w)}\\
		\bar T(\bar z)\phi(w,\bar w) \ &\sim \ \frac{\bar h}{(\bar z-\bar w)^2}+ \frac{\bar \del \phi(w,\bar w)}{(\bar z-\bar w)}\\
	\end{split}
\end{equation}
Conformal invariance in two dimensions is remarkable. The group of local conformal transformations on the complex plane is the group of holomorphic functions.
The group of global conformal transformations restricts the form of correlation functions of primary fields, e.g. the two-point functions and three-point functions to have the form
\begin{equation}
	\begin{split}
		\langle\,\phi_1(z_1,\bar z_1)\phi_2(z_2,\bar z_2)\,\rangle\ =\ &\frac{\delta_{h_1,h_2}\delta_{\bar h_1,\bar h_2}C_{12}}
		{(z_1-z_2)^{2h_1}(\bar z_1-\bar z_2)^{2\bar h_1}} \\
		\langle\,\phi_1(z_1,\bar z_1)\phi_2(z_2,\bar z_2)\phi_3(z_3,\bar z_3)\,\rangle\ =\ &\frac{C_{123}}
		{(z_1-z_2)^{\Delta_{123}}(z_2-z_3)^{\Delta_{231}}(z_1-z_3)^{\Delta_{132}}}\ \times \\
		&\frac{1}{(\bar z_1-\bar z_2)^{\bar \Delta_{123}}(\bar z_2-\bar z_3)^{\bar \Delta_{231}}(\bar z_1-\bar z_3)^{\bar \Delta_{132}}}\ \ , \\[2mm]
		\text{where}\ \Delta_{abc}\ = \ h_a+h_b-h_c\qquad&\text{and}\qquad \bar\Delta_{abc}\ = \ \bar h_a+\bar h_b-\bar h_c\, .\\
	\end{split}
\end{equation}
A two-dimensional bulk conformal field theory is completely specified by the knowledge of all two- and three-point functions of primary fields. 
\smallskip

Now, we consider a world-sheet with boundary. The simplest example is the upper-half plane, Im$z>0$ with boundary the real line $z=\bar z$. The boundary conformal field theory is in the interior of the upper half plane locally equivalent to the bulk theory. This means, that the leading singularities in the OPE of fields inserted in the interior of the upper-half plane coincide with the OPEs of the bulk theory. On the boundary we want the theory to stay conformal, which is guaranteed if the energy-momentum tensor satisfies the gluing conditions
\begin{equation}\label{eq:gluingenergymomentum}
 T(z) \ = \ \bar T(\bar z) \qquad\text{for}\qquad z\ = \ \bar z.
\end{equation}
In a two-dimensional CFT this condition ensures that there is no momentum flow across the boundary.

The additional data of a boundary conformal field theory that completely specifies the model are bulk one-point functions, bulk-boundary two-point functions and boundary three-point functions.

\subsection{Boundary states}\label{sec:bdystates}

There is an efficient concept in boundary conformal field theory, that of boundary states. In this section we follow \cite{Recknagel:1997sb} and \cite{Schomerus:2002dc}. To each boundary CFT there exists a formal state (formal, because it is not normalisable in the usual sense) in the bulk theory containing the information of the boundary conditions and of bulk one-point functions in the boundary theory.

The goal is to express correlation functions at finite temperature in the boundary theory through quantities which are completely specified in the bulk theory. In string theory terms, a boundary theory describes open strings starting and ending on some branes, while a bulk theory describes closed strings. From this point of view, boundary states are an example of a closed string to open string duality. 

Consider a CFT with energy-momentum tensor $T(z)$ and $\bar T (\bar z)$ and additional chiral fields $W(z)$ and $\bar W(\bar z)$ of half-integer conformal dimension $h_W$. In addition to preserving conformal symmetry along the boundary \eqref{eq:gluingenergymomentum}, in the boundary CFT we want to preserve half of the chiral symmetry. This means we require gluing conditions of the form
\begin{equation}\label{eq:gluingchiralfields}
	W(z)\ = \ \Omega(\bar W(\bar z))\qquad\qquad\text{for}\ z\ = \ \bar z\, .
\end{equation}
Here $\Omega$ denotes an automorphism on the space of chiral fields. In general there exists more than one boundary CFT for the gluing conditions 
\eqref{eq:gluingchiralfields}. Hence we label a boundary theory by the automorphism $\Omega$ and an additional parameter $\alpha$. In our examples this additional parameter is the transverse position of the brane which the boundary theory describes.
\smallskip

In the following the world-sheet of the boundary theory is the upper-half plane, while the bulk world-sheet is the complex plane, and we denote fields of the boundary theory by $\phibdy$ and those of the bulk model by $\phibulk$.

A correlation function at finite temperature labelled by $\beta_0$ in the boundary theory is
\begin{equation}
	\langle \phibdy_1(z_1,\bar z_1)\dots\phibdy_n(z_n,\bar z_n)\rangle^{\beta_0} \ = \ 
	\text{tr}(e^{\beta_0H^{\text{bdy}}}\phibdy_1(z_1,\bar z_1)\dots\phibdy_n(z_n,\bar z_n))	
\end{equation}
where the arguments $z_i$ are radially ordered, the trace is over the state space of the boundary theory and the Hamiltonian is $H^{\text{bdy}}=L_0^{\text{bdy}}-c/24$.
If the above fields are primaries, i.e. they scale like $\phibdy(\lambda z,\bar\lambda\bar z)=\lambda^{-h}\bar\lambda^{-\bar h}\phibdy(z,\bar z)$, then the above correlators are periodic in the time variable $t=\ln z$ up to a scale factor. We picture the above process essentially as a one-loop diagram of an open string starting and ending on our brane $(\Omega,\alpha)$.

If we exchange the role of time and space, then this process should be viewed as a closed string emitted from the brane, propagating and then being absorbed by the brane again (illustrated in figure 3.1). The interchange of space and time is realised by 
\begin{equation}
	z\ \mapsto \ \xi \ = \ \exp(\frac{2\pi i}{\beta_0}\ln z)\qquad\text{and}\qquad
	\bar z\ \mapsto \ \bar\xi \ = \ \exp(-\frac{2\pi i}{\beta_0}\ln \bar z)\, .
\end{equation}
Fields depending on $\xi$ and $\bar \xi$ should now be interpreted as bulk states, i.e. states describing closed strings. 
\begin{figure}[htb!]
	\label{fig:closedopen}
\centering%
\includegraphics[width=6cm]{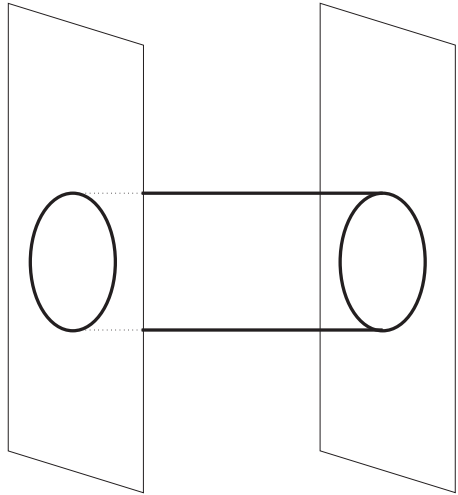}
\caption{{\em The diagram can be viewed as a closed string emitted from a brane and then being absorbed by another brane. It can also be seen as an one-loop diagram of an open string starting and ending on the two branes.}}
\end{figure}
For the primaries and the energy-stress tensor, this transformation is
\begin{equation}
	\begin{split}
		\phibulk(\xi,\bar\xi) \ &= \  \Bigl(\frac{d z}{d\xi}\Bigr)^h\, \Bigl(\frac{d\bar z}{d\bar \xi}\Bigr)^{\bar h}\,\phi^{\text{bdy}}(z,\bar z)\\                 T^{\text{bulk}}(\xi) \ &= \  \Bigl(\frac{d z}{d\xi}\Bigr)^2\,T^{\text{bdy}}(z)+\frac{c}{12}\{z,\xi\}\, ,\\
	\end{split}
\end{equation}
$\{z,\xi\}$ denotes the Schwartz derivative. The boundary state $|\alpha,\Omega\rangle$ is now defined, such that the following equation holds for any field insertions
\begin{equation}\label{eq:defboundarystate}
	\begin{split}
	\langle \phibdy_1(z_1,\bar z_1)\dots\phibdy_n(z_n,\bar z_n)&\rangle^{\beta_0} \ = \\ 
	&\langle\alpha,\Omega|e^{-\frac{2\pi^2}{\beta_0}H^{\text{bulk}}}
	\phibulk_1(\xi_1,\bar \xi_1)\dots\phibulk_n(\xi_n,\bar \xi_n)|\alpha,\Omega\rangle\, .\\
\end{split}
\end{equation}
This equation also makes sense if we replace one of the two boundary states by another boundary state $|\beta,\tilde\Omega\rangle$. Then the left-hand side describes a boundary theory with boundary conditions $(\alpha,\Omega)$ on the negative real axis and boundary conditions $(\beta,\tilde\Omega)$ on the positive real axis. We will see examples thereof in the following chapters.

There is another equivalent description of boundary states. The upper-half plane is mapped to the complement of the unit disk in the complex plane via
\begin{equation}
 z\ \mapsto \ \zeta\ = \ \frac{1-iz}{1+iz}\qquad\qquad \bar z\ \mapsto \ \bar \zeta\ = \ \frac{1+i\bar z}{1-i\bar z}\, .
\end{equation}
Under such a coordinate transformation a primary $\phi^{\text{bdy}}(z,\bar z)$ of conformal dimension $(h,\bar h)$ changes as follows
\begin{equation}
 \phi^{\text{bdy}}(z,\bar z)\ = \ \Bigl(\frac{d\zeta}{dz}\Bigr)^h\,  \Bigl(\frac{d\bar\zeta}{d\bar z}\Bigr)^{\bar h}\,\phi^{\text{bulk}}(\zeta,\bar \zeta)\, .
\end{equation}
A boundary state is a formal state $|\alpha,\Omega\rangle$
in the bulk defined such that the following equation holds for any primary $\phi$
\begin{equation}
 \langle \phi^{\text{bdy}}(z,\bar z)\rangle \ = \ 
          \Bigl(\frac{d\zeta}{dz}\Bigr)^h\, \Bigl(\frac{d\bar\zeta}{d\bar z}\Bigr)^{\bar h}\,\langle 0 |\phi^{\text{bulk}}(\zeta,\bar \zeta)|\alpha,\Omega\rangle
 \end{equation}
where the left-hand side is a one-point function evaluated in the boundary theory, while the right-hand side is evaluated in the bulk theory.
If we apply the equation above to the special case of our chiral currents inserted along the boundary $\zeta\bar\zeta=1$ and we use the gluing conditions \eqref{eq:gluingchiralfields}, then we see that the boundary state has to satisfy the Ishibashi conditions \cite{Ishibashi:1988kg}
\begin{equation}\label{eq:ishibashi}
	(W_n-(-1)^{h_W}\Omega(\bar W_{-n}))|\alpha,\Omega\rangle\ = \ 0\, .
\end{equation}
Here $W_n$ denotes the mode of the Laurent expansion of $W(z)$, i.e.
\begin{equation}
	W(z) \ = \ \sum_{n\,\in\,\Z} W_nz^{-n-h_W}\, .
\end{equation}
The analogous expression holds for $\bar W(\bar z)$.
An irreducible representation $\rho_i$ of the bulk theory allows at most one solution to the above constraint, the Ishibashi state $|i\rangle\rangle_\Omega$. It can be normalised such that
\begin{equation}\label{eq:ishibashinormalization}
   {}_\Omega\llangle i|  \tilde q^{L_0^{\text{bulk}}-\frac{c}{24}}(-1)^{F^{\text{bulk}}}|j\rrangle_\Omega\ = \ \delta_{i,j}\chi_i(\tilde q)\, ,
\end{equation}
where ${F^{\text{bulk}}}$ is a fermion number operator and $\chi_i(\tilde q)$ is the character of the representation $\rho_i$.
One feature of the superalgebras is that representations might be reducible but indecomposable, as we have seen. In that case they might possess more than one Ishibashi state and the amplitude of two Ishibashi states, in the sense of above equation, is not necessarily a true character but can also have a log $\tilde q$ dependence. We will see this in the examples in the following chapters.

The boundary state is a linear combination of the Ishibashi states
\begin{equation}
    |\alpha\rangle\ = \ \sum_{i}B^i_\alpha|i\rrangle_\Omega\, .
\end{equation}
It is very remarkable that the coefficients $B_\alpha^i$ encode all the information about the bulk one-point functions of the boundary theory $\alpha$ \cite{Lewellen:1991tb},\cite{Recknagel:1997sb}, that is 
\begin{equation}
	\langle\phi_i(z,\bar z)\rangle_\alpha\ = \ \frac{B_{\alpha,\Omega}^{i^+}}{|z-\bar z|^{h_i+\bar h_i}}
\end{equation}
where $i^+$ denotes the representation conjugate to $i$ with respect to the metric given by the two-point functions of the bulk theory.

We compute the coefficients $B_{\alpha,\Omega}^i$ using \eqref{eq:defboundarystate} without any field insertions. Then the left-hand side of \eqref{eq:defboundarystate} is the partition function of the boundary theory. For the evaluation of the right-hand side we use \eqref{eq:ishibashinormalization}, then we have
\begin{equation}
	\begin{split}
		Z_{(\alpha,\Omega);(\beta,\tilde\Omega)}(q)\ &= \ 
	\langle\alpha,\Omega|\tilde{q}^{\frac{1}{2}(L_0^\text{bulk}+\bar L_0^\text{bulk}-\frac{c}{12})}(-1)^{F^{\text{bulk}}}|\beta,\tilde\Omega\rangle\\
	&= \ \sum_i B_{\alpha,\Omega}^iB_{\beta,\tilde\Omega}^i \chi_i(\tilde q) \, .\\
\end{split}
\end{equation}
Here we introduced $q=\exp(-\beta_0)$ and $\tilde q=\exp(-\frac{2\pi^2}{\beta_0})$. The characters in terms of $\tilde q$ are linearly related to those in terms of $q$ by the modular S-matrix
\begin{equation}
	\chi_i(\tilde q) \ = \ \sum_j S_{ij} \chi_j(q)\, .
\end{equation}
Since $Z_{(\alpha,\Omega);(\beta,\tilde\Omega)}(q)$ is a true partition function, i.e. an integer combination of characters
\begin{equation}
	Z_{(\alpha,\Omega);(\beta,\tilde\Omega)}(q) \ = \ \sum_i n_{(\alpha,\Omega);(\beta,\tilde\Omega)}^i\chi_i(q)
\end{equation}
we get the following Cardy \cite{Cardy:1989ir} constraint for our coefficients
\begin{equation}
	 \sum_j B_{\alpha,\Omega}^jB_{\beta,\tilde\Omega}^jS_{ji} \ = \ n_{(\alpha,\Omega);(\beta,\tilde\Omega)}^i\, .
 \end{equation}
Often these constraints suffice to construct the boundary states. This is called modular bootstrap. 

Our goal for the following chapters is to understand in examples how to find and treat boundary states.  
\smallskip

We now turn to the specific class of models we are interested in.

\section{The bulk Wess-Zumino-Novikov-Witten model}

In this section, we introduce the bulk Wess-Zumino-Novikov-Witten (WZNW) model of a Lie supergroup. We start by introducing the action, then we show that the model possesses the affine current symmetry which allows for the Sugawara construction of the energy-momentum tensor. Finally, we state a formalism to compute correlation functions. We follow the reasoning for Lie groups, see e.g \cite{DiFrancesco:1997nk}. But in generalising to Lie supergroups one has to be carefull to include the fermions correctly.

\subsection{The bulk action}

For a field theoretic description, we need the notion of a supergroup valued field. 
Let $\Sigma$ be an orientable Riemann surface. Further let $x_a :\Sigma \rightarrow \Lambda_0(\R)$ and $\theta_b :\Sigma \rightarrow \Lambda_1(\R)$ be infinitely differentiable functions into the even, respectively odd, part of the Grassmann algebra over $\R$. By infinitely differentiable we mean a function of the form \cite{Berezin:1987wh}\vspace{1mm}
\begin{equation}
	f \ = \ f(\tau,\sigma) \ = \ \sum_{k\geq 0} \sum_{i_1,\dots,i_k}f_{i_1,\dots,i_k}(\tau,\sigma)\theta_{i_1}\dots\theta_{i_k} \ ,\vspace{1mm}
\end{equation}
where $\tau,\sigma\in \Sigma$, the $f_{i_1,\dots,i_k}(x)$ are $\R$-valued infinitely differentiable functions on $\Sigma$ and the $\theta_i$ generate the Grassmann algebra $\Lambda(\R)$. Then we introduce the local $\Lambda(\g)$-valued field on $\Sigma$\vspace{1mm}
\begin{equation}
	X(\tau,\sigma) \ = \ x_a(\tau,\sigma)t^a +\theta_b(\tau,\sigma)s^b\vspace{1mm}
\end{equation}
and the $G$-valued field $g(\tau,\sigma)=\exp X(\tau,\sigma)$.

The setup for the WZNW model is exactly as in the Lie group case. So let $\Sigma$ be the world-sheet, that is a compact Riemann surface without boundary, and $g:\Sigma\rightarrow G$ a map from the Riemann surface to the Lie supergroup $G$. Assume that there exists an extension of this map to a map $\tilde{g} : B\rightarrow G$ from a $3$-manifold $B$ with boundary $\del B=\Sigma$ to $G$. Further let $z=\tau+i\sigma$ and $\bar{z}=\tau-i\sigma$ then the kinetic term of the action is\vspace{1mm}
\begin{equation}
	S_{\text{kin}}[g] \ = \ \frac{k}{2\pi}\int_\Sigma d\tau d\sigma\ \str( g^{-1}\del g\ g^{-1}\bar{\del}g)  \vspace{1mm} 
\end{equation}
and the Wess-Zumino term is \cite{Witten:1983ar}\vspace{1mm}
\begin{equation}
	S_{\text{WZ}}[\tilde{g}] \ = \ \frac{k}{2\pi}\int_B H \ = \ 
	\frac{k}{6\pi}\int_B \str(\tilde{g}^{-1}d \tilde{g}\wedge \tilde{g}^{-1}d\tilde{g}\wedge \tilde{g}^{-1}d \tilde{g})\ .  \vspace{1mm} 
\end{equation}
The number $k$ is called the level of the model.
The full action is then \vspace{1mm}
\begin{equation}\label{eq:actionWZW}
	S[\tilde{g}] \ = \  	S_{\text{kin}}[g] + S_{\text{WZ}}[\tilde{g}] \ .\vspace{1mm}
\end{equation}
Further the variation of the action is\vspace{1mm}
\begin{equation}\label{eq:variationWZNWaction}
	\delta S \ = \ S[\tilde{g}+\delta\tilde{g}]-S[\tilde{g}]\ = \ 
	\frac{k}{\pi}\int_\Sigma d\tau d\sigma\ \str( g^{-1}\delta g\ \del(g^{-1}\bar{\del}g))\ . \vspace{1mm}
\end{equation}
Thus the bulk equations of motion tell us that we have the conserved currents $J,\bar J$ with $\delbar J=\del\bar J=0$, where
\begin{equation}
        \begin{split}
               J(z) \ &=\ -k\del g g^{-1}\qquad\qquad\text{and}\\
               \bar J(\bar z) \ &=\ kg^{-1}\bar\del g\, .\\
        \end{split}
\end{equation}
 
It is straightforward to compute the Polyakov-Wiegmann identity\vspace{1mm}
\begin{equation}
	\begin{split}\label{PolyakovWiegmann}
		S[\tilde{g}\tilde{h}] \ = \  &S[\tilde{g}]+S[\tilde{h}]	+\frac{k}{\pi}\int_\Sigma d\tau d\sigma\ \str( \del hh^{-1}\ g^{-1}\bar{\del}g) \ .\vspace{1mm}
	\end{split}
\end{equation}
The action \eqref{eq:actionWZW} is well-defined if it does not depend on the extension $\tilde{g}$ to a 3-manifold $B$.  
For type I Lie supergroup models this is done as follows \cite{Schomerus:2005bf,Gotz:2006qp,Quella:2007hr}. Type I Lie superalgebras have the distinguished $\Z$-graduation $\g=\g_-\oplus\g_0\oplus\g_+$ \eqref{eq:typeIgradiation}, where $\g_\pm$ are two irreducible representations of the bosonic subgroup $\g_0$ and the supertrace satisfies\vspace{1mm}
\begin{equation}
	\str(X_+\,Y_+)\ = \ \str(X_-\,Y_-)\ = \ 0 \qquad \text{for all}\ X_\pm\ \text{and}\ Y_\pm \ \text{in} \ \g_\pm\ .\vspace{1mm}
\end{equation}
Let $\{t^a_\pm\}$ be a basis of $\g_\pm$, then we define the fermionic fields $\theta_\pm=\theta^\pm_at^a_\pm$. Further, we parameterise a Lie supergroup element according to the distinguished $\Z$-graduation
$g=e^{\theta_-}g_0e^{\theta_+}$, where $g_0$ is an element of the bosonic subgroup.
Applying the Polyakov-Wiegmann identity \eqref{PolyakovWiegmann} twice the action becomes\vspace{1mm}
\begin{equation}\label{eq:actiontypeI}
	S[\tilde{g}] \ = \ S[\tilde{g_0}] \ + \ \frac{k}{\pi}\int_\Sigma d\tau d\sigma\ \str( \Ad(g_0)(\del\theta_+)\bar{\del}\theta_-) \ .\vspace{1mm}
\end{equation}
Thus the ambiguity in the extension of this model is the ambiguity of the Lie group WZNW model of the bosonic subgroup $G_0$ and gives well-known quantisation conditions on the level $k$ \cite{Gawedzki:1999bq}.

Type II Lie supergroups can be treated similarly \cite{Hikida:2007sz}. The distinguished $\Z$-graduation is $\g=\g_{-2}\oplus\g_{-1}\oplus\g_0\oplus\g_1\oplus\g_2$ and the supertrace satisfies
\begin{equation}
	\str(X_i\,Y_j)\ = \ 0 
\end{equation}
if $X_i$ in $\g_i$, $Y_j$ in $\g_j$ and $i+j\neq 0$.
Then parameterising a Lie supergroup element according to the distinguished $\Z$-graduation
$g=g_-e^{\theta_-}g_0e^{\theta_+}g_+$ and applying the Polyakov-Wiegmann identity \eqref{PolyakovWiegmann} four times the action becomes\vspace{1mm}
\begin{equation}
	\begin{split}
	S[\tilde{g}] \ = \ &S[\tilde{g_0}] \ + \ \frac{k}{\pi}\int_\Sigma d\tau d\sigma\ \str( \Ad(g_0)(\del\theta_+)\bar{\del}\theta_-)\ +\\
	&\frac{k}{2\pi}\int_\Sigma d\tau d\sigma\ \str( \Ad(g_0)([\theta_+,\del\theta_+]+2\del g_+g_+^{-1})
	([\bar{\del}\theta_-,\theta_-]+2g_-^{-1}\bar{\del} g_-)) \ .\vspace{1mm}
        \end{split}
\end{equation}
Thus the ambiguity in the extension in this model is the ambiguity of the Lie group WZNW model of the bosonic subgroup which corresponds to the Lie subalgebra $\g_0$. 

\subsection{The current symmetry}

Our next goal is to show, that this model possesses an affine current symmetry. This means that the modes of the currents $J(z)$ and $\bar J(\bar z)$ form two commuting copies of the affine Lie superalgebra $\hat\g$ at level $k$, where $\g$ is the underlying Lie superalgebra of the Lie supergroup $G$. Level $k$ means that the central element $K$ of the Lie superalgebra $\hat\g$ acts as the constant $k$, i.e. the WZNW model is a representation of $\hat\g$ of $K$-eigenvalue $k$. 
 We only show this symmetry for the holomorphic part, the anti-holomorphic currents are treated analogously.

Consider a variation by a holomorphic function $\omega(z)$ of the form $g\rightarrow(1+\omega(z))g$ then the Ward identity \eqref{eq:wardidentity} for this variation with $X=J$ is
\begin{equation}\label{eq:wznwwardidentity}
     \langle\,\delta_\omega J(w)\,\rangle\ = \ \langle\,[\omega,J]-k\del \omega\,\rangle \ = \ -\frac{1}{2\pi i}\oint dz\, \langle\,(\omega,J)J(w)\,\rangle\ = \ \langle\,(\delta_\omega S)J\,\rangle\, .
\end{equation}
The contour is taken counterclockwise.
The first equality sign is a direct computation as in \cite{DiFrancesco:1997nk}
\begin{equation}
	\begin{split}
		\delta_\omega J \ &=\ -k\del(\omega g)g^{-1}+k\del gg^{-1}\omega \\
		                &=\ [\omega,J]-k\del \omega\\
	\end{split}
\end{equation}
and the last equality follows from \eqref{eq:variationWZNWaction}, by using $\Ad(g)\del(g^{-1}\bar\del g)=\bar\del(\del gg^{-1})$ and changing the measure as $d^2z=2id\tau d\sigma$ and integrating by parts. 
We want to perform some explicit calculations.
We start by introducing some notation. First define the components of $J$ via (here and in the following, summation over repeated indices is understood) 
\begin{equation}
 J\ = \ \tilde J^a\kappa^{ab}t^b\qquad\qquad\text{and also}\qquad
\omega\ = \ \omega^a\kappa^{ab}t^b
\end{equation}
where $\kappa^{ab}$ denotes the invariant bilinear form $(\,t^a \, ,\,t^b \, )$ and $\{t^a\}$ a basis of our Lie superalgebra $\g$. We choose the basis such that
\begin{equation}
   \kappa^{ab}\kappa^{bc}\ = \ (-1)^{a}\delta^{ac}\, ,
\end{equation}
which is always possible as can be seen from the fundamental matrix realisations provided in example\,\ref{example:glmn}--\ref{example:ospmn}. Here and in the following we use the notation
\begin{equation}
     (-1)^{a} \ := \ (-1)^{|t^a|}\, .
\end{equation}
Since our bilinear form is non-degenerate, we can define a dual basis $\{t_a\}$
\begin{equation}
     (t^a,t_b)\ =\ \delta^a_b
\end{equation}
with dual metric $\kappa_{ab}$.
We raise and lower indices by using the metric, especially we need the formulae
\begin{equation}
	\begin{split}
   {f^{ab}}_c \ &= \ {f^{abd}}\kappa_{cd}\,,\\
   {{f^{abc}}} \ &= \  {f^{ab}}_d\kappa^{dc}\ \text{and also}\\
   \kappa_{ab}\ &= \ \kappa^{ab}\, .\\
   \end{split}
\end{equation}
Then the Ward identity \eqref{eq:wznwwardidentity} implies the following operator product expansion
\begin{equation}
  \tilde J^a(z)\tilde J^b(w)\ \sim \ \frac{k\kappa^{ba}}{(z-w)^2}-\frac{{f^{ba}}_c\tilde J^c(w)}{(z-w)}\,.
\end{equation}
This is almost an affine Lie superalgebra current symmetry, but we get some unusual signs.
We observed a similar behaviour for the invariant vector fields \eqref{eq:relationsinvariantvectorfields}. 
We get an affine current symmetry if we define
\begin{equation}
 \begin{split}
   J^a(z)\ = \ \left\{\begin{array}{cc}\ \, \tilde J^a(z) & 
\qquad\ \, \text{if}\ t^a\ \text{in}\ \g_0\oplus\g_{-1}\\
-\tilde J^a(z) & \text{if}\ t^a\ \text{in}\ \g_{1}\\
\end{array}\right. 
 \end{split}
\end{equation}
for type I Lie supergroups. In the type II case we set
\begin{equation}
 \begin{split}
   J^a(z)\ = \ \left\{\begin{array}{cc}\ \tilde J^a(z) & 
\text{if}\ t^a\ \text{in}\ \g_0\oplus\g_{-1}\\
-\tilde J^a(z) & \qquad\text{if}\ t^a\ \text{in}\ \g_{-2}\oplus\g_1\oplus\g_2\\
\end{array}\right. .
 \end{split}
\end{equation}
Now, the operator product expansion of these currents is as desired
\begin{equation}
  J^a(z) J^b(w)\ \sim \ \frac{k\kappa^{ab}}{(z-w)^2}+\frac{{f^{ab}}_c J^c(w)}{(z-w)}\, .
\end{equation}
This means that the modes of the Laurent expansion of the currents obey the relations of the affine Lie superalgebra $\hat\g$ \eqref{eq:relaffine}, i.e.
\begin{equation}
 \begin{split}
   J^a(z)\ &= \ \sum_{n\,\in\,\Z}t^a_nz^{-n-1}\\
[t^a_n,t^b_m]\ &= \ {f^{ab}}_ct^c_{n+m}+kn\delta_{n+m,0}\kappa^{ab}\, . 
 \end{split}
\end{equation}

\subsection{The Sugawara construction}

The next step is to find the Virasoro symmetry.  
We present the Sugawara construction for the holomorphic part of the energy-momentum tensor. We define the chiral field
\begin{equation}
    T(z)\ = \ \frac{(:J(z),J(z):)}{2(k+h^\vee)} \ = \ 
          \frac{:\tilde J^a(z)\kappa^{ab}\tilde J^b(z):}{2(k+h^\vee)} \ = \
 \frac{:J^a(z)\kappa^{ba} J^b(z):}{2(k+h^\vee)}
\end{equation}
and want to show that this field is the holomorphic energy-momentum tensor.
For this purpose we compute exactly as in \cite{DiFrancesco:1997nk} except for taking care of additional minus signs due to the fermions
\begin{equation}
 \begin{split}
  J^a(z):J^b\kappa^{cb}J^c:(w)\ &= \ \frac{2(k+h^\vee)J^a(w)}{(z-w)^2}\, .
 \end{split}
\end{equation}
This equation implies that the currents are primaries of conformal dimension one
\begin{equation}
	T(z)J^a(w) \ \sim \ \frac{J^a(w)}{(z-w)^2}+\frac{\del J^a(w)}{(z-w)}\, .
\end{equation}
We use this equation to show that $T(z)$ indeed satisfies the operator product expansion of an energy-momentum tensor \eqref{eq:energymomentumtensor}
\begin{equation}
	\begin{split}
		T(z)T(w)\ &\sim \ \frac{c/2}{(z-w)^4}+\frac{2T(w)}{(z-w)^2}+ \frac{\del T(w)}{(z-w)}\\
	\end{split}
\end{equation}
with central charge
\begin{equation}
	c \ = \ \frac{k\,\sdim(\g)}{k+h^\vee}\, .
\end{equation}
The superdimension is the dimension of the bosonic subalgebra minus the dimension of the fermionic part of the Lie superalgebra. 
Thus, many WZNW models on supergroups have zero or negative central charge. 

The next step is to find vertex operators of our theory. In a WZNW model of a Lie group these are the primaries of the current algebra. 

\subsection{The first order formulation}\label{section:firstorder}

The formulation provided in this section was established in \cite{Quella:2007hr}. It works only for Lie supergroups of type I, because we need the triangular decomposition \eqref{eq:typeIgradiation}.
We will sketch the procedure here.
Let $\{t^a_\pm\}$ be a basis of $\g_\pm$, then we define the fermionic fields $c=c_at^a_-$ and $\bar c=\bar c_at^a_+$. Further, we parameterise a Lie supergroup element according to the distinguished $\Z$-graduation \eqref{eq:typeIgradiation}
\begin{equation}
g\ =\ e^{c}\,g_0\,e^{\bar c}\,.
\end{equation} 
Here, $g_0$ is an element of the Lie subgroup. 
Then the action is \eqref{eq:actiontypeI} 
\begin{equation}
	S[\tilde{g}] \ = \ S[\tilde{g_0}] \ + \ \frac{k}{\pi}\int_\Sigma d\tau d\sigma\ \str( \Ad(g_0)(\del \bar c)\bar{\del}c) \ .\vspace{1mm}
\end{equation}
The idea is to find a perturbative prescription in terms of the WZNW model of the bosonic subgroup and of free fermions. For this purpose we introduce auxiliary fermionic fields $\bar b=\bar b_at^a_-$ taking values in $\g_-$ and $b=b_at^a_+$ taking values in $\g_+$. The model we want to consider is
\begin{equation}
	\begin{split}
		S \ &= \ S_0 \ + \ S_{\text{pert}} \\
		S_0 \ &= \ S[\tilde{g_0}]_{\text{ren}} \ + \ \frac{1}{2\pi}\int_\Sigma d\tau d\sigma\ \str(b\bar\del c)-\str(\bar b\del\bar c)\\
		S_{\text{pert}} \ &= \ \frac{1}{4\pi k}\int_\Sigma d\tau d\sigma\ \str( \Ad(g_0)(\bar b)b)\, .\\
	\end{split}
\end{equation}
This model is equivalent to the above WZNW model if we integrate the auxiliary fields $b,\bar b$ and if we take care about the measures. The measure of the Lie supergroup is the invariant supergroup measure \eqref{eq:Haarmeasure}, while in the model $S$, we want the invariant measure of the bosonic subgroup times the free fermionic measure. Thus we have to compute the Jacobian of change of coordinates. We will see in chapter\,\ref{chapter:gl11} in an example how this works precisely.
In general, it is shown in \cite{Quella:2007hr} that this involves the following renormalised metric for the bosonic WZNW model
\begin{equation}  
\kappa^{ij}\ \rightarrow \kappa^{ij}_{\text{ren}}\ = \ 
\kappa^{ij}-\frac{1}{k}\sum_{a,b\,\in\,\g_+}{f^{ia}}_b{f^{jb}}_a\, .
\end{equation}
The second consequence of the renormalisation is the possible appearance of a linear dilaton term coupling to the world-sheet curvature $\mathcal{R}$
\begin{equation}
 \frac{-1}{4\pi}\int d\tau d\sigma\, \mathcal{R}\ln\det\Ad(g_0)_-\, .
\end{equation}
Here, $\det\Ad(g_0)_-$ means the determinant of the adjoint action of $g_0$ on $\g_-$ (compare with \eqref{eq:Haarmeasure}. Such a term appears whenever the bosonic subgroup is not simple, i.e. in the cases $sl(n|m), gl(n|m)$ and $osp(2|2n)$.

Consider the model described by the action $S_0[g_0,c,\bar c,b, \bar b]$. It possesses the affine current symmetry of the bosonic subalgebra $\widehat{\g}_0$, e.g. the holomorphic part satisfies
\begin{equation}
J_B^i(z)J_B^j(w)\ \sim \ \frac{\kappa^{ij}_{\text{ren}}}{(z-w)^2}+\frac{{f^{ij}}_lJ_B^l(w)}{(z-w)}\, .
\end{equation}
The ghosts $b_a$ have conformal dimension 1 and the $c_a$ dimension 0. They satisfy
\begin{equation}
 b_a(z)c_b(w)\ \sim \ \frac{\kappa_{ab}}{(z-w)}\, .
\end{equation}
It turns out that one can also find an affine Lie superalgebra $\widehat{\g}$ current symmetry in the model $S_0$. 
The components of 
\begin{equation}
 \begin{split}
	 J \ &= \ J_-+J_0+J_+ \qquad\text{where} \\
J_- \ &= \ -k\del c+[c,J_B]-\frac{1}{2}:[c,:[c,b]:]:\\
J_0 \ &= \ J_B-:[c,b]:\\
J_+ \ &= \ -b
 \end{split}
\end{equation}
satisfy the relations of the $\widehat{\g}$ current algebra of level $k$
\begin{equation}
  J^a(z) J^b(w)\ \sim \ \frac{k\kappa^{ab}}{(z-w)^2}+\frac{{f^{ab}}_c J^c(w)}{(z-w)}\, .
\end{equation}
Similarly, there exists an anti-holomorphic copy of the affine current symmetry, given by the components of the following current. 
\begin{equation}
 \begin{split}
  \bar J \ &= \ \bar J_-+\bar J_0+\bar J_+  \qquad\text{where}\\
\bar J_- \ &= \ -\bar b\\
\bar J_0 \ &= \ \bar J_B+:[\bar c,\bar b]:\\
\bar J_+ \ &= \ k\bar\del\bar c-[\bar c,\bar J_B]-\frac{1}{2}:[\bar c,:[\bar c,\bar b]:]:\, .\\
 \end{split}
\end{equation}

Vertex operators are those of the model $S_0$, i.e. vertex operators of the WZNW model of the bosonic subgroup times free ghost operators. Further computations of correlation functions are performed perturbatively.

\section{The boundary WZNW model}\label{section:boundarycft}

We turn to the boundary. While bulk WZNW models on type I Lie supergroups are under good control, the boundary case has not been studied before the beginning of this thesis.  
This section is the main part of \cite{Creutzig:2008ag}. 

Boundary theory means, that the world-sheet $\Sigma$ is an orientable Riemann surface with one boundary $\del\Sigma$. Locally one usually parameterises any element in $\Sigma$ as $(\tau,\sigma)$, where the first coordinate belongs to the direction parallel to the boundary and the second one to the perpendicular direction. We also introduce complex variables $z=\tau+i\sigma$ and $\bar{z}=\tau-i\sigma$. 

Recall, that conformal invariance is preserved along the boundary if the energy-momentum tensor satisfies the boundary condition\vspace{1mm}
\begin{equation}
	T\ =\ \bar{T} \qquad\qquad \text{for} \ z\ =\ \bar{z}\ .\vspace{1mm}
\end{equation}
This is certainly satisfied if the boundary conditions of the currents are\vspace{1mm}
\begin{equation}\label{eq:gluing}
	J\ =\ \Omega(\bar{J}) \qquad\qquad \text{for} \ z\ =\ \bar{z}\ ,\vspace{1mm}
\end{equation}
where $\Om$ is an automorphism of $\g$ preserving any invariant non-degenerate supersymmetric bilinear form of $\g$.
The currents are $\Lambda(\g)$-valued fields and $\Om$ lifts to an automorphism of $\Lambda(\g)$ in the obvious way. Since these gluing conditions do not only preserve conformal symmetry but also half the current symmetry they are called maximally symmetric. 
We listed those automorphisms that preserve the metric in section \ref{automorphisms}.

\subsection{Geometry of branes on supergroups}\label{section:geometry}

The first question we ask is what kind of geometric objects the gluing conditions \eqref{eq:gluing} describe.  

For WZNW models on Lie groups the geometry of branes has been studied in detail e.g. \cite{Alekseev:1998mc,Stanciu:1999id,Felder:1999ka}. 
If a field $g$ takes values in a Lie group with definite metric, then the boundary conditions (\ref{eq:gluing}) imply that the restriction of $g$ to the boundary of the Riemann surface $\Sigma$ takes values in a twisted conjugacy class. 

The generalisation to Lie supergroups is the following. 
\begin{prp}
	Let the restriction of $g$ to the boundary of the Riemann surface $\Sigma$ take values in a subspace $N\subset G$ such that the boundary conditions (\ref{eq:gluing}) hold. We call $N$ the branes' worldvolume. If the metric restricts non-degenerately to the tangent space $\T_gN$ of the branes' worldvolume $N$ and if the tangent space of $G$ at the point $g$ decomposes in the direct sum of $\T_gN$ and its orthogonal complement $\T_gN^\perp$,\vspace{1mm}
	\begin{equation}\label{eq:orthogonal}
	\T_gG\ =\ \T_gN\oplus \T_gN^\perp\ ,\vspace{1mm}
\end{equation}
then the worldvolume $N$ is the twisted superconjugacy class\footnote{The automorphism $\Omega$ of the Lie superalgebra lifts to an automorphism of the Lie supergroup via $\Omega(\exp X)=\exp\Omega(X)$. We still denote it by $\Omega$.} \vspace{1mm}
\begin{equation} 
	C^\Omega_g\ = \ \{\ \Omega(h)gh^{-1} \ | \ h \ \in \ G \ \}\ .\vspace{1mm}
\end{equation}
\end{prp}
Since the metric is not definite, the decomposition \eqref{eq:orthogonal} is not guaranteed to hold in general. 
But for Lie supergroups with the property that the restriction of the metric
to any simple or abelian subgroup of the underlying Lie group $G_0$ is definite 
it holds for a twisted superconjugacy class that is completely delocalised in the fermionic directions, 
i.e. $\exp\Lambda(\gone)\subset C^\Omega_g$. 
This is the regular case and we call these branes typical in analogy to typical representations. 
Recall proposition \ref{proposition:conjugacy} that this is more than a mere analogy.
We will call all other branes atypical. If the gluing automorphism $\Omega$ is inner, then the above assumptions also hold 
for non-regular twisted superconjugacy classes containing a point $g$ in the bosonic Lie subgroup $G_0$ while 
they never hold for twisted superconjugacy classes containing a point $g=\exp X$ with $X$ nilpotent. 
We give an example in \cite{Creutzig:2008ag} 
of branes covering these regions.
\smallskip

Now, let us explain the above proposition.
The gluing conditions  (\ref{eq:gluing}) can be translated to boundary conditions in the tangent space $\T_g G$ tangent to the point $g\in G$, i.e.
with the help of the left and right translation
the boundary conditions read\vspace{1mm}
\begin{equation}
	\del g\ =\ -\widetilde{\Omega}_g\bar{\del}g \vspace{1mm}
\end{equation}
where $\widetilde{\Omega}_g$ is the map on the tangent space at $g$ defined as\vspace{1mm}
\begin{equation}
	\widetilde{\Omega}_g\ =\ R_g\circ\Omega\circ L_{g^{-1}}\ : \ \T_gG \rightarrow \T_gG\ .\vspace{1mm}
\end{equation}
Note that for any given tangent vector $V$ in $\T_gG$ we have $\widetilde{\Omega}_g(V)=\Omega(g^{-1}V)g$.
In terms of Dirichlet and Neumann derivatives ($2\p=\del+\bar{\del}$ and $2i\del_n=\del-\bar{\del}$) the boundary conditions are\vspace{1mm}
\begin{equation}\label{eq:gluing2}
	(1+\widetilde{\Omega}_g)\p g\ =\ -i(1-\widetilde{\Omega}_g)\del_n g\ .\vspace{1mm}
\end{equation}

We need the assumption that the metric restricts non-degenerately to $\T_gN$ and that the tangent space $\T_gG$ splits into a direct sum\vspace{1mm}
\begin{equation}
	\T_gG\ =\ \T_gN\oplus \T_gN^\perp\ ,\vspace{1mm}
\end{equation}
then equation (\ref{eq:gluing2}) identifies those vectors in $\T_gG$ which have nonzero $(1-\widetilde{\Omega}_g)$ 
eigenvalue as directions of Neumann boundary conditions, i.e. they are vectors tangent to the branes worldvolume $N$. 
Then $T_gN^\perp$ is spanned by the vectors having zero $(1-\widetilde{\Omega}_g)$ eigenvalue.

Let $V$ be
in $\T_gN^\perp$ then $\widetilde{\Om}_g(V)=V$ which is expressed in terms of $\Om$\vspace{1mm} 
\begin{equation}
	\Om(g^{-1}V)\ =\ Vg^{-1}\ .\vspace{1mm}
\end{equation}
Thus\vspace{1mm}
\begin{equation}\label{eq:step2}
	(\Om(g^{-1}V)-Vg^{-1},\Om(X))\ =\ 0\qquad \qquad \text{for all} \ X \ \text{in} \ \Lambda(\g)\ .\vspace{1mm}
\end{equation}
Since the metric is left- and right-invariant and invariant under $\Om$ (recall that $\Om$ is required to be metric preserving) \eqref{eq:step2} 
is equivalent to\vspace{1mm}
\begin{equation}
	(V,gX-\Om(X)g)\ =\ 0\ .\vspace{1mm}
\end{equation}
Hence $gX-\Om(X)g$ is orthogonal to $\T_gN^\perp$, i.e. it is tangent to the worldvolume $N$ of the brane,
but it is also tangent to the twisted superconjugacy class\vspace{1mm}
\begin{equation}
	C^{\Om}_g\ =\ \{\ \Om(h)gh^{-1}\ | \ h \ \in  G\ \}\ .\vspace{1mm}
\end{equation}
This can be seen as follows. Consider the curve $\Om(h(t))gh(t)^{-1}$ in $C^\Om_g$ through $g$, i.e. $h(0)=1$ and $\dot{h}(0)=-X$, then its tangent vector at $g$ is\vspace{1mm}
\begin{equation}
	\frac{d}{dt}\ \Om(h(t)) g h(t)^{-1}\Big|_{t=0}\ =\ gX-\Om(X)g\ .\vspace{1mm}
\end{equation}
Hence the tangent vectors of the form $gX-\Om(X)g$ are the tangent vectors of the twisted superconjugacy class $C^{\Om}_g$.
It remains to show that any tangent vector tangent to the worldvolume of the brane has the form $gX-\Om(X)g$.
Recall that those tangent vectors $V$ describe Dirichlet boundary conditions which are in the kernel
of $1-\widetilde{\Omega}_g$.
Hence the image of the adjoint operator $(1-\widetilde{\Om}_g)^\dagger$ must be $\T_gN$. 
Since $\widetilde{\Omega}_g=R_g\circ\Omega\circ L_{g^{-1}}$ is an isometry the adjoint is 
the inverse\vspace{1mm}
\begin{equation}
	(1-\widetilde{\Om}_g)^\dagger\ =\ (1-R_g\circ\Omega\circ L_{g^{-1}})^\dagger\ =\ 1-L_g\circ\Om^{-1}\circ R_{g^{-1}} \ ,\vspace{1mm}
\end{equation}
i.e. any element $W$ in $\T_gN$ can be written as \vspace{1mm}
\begin{equation}
	W\ =\ U-g\Om^{-1}(Ug^{-1})\vspace{1mm}
\end{equation}
for some $U$ in $\T_gG$. Further any vector $U$ in $\T_gG$ can be written as $U=\Om(X)g$ for some $X$ in $\Lambda(\g)$, hence
$W=\Om(X)g-gX$ for some $X$. We conclude that the worldvolume of a brane is a twisted superconjugacy class.
\smallskip

There are some remarks. 

\noindent{\bf Remark 1}

 The Lie supergroup acts on a twisted superconjugacy class by the twisted adjoint action $\Ad^\Om$\vspace{1mm}
\begin{equation}
         \begin{split}
              &\Ad^\Om(a): C^\Om_g\ \longrightarrow\ C^\Om_g \\[1mm]
              &\Ad^\Om(a)(\Om(h)gh^{-1})\ = \ \Om(a)\Om(h)gh^{-1}a^{-1} \ = \ \Om(ah)g(ah)^{-1} \\[1mm]
         \end{split}
\end{equation}
for any $a$ in $G$ and $\Om(h)gh^{-1}$ in $C^\Om_g$ . When analysing branes on a Lie supergroup one usually starts with 
its semiclassical limit, the minisuperspace \cite{Creutzig:2008ek,Quella:2007sg}.
The minisuperspace of a brane of a Lie supergroup is the quotient of the space of functions on the supergroup by those that vanish on the brane. 
The infinitesimal twisted adjoint action acts on this space. This action is the semiclassical limit of the action of 
the boundary currents on the boundary fields.
The infinitesimal twisted adjoint action can be expressed through the infinitesimal right-translation \eqref{eq:rightvectorfields}. 
Let $h$ be in $G$ and $R_X^h$ be the left-translation in the direction $X$, i.e.\vspace{1mm}
\begin{equation}
    R_X^h: G\ \longrightarrow\ T_hG\ , \qquad R_X^hh\ =\ -Xh\ .\vspace{1mm}
\end{equation}  
Further its action on $h^{-1}$ is $R_X^hh^{-1}=h^{-1}X$ since $R_X(hh^{-1})=0$, hence its action on the twisted superconjugacy class element $a=\Om(h)gh^{-1}$ is
the infinitesimal twisted adjoint action\vspace{1mm}
\begin{equation}
         R_X^h(\Om(h)gh^{-1})\ = \ -\Om(Xh)gh^{-1}+\Om(h)gh^{-1}X \ = \ -\Om(X)a+aX\ .\vspace{1mm}
\end{equation}  
\smallskip

\noindent{\bf Remark 2}

 The stabiliser of $g$ under the twisted adjoint action is the twisted supercentraliser \vspace{1mm}
\begin{equation}
	\mathcal{Z}(g,\Om) \ = \ \{ h \ \in \ G \ | \ \Om(h)g\ = \ gh \ \}\ .\vspace{1mm}
\end{equation}
Its tangent space at $g$ is the kernel of $1-\tilde{\Om}_g$. The twisted superconjugacy class can be described by the homogeneous space $G/\mathcal{Z}(g,\Om)$. 
In the regular case the twisted supercentraliser is isomorphic to the maximal set $T^\Om$ of commuting points which are pointwise 
fixed under the action of $\Om$, i.e. it is contained in a maximal torus. Whenever $\Om$ is inner $T^\Omega$ is a maximal torus. 
A maximal torus of a basic Lie superalgebra is isomorphic to the maximal torus of its bosonic Lie subalgebra.
Hence in the regular case the brane is completely delocalised  in the fermionic directions and since the metric is consistent (see Definition\ \ref{def:lsa})
the assumed orthogonal decomposition 
\eqref{eq:orthogonal} 
is true if it is true for the restriction to the Lie subgroup $G_0$.
At non regular points the brane is not necessarily completely delocalised in the fermionic directions. In these cases one has to check whether \eqref{eq:orthogonal} holds.

It certainly does not hold for superconjugacy classes containing a point $g=\exp X$ with $X$ nilpotent, since then the operator $1-\tilde{\Om}_g$ is not diagonalisable. In this case there is a new type of branes whose geometry is rather different, we give an example in \cite{Creutzig:2008ag}.  
\smallskip

\noindent{\bf Remark 3}

Gluing automorphisms must be metric preserving automorphisms of the relevant Lie algebra that is the Grassmann envelope $\Lambda(\g)$ of the Lie superalgebra $\g$. 
So far we obtained such an automorphism by lifting it from an automorphism of the Lie superalgebra $\g$. These are not all possible gluing automorphisms because conjugating by a fermionic Lie supergroup element $\theta$ is an automorphism of $\Lambda(\g)$ but not of $\g$. The above statements also hold for $\Om=\Ad(\theta)$ and a $\Ad(\theta)$ twisted superconjugacy class is simply a left translate by $\theta$ of an ordinary superconjugacy class.  

\subsection{The boundary action}\label{section:bdyaction}

In this section we will state the boundary action. 
Following \cite{Gawedzki:1999bq}, we represent a Riemann surface $\Sigma$ with boundary as $\Sigma'\backslash D$, where $\Sigma'$ is a Riemann surface without boundary and $D$ an open disc. We want to have a WZNW model based on a map $g:\Sigma\rightarrow G$ from the world-sheet with boundary to the Lie supergroup $G$. For this purpose one needs to extend the map to a $3$-manifold $B$. This is not possible for a world-sheet with boundary. Thus the idea is to first extend $g$ to a map $g':\Sigma'\rightarrow G$ then to consider the WZ term based on $g'$ and to subtract a boundary term which only depends on the restriction of $g'$ to the closure of the disc $\bar{D}$.  
This boundary term has to be such that it coincides with the restriction of the Wess-Zumino term to the disc and such that the variation of the total action vanishes provided the usual equation of motion hold in the bulk and the desired gluing condition at the boundary.

Let us introduce the action and show that it has the two properties mentioned above. 

Let $g,g',\Sigma,\Sigma'$ as above, let $\tilde{g}:B\rightarrow G$ an extension of $g'$ to a 3-manifold $B$ with boundary $\del B=\Sigma'$. Further let the restriction of $g'$ to the closure of the disc $\bar{D}$ map to a twisted superconjugacy class $C^\Om_a$ at a regular point $a$,\vspace{1mm}
\begin{equation}
	g'(\bar{D})\ \subset\ C^\Om_a \ .\vspace{1mm}
\end{equation}
Then the WZNW action for the twisted boundary conditions $J=\Om(\bar{J})$ is given by\vspace{1mm}
\begin{equation}
	S_{\Omega,a}[g] \ = \  	S_{\text{kin}}^\Sigma[g] + S_{\text{WZ}}^B[\tilde{g}]-
	\frac{k}{2\pi}\int_{\bar{D}} \omega  \, ,\vspace{1mm}
\end{equation}
where $\omega$ is (using the shorthand $\tilde{\Omega}=\Ad(g'^{-1})\circ\Omega$)\vspace{1mm}
\begin{equation}
	\omega \ = \ \frac{1}{2}\str(g'^{-1}dg'\wedge \frac{\tilde{\Om}+1 }{\tilde{\Om}-1 }g'^{-1}dg')\vspace{1mm}
\end{equation}
and $\tilde{\Om}-1$ restricted to a twisted superconjugacy class is invertible as already seen in the previous section.
If we write $g'\vert_{\bar{D}}=\Omega(h)ah^{-1}$ then \vspace{1mm}
\begin{equation}
	(\tilde{\Om}-1)^{-1}g'^{-1}dg'\big\vert_{\bar{D}} \ = \ dhh^{-1} \ .\vspace{1mm}
\end{equation}
This allows us to rewrite the  boundary term as \vspace{1mm}
\begin{equation}\label{boundaryterm}
	\frac{k}{2\pi}\int_{\bar{D}} \omega \ = \ 
	\frac{k}{2\pi}\int_{\bar{D}} d\tau d\sigma \ \str(\tilde{\Om}(\del hh^{-1})\bar{\del}hh^{-1}-\tilde{\Om}(\bar{\del}hh^{-1})\del hh^{-1})\ .\vspace{1mm}
\end{equation}
Now we can check explicitly that the proposed action has the desired properties. First the restriction of the 3-form $H$ to the twisted superconjugacy class indeed coincides with $d\omega$\vspace{1mm}
\begin{equation}
	d\omega \ = \ H\big\vert_{C^\Om_a} \ .\vspace{1mm}
\end{equation}
Furthermore the variation of the action vanishes provided the usual bulk equations of motion and the boundary equation of motions $J=\Omega(\bar{J})$ hold, \vspace{1mm}
\begin{equation}\nonumber
 \begin{split}
	\delta S_{\Omega,a}[g]\ &= \ \delta S_{\Omega,a}^{\text{bulk}}[g]+
	\frac{ik}{2\pi}\int_{\del \bar{D}} d\tau \ 
	\str\bigl(\delta h h^{-1} \bigl((1-\tilde{\Omega})\bar{\del}hh^{-1}+(\tilde{\Omega}^{-1}-1)\del hh^{-1}\bigr)\bigr)\\[2mm]
	&= \  \delta S_{\Omega,a}^{\text{bulk}}[g]+\frac{i}{2\pi}\int_{\del \bar{D}} d\tau \ \str(\Omega(\delta h h^{-1}) (J-\Omega(\bar{J})))\ .\qquad\qquad\qquad\qquad\ \end{split}
\end{equation}
A well-defined action should not depend on the extensions of the map $g$. 
In section \ref{sec:boundarygl11} the boundary GL($1|1$) model with gluing automorphism $\Omega=(-st)$ 
(see section \ref{automorphisms} for the description of $(-st)$) is studied using a triangular decomposition of the group valued field. The question is, whether this can be generalised to all type I boundary WZNW models with gluing automorphism $\Omega=(-st)$.

For general $\Om$ and any basic Lie superalgebra, there is a parameterisation of the $G$-valued field $g$ that is particularly adapted to the problem: $g=\Omega(\theta)g_0\theta^{-1}$ where $g_0$ in the bosonic subgroup $G_0$ and $\theta$ takes values in $\exp\Lambda(\gone)$.
Using the Polyakov-Wiegmann identity \eqref{PolyakovWiegmann} and the explicit form of the boundary term \eqref{boundaryterm} one can rewrite the action as\vspace{1mm}
\begin{equation}
	\begin{split}
		S_{\Omega,a}[g] \ = \ S_{\Omega\vert_{G_0},a}[g_0] +
		\frac{k}{2\pi}\int_\Sigma d\tau d\sigma\
		&\str(\theta^{-1}\del\theta\ \theta^{-1}\bar{\del}\theta )+\str(\del g_0g_0^{-1}\ \Omega(\theta^{-1}\bar{\del}\theta) )+ \\[1mm]
		-\str(\theta^{-1}&\del\theta\ g_0^{-1}\bar{\del}g_0 )-\str(\theta^{-1}\del\theta\ g_0^{-1}\Omega(\theta^{-1}\bar{\del}\theta)g_0 ) \ . \\[1mm]
	\end{split}
\end{equation}
This model then has the same quantisation conditions as the Lie group boundary WZNW model $S_{\Omega\vert_{G_0},a}[g_0]$.

Our goal is to be able to compute correlation functions in a boundary model, which we will do for \GL\ in chapter \ref{chapter:gl11}.

\chapter{Symplectic fermions}

This chapter is part of \cite{Creutzig:2008an}.
Symplectic fermions can be viewed as a simple example of a supergroup, that is a supergroup with trivial bosonic subgroup. They will turn out to be highly relevant in the \GL\ WZNW model as we will see in the next chapter. 
Symplectic fermions have been studied in detail in the bulk \cite{Kausch:2000fu}. After a short review of the bulk theory, we will give a detailed description of the boundary theory focusing on boundary states and including twisted sectors.

\section{The bulk}

Symplectic fermions are two dimension zero fermionic fields $\chi^1$ and $\chi^2$ with action
\begin{align}
    S(\chi^a)\ =\ \frac{1}{4\pi}\int_\Sigma
d^2z \ \ep_{ab}\del\chi^a\bar{\del}\chi^b\, ,
\end{align}
where the anti-symmetric symbol is defined by $\ep_{12}=-\ep_{21}=1$. This
gives the operator product expansions
\begin{align}
    \chi^a(z,\bar z)\chi^b(w,\bar w)&\sim -\ep^{ab}\ln\abs{z-w}^2\, ,
\end{align}
where $\ep^{12}=-1$.
For correlation functions, we have the requirement that a correlator is only non-vanishing if the zero-modes of $\chi^1$ and $\chi^2$ are inserted.

In view of the symplectic fermion correspondence to \GL\ twisted sectors become interesting.
A twisted sector is given, if we insert a field $\mu_\lambda$ at some point on the world-sheet, e.g. at zero. If we move the symplectic fermions around this field, they change by a phase, i.e.
\begin{equation}
	\begin{split}
    \chi^1(e^{2\pi i}z)\mu_\la(0)\ &=\ e^{-2\pi i\la}\chi^1(z)\mu_\la(0)\, ,\ \, 
    \qquad \chi^2(e^{2\pi i}z)\mu_\la(0)\ =\ e^{2\pi i\la}\chi^2(z)\mu_\la(0)\, ,\\
    \bar \chi^1(e^{-2\pi i} \bar z)\mu_\la(0)\ &=\ e^{-2\pi i\la}\bar \chi^1(\bar z)\mu_\la(0)\, , 
    \qquad \bar \chi^2(e^{-2\pi i}\bar z)\mu_\la(0)\ =\ e^{2\pi i\la}\bar \chi^2(\bar z)\mu_\la(0)\, .\\
    \end{split}
\end{equation}
$\chi^1$ and $\chi^2$ have to transform oppositely to give a symmetry
of the Lagrangian.
Then the mode expansions of the fields in these sectors are
\begin{equation}
\begin{split}
    \del\chi^1(z) \ &= \ -\sum_{n\in\Z}\chi^1_{n+\lambda}z^{-(n+\lambda)-1}\qquad{\rm and}\qquad
    \bar{\del}\bar{\chi}^1(\bar{z}) \ = \ -\sum_{n\in\Z}\bar{\chi}^1_{n-\lambda}\bar{z}^{-(n-\lambda)-1}\\
    \del\chi^2(z) \ &= \ -\sum_{n\in\Z}\chi^2_{n-\lambda}z^{-(n-\lambda)-1}\qquad{\rm and}\qquad
    \bar{\del}\bar{\chi}^2(\bar{z}) \ = \ -\sum_{n\in\Z}\bar{\chi}^2_{n+\lambda}\bar{z}^{-(n+\lambda)-1}\,.
\end{split}
\end{equation}
Since the symplectic fermions do not have any zero modes in the twisted sector, the representation in this sector is irreducible. The conformal dimension of the ground-state is
\begin{equation}
   h_\la \ = \ -\frac{\la\la^*}{2}\qquad\qquad \la^*\ = \ 1-\la\, .
\end{equation}
Correlation functions have been determined \cite{Kausch:2000fu}, they are
\begin{eqnarray}
    \langle \mu_\la(z_1,\bar z_1)\mu_{\la^*}(z_2,\bar z_2)\rangle &=& 
        -|z_{12}|^{2\la\la^*}\nonumber\\[2mm]
 \langle \mu_\la(z_1,\bar z_1)\mu_{\la^*}(z_2,\bar z_2)\normord{\chi^1\chi^2}(z_3,\bar z_3)\rangle &=&
        |z_{12}|^{2\la\la^*}\Bigl(\mathcal{Z}_\la+\ln\Bigl|\frac{z_{13}z_{23}}{z_{12}}\Bigr|^2\Bigr)
\end{eqnarray}
\begin{equation*}
\langle \mu_{\la_1}(z_1,\bar z_1)\mu_{\la_2}(z_2,\bar z_2)\mu_{\la_3}(z_3,\bar z_3)\rangle\ =\  \mathcal{C}_{\la_1\la_2\la_3}\left\{\begin{array}{cc}
\bigl|z_{12}^{\la_1\la_2}z_{13}^{\la_1\la_3}z_{23}^{\la_2\la_3}\bigr|^2 &
,\ \la_1+\la_2+\la_3=1\\[2mm]
\bigl|z_{12}^{\la_1^*\la_2^*}z_{13}^{\la_1^*\la_3^*}z_{23}^{\la_2^*\la_3^*}\bigr|^2 &
,\ \la_1+\la_2+\la_3=2
\end{array}\right.
\end{equation*}
where we take the short-hand $z_{ij}=z_i-z_j$ as usual and 
\begin{equation}
   \mathcal{C}_{\la_1\la_2\la_3}\ = \ \sqrt{
  \frac{\Gamma(\la_1)\Gamma(\la_2)\Gamma(\la_3)}
{\Gamma(\la_1^*)\Gamma(\la_2^*)\Gamma(\la_3^*)}}\, .
\end{equation}
These coefficients also appear in the \GL\ WZNW model and indicate the correspondence we will prove later on.

Let us now turn to the boundary theory.
For earlier works on boundary
models of symplectic fermions
see~\cite{Bredthauer:2002ct,Bredthauer:2002xb,Kawai:2001ur,Gaberdiel:2006pp}.
These works however do not consider all boundary conditions. 

\section{Boundary conditions}

We start our considerations by investigating possible boundary conditions.
Recall the energy momentum tensor
\begin{equation}
    T(z) \ = \  -\half\ep_{ab}\normord{\del\chi^a\del\chi^b}, \qquad \bar{T}(\bar{z}) \ = \ -\half\ep_{ab}\normord{\delbar\chi^a\delbar\chi^b}.
\end{equation}
They preserve the symplectic fermion symmetry and coincide along the boundary if
\begin{equation}\label{eq:bdyconditions}
    \begin{split}
        \del\chi \ = \ A\ \bar{\del}\chi \qquad\qquad {\rm for} \ z\ = \bar{z}\,,
\end{split}
\end{equation}
where $A=\bigl( \begin{smallmatrix}
a&b\\c&d
\end{smallmatrix} \bigr)$ is a matrix in SL(2) and for convenience we combined the two fermions in the vector $\chi = \Bigl( \begin{smallmatrix}
\chi^1\\ \chi^2
\end{smallmatrix} \Bigr)$.
In terms of Dirichlet and Neumann derivatives
($\del=\half\del_u-i\half\del_n$ and $\delbar=\half\del_u+i\half\del_n$)
the boundary conditions are
\begin{equation}
    \begin{split}
        -i\del_n \chi \ = \ \frac{A-1}{A+1}\ \del_u\chi\
\end{split}
\end{equation}
provided $1+A$ is invertible.
Then the action on the upper half-plane is
\begin{equation} \label{eq:actsymplecticfermionsbdy}
    \begin{split}
    S \ = \ &-\frac{1}{4\pi} \int d^2z\ \del\chi^t\, J\,\bar{\del}\chi\ +\
     \frac{i}{8\pi} \int\limits_{z=\bar{z}} du \ \chi^t\, J\, \frac{A-1}{A+1}\, \del_u\chi\,,
     \end{split}
  \end{equation}
where the matrix $J$ is $J=\bigl( \begin{smallmatrix}
0&-1\\ 1&0
\end{smallmatrix} \bigr)$. The variation of this action vanishes provided the above boundary conditions hold as well as the bulk
equations of motion $\del\bar{\del}\chi^\pm=0$.
  If $1+A$ is not invertible it has characteristic polynomial $\lambda^2$,
i.e. if $1+A=0$ there are Dirichlet conditions in both directions while
otherwise there is one Dirichlet and one Neumann condition. Note that
these cases resemble the atypical branes in \GL\ \cite{Creutzig:2008ag}.

\section{The Ramond sector}

We first consider the Ramond sector, ie. there are no twist fields present.
The explicit mode expansion is
\begin{equation}\label{eq:modeexpansionfermions}
     \chi^a(z,\bar{z}) \ = \ \xi^a+\chi^a_0\ln |z|^2 - \sum_{n\neq0}\frac{1}{n}\chi^a_nz^{-n}+\frac{1}{n}\bar{\chi}^a_n\bar{z}^{-n},
 \end{equation}
where the modes satisfy
\begin{equation} \label{eq:fermionrelations}
    \{ \chi^a_m,\chi^b_n\} \ = \ -m\ep^{ab}\,\delta_{m,-n} \ \ ,\ \  \{ \bar{\chi}^a_m,\bar{\chi}^b_n\} \ = \ -m \ep^{ab}\,\delta_{m,-n} \ \
    {\rm and}\ \  \{ \xi^a,\chi_0^b \} \ = \ \ep^{ab} \ \ .
\end{equation}
All other anti-commutators vanish. Note that for locality we have required $\chi_0^a=\bar\chi_0^a$.

In this section we construct the
boundary states in the Ramond sector, compute the amplitudes and construct
the corresponding open string model. We start the discussion of boundary
states by investigating Dirichlet conditions in the two fermionic
directions.

\subsection{Dirichlet conditions}

Let us first remind ourselves that if we have an extended chiral algebra
given by $W(z)$ and $\bar W(\bar z)$ we need a gluing automorphism,
$\Om$, for the boundary \eqref{eq:gluingchiralfields}
\begin{equation}\label{eq:gluingauto}
    W(z)=\Om(\bar W)(\bar z)\qquad\textrm{for } z=\bar z\,.
\end{equation}
 We now pass to closed strings via world-sheet duality. The
gluing conditions then become the following Ishibashi conditions for the
boundary states $|\al\rrangle_{\Om}$ in the CFT on the full plane \eqref{eq:ishibashi}
\begin{equation}\label{eq:ishibashigeneral}
    \left(W_n-(-1)^{h_W}\Om(\bar W_{-n})\right)|\al\rrangle_{\Om}\,,
\end{equation}
where $h_W$ is the conformal dimension of $W$.

Using~\eqref{eq:ishibashigeneral} we see that for the Dirichlet boundary
conditions ($A=-1$ in~\eqref{eq:bdyconditions}) the corresponding
Ishibashi states have to satisfy
\begin{equation}
    \begin{split}
    \left(\chi^a_n-\bar{\chi}^a_{-n}\right)|D\rrangle \ = \ 0\qquad\qquad\text{for}\ a=1,2\ ,
\end{split}
\end{equation}
note that there is no condition on $\chi^a_0$ because of the locality
constraint $\chi^a_0-\bar{\chi}^a_{0}=0$. The Ishibashi states are
explicitly constructed as
\begin{eqnarray}
    |D_0\rrangle \ &= \ \sqrt{2\pi}\exp\Bigl(\sum_{m>0}\frac{1}{m}
    \bigl(\chi^2_{-m}\bar{\chi}^1_{-m}-\chi^1_{-m}\bar{\chi}^2_{-m}\bigr)\Bigr)|0\rangle\,, \label{eq:D0}\\
    |D_\pm\rrangle \ &= \ \xi^\pm\ \exp\Bigl(\sum_{m>0}\frac{1}{m}\bigl(\chi^2_{-m}\bar{\chi}^1_{-m}-\chi^1_{-m}\bar{\chi}^2_{-m}\bigr)\Bigr)|0\rangle\,, \\
    |D_2\rrangle \ &= \ \frac{\xi^-\xi^+}{\sqrt{2\pi}}\ \exp\Bigl(\sum_{m>0}\frac{1}{m}
    \bigl(\chi^2_{-m}\bar{\chi}^1_{-m}-\chi^1_{-m}\bar{\chi}^2_{-m}\bigr)\Bigr)|0\rangle\,, \label{eq:D2}
\end{eqnarray}
where the ground state $|0\rangle$ is defined by $\chi^a_n|0\rangle=0$ for
$n\geq0$. The dual Ishibashi state is obtained by dualizing the modes
using (here $m>0$)
\begin{equation}
    {\chi^1_{-m}}^\dag \ = \ \chi^1_{m}\qquad\text{and}\qquad{\chi^2_{-m}}^\dag \ = \ -\chi^2_{m}\  .
\end{equation}
For the computation of amplitudes we need the Virasoro generators, they are
\begin{equation}
L_n\ = \ -\half\ep_{ab}\sum_{m} : \chi^a_{n-m}\chi^b_{m}:\
\end{equation}
and the central charge is $c=-2$.
Define $q=\exp 2\pi i \tau$ and $\tilde{q}=\exp( -2\pi i /\tau)$ as
usual, where $\tau$ takes values in the upper half plane. Then the
non-vanishing overlaps are
\begin{equation}\label{eq:Dirichletoverlap}
    \begin{split}
        \llangle D_0| q^{L_0^c+\frac{1}{12}}(-1)^{F^c}|D_2\rrangle \ &= \
        \llangle D_2| q^{L_0^c+\frac{1}{12}}(-1)^{F^c}|D_0\rrangle \ = \ \eta(\tau)^2, \\
        \llangle D_-| q^{L_0^c+\frac{1}{12}}(-1)^{F^c}|D_+\rrangle \ &= \
        -\llangle D_+| q^{L_0^c+\frac{1}{12}}(-1)^{F^c}|D_-\rrangle \ = \ \eta(\tau)^2, \\
                \llangle D_2| q^{L_0^c+\frac{1}{12}}(-1)^{F^c}|D_2\rrangle \ &= \ - i\tau\eta(\tau)^2\ = \ \eta(\tilde{\tau})^2\ , \\
    \end{split}
\end{equation}
where $L_0^c=L_0+\bar{L_0}$. Further $\eta(\tau)$ is the Dedekind $\eta$-function
\begin{equation}
 \eta(\tau) \ = \ q^{\frac{1}{12}}\prod_{m>0}(1-q^m)^2\, .
\end{equation}
 Its modular S-transformation is ($\tilde\tau=-1/\tau$)
\begin{equation}
 \eta(\tilde\tau)^2 \ = \ -i\tau\eta(\tau)^2\, .
\end{equation}
In section \ref{sec:bdystates} we saw that the modular transformation of an amplitude describes the spectrum of an open string, i.e. it state must be a true character. Thus only $|D_2\rangle$ makes sense as a boundary state.

\subsection{Neumann conditions}

Next we would like to display the boundary state $|A\rangle$ for our
general boundary conditions \eqref{eq:bdyconditions}. It has to satisfy
the Ishibashi condition~\eqref{eq:ishibashigeneral}
\begin{equation}
    \begin{split}
    \chi^1_n+a\,\bar{\chi}^1_{-n}+b\,\bar{\chi}^2_{-n}|A\rrangle \ = \ 0\,, \\
    \chi^2_n+c\,\bar{\chi}^1_{-n}+d\,\bar{\chi}^2_{-n}|A\rrangle \ = \ 0\,, \\
\end{split}
\end{equation}
which are satisfied by
\begin{equation}
    |A\rrangle \ = \ \mathcal{N}\ \exp\Bigl(-\sum_{m>0}\frac{1}{m}
    \bigl(a\chi^2_{-m}\bar{\chi}^1_{-m}+b\chi^2_{-m}\bar{\chi}^2_{-m}-c\chi^1_{-m}\bar{\chi}^1_{-m}-d\chi^1_{-m}\bar{\chi}^2_{-m}\bigr)\Bigr)|0\rangle\,.
\end{equation}
The dual state is
\begin{equation}
     \llangle A| \ = \ \mathcal{N}\ \llangle0| \exp\Bigl(-\sum_{m>0}\frac{1}{m}
    \bigl(-a\chi^2_{m}\bar{\chi}^1_{m}+b\chi^2_{m}\bar{\chi}^2_{m}-c\chi^1_{m}\bar{\chi}^1_{m}+d\chi^1_{m}\bar{\chi}^2_{m}\bigr)\Bigr)\ .
\end{equation}
It will turn out that the normalisation should be fixed to be
\begin{equation}\label{eq:normalization}
\mathcal{N}\ = \ \sqrt{2\pi}\, 2\, \sin\pi\mu\,,
\end{equation}
where we introduce $\mu$ via $\alpha=\exp 2\pi i\mu$ by $-\tr(A)=\alpha+\alpha^{-1}$.

Now it is straightforward to compute amplitudes between two boundary
states. Any non-zero amplitude requires the zero modes of $\chi^1$ and $\chi^2$ hence only the Dirichlet boundary state has
non-vanishing overlap with any Neumann state:
\begin{equation}\label{eq:RamondNeumannDirichletamplitude}
    \llangle A|\, q^{\frac{1}{2}L_0^c+\frac{1}{12}}\,(-1)^{F^c}\,|D_2\rrangle \ = \ \frac{\mathcal{N}}{\sqrt{2\pi}}q^{\frac{1}{12}}\prod_{m>0}(1-\alpha_{12} q^m)(1-\alpha_{12}^{-1} q^m)\,.
\end{equation}
Upon modular transformation this amplitude is the spectrum of an open
string stretching between two branes with respectively Neumann boundary
conditions given by $A$ and Dirichlet conditions. Using the formulae
provided in the appendix equation
\eqref{eq:RamondNeumannDirichletamplitude} becomes
\begin{equation}
    \begin{split}
        \frac{\mathcal{N}}{\sqrt{2\pi}}\,q^{\frac{1}{12}}\prod_{m>0}(1-\alpha q^m)(1-\alpha^{-1} q^m) \
        = \ \tilde{q}^{\frac{1}{2}(\mu-\frac{1}{2})^2-\frac{1}{24}}\,\prod_{n=0}^\infty
        \bigl(1-\tilde{q}^{n+1-\mu}\bigr)\bigl(1-\tilde{q}^{n+\mu}\bigr)\, .\\
    \end{split}
\end{equation}
Now, we construct the boundary theory of a string stretching between these
two branes and check that its spectrum is indeed given by the amplitude we
just computed, we follow~\cite{Creutzig:2006wk}. For this purpose consider the upper half plane, and demand
boundary condition $A$ for the negative real line, i.e.
\begin{equation}
    \del\chi \ = \ A\,\bar{\del}\chi    \qquad{\rm for}\ z\ = \bar{z} \qquad{\rm and}\ z+\bar{z}\ <\ 0\ ;
\end{equation}
and Dirichlet conditions for the positive real axis
\begin{equation}
    \del_u\chi \ = \ 0  \qquad{\rm for}\ z\ = \bar{z} \qquad{\rm and}\ z+\bar{z}\ >\ 0\ .
\end{equation}
Then the fields have the following SL(2) monodromy (counterclockwise)
\begin{equation}
    \del\chi(ze^{2\pi i}) \ = \ -A\del\chi(z)\,,
\end{equation}
and similar for the bared quantities. Denote by $S$ the matrix that
diagonalises the monodromy, i.e. $S(-A)S^{-1}$ is diagonal. We denote the
eigenvalues by $\alpha^{\pm1}$. Further, call the eigenvectors
$\del{\chi}^\pm$, they then have the usual mode expansion \cite{Kausch:2000fu}
\begin{equation}
    {\chi}^\pm(z) \ = \ \sum_{n\in\Z}\frac{1}{n\pm\mu}{\chi}^\pm_{n\pm\mu}z^{-(n\pm\mu)}\, .
 \end{equation}
The original fields are then explicitly
\begin{equation}
    \begin{pmatrix}\chi^1\\ \chi^2\\\end{pmatrix} \ = \ S^{-1} \begin{pmatrix}{\chi}^+\\ {\chi}^-\\\end{pmatrix}\, .
\end{equation}
Their partition function is
\begin{equation}
    \tr (\, q^{L_0-\frac{c}{24}}(-1)^F\, ) \ = \
    q^{\frac{1}{2}(\mu-\frac{1}{2})^2-\frac{1}{24}}\prod_{n=0}^\infty
        \bigl(1-q^{n+1-\mu}\bigr)\bigl(1-q^{n+\mu}\bigr)\,.
\end{equation}
The computation has been done similarly by Kausch \cite{Kausch:2000fu}. We see
that the result fits with~\eqref{eq:RamondNeumannDirichletamplitude} and
the Cardy condition is fulfilled. Thus, we nicely established our boundary
state and the open string theory it describes.

If we want to investigate amplitudes involving Neumann boundary states on
both ends, we learnt \cite{Creutzig:2006wk} that it is necessary to insert
additional zero modes in order to obtain a non-vanishing amplitude. Also
introduce $\alpha_{12}$ via
$\tr(A_1A_2^{-1})=\alpha_{12}+\alpha_{12}^{-1}$ then we get
\begin{equation}\label{eq:Ramondamplitude}
\begin{split}
    \llangle A_1|\, \chi^2\chi^1\, q^{\frac{1}{2}L_0^c+\frac{1}{12}}\,(-1)^{F^c}\,|A_2\rrangle \ &= \
\mathcal{N}_1\mathcal{N}_2\,q^{\frac{1}{12}}\prod_{m>0}(1-\alpha_{12} q^m)(1-\alpha_{12}^{-1} q^m)\\
        &= \ \mathcal{N}_{12}\,\tilde{q}^{\frac{1}{2}(\mu_{12}-\frac{1}{2})^2-\frac{1}{24}}\,\prod_{n=0}^\infty
        \bigl(1-\tilde{q}^{n+1-\mu_{12}}\bigr)\bigl(1-\tilde{q}^{n+\mu_{12}}\bigr)\,,
    \end{split}
\end{equation}
where
\begin{equation}
       \mathcal{N}_{12}\ = \ 4\pi\, \frac{\sin\pi\mu_1\, \sin\pi\mu_2}{\sin\pi\mu_{12}}\, .
\end{equation}
The open string theory is constructed almost exactly as above and again
resembles \cite{Creutzig:2006wk}. We demand boundary condition $A_1$ for
the negative real line and $A_2$ for the positive one,
\begin{equation}
    \del\chi \ = \ \left\{ \begin{array}{ll}
         A_1\,\bar{\del}\chi & \qquad{\rm if}\ z\ = \bar{z} \qquad{\rm and}\ z+\bar{z}\ <\ 0\\
         A_2\,\bar{\del}\chi & \qquad{\rm if}\ z\ = \bar{z} \qquad{\rm and}\ z+\bar{z}\ >\ 0\ .\end{array} \right.
\end{equation}
The fields have the following SL(2) monodromy
\begin{equation}
    \del\chi(ze^{2\pi i}) \ = \ A_1A_2^{-1}\,\del\chi(z)\, .
\end{equation}
Let $S$ diagonalise the monodromy, then its eigenvalues are
$\alpha_{12}^{\pm1}$ and we call the eigenvectors again $\del{\chi}^\pm$.
They have the mode expansion
\begin{equation}
    {\chi}^\pm(z) \ = \ \sqrt{\mathcal{N}_{12}}\,{\xi}^\pm+
    \sum_{n\in\Z}\frac{1}{n\pm\mu_{12}}{\chi}^\pm_{n\pm\mu_{12}}z^{-(n\pm\mu_{12})}\,,
 \end{equation}
note the extra zero mode, since the monodromy does only concern
derivatives. Its partition function with appropriate insertion is
\begin{equation}
    \tr (\chi^2\chi^1q^{L_0-\frac{c}{24}}(-1)^F ) \ = \ \mathcal{N}_{12}\,
    q^{\frac{1}{2}(\mu_{12}-\frac{1}{2})^2-\frac{1}{24}}\prod_{n=0}^\infty
        \bigl(1-q^{n+1-\mu_{12}}\bigr)\bigl(1-q^{n+\mu_{12}}\bigr)\,,
\end{equation}
and coincides with \eqref{eq:Ramondamplitude} as desired.

\section{The Neveu-Schwarz sector}

In this section we study the boundary states in the Neveu-Schwarz sector.
The states have to satisfy the usual Ishibashi condition
\begin{equation}
    \begin{split}
    \chi^1_n+a\bar{\chi}^1_{-n}+b\bar{\chi}^2_{-n}|A\rrangle_{NS} \ = \ 0 \,,\\
    \chi^2_n+c\bar{\chi}^1_{-n}+d\bar{\chi}^2_{-n}|A\rrangle_{NS} \ = \ 0 \,,
\end{split}
\end{equation}
where the modes are half-integer, i.e. $n$ in $\Z+1/2$.
The conditions are satisfied by
\begin{equation}
    |A\rrangle \ = \ \exp\Bigl(-\sum_{\substack{m>0 \\ m \, \in\, \Z+1/2}}\frac{1}{m}
    \bigl(a\chi^2_{-m}\bar{\chi}^1_{-m}+b\chi^2_{-m}\bar{\chi}^2_{-m}-c\chi^1_{-m}\bar{\chi}^1_{-m}-d\chi^1_{-m}\bar{\chi}^2_{-m}\bigr)\Bigr)|0\rangle\,.
\end{equation}
We introduce $\alpha_{12}$ as before, that is
$\tr(A_1A_2^{-1})=\alpha_{12}+\alpha_{12}^{-1}$, and get
\begin{equation}
    \begin{split}
    {}_{NS}\llangle A_1| q^{L_0^c+\frac{1}{12}}(-1)^{F_c}|A_2\rrangle_{NS} \ &= \
    q^{-\frac{1}{24}}\prod_{\substack{m>0 \\ m \, \in\, \Z+1/2}}(1-\alpha_{12} q^m)(1-\alpha_{12}^{-1} q^m)\\
    &= \    \tilde{q}^{\frac{1}{2}(\mu-\frac{1}{2})^2-\frac{1}{24}}\prod_{n>0}(1+\tilde{q}^{n-\mu})
        (1+\tilde{q}^{n-\mu^*})\,,
    \end{split}
\end{equation}
where $\alpha_{12}=e^{2\pi i\mu}$. This is the spectrum of an open string constructed similarly as in the Ramond sector,
but with antisymmetric boundary conditions in the time-direction.

\section{The twisted sectors}

Given any twisted sector we can diagonalise it and thus we can restrict to
twists that are diagonal. Call the ground state of the sector $\mu_\lambda$ on which $\chi^a$ has twists
\begin{equation}
    \chi^1 \ \longrightarrow \ e^{-2\pi i\lambda}\chi^1\qquad\text{and}\qquad\chi^2 \ \longrightarrow \ e^{2\pi i\lambda}\chi^2\,.
\end{equation}
Then recall that the mode expansions of the fields in these sectors are
\begin{equation}\label{eq:twistedmodeexpansion}
\begin{split}
    \del\chi^1(z) \ = \ -\sum_{n\in\Z}\chi^1_{n+\lambda}z^{-(n+\lambda)-1}\qquad{\rm and}\qquad
    \bar{\del}\bar{\chi}^1(\bar{z}) \ = \ -\sum_{n\in\Z}\bar{\chi}^1_{n-\lambda}\bar{z}^{-(n-\lambda)-1}\\
    \del\chi^2(z) \ = \ -\sum_{n\in\Z}\chi^2_{n-\lambda}z^{-(n-\lambda)-1}\qquad{\rm and}\qquad
    \bar{\del}\bar{\chi}^2(\bar{z}) \ = \ -\sum_{n\in\Z}\bar{\chi}^2_{n+\lambda}\bar{z}^{-(n+\lambda)-1}\,.
\end{split}
\end{equation}
Whenever $\lambda\neq1/2$ the boundary conditions are parameterised by
just one parameter $\alpha$ according to the boundary conditions
\begin{equation}
    \begin{split}
        \del \chi^1 \ = \ \alpha\bar{\del}\chi^1\qquad{\rm and}\qquad
    \del \chi^2 \ = \ \alpha^{-1}\bar{\del}\chi^2 \ .
\end{split}
\end{equation}
Only to these conditions there exist twisted Ishibashi states. The
boundary state has to satisfy the usual Ishibashi condition
\begin{equation}
    \begin{split}
        \chi^1_{n+\lambda}+\alpha\bar{\chi}^1_{-n-\lambda}|\alpha\rrangle_{\lambda} \ = \ 0\,, \\
        \chi^2_{n-\lambda}+\alpha^{-1}\bar{\chi}^2_{-n+\lambda}|\alpha\rrangle_{\lambda} \ = \ 0\,,
\end{split}
\end{equation}
and these are solved by ($\lambda^*=1-\lambda$)
\begin{equation}\label{eq:twistedIshibashistate}
    |\alpha\rrangle_{\lambda} \ = \ \mathcal{N}\exp\Bigl(-\sum_{m>0}\frac{\alpha}{m-\lambda^*}\chi^2_{-m+\lambda^*}\bar{\chi}^1_{-m+\lambda^*}
    -\frac{\alpha^{-1}}{m-\lambda}\chi^1_{-m+\lambda}\bar{\chi}^2_{-m+\lambda}\Bigr)\mu_{\lambda} \ .
\end{equation}
where we fix the normalisation to be $\mathcal{N}=e^{-2\pi i(\lambda-1/2)(\mu-1/4)}$ and $\alpha=e^{2\pi i \mu}$.
The dual boundary state is
\begin{equation}
    {}_{\lambda}\llangle\alpha | \ = \ \mathcal{\bar N}\mu_{\lambda}^\dag\exp\Bigl(\sum_{m>0}\frac{\alpha}{m-\lambda}\chi^2_{m-\lambda}\bar{\chi}^1_{m-\lambda}
    -\frac{\alpha^{-1}}{m-\lambda^*}\chi^1_{m-\lambda^*}\bar{\chi}^2_{m-\lambda^*}\Bigr)\,.
\end{equation}
Now we are prepared to compute the amplitudes (note that the conformal
dimension of the twist state is $h_\lambda=-\lambda\lambda^*/2$ and we use
the shorthand $\alpha_1\alpha_2^{-1}=e^{2\pi i\mu}$)
\begin{equation}
    \begin{split}
        {}_\lambda\langle\alpha_1| q^{L_0^c+\frac{1}{12}}(-1)^{F_c}|\alpha_2\rangle_\lambda \, &= \, \frac{q^{\frac{1}{2}(\lambda-\frac{1}{2})^2-\frac{1}{24}}}{e^{2\pi i(\lambda-\frac{1}{2})(\mu-\frac{1}{2})}}\prod_{n>0}(1-\alpha_1\alpha_2^{-1}q^{n-\lambda^*})
        (1-\alpha_2\alpha_1^{-1}q^{n-\lambda}) \\
        &= \, \tilde{q}^{\frac{1}{2}(\mu-\frac{1}{2})^2}
        \theta\Bigl(\tilde{\tau}(\frac{1}{2}-\mu)-(\lambda-\frac{1}{2}),\tilde{\tau}\Bigr)/\eta(\tilde{\tau})\\
        &= \, \tilde{q}^{\frac{1}{2}(\mu-\frac{1}{2})^2-\frac{1}{24}}\prod_{n>0}(1-u^{-1}\tilde{q}^{n-\mu})
        (1-u\tilde{q}^{n-\mu^*})\,,
    \end{split}
\end{equation}
where $u=e^{2\pi i\lambda}$. This is the character of a boundary theory
twisted by $\mu_{12}$ in an orbifold model of the symplectic fermions.
The orbifold is by an abelian subgroup $\mathcal{G}$ of SL(2),
where $\mathcal{G}$  is generated by $u$. We refer to \cite{Kausch:2000fu} for a
detailed discussion.

\chapter{The \GL\ WZNW model}\label{chapter:gl11}

We turn to the main part of the thesis, the detailed discussion of an example, the \GL\ Wess-Zumino-Novikov-Witten model. 
The bulk model has been discussed in \cite{Rozansky:1992rx} and \cite{Schomerus:2005bf}.
We start this chapter by giving an equivalent, but rather different approach (section \ref{section:gl11symplecticfermions}). We show that \GL\ is equivalent to a pair of symplectic fermions and two scalar fields. This model is far from being trivial, since we have to include twist fields for the symplectic fermions. We use the correspondence to recompute bulk correlation functions. 

Our main goal is to understand the boundary theory. 
There exist two families of boundary models. One consists of branes that are point-like in the bosonic directions and generically delocalised in the fermionic directions, while the other one consists of one volume-filling brane. The former belongs to the identity gluing automorphism and we call the corresponding branes untwisted, while the volume-filling brane will be called twisted brane.

In section \ref{sec:branes} we discuss boundary states in \GL. We compute the spectrum of strings ending on any two branes, verify Cardy's condition and observe that the structure given by amplitudes involving only untwisted branes coincides with the fusion ring.

In the last section of this chapter we solve the boundary theory of the volume-filling brane completely. For this purpose we extend the first order formulation to the boundary, this involves the introduction of an extra fermionic boundary degree of freedom.

\section{The \GL-symplectic fermion correspondence}\label{section:gl11symplecticfermions}

This section is the main result of \cite{Creutzig:2008an}.
In this section we will  show the relation between
the \GL\ WZNW model and the free scalars and symplectic fermions. Finally,
we will comment on the bulk correlation functions.

\subsection{The \GL\ WZNW model}

Our starting point for the relation between the \GL\ WZNW model and the
free theory will be the first order action for \GL\ found
in~\cite{Schomerus:2005bf}. Recall that the Lie superalgebra is generated by two
bosonic elements $E, N$ and two fermionic $\psi^{\pm}$ which have the
following non-zero (anti)commutator relations
\begin{align}\label{eq:commu}
    [N,\psi^{\pm}]=\pm\psi^{\pm},\quad \{\psi^-,\psi^+\}=E.
\end{align}
Further, we have a family of supersymmetric bilinear forms, but below we
will always work with
\begin{align}\label{eq:supertrace}
    \str(NE)=\str(\psi^+\psi^-)=-1.
\end{align}
For the \GL\ supergroup we choose a Gauss-like decomposition of the form
$$ g \ = \ e^{c_- \psi^-} \, e^{XE + YN} \, e^{-c_+
\psi^+}. $$
The WZNW model thus has two bosonic fields $X(z,\bar z),Y(z, \bar z)$ and two fermionic
fields $c_\pm(z, \bar z)$, and the action takes the form
\begin{equation}\label{eq:SWZW}
    \begin{split}
S_{\text{WZNW}}[g]\ &=\ \frac{k}{4\pi }\int_\Sigma
d^2z\
  \left( -\del X\bar{\del}Y-\del
  Y\bar{\del}X+2e^{Y}\del\cp\bar{\del}\cm\right)\, ,
  \end{split}
\end{equation}
where $k$ is the level. Variation of the action leads to the usual bulk
equations of motion \cite{Creutzig:2007jy}.

The holomorphic current of the \GL\ WZNW model is in our notation given by
$k\del g g^{-1}$. The components corresponding to the generators are
\begin{align}
    J^E &= -k\del Y, &
    J^N &= -k\del X +k\cm \del\cp \, e^{Y} \ ,\nonumber \\[2mm]
    J^- &= ke^{Y}\del\cp, &
    J^+  &= -k\del\cm -k\cm \del Y \ ,\label{eq:holomorphiccurrents}
\end{align}
Similarly, for the anti-holomorphic current $-kg^{-1}\delbar g$ the
components are
\begin{align}
        \bar{J}^E &= k\bar{\del} Y,  &
        \bar{J}^N &= k\bar{\del} X -k \bar{\del}\cm\, \cp\, e^{Y} \ , \nonumber
        \\[2mm]
            \bar{J}^+&= ke^{Y}\bar{\del}\cm, &
        \bar{J}^-&= -k\bar{\del}\cp-k\cp\bar{\del} Y \ .\label{eq:antiholomorphiccurrents}
\end{align}
Let us also mention that the modes of this affine algebra satisfy
\begin{align}\label{eq:affine}
[J^E_n,J^N_m]=-km\de_{n+m},\quad[J^N_n, J^{\pm}_m]=\pm J^{\pm}_{n+m},\quad \{J^-_n,J^+_m\}=J^E_{n+m}+km\de_{n+m},
\end{align}
where we note that the modes can be rescaled such that the algebra is
independent of the level $k$. Equation~\eqref{eq:affine} corresponds to the
OPE
\begin{equation}\label{eq:currentope}
    J^A(z) J^B (w)\sim -k\frac{\str(AB)}{(z-w)^2}+\frac{[A,B\}}{z-w}.
\end{equation}

\subsection{First order formulation}

Following section \ref{section:firstorder} we will now pass to a first order
formalism by introducing two additional fermionic auxiliary fields $b_\pm$
of weight $\Delta(b_\pm) = 1$. Naively, the action would be
\begin{equation}
        \frac{1}{4\pi}\int_\Sigma d^2z\ \left(- k\del X\bar{\del}Y-
        k\del Y\bar{\del}X+2\bp\del\cp+2\bm\bar{\del}\cm+\frac{2}{k}e^{-Y}\bm\bp\right).
\end{equation}
This reduces to~\eqref{eq:SWZW} if we integrate out $b_\pm$ using their
equations of motion
\begin{align}\label{eq:polbb}
   b_- = k \del c_+ \exp
Y,\qquad b_+ = - k \bar \del c_- \exp Y.
\end{align}
However, we get a quantum correction in going from the \GL\ invariant
measure used for the action in~\eqref{eq:SWZW} to the free-field measure
$\mathcal{D}X\mathcal{D}Y
      \mathcal{D}\cm\mathcal{D}\cp\mathcal{D}\bm\mathcal{D}\bp$ that we
want to use for our first order formalism. In analogy with
\cite{Gerasimov:1990fi} the correction is
\begin{align}\label{eq:polmeas}
    \ln\det\left(|\rho|^{-2}e^{-Y}\del e^{Y}\bar \del\right)=\frac{1}{4\pi}\int d^2 z\left(\del Y\bar\del Y+\frac{1}{4}\sqrt{G}\mathcal{R}Y\right).
\end{align}
Here $G$ is the determinant of the world-sheet metric and $\mathcal{R}$
its Gaussian curvature. $|\rho|^2$ is the metric and we have the relation
$\sqrt{G}\mathcal{R} \ = \ 4 \del\bar{\del}\log |\rho|^2$. We thus get a
correction to the kinetic term and a background charge for $Y$. The first
order action including the correction is
\begin{multline}\label{eq:pol1st}
    S(X,Y,b_{\pm},c_\pm) = \frac{1}{4\pi}\int_\Sigma
d^2z \bigg(-k\del X\bar{\del}Y-k\del
  Y\bar{\del}X+\del
  Y\bar{\del}Y+\frac{1}{4}\sqrt{G}\mathcal{R}Y \\ +2\cp\del\bp+2\cm\bar{\del}\bm+\frac{2}{k}e^{-Y}\bm\bp\bigg).
\end{multline}

We also get a quantum correction to the current. This will happen where we
have to choose a normal ordering of the terms in the
current~\eqref{eq:holomorphiccurrents}. We fix this by demanding that the
currents obey the OPEs~\eqref{eq:currentope}. Indeed, we have to add $\del
Y$ to $J^N$ to ensure that it has a regular OPE with itself. Thus the
holomorphic currents in the free field formalism are
\begin{align*}
    J^E &= -k\del Y ,  &
    J^N &= -k\del X +\cm b_-+\del Y \ , \\[2mm]
    J^- &= b_-,  &
    J^+ &=  -k\del\cm -k\cm \del Y \ ,
\end{align*}
where we suppress the normal ordering. We get
similar expressions for the anti-holomorphic currents.

\subsection{The correspondence}

If we integrate out $b_\pm$ in~\eqref{eq:pol1st} we simply obtain
the original \GL\ WZNW model. We will now show that if we instead bosonize the $bc$
system to obtain a system of three scalars, it is possible to perform a
field redefinition such that one of the scalars decouples. We can then
return to a new $b'c'$ formalism and integrate out $b'_\pm$ to arrive at a
decoupled theory of two scalars and a set of symplectic fermions.

In this process the current becomes more symmetric and simple. It can be
seen as a guideline for which transformations to perform and we will
therefore explicitly follow the transformation of the current in each
step.

We will start by only discussing the transformation of the action and the
current. The map of the vertex operators will be determined in the next
subsection.

To begin we bosonize the $bc$ system in~\eqref{eq:pol1st} in the standard
way~\cite{Friedan:1985ge}
\begin{align}
    c_\pm =e^{\rho^{R,L}},&\qquad b_\pm=e^{-\rho^{R,L}},\nonumber \\
    \cp\del\bp+\cm\bar{\del}\bm &= -\frac{1}{2}\del\rho\bar\del\rho+\frac{1}{8}\sqrt{G}\mathcal{R}\rho,\nonumber \\
    b_- c_-&=-\del\rho^L, \label{eq:polbos}
\end{align}
where we denote left and right components of scalars by superscripts
$L,R$. In the currents we likewise have to introduce left and right
indices and the holomorphic currents then become
\begin{align}
    J^E &=\ -k\del Y^L,  &
    J^N &=\ -k\del X^L +\del\rho^L+\del Y^L \ ,\nonumber \\[2mm]
    J^- &=\ e^{-\rho^L},  &
    J^+ &= \ -k\del(\rho^L + Y^L)e^{\rho^L} \ ,
\end{align}
and the action is
\begin{equation}\label{eq:polbosact}
\begin{split}
    S(X,Y,b_{\pm},c_\pm) \ = \ \frac{1}{4\pi}\int_\Sigma
                  d^2z \Bigl(-k\del X\bar{\del}Y&-k\del  Y\bar{\del}X+\del   Y\bar{\del}Y+\Bigr. \\
                         \Bigl.&-\del\rho\bar\del\rho+\frac{1}{4}\sqrt{G}\mathcal{R}(Y+\rho)+\frac{2}{k}e^{-Y-\rho}\Bigr)\, .\\
  \end{split}
\end{equation}

We observe, both from the current and the action, that it is very
natural to go to variables $Y, Z, \rho'$ where
\begin{align}\label{eq:polphi1phi2}
    \rho'=Y+\rho,\qquad Z=kX-\rho-Y=kX-\rho'.
\end{align}
The currents and the action in these variables are
\begin{align}
    J^E &= -k\del Y^L,& J^N&=-\del Z^L,\nonumber \\
    J^- &= e^{Y^L-\rho'^L},& J^+ &= -k\del\rho'^Le^{\rho'^L-Y^L}
\end{align}
and
\begin{equation}
    S(Z,Y,\rho')\ =\ \frac{1}{4\pi}\int_\Sigma
d^2z \left(-\del Z\bar{\del}Y-\del
  Y\bar{\del}Z-\del\rho'\bar\del\rho'+\frac{1}{4}\sqrt{G}\mathcal{R}\rho'+\frac{2}{k}e^{-\rho'}\right).\label{eq:poltransact}
\end{equation}
Hence we got a theory of two scalars decoupled from another scalar with
screening charge and linear dilaton term. For calculation of correlation functions this is a very
efficient formulation of the theory. We will, however, go one step further
and rewrite the screened Coulomb gas in terms of symplectic fermions.

We thus return to a $b'c'$ system using again~\eqref{eq:polbos}, but now
for the field $\rho'$. This gives us the following simple expressions
\begin{align}
    J^E &= -k\del Y^L,& J^N &=-\del Z^L,\nonumber \\
    J^- &= e^{Y^L}b'_-, & J^+ &= -ke^{-Y^L}\del c'_-\, ,
\end{align}
for the currents and for the action it becomes
\begin{equation}
    S(Z,Y,b'_{\pm},c'_\pm) = \frac{1}{4\pi}\int_\Sigma
d^2z \left(-\del Z\bar{\del}Y-\del
  Y\bar{\del}Z+2\cp'\del\bp'+2\cm'\bar{\del}\bm'+\frac{2}{k}b'_- b'_+\right).\label{eq:polfinal}
\end{equation}
We can now integrate out the fields $b'_\pm$ getting the equations of
motion
\begin{align}\label{eq:polbs}
    b'_+=-k\bar\del c'_-,\quad b'_-=k\del c'_+\, ,
\end{align}
and arrive at
\begin{align}
    S(X,Y,c_\pm) = \frac{1}{4\pi}\int_\Sigma
d^2z \left(-\del Z\bar{\del}Y-\del
  Y\bar{\del}Z+2k\del\cp'\bar{\del}\cm'\right)\, .\label{eq:polfinalsymplecticfermion1}
\end{align}
Of course, we have to be careful when the vertex operators depend on $b'$. As
we will see below, the vertex operators for typical representations will
be twist operators which we interpret as not containing $b$.

To remove the dependence on the level $k$ in the action we introduce
$\chi^a$ by
\begin{align}\label{eq:rescaling}
    \sqrt{k}c'_+=\chi^1,\quad \sqrt{k}c'_-=\chi^2,
\end{align}
and the currents and action are then
\begin{align}
    J^E &= -k\del Y^L,& J^N &=-\del Z^L,\nonumber \\
    J^- &= \sqrt{k}e^{Y^L}\del \chi^1, & J^+ &= -\sqrt{k}e^{-Y^L}\del \chi^2,\label{eq:finalcurrent}
\end{align}
\begin{align}
    S(X,Y,\chi^a) = \frac{1}{4\pi}\int_\Sigma
d^2z \left(-\del Z\bar{\del}Y-\del
  Y\bar{\del}Z+\ep_{ab}\del\chi^a\bar{\del}\chi^b\right)\, .\label{eq:polfinalsymplecticfermion2}
\end{align}
where the anti-symmetric symbol is defined by $\ep_{12}=-\ep_{21}=1$. This
gives the OPEs
\begin{align}
    \chi^a(z,\bar z)\chi^b(w,\bar w)&\sim -\ep^{ab}\ln\abs{z-w}^2,\nonumber \\
    Z(z,\bar z) Y(w,\bar w)&\sim\ln\abs{z-w}^2.\label{eq:polopesympl}
\end{align}
where $\ep^{12}=-1$. This is the action and current that was constructed
in~\cite{LeClair:2007aj}. In that reference it was also found that the
action has an enlarged OSp$(2|2)$ symmetry.

For future reference, let us sum up the correspondence between the
symplectic fermions and the underlying $b',c'$ system. We have
\begin{align}
    \delbar\chi^1&=\sqrt{k}\delbar c'_+, & \delbar\chi^2&=\sqrt{k}\delbar c'_-=-\frac{1}{\sqrt{k}}b'_+,\nonumber \\
    \del\chi^1&=\sqrt{k}\del c'_+=\frac{1}{\sqrt{k}}b'_-, & \del\chi^2&=\sqrt{k}\del c'_-,\label{eq:polderivdict}
\end{align}
which will be useful in the next section where we study what happens to
the vertex operators.

\subsection{Mapping of the vertex operators}\label{section:voa}

We now consider the mapping of the \GL\ vertex operators under the
transformation that we found in the last subsection. The basis of vertex
operators to be used with the first order action~\eqref{eq:pol1st} were
found in~\cite{Schomerus:2005bf} by a minisuperspace analysis.  We will
here use the notation of~\cite{Creutzig:2008ek} and write the operators as
\begin{equation}\label{eq:polvertexope1}
    V_{\<-e,-n+1\>}=\normord{e^{eX+nY}}\begin{pmatrix}
                           1 & c_-\\
                           c_+ & c_-c_+ \\
                         \end{pmatrix},
\end{equation}
and the conformal dimension is
\begin{equation}\label{eq:polconfdim}
    \De_{(e,n)}=\frac{e}{2k}(2n-1+\frac{e}{k}).
\end{equation}
For $e\neq mk$, where $m$ is an integer, the columns of this matrix will
correspond to the two-dimensional representation $\<-e,-n+1\>$ for the left-moving currents while the rows correspond to
the representation $\<e,n\>$ under the right-moving currents.

Let us first consider the transformation giving us~\eqref{eq:poltransact}:
\begin{gather}
    X=\frac{1}{k}(\rho'+Z),\nonumber\\
    c_-=e^{\rho'^L_1-Y^L},\quad b_-=e^{-\rho'^L_1+Y^L}.\label{eq:poltranstocol}
\end{gather}
This maps the vertex operators to
\begin{equation}\label{eq:polvertexope2}
    V_{\<-e,-n+1\>}=\normord{e^{\frac{e}{k}\rho'+\frac{e}{k}Z+nY}\begin{pmatrix}
                           1 & e^{\rho'^L-Y^L}\\
                           e^{\rho'^R-Y^R} & e^{\rho'-Y} \\
                         \end{pmatrix}}.
\end{equation}
Here we generally split scalar fields into the left and right handed part
as $\rho'=\rho'^L+\rho'^R$. Some comments are in order here: Firstly,
rather than thinking of e.g. $c_-$ in~\eqref{eq:polvertexope1} as a
function to be evaluated under the path integral, we have here used
bosonization and will think about the vertex operators in the operator
formalism. This means that $c_-$ is a holomorphic operator. Secondly, for
the $YZ$ system the vertex operators are
\begin{equation}\label{eq:polvertexope2.1}
V_{\<-e,-n+1\>}^{\textrm{B}}=\begin{pmatrix}
  \normord{e^{\frac{e}{k}Z+nY}} & \normord{e^{\frac{e}{k}Z+(n-1)Y^L+nY^R}} \\
  \normord{e^{\frac{e}{k}Z+nY^L+(n-1)Y^R}} & \normord{e^{\frac{e}{k}Z+(n-1)Y}} \\
\end{pmatrix},
\end{equation}
whereas for the $\rho'$ system they are
\begin{equation}\label{eq:polvertexope2.2}
    V_{\<-e,-n+1\>}^{\textrm{F}}=\begin{pmatrix}
    \normord{e^{\frac{e}{k}\rho'}} & \normord{e^{(\frac{e}{k}+1)\rho'^L+\frac{e}{k}\rho'^R}}\\
    \normord{e^{\frac{e}{k}\rho'^L+(\frac{e}{k}+1)\rho'^R}} & \normord{e^{(\frac{e}{k}+1)\rho'}} \\
                         \end{pmatrix}.
\end{equation}
Thus in the off-diagonal terms, the splitting into holomorphic and
anti-holomorphic parts means that the correlation functions calculated in
respectively the $YZ$ system and the $\rho'$ system are not separately
real, but only the combined correlation function can be expressed in the
absolute values of the insertions $z_i$. Also, we see that around the
off-diagonal terms in the operator~\eqref{eq:polvertexope2.1} the field
$Z$ gets an additive twist. The overall twist vanishes due to charge
conservation for $Y$.

Since $\rho'$ now appears with non-integer momenta, we see that in going to
the $b',c'$ system with action~\eqref{eq:polfinal} we get twist operators.
Precisely, the vertex operator~\eqref{eq:polvertexope2.2} maps into
\begin{align}\label{eq:polvertexopebc}
V_{\<-e,-n+1\>}^{\textrm{F}}=\begin{pmatrix}
  \tilde{\mu}^L_{e/k}\tilde{\mu}^R_{e/k} & \tilde{\mu}^L_{e/k+1}\tilde{\mu}^R_{e/k} \\
  \tilde{\mu}^L_{e/k}\tilde{\mu}^R_{e/k+1} & \tilde{\mu}^L_{e/k+1}\tilde{\mu}^R_{e/k+1} \\
\end{pmatrix},
\end{align}
where the twist states are defined by
\begin{align}\label{eq:poltwist}
    c'_-(e^{2\pi i}z)\tilde{\mu}^L_{\la}(0)=e^{2\pi i\la}\tilde{\mu}^L_{\la}(0).
\end{align}
This is solved by
\begin{align}\label{eq:poltwisttoexp}
    \tilde{\mu}^L_{\la}\equiv\normord{e^{\la\rho'^L}},
\end{align}
but only uniquely in $\la$ modulo integers and, naturally, up to a
normalisation. The conformal dimension is $-\half\la(1-\la)$ so the ground
states have $0<\la<1$. We can step $\la$ up and down with respectively
$c'_-$ and $b'_-$ e.g.
\begin{align}\label{eq:poltwist2}
    c'_-(z)\tilde{\mu}^L_{\la}(0)\sim\frac{1}{z^{-\la}}\tilde{\mu}^L_{\la+1}(0).
\end{align}
Also note that
\begin{align}
    \tilde{\mu}^R_{\la}\equiv\normord{e^{\la\rho'^R}},
\end{align}
fulfils
\begin{align}\label{eq:poltwist3}
    c'_+(e^{-2\pi i}{\bar z})\tilde{\mu}^R_{\la}(0)=e^{-2\pi i\la}\tilde{\mu}^R_{\la}(0).
\end{align}

To obtain the symplectic fermions requires integrating out $b'$. This
means that the anti-holomorphic part of $c'_-$ is non-trivial in the OPEs.
As an example, $c'_+$ and $c'_-$ with
action~\eqref{eq:polfinalsymplecticfermion1} have a singular OPE that is
$\sim\frac{1}{k}\ln\abs{z-w}^2$. However, using
equations~\eqref{eq:polderivdict} we get the mapping of $\del c_-'$ and
$b_-'$ to the holomorphic operators $\del\chi^2$ and $\del\chi^1$.
Likewise, $\delbar c_+'$ and $b_+'$ will correspond to the
anti-holomorphic operators $\delbar\chi^1$ and $\delbar\chi^2$.

One has to be careful since we in principle can not integrate out $b'$
when the vertex operators depend on $b'_-b'_+$. However, for the twist
operators it seems plausible since, at least for $\la>0$, we can naively
think of $\mu_\la$ as $c'^\la$. To check this we will in the next section
compare the correlation functions to the already known calculation for the
symplectic fermions. The twist fields in the $b',c'$ system then directly
translates into twist fields of the symplectic fermions. The symplectic
fermion twist fields are defined by~\cite{Kausch:2000fu}
\begin{align}
    \chi^1(e^{2\pi i}z)\mu_\la(0)&=e^{-2\pi i\la}\chi^1(z)\mu_\la(0), & \chi^2(e^{2\pi i}z)\mu_\la(0)&=e^{2\pi i\la}\chi^2(z)\mu_\la(0),\nonumber\\
    \bar \chi^1(e^{-2\pi i} \bar z)\mu_\la(0)&=e^{-2\pi i\la}\bar \chi^1(\bar z)\mu_\la(0), & \bar \chi^2(e^{-2\pi i}\bar z)\mu_\la(0)&=e^{2\pi i\la}\bar \chi^2(\bar z)\mu_\la(0),\label{eq:polsympltwist}
\end{align}
where $\chi^1$ and $\chi^2$ has to transform oppositely to give a symmetry
of the Lagrangian. Here we have split the symplectic fermions into their
chiral and anti-chiral parts
$\chi^a(z,\bar{z})=\chi^a(z)+\bar{\chi}^a(\bar{z})$. The anti-holomorphic
part must transform in the same way under $\bar{z}\mapsto e^{-2\pi
i}\bar{z}$, but importantly $\la$ can differ by an integer between the
holomorphic and anti-holomorphic sector. The
condition~\eqref{eq:polsympltwist} is fulfilled by
$\tilde{\mu}_\la^L\tilde{\mu}_\la^R$ and the other operators
in~\eqref{eq:polvertexopebc}. However, we have done the
rescaling~\eqref{eq:rescaling} so if we think of the twist operator as
$(c'_-)^\la$ we should choose the following normalisation:
\begin{align}\label{eq:newtwistfield}
    \mu^L_\la=\sqrt{k}^\la\tilde{\mu}^L_\la=\sqrt{k}^\la\normord{e^{\la\rho'^L}},
\end{align}
and similarly for the anti-holomorphic part. Thus the vertex
operator~\eqref{eq:polvertexopebc} maps into
\begin{align}\label{eq:polvertexopesymplectic}
V_{\<-e,-n+1\>}^{\textrm{F}}\mapsto k^{-\frac{e}{k}}\begin{pmatrix}
  {\mu}^L_{e/k}{\mu}^R_{e/k} & \frac{1}{\sqrt{k}}{\mu}^L_{e/k+1}{\mu}^R_{e/k} \\
  \frac{1}{\sqrt{k}}{\mu}^L_{e/k}{\mu}^R_{e/k+1} & \frac{1}{k}{\mu}^L_{e/k+1}{\mu}^R_{e/k+1} \\
\end{pmatrix}.
\end{align}
A notation with splitting into left and right part, like in the $b'c'$
system, turns out to be useful. The twist values can be stepped up and
down using the following OPEs:
\begin{align}\label{eq:polsympltwistope1}
    \del\chi^1(z)\mu^L_\la(0)&\sim\frac{1}{z^\la}\mu^L_{\la-1}(0),\quad \del\chi^2(z)\mu^L_{\la}(0)\sim\frac{\la}{z^{1-\la}}\mu^L_{\la+1}(0),
\end{align}
and correspondingly
\begin{align}\label{eq:polsympltwistope2}
    \delbar\bar \chi^1(\bar z)\mu^R_\la(0)&\sim\frac{\la}{\bar z^{1-\la}}\mu^R_{\la+1}(0), \quad \delbar\bar \chi^2(\bar z)\bar \mu^R_\la(0)\sim-\frac{1}{\bar z^{\la}}\mu^R_{\la-1}(0).
\end{align}
We note here again that up to a sign the anti-holomorphic side is
understood by seeing $\mu^R_\la$ as $\mu^L_{1-\la}$.

To conclude, the total vertex operator $V_{\<-e,-n+1\>}$ in the $YZ$ and
symplectic fermion system with
action~\eqref{eq:polfinalsymplecticfermion2} takes the form
\begin{equation}\label{eq:finalvertexoperator}
 \begin{split}
   &V_{\<-e,-n+1\>}\mapsto\\ &k^{-\frac{e}{k}} \begin{pmatrix}
  \normord{e^{\frac{e}{k}Z+nY}}{\mu}^L_{e/k}{\mu}^R_{e/k} & \frac{1}{\sqrt{k}}\normord{e^{\frac{e}{k}Z+(n-1)Y^L+nY^R}}{\mu}^L_{e/k+1}{\mu}^R_{e/k} \\
  \frac{1}{\sqrt{k}}\normord{e^{\frac{e}{k}Z+nY^L+(n-1)Y^R}}{\mu}^L_{e/k}{\mu}^R_{e/k+1} & \frac{1}{k}\normord{e^{\frac{e}{k}Z+(n-1)Y}}{\mu}^L_{e/k+1}{\mu}^R_{e/k+1} \\
\end{pmatrix}
\end{split}
\end{equation}
We note that equations~\eqref{eq:polsympltwistope1} can be used to check
that the columns of this operator transform in the \mbox{$\<-e,-n+1\>$}
representation of \GL\ under the left-moving
currents~\eqref{eq:finalcurrent}. These operators are indeed close to the
operators found in~\cite{LeClair:2007aj}. Let us now check the correlation functions of these vertex operators.

\subsection{Bulk correlation functions}

We will now compare the correlation functions of the primary
fields~\eqref{eq:polvertexope1} obtained in the \GL\ model to the
calculations done for the symplectic fermions in~\cite{Kausch:2000fu}. The
similarity was already noted in~\cite{Schomerus:2005bf}.

Let us first note that from equations~\eqref{eq:polvertexope2.1}
and~\eqref{eq:polvertexope2.2} the vertex
operators~\eqref{eq:polvertexope1} in the $Y,Z,\rho'$
picture~\eqref{eq:poltransact} takes the form
\begin{align}\label{eq:vertexwithindices}
    {V_{\<-e,-n+1\>}}_\si^{\bar\si}=\normord{e^{\frac{e}{k}Z+(n-\si)Y^L+(n-\bar\si)Y^R}e^{(\frac{e}{k}+\si)\rho'^L+(\frac{e}{k}+\bar\si)\rho'^R}},
\end{align}
where $\si,\bar\si\in\{0,1\}$ labels respectively the columns and the
rows.

We consider the three-point function
\begin{align}\label{eq:3pt1}
    A=\expect{{V_{\<-e_1,-n_1+1\>}}_{\si_1}^{\bar\si_1}(z_1){V_{\<-e_2,-n_2+1\>}}_{\si_2}^{\bar\si_2}(z_2){V_{\<-e_3,-n_3+1\>}}_{\si_3}^{\bar\si_3}(z_3)}.
\end{align}
The correlation function splits into a $YZ$ and a $\rho'$ part,
$A=A^{\textrm{B}}A^{\textrm{F}}$. The $YZ$ part is easily evaluated to be
\begin{equation}\label{eq:3pt2}
	\begin{split}
    A^{\textrm{B}}\ = \ \delta\big(\sum_i \frac{e_i}{k}\big)\delta&\big(\sum_i(n_i-\si_i)\big)\delta\big(\sum_i(n_i-\bar\si_i)\big)\, \times\\
    &\times\,\prod_{i<j}(z_i-z_j)^{\frac{e_i}{k}(n_j-\si_j)+\frac{e_j}{k}(n_i-\si_i)}(\bar z_i-\bar
    z_j)^{\frac{e_i}{k}(n_j-\bar\si_j)+\frac{e_j}{k}(n_i-\bar\si_i)}\, ,\\
    \end{split}
\end{equation}
where the indices run from 1 to 3. The $\de$-functions follow directly
from the $J^E$ and $J^N$ currents. The $\rho'$ part is also easily
evaluated. Here one has to remember that the overall $\rho'$ charge has to
sum to one due to the background charge of $\rho'$. This means that we can
maximally have two insertions of the interaction term of the
action~\eqref{eq:poltransact}. However, as was commented
in~\cite{Schomerus:2005bf}, the part with two interaction terms vanish.
The part with one interaction term is calculated using the Dotsenko-Fateev
like integral used in~\cite{Schomerus:2005bf}. We get
\begin{equation}
	\begin{split}
    A^{\textrm{F}}\ &=\  A^{\textrm{F}}_1+ A^{\textrm{F}}_2\\
     A^{\textrm{F}}_1 \ &= \ \delta\big(\sum_i\si_i -1\big)\delta\big(\sum_i\bar\si_i -1\big)
                              \prod_{i<j}(z_i-z_j)^{(\frac{e_i}{k}+\si_i)(\frac{e_j}{k}+\si_j)}
			      (\bar z_i-\bar z_j)^{(\frac{e_i}{k}+\bar \si_i)(\frac{e_j}{k}+\bar \si_j)}\\
  A^{\textrm{F}}_2\ &= \ -\frac{1}{k}\delta\big(\sum_i\si_i -2\big)\delta\big(\sum_i\bar\si_i -2\big)(-1)^{\si_3+\bar\si_3}\, \times\\
  &\qquad\qquad\qquad\times\,\frac{\Ga(1-\frac{e_1}{k}-\si_1)\Ga(1-\frac{e_2}{k}-\si_2)\Ga(1-\frac{e_3}{k}-\bar\si_3)}
             {\Ga(\frac{e_3}{k}+\si_3)\Ga(\frac{e_1}{k}+\bar \si_1)\Ga(\frac{e_2}{k}+\bar\si_2)}\,\times\\
    &\qquad\qquad\qquad\times\,\prod_{i<j}(z_i-z_j)^{(\frac{e_i}{k}+\si_i-1)(\frac{e_j}{k}+\si_j-1)}(\bar z_i-\bar z_j)^{(\frac{e_i}{k}+\bar \si_i-1)(\frac{e_j}{k}+\bar \si_j-1)}\label{eq:3pt3},
\end{split}
\end{equation}
where the first part $A^{\textrm{F}}_1$ corresponds to no interaction term and the second part $A^{\textrm{F}}_2$ to
one interaction term. We have here used that $\sum_i e_i=0$ due to the
delta-function from the $YZ$ part of the correlation function
in~\eqref{eq:3pt2}.

If we combine the two parts in~\eqref{eq:3pt2} and~\eqref{eq:3pt3} the
symmetry between the holomorphic and anti-holomorphic sector is restored
and we arrive at
\begin{multline}\label{eq:3pt4}
    A=\delta\big(\sum_i \frac{e_i}{k}\big)\delta\big(\sum_i(n_i-\si_i)\big)\delta\big(\sum_i(n_i-\bar\si_i)\big)\\
    \bigg(\delta\big(\sum_i\si_i -1\big)\delta\big(\sum_i\bar\si_i -1\big)\prod_{i<j}\abs{z_i-z_j}^{2\frac{e_i}{k}n_j+2\frac{e_j}{k}n_i+2\frac{e_ie_j}{k^2}} \\
    -\frac{1}{k}\delta\big(\sum_i\si_i -2\big)\delta\big(\sum_i\bar\si_i -2\big)(-1)^{\si_3+\bar\si_3}\frac{\Ga(1-\frac{e_1}{k}-\si_1)\Ga(1-\frac{e_2}{k}-\si_2)\Ga(1-\frac{e_3}{k}-\bar\si_3)}{\Ga(\frac{e_3}{k}+\si_3)\Ga(\frac{e_1}{k}+\bar \si_1)\Ga(\frac{e_2}{k}+\bar\si_2)} \\
    \times \prod_{i<j}\abs{z_i-z_j}^{2\frac{e_i}{k}(n_j-1)+2\frac{e_j}{k}(n_i-1)+2\frac{e_ie_j}{k^2}}
    \bigg),
\end{multline}
as was derived in~\cite{Schomerus:2005bf}. This indeed supports the
validity of our decoupling of the \GL\ WZNW model into a set of free
scalars and the $\rho'$ system with action~\eqref{eq:poltransact}. The
result may not look local, e.g. does not seem to be symmetric in
interchanging operator 2 and 3, due to the asymmetric-looking $\Ga$
functions. However, these can be rewritten in the following symmetric form
\begin{align}\label{eq:3pt4.5}
    (-1)^{\si_3+\bar\si_3}\frac{\Ga(1-\frac{e_1}{k}-\si_1)\Ga(1-\frac{e_2}{k}-\si_2)\Ga(1-\frac{e_3}{k}-\bar\si_3)}{\Ga(\frac{e_3}{k}+\si_3)\Ga(\frac{e_1}{k}+\bar \si_1)\Ga(\frac{e_2}{k}+\bar\si_2)}=\prod_i\frac{\Ga(1-\frac{e_i}{k})}{\Ga(\frac{e_i}{k})}\left(\frac{-e_i}{k}\right)^{-\si_i-\bar\si_i}.
\end{align}

As we see from the result~\eqref{eq:3pt4} one has to be careful in the
limit when $e_i$ is an integer multiple of $k$. As was shown
in~\cite{Schomerus:2005bf} this gives logarithmic correlation functions.
For now let us not consider these limits. Thus we get genuine twist
operators when going to the symplectic fermions and the twists are
$\la_i=e_i/k+\si_i$ in the holomorphic sector and
$\bar\la_i=e_i+\bar\si_i$ in the anti-holomorphic sector when we compare
equation~\eqref{eq:vertexwithindices}
with~\eqref{eq:polvertexopesymplectic}. As we see from the vertex
operators in~\eqref{eq:polvertexopesymplectic}, the results that we expect
from the symplectic fermions to comply with correlation
function~\eqref{eq:3pt3} are
\begin{multline}\label{eq:3pt5}
    \expect{\mu^L_{\la_1}(z_1)\mu^R_{\bar \la_1}(\bar z_1)\mu^L_{\la_2}(z_2)\mu^R_{\bar \la_2}(\bar z_2)\mu^L_{\la_3}(z_3)\mu^R_{\bar \la_3}(\bar z_3)}_{\mathrm{SF}}=\prod_{i<j}(z_i-z_j)^{\la_i\la_j}(\bar z_i-\bar z_j)^{\bar \la_i\bar
    \la_j}
\end{multline}
for $\sum_i\la_i=\sum_i\bar\la_i=1$, and
\begin{multline}\label{eq:3pt6}
    \expect{\mu^L_{\la_1}(z_1)\mu^R_{\bar \la_1}(\bar z_1)\mu^L_{\la_2}(z_2)\mu^R_{\bar \la_2}(\bar z_2)\mu^L_{\la_3}(z_3)\mu^R_{\bar \la_3}(\bar z_3)}_{\mathrm{SF}}\\
    =-(-1)^{\la_3-\bar\la_3}\frac{\Ga(\la_1^*)\Ga(\la_2^*)\Ga(\bar\la_3^*)}{\Ga(\bar\la_1)\Ga(\bar\la_2)\Ga(\la_3)}\prod_{i<j}(z_i-z_j)^{\la^*_i\la^*_j}(\bar z_i-\bar z_j)^{\bar \la^*_i\bar
    \la^*_j}
\end{multline}
for $\sum_i\la_i=\sum_i\bar\la_i=2$, where $\la^*=1-\la$ and the subscript SF means that the expectation value
is calculated using the symplectic fermion part of the
action~\eqref{eq:polfinalsymplecticfermion2}. Here $\mu_\la$ are the twist
operators defined in eq.~\eqref{eq:polsympltwist}. We have also used that
in going to this expectation value under the
rescaling~\eqref{eq:rescaling} we have to multiply the correlation
functions with an overall factor of $k$. This is because the correlation
function normalisation is relative to the correlator of $\bar\chi^1\chi^2$
or $c'_+c'_-$ in the $b'c'$ system in eq.~\eqref{eq:polfinal}. This simply
means that the dependence on $k$ disappears due to the normalisation in
eq.~\eqref{eq:newtwistfield} as is expected.

We want to compare this to the calculation of bulk twist correlators done
by Kausch in~\cite{Kausch:2000fu}. In that paper, of course, only twist
fields with identical twist in the holomorphic and anti-holomorphic sector
are treated so we take $\la_i=\bar\la_i$. Further, we have to remember
that the twist fields are only defined up to normalisation. To compare
with Kausch we use one of the equations~\eqref{eq:3pt5},~\eqref{eq:3pt6}
to fix the normalisation and can then compare to the second one. The
normalisation is fixed by defining
\begin{align}\label{eq:3pt7}
    \mu^L_{\la}\mu^R_{\la}=-\sqrt{\frac{\Ga(\la^*)}{\Ga(\la)}}\mu_{\la}.
\end{align}
Then we get
\begin{align}
    \expect{\mu_{\la_1}(z_1,\bar z_1)\mu_{\la_2}(z_2,\bar z_2)\mu_{\la_3}(z_3,\bar z_3)}_{\mathrm{SF}}&
    =\prod_{i}\sqrt{\frac{\Ga(\la_i)}{\Ga(\la_i^*)}}\prod_{i<j}\abs{z_i-z_j}^{2\la_i\la_j}\quad\textrm{for }\sum_i\la_i=1,\nonumber\\
    &=\prod_{i}\sqrt{\frac{\Ga(\la^*_i)}{\Ga(\la_i)}}\prod_{i<j}\abs{z_i-z_j}^{2\la^*_i\la^*_j}\quad\textrm{for }\sum_i\la_i=2,\label{eq:3pt8}
\end{align}
which is exactly as in~\cite{Kausch:2000fu}. We can also compare with the
two-point function which is easily calculated and also get a match here.
Note, however, that in~\cite{Kausch:2000fu} only ground state twist fields
with $0<\la<1$ are considered. Our results thus compare precisely in this
range, and are the analytic continuation of the twists $\la$ for the
results in~\cite{Kausch:2000fu}.

In the case where we allow the $e_i$ to be zero or an integer multiple of
$k$, we have to take into account the zero modes of the symplectic
fermions. This gives four different ground states in the symplectic model
- two fermionic and two bosonic, where the last two span a Jordan block
for $L_0$. The result is that we get logarithmic branch cuts in the
correlation functions. This can be seen from the \GL\ side where the $\Ga$
functions diverge when $\la$ becomes integer~\cite{Schomerus:2005bf}. Thus
we also get agreement from the two sides of the correspondence here.
\smallskip

Now, having established the correspondence, we want to apply it. There are
two apparent applications. For point-like branes in the \GL\ WZNW model,
so far it could be argued that correlators containing only boundary fields
behave like untwisted symplectic fermions see section \ref{sec:coratyp}, but it
was not possible to handle insertions of bulk fields. Now, we are in a
position to approach the problem of computing correlation functions
involving bulk and boundary fields. We will refrain from this problem for
now, but keep it in mind for future research. Instead, we consider the
study of boundary states in section \ref{sec:branes}.

\section{Branes}\label{sec:branes}

The aim of this section is to initiate a systematic study of boundary
conditions in WZNW models on supergroups based on the example of \GL.%
\footnote{Spectra of supersymmetric coset models with open boundary 
conditions were also studied previously, in particular in 
\cite{Read:2001pz,Read:2007qq}.}
Let us list the main results of this section in more detail. Recall
that maximally symmetric boundary conditions in conformal field
theories carry two labels. The first one refers to the choice of a
gluing condition between left and right moving chiral fields. The
second label parametrises different boundary conditions
associated with the same gluing condition. In uncompactified free
field theory, for example, the two labels correspond to the dimension
of the brane and its transverse position. The relation between these
labels and the branes' geometry becomes more intricate when the
world-sheet theory is interacting.
\smallskip
  
Recall that \gl\ possesses two different gluing automorphisms (section \ref{section:automorphismsgl11}).
Those branes corresponding to the trivial gluing automorphism will be called untwisted, while the other we call twisted. 
After a detailed study of
the branes' geometry we shall provide exact boundary states for
generic and non-generic untwisted branes on \GL\ in section \ref{section:bdystateuntwisted}.
There, we shall also discuss what happens when a generic brane is
moved onto one of the lines $y_0 = 2\pi s$: It turns out to split
into a pair of non-generic branes with a transverse separation
that is proportional to the level of the WZNW model. Section \ref{section:comparisoncardy}
contains a detailed discussion of the relation between our
findings for boundary conditions in a local logarithmic conformal
field theory and the usual Cardy case of unitary rational models
\cite{Cardy:1989ir}.
We shall see that in both cases branes are parameterised by
irreducible representations of the current algebra. Furthermore,
the spectra between any two branes can be determined by fusion.
Similar results for the $p=2$ triplet model have been obtained 
in \cite{Gaberdiel:2006pp}. In the case of \GL\ WZNW model we will 
establish that most of the boundary spectra are not fully 
reducible. This applies in particular to the spectrum of 
boundary operators on a single generic brane.

\subsection{Untwisted branes: Geometry and particle limit}

This section is devoted to the geometry of branes associated with
the trivial gluing automorphism. We shall show that such branes
are localised at a point $(x_0,y_0)$ on the bosonic base of \GL.
For generic choices $y_0$, they stretch out along the fermionic
directions, i.e.\ the fermionic fields obey Neumann type boundary
conditions. When $y_0 = 2\pi s, s \in \mathbb{Z}$, on the other
hand, the corresponding branes are point-like. These geometric
insights from the first part of the section are then used in the
second part to study branes in the particle limit in which the
level $k$ is sent to infinity. Most importantly, we shall
provide minisuperspace analogues of the boundary states for
both generic and non-generic untwisted branes, see eqs.\ %
\eqref{eq:MSSbst1} and \eqref{eq:MSSbst2}, respectively.

Recall that a boundary condition is said to be
maximally symmetric if left and right moving currents can be
identified along the boundary, up to the action of an automorphism
$\Omega$,
\begin{equation} \label{eq:glue}
    J^a(z)\ =\ \Omega\bigl(\bar{J^a}(\bar{z})\bigr) \ \ \text{for} \
                                      z = \bar{z}
    \ .
\end{equation}
where $J^a = E,N,\Ppm$ when we deal with the \GL\ model. For
$\Omega$ we can insert any of the automorphisms of \gl (section \ref{section:automorphismsgl11}).
\smallskip

It will be convenient to rewrite the gluing conditions
\eqref{eq:glue} in terms of those fields that appear in the action
of the \GL\  WZNW model. In principle, there exist various choices
that come with different parameterisations of the supergroup \GL.
One possible set of coordinate fields is introduced through
\begin{equation}\label{eq:par1}
    g\ =\ e^{i\cm\Pm}\, e^{iXE+iYN}\, e^{i\cp\Pp}\ \ .
\end{equation}
The fields $X$ and $Y$ are bosonic while $\cpm$ are fermionic.
Inserting our specific choice of the parameterisation
\eqref{eq:par1}, the currents take the following form
\begin{equation} \label{eq:J}
	\begin{split}
    \bar{J} \ &= \ k g^{-1}\bar{\del} g\\
&= \ kie^{iY}\bar{\del}\cm\Pm + k\bigl(i\bar{\del}X - 
     (\bar{\del}\cm)\cp e^{iY}\bigr) E +ki\bar{\del}Y N +
     k(i\bar{\del} \cp- \cp \bar{\del} Y)\Pp\\
     \end{split}
\end{equation}
and
\begin{equation} \label{eq:bJ}
	\begin{split}
    J \ &= \ -k\del g g^{-1}\\
&=\ -k(i\del\cm-\cm\del Y)\Pm -k\bigl(i\del X -
    \cm (\del \cp) e^{iY}\bigr)
    E -k i\del Y N -k ie^{iY}\del\cp\Pp.\\
    \end{split}
\end{equation}

\subsubsection{Geometric interpretation of untwisted branes}

In the previous section we have made a number of general statements
concerning the geometry of maximally symmetric branes on (super-)group
target spaces. Here, we want to step back a bit and work out the
precise form of the boundary conditions for coordinate fields. We
shall continue to use the specific parameterisation \eqref{eq:par1}
of \GL. Insertion of our explicit formulae \eqref{eq:J} and
\eqref{eq:bJ} for left and right moving currents into the gluing
condition \eqref{eq:glue} with $\Omega= \mathbb{I}$ gives
\begin{equation} \label{eq:BUG}
    \begin{split}
        &\p Y \ = \ 0 \qquad , \qquad \p Z\  = \ 0 \qquad ,
         \qquad \text{for} \ z\, =\, \bar{z} \ , \\[2mm]
        &\text{where} \qquad Z=X+i\cm\cp(e^{-iY}-1)^{-1} \  \\
        \end{split}
\end{equation}
and $\partial_p$ denotes the derivative along the boundary. In
other words, both bosonic fields $Y$ and $Z$ satisfy Dirichlet
boundary conditions. Untwisted branes in the \GL\ WZNW model are
therefore parameterised by the constant values $(y_0,z_0)$ the two
bosonic fields $Y,Z$ assume along the boundary. For the two basic
fermionic fields we obtain similarly
\begin{equation} \label{eq:FUG}
    \begin{split}
        &\pm 2 \sin^2(Y/2)\n \dpm \ = \ \sin(Y)\,\p \dpm \qquad ,
         \qquad \text{for} \ z\, =\, \bar{z}\ , \\[2mm]
        &\text{where} \qquad\dpm\ =\ \cpm e^{iY/2}\sin^{-1}(Y/2)/2i\
        .\\
         \end{split}
\end{equation}
Thereby, the fermionic directions are seen to satisfy Neumann
boundary conditions with a constant B-field whose strength
depends on the position of the brane along the bosonic base. We
shall provide explicit formulae below. For the moment let us point
out that the condition \eqref{eq:FUG} degenerates whenever the
value $y_0$ of the bosonic field $Y$ on the boundary approaches an
integer multiple of $2\pi$. In fact, when $y_0 = 2\pi s, s \in
\mathbb{Z}$ we obtain Dirichlet boundary conditions in all
directions, bosonic and fermionic ones,
\begin{equation}
    \p Y\ =\ \p Z=\p \dpm \ =\ 0 \ \ \text{for} \ z\, =\, \bar{z}.
\end{equation}
In the following, we shall refer to the branes with parameters
$(z_0,y_0 \neq 2\pi s)$ as {\em generic (untwisted)
branes}. These branes are localised at the point $(z_0,y_0)$ of
the bosonic base and they stretch out along the fermionic
directions. A localisation at points $(z_0,2\pi s),
s \in \mathbb{Z}$, implies Dirichlet boundary conditions for
the fermionic fields. We shall refer to the corresponding
branes as {\em non-generic (untwisted) branes}.
\smallskip

We have seen in the description of our gluing conditions that it
was advantageous to introduce fields $Z$ and $d_\pm$ instead of
$X$ and $\cpm$. They correspond to a new choice of coordinates on
the supergroup \GL\
\begin{equation}\label{eq:par}
    g\ =\ e^{i\cm\Pm}e^{ixE+iyN}e^{i\cp\Pp}\ =\
     e^{i\dm\Pm}e^{-i\dpp\Pp}e^{izE+iyN}e^{i\dpp\Pp}e^{-i\dm\Pm}
\end{equation}
that is particularly adapted to the description of untwisted branes.
In fact, we recall from our general discussion that untwisted branes
are localised along conjugacy classes. It is therefore natural to
introduce a parameterisation in which supergroup elements $g$ are
obtained by conjugating bosonic elements $g_0 = \exp (iz_0E +iy_0N)$
with exponentials of fermionic generators. From equation \eqref{eq:par}
it is also easy to read off that conjugacy classes containing a bosonic
group element $g_0$ contain two fermionic directions as long as $y_0
\neq 2\pi s$. In case $y_0 = 2\pi s$, conjugation of $g_0$ with the
fermionic factors is a trivial operation and hence the conjugacy
classes consist of points only.
\smallskip

It is instructive to work out the form of the background metric
and B-field in our new coordinates. To this end, let us recall
that
\begin{equation} \label{eq:sold}
ds^2\ =\ \str\bigl((g^{-1}dg)^2\bigr)\ =\ 2dxdy-2e^{iy}d\etam d\etap\ \ .
\end{equation}
Here, the super-coordinates $x,y,\etapm$ correspond to our
coordinate fields $X,Y,c_\pm$. Similarly, the Wess-Zumino
3-form on the supergroup \GL\ is given by
\begin{equation}\label{Hold}
H\ =\ \frac{2}{3}\; \str(g^{-1}dg)^{\wedge 3}\ =\
2ie^{iy}d\etam\wedge d\etap\wedge dy\ \ .
\end{equation}
After the appropriate change of coordinates from $(x,y,\etapm)$
to $(z,y,\zpm)$, the metric reads
\begin{equation} \label{eq:snew}
    ds^2\ =\ 2dzdy+8\sin^2(y/2)d\zm d\zp
\end{equation}
and the $H$ field becomes
\begin{equation} \label{eq:Hnew}
H \ = \ 4 i \bigl(\cos(y)-1\bigr) d\zm\wedge d\zp \wedge d
y \ \ .
\end{equation}
It is easy to check that $H = dB$ possesses a 2-form potential
$B$ given by
\begin{equation} \label{Bnew}
    B\ =\ 4i\sin(y)\,d\zm\wedge d\zp\ + 2i\zp d\zm\wedge dy -
         2i\zm d\zp \wedge dy \ \ .
\end{equation}
Upon pull back to the untwisted branes we can set $dy=0$ and the
B-field becomes
\begin{equation}
 \pi^\ast_{\text{brane}}\, B\ = \ 4i\sin(y)\,d\zm\wedge d\zp\ \ .
\end{equation}
This expression together with our formula \eqref{eq:snew} for the
metric allow to recast the boundary conditions \eqref{eq:FUG} for
the fermionic fields in theories with generic untwisted boundary
conditions in the familiar form (section \ref{section:bdyaction}.

\subsubsection{Boundary states in the minisuperspace theory}

As in the analysis of the bulk \GL\ model \cite{Schomerus:2005bf}
it is very instructive to study the properties of untwisted branes
in the so-called particle or minisuperspace limit. Thereby we
obtain predictions for several field theory quantities in the
limit where the level $k$ tends to infinity. Our first aim is
to present formulae for the minisuperspace analogue of Ishibashi
states. Using our insights from the previous subsection we shall
then propose candidate boundary states for the particle limit
and expand them in terms of Ishibashi states.
\smallskip

Let us begin by recalling a few basic facts about the supergroup
\GL\ or rather the space of functions $\L2$ it determines, see
\cite{Schomerus:2005bf}. The latter is spanned by the elements
\begin{equation} \label{eq:basis}
e_0(e,n)\ =\ e^{iex+iny}\ , \ \ e_\pm(e,n)\ =\ \etapm e_0(e,n)\, \
\ \  e_2(e,n)\ =\ \etam\etap e_0(e,n) \ .
\end{equation}
where the coordinates are the same as in the previous subsection.
Right and left invariant vector fields take the following form
\begin{equation}
R_E\ =\ i\del_x \ , \ \ R_N\ =\ i\del_y+\etam\del_- \ , \ \ R_+\
=\ -e^{-iy}\del_+-i\etam\del_x \ , \ \ R_-\ =\ -\del_-\ ,
\end{equation}
and
\begin{equation}
L_E\ =\ -i\del_x \ , \ \ L_N\ =\ -i\del_y-\etap\del_+ \ , \ \ L_-\
\ =\ e^{-iy}\del_- -i\etap\del_x \ , \ \ L_+\ =\ \del_+\ ,
\end{equation}
These vector fields generate two \mbox{(anti-)}commuting copies of the
underlying Lie superalgebra \gl. For the reader's convenience we
also wish to reproduce the invariant Haar measure on \GL,
\begin{equation} d\mu \ = \ e^{-iy}dxdyd\etap d\etam\ \ .
\end{equation}
The decomposition of $\L2$ with respect to both left and right
regular action was analysed in \cite{Schomerus:2005bf}. Here, we
are most interested in properties of the adjoint action $\ad_X =
R_X + L_X$ since it is this combination of the symmetry generators
that is preserved by the untwisted D-branes.
\smallskip

In the last subsection we saw that the coordinates $(z,y,\zpm)$ which describe conjugacy classes are particularly adapted to the description of untwisted branes. When we use these coordinates the adjoint action takes the following simple form 
\begin{equation}
	\label{eq:adjointaction}
	\ad_E\ = \ 0 \ , \ \ \ad_N\ = \ \zm\del_--\zp\del_+ \ , \ \ \ad_+ \ = \ \del_+ \ , \ \ \ad_- \ = \ - \del_- \ . 
\end{equation}
Here $\del_-$ and $\del_+$ denote the derivatives with respect to $\zm$ and $\zp$. 
The space
of functions ${\cal N}_{z_0,y_0}$ vanishing along
the brane at $(z_0,y_0)$ is spanned by 
\begin{equation}
	e^{iez+iny}-e^{iez_0+iny_0}\ , \ \zpm(e^{iez+iny}-e^{iez_0+iny_0})\ , \ \zm\zp(e^{iez+iny}-e^{iez_0+iny_0})\ . \
\end{equation}
Clearly, the adjoint action
may be restricted to the space ${\cal N}_{z_0,y_0}$. From
now on we shall consider ${\cal N}_{z_0,y_0}$ as a \gl\
submodule of  $\L2$. The space of functions on the brane may be constructed as a quotient
of the space of functions on the
supergroup by the submodule ${\cal N}_{z_0,y_0}$ of functions vanishing
along the brane. This quotient is represented by the functions $1,\zm, \zp$ and $\zm\zp$. Under the adjoint action, these functions
transform in a 4-dimensional indecomposable representation ${\cal P}_0$
of \gl . The latter is known as the projective cover of the trivial
representation. Thus, we have shown that the space of functions on a
generic brane transforms in a projective module ${\cal P}_0$. According
to the usual rules, functions on the brane are the minisuperspace model
for boundary operators in the full field theory.  
\smallskip

The next aim is to construct a canonical basis in the space of
(co-)invariants. By definition, a (co-)invariant $|\psi\rrangle$
($\llangle \psi |$) is a state (linear functional) satisfying
\begin{equation} \label{eq:MSSinv}
 \ad_X |\psi \rrangle \ = \ (R_X + L_X) |\psi\rrangle \ = \ 0 \ \ \ , \ \
\ \llangle \psi | \,\ad_X \ = \ \llangle \psi| (R_X + L_X) \ = \ 0 \
\ .
\end{equation}
These two linear conditions resemble the so-called
Ishibashi conditions in boundary conformal field theory. In the
minisuperspace theory, it is easy to describe the space of
solutions. One may check by a short computation that a generic
invariant takes the form
\begin{equation}
|e,n\rrangle_0 \ = \ \frac{1}{2\pi \sqrt e} \bigl( e_0(e,n)- e_0(e,n-1)
+ee_2(e,n)\bigr)  \ \ .
\end{equation}
The pre-factor $1/2\pi \sqrt e$
is determined by a normalisation condition to be spelled out
below. We note that the function $|e,n\rrangle_0$ is obtained by
taking the super-trace of supergroup elements in the typical
representation $\langle e,n\rangle$.\footnote{Our conventions for
the representation theory of \gl\ are the same as in
\cite{Gotz:2005jz}. In particular, $\langle e,n\rangle$ denotes a
2-dimensional graded representation of \gl. Let us agree to
consider the state with smaller $N$-eigenvalue as even (bosonic).
The same representation with opposite grading shall receive an
additional prime, i.e.\ it is denoted by $\langle e,n\rangle'$.}
To each of the invariants $|e,n\rrangle_0$ we can assign a
co-invariant $_0\llangle e,n|:\L2 \rightarrow \mathbb{C}$ through
\begin{equation}
\ _0\llangle e,n| \ = \ \int d\mu \,\frac{1}{2 \pi \sqrt e} \bigl(
e_0(-e,-n+1)- e_0(-e,-n) -ee_2(-e,-n+1)\bigr)\ .
\end{equation}
Our normalisation of both $|e,n\rrangle_0$ and the dual invariant
$_0\llangle e,n|$ ensures that
$$\ _0\llangle e,n| (-1)^F u_1^{\frac12(L_E-R_E)} u_2^{\frac12 (L_N-R_N)}
 |e',n'\rrangle_0 \ = \ \delta(n'-n) \, \delta(e'-e)
 \, \chi_{\langle e,n\rangle} (u_1,u_2)$$
where $ \chi_{\langle e,n\rangle} (u_1,u_2) =
u_1^e\left(u_2^{n-1}- u_2^{n}\right)$ is the super-character of
the typical representation $\langle e,n\rangle$ of \gl. If we
re-scale the invariants $|e,n\rrangle_0$ and then send $e$ to zero
we obtain another family of invariants,
\begin{equation}
|0,n\rrangle_0 \ := \  \lim_{e\rightarrow 0} \sqrt{e}\,
|e,n\rrangle_0 \ = \ e_0(0,n) - e_0(0,n-1) \ \ .
\end{equation}
Similarly, we define the dual $\ _0\llangle 0,n|$ as a limit of $\
_0\llangle -e,-n+1| \sqrt{e}$. By construction, the states
$|0,n\rrangle_0$ and the associated linear forms possess
vanishing overlap with each other and with the states
$|e,n\rrangle_0$,
\begin{equation}
\ _0\llangle 0, n| u_1^{\frac12(L_E-R_E)} u_2^{\frac12 (L_N-R_N)}
 |e',n'\rrangle_0 \ = \ 0
\end{equation}
for all $e'$, including $e'=0$. This does certainly not imply that
$\ _0\llangle 0,n|$ acts trivially on the space of functions.
\smallskip

It is easy to see that the functions $|0,n\rrangle_0$ do not yet
span the space of invariants. What we are missing is a family of
additional states $|n\rrangle_0$  which is given by
$$ |n\rrangle _0 \ = \ \frac{1}{2\pi}\,  e_0(0,n) \ \ \ \mbox{ for } \ \ \
   n \ \in \ [0,1[ \ \ . $$
The corresponding dual co-invariants are given by the prescription
\begin{equation} \label{eq:nd}
\ _0\llangle n| \ = \ \frac{1}{2\pi} \, \int d\mu \, 
  \sum_{m \in \mathbb{Z}} e_2(0,-n+m+1) \ \ . 
\end{equation}
Our normalisation ensures that
\begin{equation}
\ _0\llangle n| (-1)^F u_1^{\frac12(L_E-R_E)} u_2^{\frac12
(L_N-R_N)} |n'\rrangle_0 \ = \ \delta(0) \, \delta(n'-n)\, 
\chi_{\langle n\rangle} (u_1,u_2)
\end{equation}
where $\chi_{\langle n \rangle}(u_1,u_2) = u_2^n$. The divergent 
factor $\delta(0)$ stems from the infinite volume of our target 
space and it could absorbed into the normalisation of the 
Ishibashi state. Let us observe that the co-invariants $ _0 
\llangle n|$ may be obtained by a limiting procedure from 
$\ _0\llangle e,n|$,
\begin{equation} \label{eq:nlimit}
\ _0 \llangle n | \ = \ - \lim_{e\rightarrow 0}\
\frac{1}{\sqrt{e}} \ \sum_m \, _0\llangle e,n+m|\ \ .
\end{equation}
A similar construction can be performed with the Ishibashi states
$|e,n\rrangle_0$ to give the formal invariants $\sum_m e_2(0,n+m)$.
They are formally dual to co-invariants given by $\int d\mu
e_0(0,-n+1)$. In our discussion, and in particular when we wrote
eq.\ \eqref{eq:nd}, we have implicitly equipped $\L2$ with a
topology that excludes to consider $\sum_m e_2(0,n+m)$ as
a true function. While the dual functional $\int d\mu
e_0(0,-n+1)$ does not suffer from any such problem, it so happens
not to appear in the construction of boundary states. This is why
we do not bother giving it a proper name.
\medskip

It is our aim now to determine the coupling of bulk modes to
branes in the minisuperspace limit. In the particle limit, the
bulk 1-point functions are linear functionals $f \mapsto \langle f
\rangle$ on the space $\L2$ of functions such that $\langle \ad_X
f \rangle = 0$, i.e.\ they are co-invariants. The first family of
co-invariants we shall describe corresponds to branes in generic
positions $(z_0,y_0)$. Since these are localised at a point
$(z_0,y_0)$ on the bosonic base and delocalised along the
fermionic directions, their density is given by
\begin{equation}
  \begin{split}
\rho_{(z_0,y_0)} & =\ -2i \sin(y_0/2)\, \delta(y-y_0)
\, \delta(z-z_0) \\[2mm]
& =\ - 2 i \sin(y_0/2)\,  \delta(y-y_0) \, \delta\bigl(x - i
\etam\etap (1-e^{-iy})^{-1} - z_0\bigr) \ \ .
  \end{split}
\end{equation}
The constant prefactor $-2i\sin(y_0/2)$ was chosen simply to 
match the normalisation of our boundary states below.  
Obviously, the density $\rho_{(z_0,y_0)}$ is invariant under the
adjoint action. It gives rise to a family of co-invariants through
the prescription
\begin{equation} \label{eq:coinv}
 f \mapsto \langle f \rangle_{\rho}  \ :=\
   \int d\mu \, \rho(x,y,\etapm) \, f(x,y,\etapm)\ \ .
\end{equation}
Geometrically, the integral computes the strength of the coupling
of a bulk mode $f$ to a brane with mass density $\rho$. It is not
difficult to check that our functional $\langle \cdot
\rangle_{(z_0,y_0)}$ admits an expansion in terms of dual
Ishibashi states as follows,
\begin{equation} \label{eq:MSSbst1}
\begin{split}
\langle\  \cdot\  \rangle_{(z_0,y_0)} & \equiv \ _0\langle z_0, y_0| \ = \
   \int dedn \, \sqrt{e} \,
e^{i(n-1/2) y_0 +iz_0e} \ _0 \llangle e,n|\\[2mm]
& \hspace*{-2cm} = \ \int_{e\neq 0} dedn\,\sqrt{e}\,e^{i(n-1/2) y_0
+iz_0e} \ _0 \llangle e,n| + \int dn \, e^{i(n-1/2) y_0} \
_0\llangle 0,n|\ \ .
\end{split}
\end{equation}
In the second line of this formula we have separated typical and
atypical contributions to the boundary state. Considering that the
state $_0\llangle 0,n|$ is obtained through the limiting procedure
$_0\llangle 0,n| = \lim_{e\rightarrow 0} \sqrt {e} \
_0\llangle e,n|$, the second term is the natural continuation of
the first. In this sense, we may drop the condition $e\neq 0$ in
the first integration and combine typical and atypical terms into
the single integral appearing in the first line. We observe that all
$\langle \cdot\rangle_{(z_0,y_0)}$ vanish on functions $e_0(e,n)$ with
$e=0$.
\smallskip

Let us now turn to the non-generic branes. These are localised
also in the fermionic directions. Hence, their density takes the
form
\begin{equation}
\rho^s_{z_0} \ = \ \, (-1)^s\,\delta(y-2\pi s)\,
 \delta(x-z_0)\, \delta (\etap)\, \delta(\etam)
\end{equation}
where $s$ is an integer. When this density is inserted into the
general prescription \eqref{eq:coinv}, we obtain another family of
co-invariants. Its expansion in terms of Ishibashi states reads
\begin{equation} \label{eq:MSSbst2}
\begin{split}
\langle\  \cdot\  \rangle^s_{z_0}& = \ _0\langle z_0;s| \ = \
  \int dedn
\,\frac{1}{\sqrt{e}}\, e^{2\pi i(n-1/2)s + iez_0} \ _0\llangle e,n|
 \\[2mm]
& \hspace*{-2cm} = \ \int_{e \neq 0} de dn \,\frac{1}{\sqrt{e}}\,
e^{2\pi i(n-1/2) s + iez_0} \  _0\llangle e,n| - \int_0^1 dn\,
e^{2\pi i (n-1/2) s} \  _0\llangle n|\ \ .
\end{split}
\end{equation}
Once more, the second line displays typical and atypical
contributions to the boundary state separately. In passing from
the first to the second line, we exploited $s \in \mathbb{Z}$
along with our observation \eqref{eq:nlimit}.
\smallskip

The two families $\langle \cdot \rangle_{(z_0,y_0)}$ with $y_0
\neq 2\pi s$ and $\langle \cdot \rangle^s_{z_0}$ are not entirely
independent. In fact, we note that boundary states from the
generic family may be `re-expanded' in terms of members from
the non-generic family when the paremeter $y_0$ tends to $2\pi
s$. The precise relation is
\begin{equation} \label{eq:branerel}
\lim_{y_0 \rightarrow 2\pi s} \, \langle f \rangle_{(z_0,y_0)} \ =
\ \frac{1}{i} \frac{\partial}{\partial z_0}
   \langle f \rangle^s_{z_0}\ \
\end{equation}
for all elements $f \in \L2$. We shall find that both
families of co-invariants can be lifted to the full field theory.
An analogue of relation \eqref{eq:branerel} also holds in the
field theory. It tells us that, for special values of the
parameters, branes from the generic family decompose into a
superposition of two branes from the non-generic family. Their
distance is finite for finite level but tends to zero as $k$ is
sent to infinity.

\subsection{Untwisted boundary states and their spectra}\label{section:bdystateuntwisted}

We are now prepared to spell out the boundary states and boundary
spectra for maximally symmetric branes with trivial gluing
conditions. As we have argued in the previous subsection, they come
in two different families. After a few comments on the relevant
Ishibashi states, we construct the boundary states for branes in
generic positions in the second subsection. Branes in non-generic
position are constructed in the third part of this section.

\subsubsection{Characters and Ishibashi states}

In this subsection we shall provide a list of untwisted Ishibashi
states from which the boundary states of the \GL\ WZNW model will
be built in consecutive subsections. By definition, an untwisted
Ishibashi state is a solution of the following set of linear
relations
\begin{equation}
\left( X_n +\bar X_{-n} \right) |\Psi\rrangle\ = \ 0 \ \ \ \
\text{for} \ \ \ X = E,N,\Psi^\pm \ \ .
\end{equation}
These relation lift our invariance conditions \eqref{eq:MSSinv} from
the particle model to the full field theory. It is obvious that
solutions must be in one-to-one correspondence to invariants in
the minisuperspace theory.
\smallskip

We now construct the Ishibashi states using our symplectic fermion
correspondence. Recall that the currents take the
form~\eqref{eq:finalcurrent}
\begin{align}
        J^E &= -k\del Y, & J^N &= -\del Z, &  J^- &= \sqrt{k}e^{Y^L}\del\chi^1, &   J^+ &= -\sqrt{k}e^{-Y^L}\del\chi^2, \\
        \bar{J}^E &= k\bar{\del} Y, & \bar{J}^N &=  \delbar Z, &
        \bar{J}^- &= -\sqrt{k}e^{-Y^R}\delbar\chi^1, &  \bar{J}^+ &= \sqrt{k}e^{Y^R}\delbar\chi^2.
\end{align}
Further, the fermions have mode expansion as in equation
\eqref{eq:modeexpansionfermions} and relations \eqref{eq:fermionrelations}
(or the twisted versions thereof) while the two scalars have expansion
\begin{equation}
    \begin{split}
        Y^L(z) \ &= \ Y_0^L + p_Y^L\, \ln z - \sum_{n\neq0}\, \frac{1}{n}\,  Y^L_n\, z^{-n},  \\
        Y^R(z) \ &= \ Y_0^R + p_Y^R\, \ln \zbar - \sum_{n\neq0}\, \frac{1}{n}\,  Y^R_n\, \zbar^{-n},  \\
        Z^L(z) \ &= \ Z_0^L + p_Z^L\, \ln z - \sum_{n\neq0}\, \frac{1}{n}\,  Z^L_n\, z^{-n},  \\
        Z^R(z) \ &= \ Z_0^R + p_Z^R\, \ln \zbar - \sum_{n\neq0}\, \frac{1}{n}\,  Z^R_n\, \zbar^{-n},
    \end{split}
\end{equation}
and relations
\begin{equation}
    [Y_n^{L,R},Z_m^{L,R}]\ = \ -m\delta_{n,-m}\qquad{\rm and}\qquad [Z_0^{L,R},p_Y^{L,R}]\ = \ [Y_0^{L,R},p_Z^{L,R}]\ = \ -1\, .
\end{equation}
To ensure locality we have $p_Y^L=p_Y^R$ and also $Z_0^L=Z_0^R$ for the
conjugate modes. However, we will not demand $p_Z^L=p_Z^R$ and
correspondingly not $Y_0^L=Y_0^R$ since $Z$ has an additive twist around
our winding states~\eqref{eq:finalvertexoperator}.

The energy momentum tensor is
\begin{equation}
    T(z)\ = \ \del Y\del Z -\half\epsilon_{ab}\del\chi^a\del\chi^b\qquad{\rm and}\qquad
    \bar{T}(\zbar)\ = \ \delbar Y\delbar Z -\half\epsilon_{ab}\delbar\chi^a\delbar\chi^b\, ,
\end{equation}
and thus the Virasoro modes are
\begin{equation}
    \begin{split}
    L_n \ = \ &-\sum_{m\, \in\,\Z}\, \normord{\chi^1_{n-m}\chi^2_m} +  \sum_{m\,\neq\,0,n} \normord{Y^L_{n-m}Z^L_m}+\\
            &+\sum_{m\,\neq\,0}(\normord{p_Y^L Z^L_m}+\normord{p_Z^L Y^L_m})+\de_{n,0}\,p^L_Yp^L_Z\,,\\
    \bar{L}_n \ = \ &-\sum_{m\, \in\,\Z}\, \normord{\bar\chi^1_{n-m}\bar\chi^2_m} +  \sum_{m\,\neq\,0,n} \normord{Y^R_{n-m}Z^R_m}+\\
            &+\sum_{m\,\neq\,0}(\normord{p_Y^R Z^R_m}+\normord{p_Z^R Y^R_m})+\de_{n,0}\,p^R_Yp^R_Z \,.\\
\end{split}
\end{equation}
We also need the zero modes of the currents corresponding to the Cartan generators $J^E$ and $J^N$:
\begin{equation}
    E_0 \ = \ -kp_Y^L,\qquad\bar{E}_0 \ = \ kp_Y^R,\qquad N_0 \ = \ -p_Z^L ,\qquad \bar{N}_0 \ = \ p_Z^R\, .
\end{equation}
Let us now consider the Ishibashi states. We start by spelling out the
Ishibashi conditions for the untwisted case. As noted above, the gluing
condition $J=\bar{J}$ means that the bosonic fields simply satisfy
Dirichlet conditions
\begin{equation}
    \del_u Y \ = \ \del_u Z \ = \ 0\, .
\end{equation}
Using these Dirichlet conditions for the field $Y=Y^L+Y^R$ the fermionic ones can be written as follows
\begin{equation}
 e^{Y_0^L}\del\chi^1 \ = \ -e^{-Y_0^R}\delbar\chi^1
    \qquad{\rm and}\qquad e^{-Y_0^L}\del\chi^2 \ = \ -e^{Y_0^R}\delbar\chi^2\, .
\end{equation}
Then correspondingly the Ishibashi conditions for the bosonic fields are
\begin{equation}
\begin{split}
    \left(Y^L_n-Y^R_{-n}\right)\, |\, I \,\rrangle \ &= \ \left(Z^L_n-Z^R_{-n}\right)\, |\, I \,\rrangle \ = \ 0\qquad\ n\neq0\\
    \left(p^L_Z-p^R_{Z}\right)\, |\, I \,\rrangle \ &= \left(p^L_Y-p^R_{Y}\right)\, |\, I \,\rrangle \ = \ 0\, ,\\
    \end{split}
\end{equation}
note that there is no conditions on the zero modes $Y^L_0$ and $Y^R_0$.
Further, the conditions for the fermionic ones are
\begin{equation}
    \big(e^{Y_0^L}\chi^1_n-e^{-Y_0^R}\bar\chi^1_{-n}\big)\, |\, I \,\rrangle \ = \
    \big(e^{-Y_0^L}\chi^2_n-e^{Y_0^R}\bar\chi^2_{-n}\big)\, |\, I \,\rrangle \ = \ 0 \, .
\end{equation}
The Ishibashi states clearly factorises into a bosonic and a fermionic part
and are easily constructed as follows. The typical primary of \GL,
$\<e,n\>_R$, is the representation with ground state
$|n,\mu_\lambda\rangle$ where $\lambda=e/k$ satisfying
\begin{equation}
    \begin{split}
    p_Z^L|n,\mu_\lambda\rangle \ &= \ p_Z^R|n,\mu_\lambda\rangle \ = \ n|n,\mu_\lambda\rangle\,, \\
    p_Y^L|n,\mu_\lambda\rangle \ &= \ p_Y^R|n,\mu_\lambda\rangle \ = \ \lambda |n,\mu_\lambda\rangle \, .
\end{split}
\end{equation}
Further, recall that the fermions have the mode expansion in the presence
of the ground state $\mu_\lambda$ \eqref{eq:twistedmodeexpansion}
\begin{equation}
    \begin{split}
        \chi^1(z,\bar{z})\ &= \ \sum_{n\,\in\,\Z+\lambda}\,\frac{1}{n}\,\chi^1_{n}\,z^{-n}\,+\,
                                \sum_{n\,\in\,\Z+\lambda^*}\,\frac{1}{n}\,\bar{\chi}^1_{n}\,\zbar^{-n}\,,\\
        \chi^2(z,\bar{z})\ &= \ \sum_{n\,\in\,\Z+\lambda^*}\,\frac{1}{n}\,\chi^2_{n}\,z^{-n}\,+\,
                                \sum_{n\,\in\,\Z+\lambda}\,\frac{1}{n}\,\bar{\chi}^2_{n}\,\zbar^{-n}\,,\\
    \end{split}
\end{equation}
where $\lambda^*=1-\lambda$.
Then the bosonic Ishibashi state is
\begin{equation}
    |n,e\rrangle_B \ = \  \exp\Bigl(\sum_{m>0}\frac{1}{m}\bigl(Y^L_{-m}Z^R_{-m}+Z^L_{-m}Y^R_{-m}\bigr)\Bigr)|n,\mu_\lambda\rangle_B\,, \\
\end{equation}
and the fermionic one is computed as \eqref{eq:twistedIshibashistate}
\begin{equation}\label{eq:twistedGLIshibashistate}
    |n,e\rrangle_F \ = \ \exp\Bigl(-\sum_{m\,>\,0}
    \frac{e^{Y_0^L+Y_0^R}}{m-\lambda}\chi^1_{-m+\lambda}\bar{\chi}^2_{-m+\lambda}
    -\frac{e^{-Y_0^L-Y_0^R}}{m-\lambda^*}\chi^2_{-m+\lambda^*}\bar{\chi}^1_{-m+\lambda^*}\Bigr)|n,\mu_\lambda\rangle_F \ .
\end{equation}
and the Ishibashi state is then the product of the two. The following
simple computations are crucial
\begin{equation}
    \begin{split}
        q^{L_0}e^{\pm Y^L_0} \ &= \ e^{\pm Y_0^L}q^{L_0\mp \frac{E_0}{k}} \qquad , \qquad
        Z^{N_0}e^{\pm Y^L_0} \ = \ e^{\pm Y^L_0}Z^{N_0\mp1}\,, \\
        q^{\bar{L}_0}e^{\pm Y^R_0} \ &= \ e^{\pm Y_0^R}q^{\bar{L}_0\pm \frac{\bar{E}_0}{k}} \qquad , \qquad
        Z^{\bar{N}_0}e^{\pm Y^R_0} \ = \ e^{\pm Y^R_0}Z^{\bar{N}_0\pm1}\,, \\
    \end{split}
\end{equation}
Introduce $L_0^c=\frac{1}{2}(L_0+\bar{L}_0)$ and
$N_0^c=\frac{1}{2}(N_0-\bar{N}_0)$ as usual. Then we get the fermionic contribution
of the overlap, that is
\begin{equation}
    _F\llangle n,e|q^{L_0^c+\frac{1}{12}}z^{N_0^c}(-1)^{F^c} |n,e\rrangle_F \ = \
    z^{n}(1-z^{-1})q^{\frac{1}{2}(\lambda-\frac{1}{2})^2-\frac{1}{24}}\prod_{n>0}(1-z^{-1}q^n)(1-zq^n)\,,
\end{equation}
and the bosonic
 \begin{equation}
    _B\llangle n,e|q^{L_0^c-\frac{1}{12}}z^{N_0^c}(-1)^{F^c} |n,e\rrangle_B \ = \
    -\frac{q^{n\lambda}}{\eta(\tau)^2}\,,
\end{equation}
where we normalised the dual state such that we get the minus sign. Then
in total, we arrive at
\begin{equation}
    \begin{split}
        \llangle n,e|q^{L_0^c}z^{N_0^c}(-1)^{F^c}|n,e\rrangle \ &= \
    z^{n-1}(1-z)\frac{q^{n\lambda+\frac{1}{2}(\lambda-\frac{1}{2})^2-\frac{1}{24}}}{\eta(\tau)^2}\prod_{n>0}(1-z^{-1}q^n)(1-zq^n) \\
    &= \ \hat\chi_{<e,n>}(z,\tau)\,.
    \end{split}
\end{equation}
So far we assumed $0<\lambda<1$, whenever $\lambda$ becomes zero our
Dirichlet symplectic fermion boundary states come into the game. There are
four of them. Denote by $|n,0\rangle$ the ground state with $N_0$
eigenvalue $n$, i.e.
\begin{align}
    N_0|n,0\rangle&=n|n,0\rangle,\qquad  E_0|n,0\rangle=0\,,\nonumber \\
    Y_m|n,0\rangle&=Z_m|n,0\rangle=\chi^a_m|n,0\rangle=\chi^a_0|n,0\rangle=0,\quad\textrm{for }m>0.
\end{align}
Then the Ishibashi states are
\begin{alignat}{3}\nonumber
        |n_0\rrangle \ &=& \
    \exp\Bigl(\sum_{m>0}\frac{1}{m}\bigl(Y^L_{-m}Z^R_{-m}+Z^L_{-m}Y^R_{-m}-
    e^{Y^L_0+Y_0^R}\chi^1_{-m}\bar{\chi}^2_{-m}+e^{-Y^L_0-Y^R_0}\chi^2_{-m}\bar{\chi}^1_{-m}\bigr)\Bigr)|n,0\rangle\\ 
    |n_\pm\rrangle \ &=& \ \xi^\pm|n_0\rrangle\quad\qquad\qquad\qquad\qquad\qquad\qquad\qquad\qquad\qquad\qquad\qquad\qquad\qquad\qquad\qquad\\
    |n\rrangle \ &=& \ \xi^-\xi^+|n_0\rrangle\qquad\qquad\qquad\qquad\qquad\qquad\qquad\qquad\qquad\qquad\qquad\qquad\qquad\qquad\qquad\nonumber
\end{alignat}
and we arrive at the following amplitudes
\begin{equation}\label{eq:ishiamplin}
    \begin{split}
        \llangle n_0|q^{L^c_0}z^{N^c_0}(-1)^{F^c}|n\rrangle \ &= \ \chi_0(\mu,\tau), \\
                \llangle n|q^{L^c_0}z^{N^c_0}(-1)^{F^c}|n_0\rrangle \ &= \ -\chi_0(\mu,\tau), \\
                \llangle n_\pm|q^{L^c_0}z^{N^c_0}(-1)^{F^c}|n_\mp\rrangle \ &= \ -\chi_0(\mu,\tau), \\
                \llangle n|q^{L^c_0}z^{N^c_0}(-1)^{F^c}|n\rrangle \ &= \ -2\pi i\tau\chi_0(\mu,\tau), \\
    \end{split}
\end{equation}
where
\begin{equation}
    \chi_0(\mu,\tau)=z^{n-1}q^{\frac{1}{12}}\prod_{n>0}(1-z^{-1}q^n)(1-zq^n)/\eta(\tau)^2.
\end{equation}
All other amplitudes vanish unless zero modes are inserted.

Let us now consider twist states $\mu_{\tilde\lambda}$ where
$\tilde\lambda\not\in\ ]0,1[\,$. We saw in section \ref{section:voa} that
such states are simply descendants of $\mu_\lambda$ where $\tilde\lambda
=\lambda+m$ for some integer $m$ and $\la\in\,]0,1[$. The state
$|n,\mu_{\tilde\lambda}\rangle$ satisfies the following conditions
\begin{equation}
      N_0|n,\mu_{\tilde\lambda}\rangle\ = \ n|n,\mu_{\tilde\lambda}\rangle\qquad \text{and} \qquad E_0|n,\mu_{\tilde\lambda}\rangle \ = \ k(\lambda+m)|n,\mu_{\tilde\lambda}\rangle\,.
\end{equation}
The Ishibashi state $|e,n\rrangle$ (with $e/k=\tilde\lambda=\lambda+m$) in this representation is obtained from the previously constructed ones as
\begin{equation}
        |n,e\rrangle \ = \ e^{m(Z_0^L-Z_0^R)}e^{m(Y_0^L+Y_0^R)}|n,e-mk\rrangle\, .
\end{equation}
The amplitude is computed using
\begin{equation}
        q^{L_0^c}e^{m(Z_0^L-Z_0^R)}\ =\ e^{m(Z_0^L-Z_0^R)}q^{L_0^c-mN_0^c}\,,
\end{equation}
and the spectral flow formulae provided in appendix~\ref{sc:Reps}
\begin{equation}
    \begin{split}
        \llangle n,e|q^{L_0^c}z^{N_0^c}(-1)^{F^c}|n,e\rrangle \ &= \ \hat\chi_{<e-mk,n+m>}(z-m\tau,\tau) \ = \ (-1)^m\hat\chi_{<e,n>}(z,\tau)\, .
    \end{split}
\end{equation}
A similar construction holds also for the atypical part.

\subsubsection{The generic boundary state}\label{section:generic}

In this section, we propose the boundary state corresponding to a
generic brane localised at $(z_0,y_0)$ with $y_0 \neq 2\pi s$ and perform a
non-trivial Cardy consistency check \cite{Cardy:1989ir}. For this purpose,
we need to know the modular properties of the characters. They are
easily computed with the help of \cite{Mumford} and we list them in
appendix \ref{sc:Mods}.

\begin{prp}{\rm (Generic boundary state)} The boundary state of branes
associated with generic position parameters $z_0$, $y_0$ is
\begin{equation}\label{eq:bst1}
	\begin{split}
    |z_0,y_0\rangle\ =\ &\sqrt{\frac{2i}{k}}\int_{\substack{e\neq mk\\ m\, \in \, \Z}} de dn\
     \exp\bigl(i(n-1/2)y_0+iez_0\bigr)\ \sin^{1/2}(\pi e/k)\ |n,e\rrangle \ - \\
     &\frac{\sqrt{2\pi i}}{k}\sum_{m\,\in\,\Z}\int dn\
     \exp\bigl(i(n-1/2)y_0+imkz_0\bigr)\ |n_0\rrangle^{(m)}\ .
     \end{split}
\end{equation}
We shall argue below that these boundary states give rise to
elementary branes if and only if the parameter
  $y_0 \not\in 2\pi\mathbb{Z}$.
\end{prp}
Before we show that our Ansatz for the generic boundary states
produces the expected boundary spectrum, let us make a few
comments. To begin with, it is instructive to compare the
coefficients of the Ishibashi states in $|z_0,y_0\rangle$ with the
minisuperspace result eq.\ \eqref{eq:MSSbst1}. If we send $k$ to
infinity, the factor $\sin^{1/2}(\pi e/k)$ is proportional to the
factor $\sqrt e$ that appears in the 1-point coupling of bulk
modes in the minisuperspace theory. The replacement $\sqrt e
\rightarrow \sin^{1/2} (\pi e/k)$ is necessary to ensure that the
field theory couplings are invariant under spectral flow \eqref{eq:SFKac}. 
\smallskip

In order to check the consistency of our proposal for the boundary
states with world-sheet duality, we compute the spectrum between a
pair of generic branes,
\begin{eqnarray}
\nonumber \langle
z_0,y_0|(-1)^{F^c}\tilde{q}^{L_0^c}\tilde{z}^{N_0^c}|z'_0,y'_0\rangle
\!\!&\ =\ \!\!\frac{2i}{k}\int de' dn'
e^{i(n'-\frac12)(y'_0-y_0)+ie'(z'_0-z_0)}\sin(\pi e'/k)
   \hat{\chi}_{\langle e',n'\rangle }(\tilde{\mu},\tilde{\tau})\\[2mm]
&\ =\ \hat{\chi}_{\langle e,n\rangle }(\mu,\tau)\ -
                \hat{\chi}_{\langle e,n+1\rangle }(\mu,\tau)
\label{eq:gg}
\end{eqnarray}
where the momenta $e,n$ are related to the coordinates of the
branes according to
$$ e\ =\ \frac{k(y'_0-y_0)}{2\pi} \ \ \ \ , \ \ \
n\ =\ \frac{k(z'_0-z_0)}{2\pi}-\frac{y'_0-y_0}{2\pi}\ .$$
To begin
with, the result is a combination of characters with integer
coefficients. Hence, it can be consistently interpreted as the
partition function for open strings that stretch in between the
two branes. If we put both branes into the same position
$(z_0,y_0)$, then the result specialises to
\begin{equation} \label{eq:gbspec}\langle
z_0,y_0|(-1)^{F^c}\tilde{q}^{L_0^c}\tilde{u}^{N_0^c}
|z_0,y_0\rangle \ = \ \hat{\chi}_{\langle 0,0\rangle }(\mu,\tau)\
         -  \hat{\chi}_{\langle 0,1\rangle }(\mu,\tau)
                \ = \ \hat{\chi}_{{\mathcal P}_0}(\mu,\tau).
\end{equation}
In the last step we have observed that the super-characters of the
representation spaces over the two atypical Kac modules $\langle
0,0\rangle$ and $\langle 0,1\rangle'$ combine into the character of
the representation that is generated from the projective cover
${\mathcal{P}}_0$. This outcome was expected: it signals that the
state space of open strings on a generic branes contains no
bosonic zero modes and two fermionic ones. The latter give rise to
the four ground states of the projective cover. This is in
agreement with the fact that generic branes stretch out along the
fermionic directions.
\smallskip

There is one important subtlety in our interpretation of the
result \eqref{eq:gbspec} that we do not want to gloss over. While
the character of the projective cover $\hat{\mathcal{P}}_0$ is
the same as that of the two affine Kac modules, the corresponding
representations are not. The characters are blind against the
nilpotent parts in $L_0$ and hence they cannot distinguish between
an indecomposable and its composition series. But for the
conformal field theory, the difference is important. In
particular, the generator $L_0$ is diagonalisable on all Kac
modules, atypical or not, but it has a nilpotent contribution in
the \agl-module over $\mathcal{P}_0$. Hence, if the boundary
spectrum does transform in $\hat{\mathcal{P}}_0$, then some
boundary correlators are guaranteed to display logarithmic
singularities when two boundary coordinates come close to each
other. The information we obtained from the boundary states using
world-sheet duality alone is not sufficient to make any rigorous
statements on the existence of such logarithms. But in the
minisuperspace limit $k \rightarrow \infty$ we have clearly identified
the projective cover $\mathcal{P}_0$ as the relevant structure.
Since $L_0$ is not diagonalisable in that limit, it cannot be so
for finite level $k$.

\subsubsection{Non generic point-like branes}

Let us now turn to the boundary states of non-generic untwisted
branes in the \GL\ WZNW model. From our discussion of the geometry
we expect them to be parameterised by a single real modulus $z_0$
and to possess a spectrum without any degeneracy of ground states.
These expectations will be met. Let us begin by spelling out the
formula for the non-generic boundary states.

\begin{prp}{\rm (Non-generic boundary states)}
The boundary states of elementary bra\-nes associated with
non-generic position parameters $z_0$ and $y_0=2\pi s, s \in
\mathbb{Z},$ are given by
\begin{equation}
    \begin{split}
        |z_0;s\rangle & =\ \frac{1}{\sqrt{2ki}}
        \int_{e \neq mk} dedn\ \exp\bigl(2\pi i (n-1/2) s +
   iez_0\bigr)\  \sin^{-1/2}(\pi e/k)\
         |n,e\rrangle\  \\[2mm] & \hspace*{2cm} - \,
       \frac{1}{\sqrt{2\pi i}}
       \, \sum_{m\in\Z} \int dn\ \exp\bigl(2\pi i (n-1/2) s +
   imk z_0\bigr)\ |n\rrangle^{(m)}\ \ .
    \end{split}
\end{equation}
\end{prp}
If we send the level $k$ to infinity in the boundary states
$|z_0;s\rangle$, then the coefficient of the Ishibashi
state $|e,s\rrangle$ gets replaced by $1/\sqrt{e}$ and thereby it
reproduces the coupling \eqref{eq:MSSbst2} of bulk modes in the
minisuperspace theory. Once more, the replacement
$1/\sqrt{e}\mapsto\sin^{-1/2}(\pi e/k)$ is necessary to ensure
spectral flow symmetry of the field theoretic couplings.%
\smallskip

Note that the non-generic boundary states only involve to the 
special family $|n\rrangle^{(m)}$ of atypical Ishibashi states. 
In case of generic boundary states, we had found non-vanishing 
couplings to the regular atypical Ishibashi states $|n_0\rrangle^{(m)}$. 
\smallskip 

Let us verify that the proposed boundary states produce a
consistent open string spectrum. In order to do so, we investigate
the overlap between two non-generic boundary states
$|z_0;s\rangle$ and $|z_0';s'\rangle$,
\begin{eqnarray} \nonumber
\langle
z_0;s|(-1)^{F^c}\tilde{q}^{L_0^c}\tilde{z}^{N_0^c}|z'_0;s'\rangle
& = & \int  \frac{de'dn'}{2ki}\, \frac{e^{2\pi i(n'-1/2)(s'-s) +
ie'(z_0'-z_0)}}{\sin(\pi e'/k)} \ \hat{\chi}_{\langle
e',n'\rangle} (\tilde{\mu},\tilde{\tau})\\[2mm]
& = & \, \hat{\chi}_{\langle n \rangle}^{(m)}(\mu,\tau)
\label{eq:ngng}
\end{eqnarray}
where the labels $n$ and $m$ in the
character are related to the branes' parameters through
\begin{equation}
n \ = \ \frac{k(z'_0-z_0)}{2\pi} + s-s' \ \ \ \ , \ \ \ m = s'-s\ \ .
\end{equation}
$\hat{\chi}_{\langle n \rangle}^{(m)}$ are characters of
atypical irreducible representation of \agl. For $m=0$ the
corresponding representations are generated from the 1-dimensional
irreducible atypical representations $\langle n\rangle$ of the
finite-dimensional Lie superalgebra \gl\ by application of current
algebra modes. The representations with $m \neq 0$ are obtained
{}from those with $m=0$ by spectral flow (see Appendix A).
\smallskip

The following limit for $t$ any integer shows that in equation ~\eqref{eq:ngng} is indeed a hidden $\tau$-dependence
\begin{equation}
    \lim_{e\rightarrow mk}\frac{1}{2ki}\int dn\ \frac{e^{2\pi itn}}{\sin(\pi e/k)} \ \hat{\chi}_{\langle
    e,n\rangle} (\tilde{\mu},\tilde{\tau}) \ = \ \int dn\ \tau\, e^{2\pi itn}\hat{\chi}_{\langle n\rangle}^{(m)} (\tilde{\mu},\tilde{\tau})\,.
\end{equation}
Thus we observe that the Ishibashi state $|n\rrangle$ \eqref{eq:ishiamplin} with its $\tau$-dependence is the natural atypical Ishibashi state contributing to the atypical boundary state.
\smallskip

We also want to look at the spectrum of boundary operators that
can be inserted on a boundary if we impose non-generic boundary
conditions with parameters $z_0$ and $s$. Specialising eq.\
\eqref{eq:ngng} to the case with $z'_0 = z_0$ and $s'=s$ we find
\begin{equation} \nonumber
\langle
z_0;s|(-1)^{F^c}\tilde{q}^{L_0^c}\tilde{u}^{N_0^c}|z_0;s\rangle
  \ =\ \hat{\chi}_{\langle 0 \rangle}^{(0)}(\mu,\tau) \ \ .
\end{equation}
Hence, the spectrum consists of states that are generated from a
single invariant ground state $|0\rangle$ by application of
current algebra modes with negative mode indices. In particular,
the zero modes of the fermions act trivially on ground states.
This is in agreement with our geometric insights according to
which non-generic branes are localised in all directions,
including the two fermionic ones.
\smallskip

Further, the overlap between a generic and a non-generic state is
\begin{equation}
    \begin{split}
       \langle z_0,y_0|(-1)^{F^c}\tilde{q}^{L_0^c}\tilde{z}^{N_0^c}|z'_0;s\rangle
\ &= \ \int  \frac{de'dn'}{k}\, e^{ i(n'-1/2)(2\pi s-y_0) + ie'(z_0'-z_0)} \ \hat{\chi}_{\langle
e',n'\rangle} (\tilde{\mu},\tilde{\tau})\\[2mm]
 &= \ \hat{\chi}_{\langle e,n \rangle}(\mu,\tau)\,, \\
\end{split}
\end{equation}
where
\begin{equation}
    n \ = \ \frac{k(z'_0-z_0)}{2\pi} + \frac{y_0}{2\pi}-s+\frac{1}{2} \ \ \ \ , \ \ \ \frac{e}{k}\ = \ s-\frac{y_0}{2\pi}\ \ .
\end{equation}

We may now ask what happens if we send the parameter $y_0$ of the
generic brane to $y_0=2\pi s$. From our formulae for boundary
states we deduce that
\begin{equation*}
|z_0,2\pi s\rangle\ =\ %
\int \frac{dedn}{\sqrt{2ki}}\, \frac{e^{ie(z_0+\frac{\pi}{k})}
     - e^{ie(z_0-\frac{\pi}{k})}}{\sin^{1/2}(\pi e/k)} \,
e^{2\pi i(n-1/2)s}\, |e,n\rrangle
\ =\ |z_0+\pi/k;s\rangle -
|z_0  - \pi/k;s\rangle\ .
\end{equation*}
In other words, when a generic brane is moved onto one of the
special lines $y_0 = 2\pi s$, it decomposes into a
brane-anti-brane pair. Its constituents sit in positions $z_0\pm
\pi/k$ and possess the same discrete parameter $s$. This relation
between non-generic branes and generic branes in non-generic
positions is a field theoretic analogue of the equation
\eqref{eq:branerel} we discovered in the minisuperspace
theory.%

\subsection{Comparison with Cardy's theory}\label{section:comparisoncardy}

Let us recall a few rather basis facts concerning branes in
rational unitary conformal field theory. For simplicity we shall
restrict to cases with a charge conjugate modular invariant and a
trivial gluing automorphism $\Omega$ (the so-called `Cardy case').
This will allow a comparison with the results of the previous
subsections. In the Cardy case, elementary boundary conditions
turn out to be in one-to-one correspondence with the irreducible
representations of the chiral algebra \cite{Cardy:1989ir}.
Let us label these by $J$,
with $J=0$ being reserved for the vacuum representation. The
boundary condition with label $J=0$ has a rather simple spectrum
containing only the vacuum representation $\mathcal{H}_0$. More
generally, if we impose the boundary condition $J=0$ on one side of
the strip and any other elementary boundary condition on the
other, the spectrum consists of a single irreducible ${\cal H}_J$.
Finally, the spectrum between two boundary conditions with label
$J_1$ and $J_2$ is determined by the fusion of $J_1$ and $J_2$. We
shall now discuss that all these statements carry over to
untwisted branes in the \GL\ WZNW model. The fusion procedure,
however, can provide spectra containing indecomposables that are
not irreducible.
\smallskip

\subsubsection{Brane parameters and representations}

We proposed that the \GL\ WZNW model possesses two families of
elementary branes. The first one is referred to as the generic
family and its members are parameterised by $(z_0,y_0)$ with $y_0
\neq 2\pi s, s \in \mathbb{Z}$. Boundary states for the generic
branes were provided in subsection \ref{section:generic}. These are also defined for
integer $y_0/2\pi$ but we have argued that the corresponding
branes are not elementary. They rather correspond to
superpositions of branes from the second family. This second
family consists of branes with only one continuous modulus $z_0$
and a discrete parameter $s$. Their boundary states can be found
in subsection 4.3.
\smallskip

There is one distinguished brane in this second family with
$z_0=0$ and $s=0$. We propose that it plays the role of the $J=0$
brane in rational conformal field theory. In order to confirm this
idea, we compute the spectrum of open strings stretching between
$z_0=0,s=0$ and any of the other elementary branes. If the second
brane is non-generic with parameters $z_0,s$, the relative
spectrum reads
\begin{equation}
\langle 0;
0|(-1)^{F^c}\tilde{q}^{L_0^c}\tilde{u}^{N_0^c}|z_0;s\rangle\ =\
\hat{\chi}_{\langle n\rangle}^{(m)}(\mu,\tau)\
\end{equation}
where the parameter $n$ on the character is
\begin{equation} \label{eq:posrep1}
 n \ = \ n(z_0;s) \ =\ \frac{kz_0}{2\pi} - s \ \ \ , 
\ \ \ m \ =\ m(z_0;s) \ = \ s\ .
\end{equation}
Indeed, we see that the open string spectrum corresponds to a
single irreducible atypical module of \agl, in agreement with the
expectations from rational conformal field theory.%
\smallskip

Let us now consider the case in which the second brane is located
in a generic position $(z_0,y_0)$. From the boundary state we find
\begin{equation}
\langle0;
0|(-1)^{F^c}\tilde{q}^{L_0^c}\tilde{u}^{N_0^c}|z_0,y_0\rangle \ =
\ \hat{\chi}_{\langle e,n\rangle }(\mu,\tau)\ ,
\end{equation}
where the parameters of the character on the right hand side are
\begin{equation} \label{eq:posrep2}
 e \ = \ e(z_0,y_0) \ = \ \frac{ky_0}{2\pi}  \ \ \ , \ \ \
 n \ = \ n(z_0,y_0) \ =\ \frac{k z_0}{2\pi}-\frac{y_0}{2\pi}+ \frac12\ \ .
\end{equation}
As long as $y_0/2\pi$ is not an integer, $e$ is not a multiple of
the level and hence, $\hat \chi_{\langle e,n\rangle}$ is the
character of a single irreducible representation of \agl.
\smallskip

At this point we have found that all our elementary branes are
labelled by irreducible representations of \agl. In case of the
elementary generic branes, the relation between the position
moduli $(z_0,y_0), y_0 \neq 2 \pi m,$ and representation labels
$\langle e,n\rangle, e \neq mk,$ is provided by eq.\
\eqref{eq:posrep2}. All typical irreducible representations of
\agl\ appear in this correspondence. For the non-generic branes the
relation between their parameters $(z_0;s)$ and the representation
labels of an atypical irreducible can be found in eq.\
\eqref{eq:posrep1}. Once more, all atypical irreducibles appear in
this correspondence. Hence, branes in the \GL\ WZNW model are in
one-to-one correspondence with irreducible representations of the
current superalgebra \agl, just as in rational conformal field
theory.

\subsection{Brane spectra and fusion}

Let us now analyse whether we can find the spectrum between a pair
of elementary branes through fusion of the corresponding current
algebra representations. For the convenience of the reader we have
listed the relevant fusion rules for irreducible representations
of the current superalgebra \agl\ in Appendix~\ref{sc:AffFus}.
\smallskip

The spectrum between two typical branes with parameters
$(z_0,y_0)$ and $(z'_0,y'_0)$ has been computed in eq.\
\eqref{eq:gg}. We can convert the brane parameters into
representation labels with the help of eq.\ \eqref{eq:posrep2} and
then exploit the known fusion product of the corresponding
representations. In case $y'_0-y_0 \neq 2\pi \mathbb{Z}$ we
find
\begin{eqnarray} \label{fus1} & &  \Bigl\langle
\frac{ky_0}{2\pi},\frac{kz_0}{2\pi}-\frac{y_0}{2\pi} + \frac12
\Bigr\rangle^\ast \otimes_F \Bigl\langle
\frac{ky'_0}{2\pi},\frac{kz'_0}{2\pi}-\frac{y'_0}{2\pi} +
\frac12
\Bigr\rangle \\[4mm] & & \hspace*{.5cm} \cong \ \Bigl\langle \frac{k(y'_0-y_0)}{2\pi},
\frac{k(z'_0-z_0)}{2\pi} - \frac{y'_0-y_0}{2\pi} + 1\Bigr\rangle \,
\oplus \, \Bigl\langle \frac{k(y'_0-y_0)}{2\pi},
\frac{k(z'_0-z_0)}{2\pi} - \frac{y'_0-y_0}{2\pi}\Bigr\rangle' \nonumber
\end{eqnarray}
Here, $\otimes_F$ denotes the fusion product and we used the rule
$\langle e,n\rangle^\ast=\langle -e,-n+1\rangle'$ for the
conjugation of representations. Then we inserted the known fusion
rules while keeping track of whether the representation is
fermionic or bosonic. The result agrees nicely with the true
spectrum we computed earlier.
\smallskip

When the difference $(y'_0-y_0)/2\pi = m$ is an integer, the
fusion of the two representations on the left hand side of
\eqref{fus1} results in a single indecomposable. It is the image
of the affine representation over the projective cover
$\hat{\mathcal{P}}_{(k(z'_0-z_0)-(y'_0-y_0))/2\pi}$ under $m$ units of
spectral flow, i.e.\ %
\begin{equation} \label{fus2} \Bigl\langle
\frac{ky_0}{2\pi},\frac{kz_0}{2\pi}-\frac{y_0}{2\pi} + \frac12
\Bigr\rangle^\ast \otimes_F \Bigl\langle
\frac{ky'_0}{2\pi},\frac{kz'_0}{2\pi}-\frac{y'_0}{2\pi} +
\frac12 \Bigr\rangle \ = \ \Bigl(\mathcal{P}^{(m)}_{(k(z'_0-z_0)-(y'_0-y_0))/2\pi}\Bigr)'
\end{equation}
where $m = (y'_0-y_0)/2\pi$. Our minisuperspace theory along with
the boundary states confirm this result in the case $y_0 = y_0'$
and $z_0 = z_0'$ (see our discussion at the end of section 4.2).
For other choices of the parameters, we only see that the fusion
rules provide a representation with the correct character. Whether
the true state space is given by a single indecomposable or by a
sum of Kac modules or even irreducibles cannot be resolved
rigorously with the methods we have at our disposal. Nevertheless,
it seems very likely that the projective cover is what appears
since this is the only result which is also consistent with
spectral flow symmetry.
\smallskip

The fusion between atypical irreducibles is rather simple. It
leads to a prediction  for the spectrum between two non-generic
branes that should be checked against our earlier result
\eqref{eq:ngng},
$$ \Bigl(\,\Bigl\langle \frac{kz_0}{2\pi} -s\Bigr\rangle^{(s)}\Bigr)^\ast \otimes_F
   \,\Bigl\langle \frac{kz'_0}{2\pi}-s'\Bigr\rangle^{(s')} \ \cong \
   \,\Bigl\langle \frac{k(z'_0-z_0)}{2\pi}+s-s'\Bigr\rangle^{(s'-s)}\ \ .
$$
Once more, the findings from world-sheet duality are consistent
with the fusion prescription. There is one final check to be
performed. It concerns the spectrum between a non-generic brane
with parameters $(z_0;s)$ and a generic one with moduli
$(z_0,y_0)$. From the fusion we find
\begin{equation}
\Bigl(\Bigl\langle \frac{kz_0}{2\pi} -s \Bigr\rangle^{(s)}\Bigr)^\ast
  \otimes_F\Bigl\langle
\frac{ky'_0}{2\pi},\frac{kz'_0}{2\pi}-\frac{y'_0}{2\pi} + \frac12
\Bigr\rangle \ = \ \Bigl\langle -s k + \frac{ky'_0}{2\pi},
\frac{k(z'_0-z_0)}{2\pi} - \frac{y_0'}{2\pi}+s+\frac12\Bigr\rangle
\ \ .
\end{equation}
It may not come as a big surprise that this
fusion rule correctly predicts the spectrum between a generic and
a non-generic brane. In fact, from our formulae for boundary
states and modular transformation we find
\begin{equation}
  \begin{split}
& \bigl\langle z_0;s\bigr| (-1)^{F^c}\tilde q^{L_0^c} \tilde
u^{N^c_0} \bigl|z'_0,y_0'\bigr\rangle \ = \ \hat \chi_{\langle e
,n \rangle}(\mu,\tau)
\\[4mm] \text{where}  \ &\ \ \ e \ = \ - k s +
\frac{ky_0'}{2\pi}\ \ , \ \ n \ =  \frac{k(z'_0-z_0)}{2\pi} -
\frac{y_0'}{2\pi}+s + \frac12\ .
  \end{split}
\end{equation}
In conclusion we found that the spectra between any pair of
elementary branes may be determined by the fusion of the
corresponding irreducible representations. It is important to
stress that the fusion product of irreducible representations can
produce representations that are not fully reducible.

\subsection{Twisted boundary state}

The group of outer automorphisms of the Lie superalgebra \gl\ is of order 2. We already discussed the boundary states belonging to the trivial one.
The non-trivial one defines the following gluing conditions on the currents
\begin{equation}
    J^{E}\ = \ -\bar{J}^E \ \ \ , \ \ \ J^{N}\ = \ -\bar{J}^N \ \ \ , \ \ \
    J^{+}\ = \ -\bar{J}^- \ \ \ , \ \ \ J^{-}\ = \ \bar{J}^+\qquad{\rm for}\ z\ = \ \zbar\, .
\end{equation}
This translates into Neumann conditions for the bosonic and the fermionic fields, that is
\begin{equation}
    \del_n Y\ = \ \del_n Z \ = \ 0\qquad{\rm for}\ z\ = \ \zbar\,
\end{equation}
implying especially that the left movers of $Y$ coincide with its right movers up to the zero modes
\begin{equation}
    Y^L - Y^R \ = \ Y^L_0 - Y^R_0 \qquad{\rm for}\ z\ = \ \zbar\, .
\end{equation}
Thus the gluing conditions for the fermions are
\begin{equation}
    e^{Y_0^L}\del\chi^1\ = \ e^{Y_0^R}\delbar\chi^2\qquad,\qquad e^{-Y_0^L}\del\chi^2\ = \ -e^{-Y_0^R}\delbar\chi^1 \qquad{\rm for}\ z\ = \ \zbar\, .
\end{equation}

The boundary state $|\,\Omega\rrangle$ is easily constructed as before. It
has to satisfy
\begin{equation} \label{gluetwisted}
    \begin{split}
    (Y_n^L + Y^R_{-n})\, |\, \Omega\rrangle \ =(p_Y^L + p^R_{Y})\, |\, \Omega\rrangle \ =\ 0\,, \\
    (Z_n^L + Z^R_{-n})\, |\, \Omega\rrangle \ =(p_Z^L + p^R_{Z})\, |\, \Omega\rrangle \ =\ 0\,, \\
    (e^{Y_0^L}\chi^1_n + e^{Y_0^R}\bar{\chi}^2_{-n})\, |\, \Omega\rrangle \ =\ 0\,, \\
    (e^{-Y_0^L}\chi^2_n - e^{-Y_0^R}\bar{\chi}^1_{-n})\, |\, \Omega\rrangle \ =\ 0 \,,\\
\end{split}
\end{equation}
which can be computed to be
\begin{equation} \label{boundtwisted}
                |\, \Omega\,\rrangle \ = \ \sqrt{\pi/i}\exp \Bigl( \sum_{n=1}^\infty
\frac{1}{n} \bigl(Y^L_{-n}Z^R_{-n} + Z^L_{-n}Y^R_{-n}
- e^{Y^R_0-Y^L_0}\chi^2_{-n} \bar{\chi}^2_{-n} + e^{Y^L_0-Y^R_0}\chi^1_{-n} \bar{\chi}^1_{-n}\bigr)\Bigr)|0,0\,\rangle \, .
\end{equation}
Here, $|0,0\,\rangle$ denotes the vacuum defined by
$\chi_n^a|0,0\,\rangle=0$ for $n\geq0$ and
$Z^{L,R}_n|0,0\,\rangle=Y^{L,R}_n|0,0\,\rangle=p_Y^{L,R}|0,0\,\rangle=p_Z^{L,R}|0,0\,\rangle=0$
for $n>0$. The dual boundary state is constructed analogously.

Our main aim now is to compute some non-vanishing overlap of the twisted
boundary state $|\Omega\rrangle$. This requires the insertion of the
invariant bulk field $\chi^1\chi^2$, i.e.
\begin{equation}
    \begin{split}\label{eq:amplitude}
 \llangle\Omega\, |\, \tilde q^{L_0^c}(-1)^{F^c}\,
   \tilde z^{N_0^c}\, \chi^1\chi^2
         \, |\, \Omega\rrangle\, =\ \frac{\pi}{2k}\int de dn \ \frac{\hat{\chi}_{\langle e,n\rangle}
        (\tau,\mu)} {\sin(\pi e/k)} \ .
    \end{split}
\end{equation}
where $L_0^c = (L_0+\bar{L}_0)/2$ and $N_0^c = (N_0+\bar{N}_0)/2$ are
obtained from the zero modes of the Virasoro field and the current $N$.
Here the normalisation in~\eqref{boundtwisted} by $\sqrt{\pi/i}$ was
important.
This amplitude will be tested in section \ref{sec:coratyp}.

\subsection{Mixed amplitudes and their open strings}

The \GL-symplectic fermion correspondence allowed us to construct
boundary states explicitly. The new explicit formulation also allows
us to compute new quantities such as overlaps for atypicals
\begin{equation}
    \begin{split}\label{eq:mixedatypicalamplitude}
  \llangle\Omega\, |\, \tilde q^{L_0^c}(-1)^{F^c}\,  \tilde z^{N_0^c}\,     \, |\, z_0;s\rrangle\  &=\
                  \sqrt{\frac{1}{2}}(-1)^s\, \prod\limits_{n=0}^\infty\, \frac{(1-\tilde q^n)}{(1+\tilde q^n)}\\
  &= \ (-1)^s\,q^{\frac{1}{32}}\,\prod\limits_{n=0}^\infty\, \frac{(1-q^{n+\frac{1}{4}}) (1-q^{n+\frac{3}{4}})}
          {(1-q^{n+\frac{1}{2}})^2}\,.\\
    \end{split}
\end{equation}
Note the independence on $z$, no matter whether we take $N_0^c$ as in the previous section or as in the untwisted case ($N_0^c=(N_0\pm\bar N_0)/2$), which is natural since there does not exist a distinguished choice for $N_0^c$ for mixed amplitudes.

The corresponding open string theory is easily constructed using our
previous experience. That is, we demand untwisted gluing conditions on the
negative real line
\begin{equation}
    \begin{split}
        \del_u Y\ &= \ \del_u Z \ = \ 0\ \ , \\
        e^{Y_0^L}\del\chi^1 \ &= \ -e^{-Y_0^R}\bar{\del}\chi^1\ \ , \\
        e^{-Y_0^L}\del\chi^2 \ &= \ -e^{Y_0^R}\bar{\del}\chi^2  \qquad\qquad{\rm for}\ z\ = \ \bar{z} \qquad{\rm and}\ z+\bar{z}\ <\ 0\ ;\\
    \end{split}
\end{equation}
and twisted on the positive one
\begin{equation}
      \begin{split}
          \del_n Y\ &= \ \del_n Z \ = \ 0\ \ , \\
        e^{Y_0^L}\del\chi^1 \ &= \ e^{Y_0^R}\bar{\del}\chi^2\ \ , \\
        e^{-Y_0^L}\del\chi^2 \ &= \ -e^{-Y_0^R}\bar{\del}\chi^1 \qquad\qquad{\rm for}\ z\ = \ \bar{z} \qquad{\rm and}\ z+\bar{z}\ >\ 0\ ,\\
    \end{split}
\end{equation}
Then the fermions have a monodromy of order four around the origin
\begin{equation}
    \del\chi^1(ze^{2\pi i}) \ = \ i\,\del\chi^1(z)\qquad,\qquad \del\chi^2(ze^{2\pi i}) \ = \ -i\,\del\chi^2(z)\,,
\end{equation}
and the bosons a monodromy of order two
\begin{equation}
    \del Y(ze^{2\pi i}) \ = \ -\del Y\qquad,\qquad \del Z(ze^{2\pi i}) \ = \ -\del Z(z)\, .
\end{equation}
Thus the fermions have mode expansion
\begin{equation}
    \begin{split}
    \chi^1(z) \ &= \  \sum_{n\,\in\,\Z+\frac{3}{4}}\frac{1}{n}\chi^1_nz^{-n}\,,\\
    \chi^2(z) \ &= \  \sum_{n\,\in\,\Z+\frac{1}{4}}\frac{1}{n}\chi^2_nz^{-n}\,,\\
\end{split}
\end{equation}
and the bosons
\begin{equation}
    \begin{split}
        Y(z) \ &= \ \sum_{n\,\in\,\Z+\frac{1}{2}}\, \frac{1}{n}\,  Y_n\, z^{-n}\,,  \\
        Z(z) \ &= \ \sum_{n\,\in\,\Z+\frac{1}{2}}\, \frac{1}{n}\,  Z_n\, z^{-n}\,,  \\
    \end{split}
\end{equation}
We define the ground state to be bosonic if $s$ (the position parameter of
the non-generic brane) is even and fermionic if it is odd. The partition
function is then
\begin{equation}
    \tr(q^{L_0}(-1)^F) \ = \ (-1)^s\,q^{\frac{1}{32}}\,\prod\limits_{n=0}^\infty\, \frac{(1-q^{n+\frac{1}{4}}) (1-q^{n+\frac{3}{4}})}
          {(1-q^{n+\frac{1}{2}})^2}\, .
  \end{equation}
 The amplitude involving typical fields requires as usual zero mode insertions, i.e.
\begin{equation}
    \begin{split}\label{eq:mixedtypicalamplitude}
  \llangle\Omega\, |\, \tilde q^{L_0^c}(-1)^{F^c}\,
  \tilde z^{N_0^c}\, \chi^1\chi^2   \, |\, z_0,y_0\rrangle\  &=\  \frac{\sqrt{2}\pi}{k}e^{-iy_0/2}\frac{\prod\limits_{n=0}^\infty (1-\tilde q^n)}{\prod\limits_{n=0}^\infty (1+\tilde q^n)}\\
  &= \ \frac{2\pi}{k}e^{-iy_0/2}\, q^{\frac{1}{32}}\,\prod\limits_{n=0}^\infty\, \frac{(1-q^{n+\frac{1}{4}}) (1-q^{n+\frac{3}{4}})}
          {(1-q^{n+\frac{1}{2}})^2}\,,\\
    \end{split}
\end{equation}
and its open string spectrum can be constructed as in the symplectic fermion case.

In summary, we have been able to give a complete discussion of Cardy
boundary states in the \GL\ WZNW model. This was only possible due to the
new formulation in terms of symplectic fermions. As a result, we saw that
indeed also for the Lie supergroup \GL\ Cardy's condition holds, i.e. any
amplitude of two boundary states indeed describes an open string spectrum. Further, we saw that 
the overlap between two boundary states with trivial gluing conditions is given by fusion.
The twisted boundary state then gives a one-dimensional extension of the fusion ring.

\subsection{Conclusions}

In this section we have studied maximally symmetric branes in the
WZNW model on the simplest supergroup \GL . Following previous
reasoning for bosonic models \cite{Alekseev:1998mc} we have shown
that such branes are localised along (twisted) super-conjugacy
classes, an insight that generalises to other
supergroup target spaces (section \ref{section:geometry}). As in the case of the $p=2$ triplet 
theory \cite{Gaberdiel:2006pp}, untwisted branes turn out 
to be in one-to-one correspondence with irreducible representations 
of the current algebra. This correspondence relies on the existence 
of an `identity' brane whose spectrum consists of the irreducible 
vacuum representation only. The spectrum between the identity and any
other elementary brane is built from a single irreducible of \agl\
and any such irreducible appears in this way. Moreover, one can
compute the spectrum between any two elementary branes by fusion
of affine representations. What we have just listed are
characteristic features of Cardy's theory for rational
non-logarithmic conformal field theories. Our work proves that
they extend at least to one of the simplest logarithmic field
theory and it seems very likely that they hold more generally in
all WZNW models on (type I) supergroups, see also 
\cite{Gaberdiel:2006pp} for related findings in the $p=2$ 
triplet theory. 
\smallskip

In spite of these parallels to bosonic WZNW models, branes on
supergroups possess a much richer spectrum of possible geometries.
Whereas Dirichlet branes on a purely bosonic torus, for example,
are all related by translation, we discovered the existence of
atypical lines on the bosonic base of the \GL\ WZNW model. The
distance between any two such neighboring parallel lines is
controlled by the level $k$. When a typical untwisted brane is
moved onto one of these lines, it splits into two atypical ones.
Individual atypical branes possess a single modulus that describes
their dislocation along the atypical lines. In order for them to
leave an atypical line they must combine with a second atypical
brane. Processes of this kind model the formation of long
multiplets from shorts. Hence, on more general group manifolds,
more than just two atypical branes may be required to form a
generic brane. Let us stress, however, that the notions of long
(typical) and short (atypical) multiplets which are relevant for
such processes derive directly from the representation theory of the affine
Lie superalgebra. Thereby, all spectral flow symmetries are
built into our description. We also wish to point out the obvious
similarities with so-called fractional branes at orbifold singularities, 
see e.g.\ the discussions in section 4.3 of \cite{Recknagel:1998ih}.  
\smallskip

Another interesting and new feature of branes on \GL\ is the
occurrence of boundary spectra that cannot be decomposed into a
direct sum of irreducibles. In particular we have shown that the
spectrum of boundary operators on a single generic brane is
described by the projective cover of the vacuum module.
For more general group manifolds, we expect the corresponding
projective cover to be present as well, though along with
additional stuff. The generator $L_0$ of dilatations is not 
diagonalisable on projective covers, see e.g.\ 
\cite{Schomerus:2005bf}. According to the usual reasoning, this 
implies the existence of logarithmic singularities in boundary 
correlation functions on branes in generic positions. As we have 
remarked before, the modular bootstrap alone did not allow for such a
strong conclusion as it is blind to all nilpotent contributions
within $L_0$. But in addition to the standard conformal field
theory analysis, our investigation of the \GL\ WZNW model also
draws from the existence of the geometric regime at large level
$k$. The presence of projective covers is easily understood in the
minisuperspace theory and it must persist when field theoretic
corrections are taken into account. \smallskip

We would also like to note, that there is a related paper \cite{Gaberdiel:2007jv}
which discusses branes in triplet models with $p \geq 2$. 
The results of Gaberdiel and Runkel show that branes in triplet 
models share many features with the outcome of our analysis. In 
particular, for trivial gluing automorphism, branes in both models 
are labelled by irreducible representations of the chiral algebra. 
Also the labels for relevant Ishibashi states follow the same 
pattern: We have found one `generic' Ishibashi state for each 
Kac module and an exceptional family with members being associated 
to atypical blocks. When the same rules are applied to the triplet 
models, we obtain a set of Ishibashi states that seems closely 
related to those used in \cite{Gaberdiel:2007jv}. Furthermore, 
Gaberdiel and Runkel also find that the partition function
for any pair of boundary conditions may be determined by 
fusion of representations. The existence of a geometric 
regime for the \GL\ WZNW model allows us to go one step 
further. It gives us full control over the structure of the 
state space and thereby also over the nilpotent contributions 
to $L_0$ which are not visible in partition functions. Fusion 
of \agl\ representations was shown to correctly reproduce the 
state spaces of boundary theories in the \GL\ WZNW model. Let 
us stress, however, that the triplet and the \GL\ WZNW model 
are close cousins (see e.g. the discussion in \cite{Quella:2007hr}). 
It would therefore be somewhat premature to claim that all these 
structures will be present in more general logarithmic 
conformal field theories.

\section{The boundary \GL\ WZNW model}\label{sec:boundarygl11}

This section gives a complete discussion of volume filling branes in the \GL\ WZNW model. We compute those correlation functions which specify the boundary theory completely, these are the bulk one-point functions, the bulk-boundary two-point functions and boundary three-point functions. The results are those of \cite{Creutzig:2008ek}.

\subsection{Volume filling brane: The classical action}

Our aim in this section is to discuss the classical
description of volume filling branes in the \GL\ WZNW model. To
begin with, we spell out the standard action of the WZNW model
with so-called twisted boundary conditions. Their geometric
interpretation as volume filling branes with a non-zero B-field is
recalled briefly. In order to set up a successful computation
scheme for the quantum theory later on, we shall need a different
formulation of the theory. As in the bulk theory, computations of
correlations functions require a Kac-Wakimoto like representation
of the model \cite{Schomerus:2005bf}. Finding such a first order
formalism for the boundary theory is not entirely straightforward.
We shall see that it requires introducing an additional fermionic
boundary field.

\subsubsection{The boundary WZNW model}

Following our earlier work on WZNW models for type I supergroups,
we parametrise the supergroup \GL\ through a Gauss-like
decomposition of the form
$$ g \ = \ e^{i \eta_- \psi^-} \, e^{ixE + iyN} \, e^{i \eta_+
\psi^+} $$ where $E,N$ and $\psi^\pm$ denote bosonic and fermionic
generators of \gl, respectively. In the WZNW model, the two even
coordinates $x,y$ become bosonic fields $X,Y$ and similarly, two
fermionic fields $c_\pm$ come with the odd coordinates $\eta_\pm$.
Let us now consider a boundary WZNW model with the action
\begin{equation}\label{eq:bdySWZW}
    \begin{split}
S_{\text{WZNW}}(X,Y,c_\pm)\ =\  &-\frac{k}{4\pi i}\int_\Sigma
d^2z\
  \left( \del X\bar{\del}Y+\del
  Y\bar{\del}X+2e^{iY}\del\cp\bar{\del}\cm\right)\,  + \\[2mm]
        & +\frac{k}{8\pi i}\int du \ e^{iY}(\cp+\cm)\del_u(\cp+\cm)\, ,
    \end{split}
\end{equation}
where $u$ parametrises the boundary of the upper half plane.
Variation of the action leads to the usual bulk equations of
motion along with the following set of boundary conditions\\[0mm]
\begin{equation}\label{eq:bdyeom}
    \begin{split}
      \del_v Y\ = \ 0 \ \ \ \ & , \ \ \ \
         2 \del_v X \ = \ e^{iY}(\cp+\cm)\,  \del_u(\cp+\cm)\ ,
         \\[3mm]
      \pm 2  \del_v c_\pm\, & = \ 2 i \del_u c_\mp - (\cm+\cp)\,
       \del_u Y \ . \\[3mm]
    \end{split}
\end{equation}
Here, we have used the derivatives $\del_u = \del + \bar{\del}$
and $\del_v = i (\del - \bar{\del})$ along and perpendicular to
the boundary. The equations \eqref{eq:bdyeom} imply Neumann
boundary conditions for all four fields of our theory, i.e.\ we
are dealing with a volume filling brane. Since the normal
derivatives of the fields $X$ and $c_\pm$ do not vanish, our brane
comes equipped with a B-field. A more detailed discussion of the
brane's geometry can be found in our recent paper
\cite{Creutzig:2007jy}.

In order to see that our boundary conditions preserve the full
chiral symmetry, we recall that the holomorphic currents of the
\GL\ WZNW model take the form
\begin{equation*}
    \begin{split}
    J^E\, =\ ik\del Y \ \ \ \ \ \ \ \ \ & , \ \ \ \
    J^N\, =\ ik\del X -k\cm \del\cp \, e^{iY} \ , \\[2mm]
    J^-\, =\ ike^{iY}\del\cp\  \ \  & , \ \ \ \
    J^+ \, = \ ik\del\cm -k\cm \del Y \ , \\[1mm]
\end{split}
\end{equation*}
and similarly for the anti-holomorphic currents,
\begin{equation*}
    \begin{split}
        \bar{J}^E \, =\ -ik\bar{\del} Y \ \ \ \ \ \ \ \  & , \ \ \ \
        \bar{J}^N\, =\ -ik\bar{\del} X +k \bar{\del}\cm\, \cp\, e^{iY} \ ,
        \\[2mm]
            \bar{J}^+\, =\ ike^{iY}\bar{\del}\cm \  \
            \ \ & , \ \ \ \
        \bar{J}^-\, =\ ik\bar{\del}\cp-k\cp\bar{\del} Y \ . \ \
        \\[1mm]
\end{split}
\end{equation*}
If we plug the boundary conditions \eqref{eq:bdyeom} into these
expressions for chiral currents, we obtain the gluing condition
$ J^X(z) = \Omega \bar J^X(\bar z)$ for $X=E,N,\pm$ and all along
the boundary at $z= \bar z$. Here, the relevant gluing automorphism
$\Omega$ is obtained by lifting the automorphism
\begin{equation}
    \label{eq:twisted}
    \Omega(E)\, = \,  - E, \ \ \Omega(N)\, =\,  - N
    , \ \ \Omega(\psi^+) \, =\,  - \psi^-, \ \ \Omega(\psi^-)\,
   = \, \psi^+\  
\end{equation}
from the finite dimensional superalgebra \gl\ to the full affine
symmetry. In \cite{Creutzig:2007jy} we called these gluing
conditions {\em twisted} and showed that there is a unique brane
corresponding to this particular choice of $\Omega$.

\subsubsection{First order formulation}

Computations of bulk and boundary correlators in the presence of
twisted D-branes shall be performed in a first order formalism. In
the bulk, it is well-known how this works \cite{Schomerus:2005bf}.
There, the bulk action is built of a free field theory involving
two additional fermionic auxiliary fields $b_\pm$ of weight
$\Delta(b_\pm) = 1$ along with the original fields $X,Y$ and
$c_\pm$,
\begin{equation}
    \begin{split}
        S_{0;\text{cl}}^{\text{bulk}}[X,Y,c_\pm,b_\pm] \ = \
        &-\frac{k}{4\pi i}\int_\Sigma d^2z\ \left( \del X\bar{\del}Y+
        \del Y\bar{\del}X\right)
        \\[2mm]
         &-\frac{1}{2\pi i}\int_\Sigma d^2z\
         \left(\bp\del\cp+\bm\bar{\del}\cm\right)  \ . \\
        \end{split}
\end{equation}
We placed a subscript `cl' on the action to distinguish it from the
action we shall use in our path integral computations later on. If
the following bulk marginal interaction term is added to the free
field theory,
\begin{equation}
    \label{eq:Sbulk}
    S^{\text{bulk}}_{\text int}[X,Y,c_\pm,b_\pm] \ = \
    -\frac{1}{2k\pi i}\int_\Sigma d^2z\ e^{-iY}\bm\bp \ 
\end{equation}
the equations of motion for $b_\pm$ read $b_- = k \del c_+ \exp
iY$ and $b_+ = - k \bar \del c_- \exp iY$ so that we recover the
bulk WZNW-model upon insertion into the first order action. In
extending this treatment to the boundary sector, we are tempted to
add the ``square root'' of the bulk interaction as a boundary
term. This is indeed what happens for the closely related $AdS_2$
branes in $AdS_3$ \cite{Fateev:2007wk}. Here, however, it cannot
possibly be the right answer, at least not without a proper notion
of what we mean by taking the square root. In fact, the naive
square root of $\bm\bp\exp(-iY)$ is something like
$b_\pm\exp(-iY/2)$, i.e.\ a fermionic operator. It makes no sense
to add such an object to the bulk theory. In order to take a
bosonic square root of the bulk interaction, we introduce a new
fermionic boundary field $C$ of weight $\Delta(C) = 0$ and add
the following terms to the bulk theory,
\begin{eqnarray}
    S^{\text{bdy}}_0[X,Y,c_\pm,b_\pm,C] & = &
    \frac{1}{8\pi i}\int du \ \left( kC\del_uC+4(\cp+\cm)\bp\right)  \\[2mm]
    S^{\text{bdy}}_{\text{int}}[X,Y,c_\pm,b_\pm,C]
     & = & -\frac{1}{2\pi i}\int du \ e^{-iY/2}\bp C \ .
\end{eqnarray}
The idea to involve an additional fermionic boundary field in the
action of supersymmetric brane configurations is not new. It was
initially proposed in \cite{Warner:1995ay} and has been put to use
more recently \cite{Kapustin:2003ga,Brunner:2003dc} in the context
of matrix factorisations. Our boundary action resembles the one
Hosomichi employed to treat branes in $N=2$ super Liouville theory
\cite{Hosomichi:2004ph}. The full \gl\ boundary theory now takes
the form
\begin{equation}
    S[X,Y,\cpm,b_\pm,C] \ = \ S_{0,\text{cl}}^{\text{bulk}} +
      S_0^{\text{bdy}} + S^{\text{bulk}}_{\text{int}}+
       S^{\text{bdy}}_{\text{int}} \ = \ S_{0,\text{cl}} + S_{\text{int}}
\end{equation}
where
\begin{equation}
 \begin{split}
         S_{0,\text{cl}} \ = & \ -\frac{k}{4\pi i}\int_\Sigma d^2z\ \left(\del X\bar{\del}Y+
             \del Y\bar{\del}X \right)   \\[2mm]
         & - \frac{1}{2\pi i}\int_\Sigma d^2z\
         \left(\cp\del\bp+\cm\bar{\del}\bm\right)
         +\frac{1}{8\pi i}\int du \ kC\del_uC \ ,\\[3mm]
     S_{\text{int}} \ =  & \ -\frac{1}{2k\pi i}\int_\Sigma d^2z\ e^{-iY}\bm\bp \
     - \frac{1}{2\pi i}\int du \ e^{-iY/2}\bp C\ \ .  \\
     \end{split}
\end{equation}
Here, we have performed a partial integration on the kinetic term
for the bc-system, thereby absorbing the contribution $b_+(c_- +
c_+)$ from the boundary action. This is similar to the
case of $AdS_2$ branes in $AdS_3$ \cite{Fateev:2007wk}. In order to
complete the description of the classical action, we add the
following Dirichlet boundary condition for the fields $b_\pm$,
\begin{equation}
b_+(z) + b_-(\bar z) \ = \ 0 \ \ \ \mbox{ for } \ \ z \ = \ \bar z
\ .
\end{equation}
If the action is varied with this boundary condition, we recover
the boundary equations of motion \eqref{eq:bdyeom}. More
precisely, we obtain four equations among boundary fields. Two of
these can be used to determine the boundary fields $C$ and $b_+ =
- b_-$ through $X,Y$ and $c_\pm$,
\begin{equation}\label{Cbpm}
 C \ = \ \, e^{iY/2} \, (\cp + \cm) \ \ \ , \ \ \ \pm2 b_\pm \ = \
 k\, e^{iY/2} \partial_u C \ .
\end{equation}
The four equations among boundary fields along with the bulk
equations motion for $b_\pm$ imply the eqs. \eqref{eq:bdyeom}. We
leave the details of this simple computation to the reader.

We have now set up a first order formalism for the twisted brane
on GL(1$|$1). Let us stress again that is was necessary to
introduce an additional fermionic field $C$ on the boundary of the
world-sheet. Above we have motivated this new degree of freedom by
our desire to take a bosonic square root of the bulk interactions.
But there is another, more geometric, way to argue for the
additional field $C$. We mentioned before that the
first order formalism for the GL(1$|$1) WZNW model is very similar
to that for the Euclidean $AdS_3$, only that the bosonic
coordinates $\gamma, \bar \gamma$ of the latter are replaced by
fermionic ones. The first order formalism for $AdS_2$ branes in
$AdS_3$ was set up in \cite{Fateev:2007wk} and it describes a brane
that is localised along a 1-dimensional subspace of the
$\gamma\bar \gamma$ plane. Correspondingly, only a single $\gamma$
zero mode remains after imposing the boundary conditions. The
brane on GL(1$|$1) we are attempting to describe, however, is
volume filling and therefore it extends in both fermionic
directions. Therefore, we need two independent fermionic zero
modes. These are provided by the zero modes of the three fields
$c_\pm$ and $C$. Note that these fields are related by equation
\eqref{Cbpm}.

\subsection{Volume filling branes: The quantum theory}

Our next step is to develop a computational scheme for correlation
functions in the boundary WZNW model with twisted boundary
conditions. We shall use the first order formulation of section
2.2 as our starting point and consider the full WZNW model as a
deformation of a free field theory involving the fields
$X,Y,c_\pm,b_\pm$ and the fermionic boundary field $C$. This free
field theory will be described in more detail in the first
subsection. The definition of vertex operators and their
correlation functions in the WZNW model is the subject of
subsection 3.2.

\subsubsection{The free theory and its correlation functions}

Our strategy is to employ the first order formulation we set up in
the previous section. In order to do so, we have to add a few
comments on the measures we are using in the path integral
treatment. To begin with, the supergroup invariant measure of the
WZNW model is given by
\begin{equation} \label{WZNWmeasure}
    d\mu_{\text{WZNW}} \ \sim \ \, \mathcal{D}X\mathcal{D}Y
      \mathcal{D}(e^{iY/2} \cm)\mathcal{D}(e^{iY/2}\cp)\ .
\end{equation}
This gets multiplied with ${\cal D}b_+ {\cal D}b_- {\cal D} C$
when we pass to the first order formalism. But in the following we
would like to employ the standard free field measure
$$ d \mu_{\text{free}}\ \sim \ \mathcal{D}X\mathcal{D}Y
      \mathcal{D}\cm\mathcal{D}\cp\ .
      $$
The two measures are related by a Jacobian of the form (see
e.g.\ \cite{Gerasimov:1990fi} for similar computations)
\begin{equation}
 \begin{split}\label{measure}
    d\mu_{\text{WZNW}} \ &=  \  \left(\sdet(
  G^{ab} e^{iY}{\del_a}e^{-iY}\del_b)\right)^{-1} \ d\mu_{\text{free}} \\[3mm]
  \ &=  \ e^{\frac{1}{8\pi}\int dudv\ \sqrt{G}
   (-G^{ab}\del_a\  Y{\del}_bY+i\mathcal{R}Y) +\frac{1}{8\pi}\int du\
   i\sqrt{G}\mathcal{K}Y } \ d\mu_{\text{free}}  .\\
\end{split}
\end{equation}
Here, $G_{ab}$ is the metric on the world-sheet, ${\cal R} =
\del_a \del^a \log G $ and ${\cal K} = \frac{1}{2i}\partial_v \log G $
are its Gaussian and geodesic curvature, respectively. These two
quantities feature in the Gauss-Bonnet theorem for surfaces with
boundary,
\begin{equation}
 \frac{1}{4\pi}\int_\Sigma dudv\ \sqrt{G} \mathcal{R}+\frac{1}{4\pi}\int du\
 \sqrt{G}\mathcal{K}  \ = \ \chi(\Sigma) \ = \ 1 \ ,
\end{equation}
where $\chi(\Sigma) = 1$ is the Euler characteristic of the disc.
We can now pass to the upper half plane again where all curvature
is concentrated at infinity. The effect of the curvature terms in
the WZNW measure is to insert a background charge $Q_Y =
\chi(\Sigma)/2 = 1/2 $ for the field $Y$ at infinity. In addition,
the measure \eqref{measure} also contains a term that is quadratic
in $Y$. We simply add this to the free part of our action, i.e.\
we define
\begin{equation}
  \begin{split} \label{qS0}
   S_0 \ = &\  -\frac{1}{4\pi i}\int_\Sigma d^2z\ \left( k\, \del X\bar{\del}Y+
             k \, \del  Y\bar{\del}X
         -  \del Y\bar{\del}Y\right)  \\[2mm]
         & - \frac{1}{2\pi i}\int_\Sigma d^2z\
         \left(\cp\del\bp+\cm\bar{\del}\bm\right)
         +\frac{1}{8\pi i}\int du \ kC\del_uC \ ,
  \end{split}
\end{equation}
Note, that the new term in the actions modifies the formula for
the current $J^N$ by adding an additional $\del Y$ and similarly
for the anti-holomorphic partner.

In our path integral we now integrate with the free field theory
measure $d \mu_{\text{free}}$ over all fields subject to the
boundary condition $b_+ + b_- = 0$. Configurations for the other
fields are not constrained in the path integral. In the free
quantum field theory, they satisfy the linear (``Neumann'')
boundary conditions
\begin{equation}
    \begin{split}
       \del_v Y \ = & \ 0 \ \ \ \ , \ \ \ \ \ \ \ \,
       \del_v  X \ = \ 0  \ , \\[2mm]
       \del_u C \ = & \ 0 \ \ \ \ , \ \ \
       \cp+\cm\ = \ 0 \ \ . \\
       \end{split}
\end{equation}
These equations are satisfied in all correlation functions or,
equivalently, as operator equations on the state space of the free
field theory. Note that, according to the last equation, the zero
modes of $c_+$ and $c_-$ coincide in our free boundary theory. The
necessary second fermionic zero mode is exactly what is provided
by the field $C$.

Arbitrary correlation functions in the free field theory can now
easily be computed with the help of Wick's theorem. All we need to
use is the following list of operator product expansions
\begin{equation}
    \begin{split}
         X(z,\bar{z})Y(z,\bar{z})  \ \sim \ &
          \frac{1}{k}\ln|z-w|^2+\frac{1}{k}\ln|z-\bar{w}|^2
          \\[2mm]
         \cm(z)b_-(w)  \ \sim \ \ \frac{1}{w-z} &  \ \
        \ \
         \cp(\bar{z})b_+(\bar{w})  \ \sim \
         \frac{1}{\bar{w}-\bar{z}}\\[2mm]
         \cm(z)b_+(\bar{w}) \ \sim \
         \frac{1}{z-\bar{w}} & \ \ \ \
         \cp(\bar{z})b_-(w)  \ \sim  \
         \frac{1}{\bar{z}-w}\\[2mm]
         C(v)C(u) \ &\ \sim \  \frac{2\pi i}{k}\sign(v-u)\ \ \ .
    \end{split}
\end{equation}
Let us remark that a non-vanishing correlation function in the
free field theory requires that the fields $c$ outnumber the
insertions of $b$ by one. Furthermore, $C$ must be inserted an odd
number of times. We also recall that there is a non-vanishing
background charge $Q_Y = 1/2$ for the field $Y$. On the disk, the
corresponding U(1) charges of all tachyon vertex operators must
add up to $Q_Y \chi(\Sigma) = 1/2$ in order for the correlator to
be non-zero. These rules imply that the 1-point function of the
bulk identity field vanishes. In order to normalise the vacuum
expectation value, we require that
\begin{equation}
    \langle \, \left(\cm(z)-\cp(\bar{z})\right) \, C(u) \,
     e^{ieX(z,\bar z) + inY(z,\bar{z})}\, \rangle_0 \ = \
      \delta(e) \delta(n-1/2)  \ .
\end{equation}
Note that the product of fields in brackets is the simplest
expression that meets all our requirements: The U(1)$_Y$ charge of
the tachyon vertex operators is $m = 1/2$, we inserted one $c_\pm$
and no field $b_\pm$ and multiplied with a single $C$ in order to
make the total insertion bosonic again.

\subsubsection{Correlation functions in boundary WZNW model}

Now that we have learnt how to perform computations in the free
field theory described by the action \eqref{qS0}, we would like
to add our interaction term
\begin{equation}\label{Sint}
    S_{\text{int}} \ = \
    -\frac{1}{2k\pi i}\int_\Sigma d^2z\ e^{-iY}\bm\bp \ -
    \frac{1}{2\pi i}\int du \ e^{-iY/2}\bp C \ .
\end{equation}
The idea is to calculate correlators of the full boundary
WZNW model perturbatively, i.e.\ by expanding the exponential of
the interaction in a power series. Even though there is a priori
an infinite number of terms to be considered, only finitely many
contribute to our perturbative expansion. This is very similar to
what has been observed in the bulk model \cite{Schomerus:2005bf}.

Before we can spell out precise formulae for the quantities we
want to compute, we need to explain how to associate free field
theory vertex operators to the fields of the interacting WZNW
model. The latter are in one-to-one correspondence with functions
on the supergroup GL(1$|$1) and they may be characterised by
their behaviour with respect to global gl(1$|$1) transformations.
We shall first recall from \cite{Schomerus:2005bf} how this works
for bulk fields.

Let us begin by collecting a few basic facts about the space of
functions on the supergroup \GL\ \cite{Schomerus:2005bf}. As for
any other group or supergroup, $\L2$ carries two graded-commuting
actions of the Lie superalgebra gl(1$|$1). These are generated by
the following right and left invariant vector
fields\\[-2mm]
\begin{equation}
\begin{array}{llll}
 R_E = i\del_x \ , & R_N  = i\del_y+\etam\del_- \ ,
& R_+ = -e^{-iy}\del_+-i\etam\del_x \ ,  & R_- = -\del_-\ , \\[4mm]
 L_E = -i\del_x \ ,  & L_N = -i\del_y-\etap\del_+ \ ,
& L_- = e^{-iy}\del_- -i\etap\del_x \ ,  & L_+ = \del_+\ .\\[3mm]
\end{array}
\end{equation}
A typical irreducible multiplet for \gl\ is 2-dimensional. Hence,
typical irreducible multiplets of the combined left and right
action are spanned by four functions in the supergroup. As in
\cite{Schomerus:2005bf} we shall combine these functions into a
$2\times 2$ matrix of the form
\begin{equation}
    \begin{split}\label{phi}
        \varphi_{\<-e,-n+1\>}\ = \ e^{iex+iny}
        \left( \begin{array}{cc}
            1& \etam    \\
                 \etap &e^{-1}e^{-iy}+\etap\etam \\
                        \end{array}
                           \right)
        \end{split}
\end{equation}
The rows span the typical irreducibles $\<-e,-n+1\>$ of the right
regular action. Columns transform in the  representations
$\<e,n\>$ of the left regular action. Note that $\varphi_{\<
e,n\>}$ is only well defined for $e\neq0$, i.e.\ in the typical
sector of the minisuperspace theory.

Following \cite{Schomerus:2005bf}, the bulk vertex operators in
the free field theory are modelled after the matrices
$\varphi_{\<e,n\>}$. More precisely, let us introduce typical bulk
operators through
\begin{equation}
    \begin{split}\label{vertV}
        V_{\<-e,-n+1\>}(z,\bar{z})\ = \ e^{ieX+inY}
        \left( \begin{array}{cc}
                     1& \cm  \\
                 \cp & \cp\cm \\
                        \end{array}
                           \right)
        \end{split}
\end{equation}
Since the weight of the fermionic fields $c_\pm$ vanishes,
all four fields in this matrix possess the same conformal
dimension,
\begin{equation}
    \Delta_{(e,n)} \ = \ \frac{e}{2k}(2n-1+\frac{e}{k})\ .
\end{equation}
Note that one of the terms in the lower left corner of the
minisuperspace matrix $\varphi_{\<e,n\>}$ has no analogue on the
vertex operator $V_{\<-e,-n+1\>}$. We consider this term as
`subleading'. It is reconstructed when we build correlation
functions of the interacting WZNW model (see
\cite{Schomerus:2005bf} and \cite{Gotz:2006qp} for more details).
\smallskip

Let us now repeat the previous analysis for the boundary fields.
Since our twisted brane is volume filling, the relevant space of
minisuperspace wave functions is again the space $\L2$ of all
functions on the supergroup \GL . But this time, it comes equipped
with a different action of the Lie superalgebra \gl. In fact,
minisuperspace wave functions as well as boundary vertex operators
are now distinguished by their transformation under a single twisted
adjoint action $\ad^{\Omega}_X = R_X + L^\Omega_X$ of \GL\ on
$\L2$. Explicitly, the generators of \gl\ transformations are
given by
\begin{equation}
    \begin{split}\label{eq:twad}
      \ad_E^\Omega\ &=\ 2i\del_x \qquad\qquad , \ \
      \ad_N^\Omega\ =\ 2i\del_y+\etap\del_++\etam\del_- \ , \\
      \ad_-^\Omega\ &=\ \del_+-\del_- \qquad , \ \
      \ad_+^\Omega\ =\ -e^{-iy}(\del_-+\del_+)+i(\etap-\etam)\del_x \ \ . \\
    \end{split}
\end{equation}
Under the twisted adjoint action of \gl\ on $\L2$, each typical
multiplet appears with two-fold multiplicity
\cite{Creutzig:2007jy}. Once more, we propose to assemble the
corresponding four functions into a $2\times 2$ matrix of the form
\begin{equation}
    \begin{split} \label{eq:bdrmat}
        \psi_{\<-2e,-2n+1\>}\ = \ e^{iex+iny}
        \left( \begin{array}{cc}
                             1& \etap-\etam \\
                 \eta &2e^{-1}e^{-iy/2}+(\etap-\etam)\eta \\
                        \end{array}
                           \right)
        \end{split}
\end{equation}
where we introduced the shorthand $\eta=e^{iy/2}(\etam+\etap)$.
The reader is invited to check that the two rows of this matrix
each span the 2-dimensional typical irreducible $\<-2e,-2n+1\>$
under the twisted adjoint action \eqref{eq:twad} of the
superalgebra \gl.

Boundary vertex operators are modelled after the matrices $
\psi_{\<-2e,-2n+1\>}$ more or less in the same way as in the case
of bulk fields,
\begin{equation}
    \begin{split}\label{vertU}
        U_{\<-2e,-2n+1\>}(u)\ = \ e^{ieX+inY}
        \left( \begin{array}{cc}
                     1& \cp-\cm   \\
                 C & (\cp-\cm)C \\
                        \end{array}
                           \right)\ \ .
        \end{split}
\end{equation}
Again, we dropped the $y$-dependent term in the lower right corner
of the matrix \eqref{eq:bdrmat}. Eventually, we will see how this
term is recovered in boundary correlation functions. The main new
aspect of the prescription \eqref{vertU}, however, concerns the
appearance of the fermionic boundary field $C$ that we inserted in
place of the function $\eta$. This substitution is motivated by
the classical equation of motion \eqref{Cbpm}.
\smallskip

After this preparation we are able to spell out how correlation
functions of bulk and boundary fields can be computed for the
interacting WZNW model. More precisely, we define,
\begin{equation}
    \begin{split} \label{corrdef}
  \left\langle \prod_{\nu=1}^m\Phi_{\langle
        e_\nu,n_\nu\rangle}
         (z_\nu,\bar{z}_\nu)\right.&\left.\prod_{\mu=1}^{m'}
          \Psi_{\langle e_\mu,n_\mu\rangle}(u_\mu)
          \right\rangle \ =
          \\[2mm]
        & \sum_{s=0}^\infty\, \frac{(-1)^s}{s!}\ \left\langle
        \,  \left(S_{\text{int}}\right)^s \
        \prod_{\nu=1}^m V_{\langle e_\nu,n_\nu\rangle}(z_\nu,\bar{z}_\nu)
        \prod_{\mu=1}^{m'}
U_{\langle e_\mu,n_\mu\rangle}(u_\mu)\right\rangle_0 \ . \\
    \end{split}
\end{equation}
Here, $S_{\text{int}}$ is the interaction \eqref{Sint} and all
correlation functions on the right side are to be computed in the
free field theory \eqref{qS0}. The relevant vertex operators $V$
and $U$ were introduced in equations \eqref{vertV} and
\eqref{vertU} above. For later use we also note that bosonic
correlators can be determined by means of the following standard
formula,
\begin{eqnarray} \nonumber
        & &\hspace*{-5mm} \left\langle\prod_{\nu=1}^m V_{(e_\nu,n_\nu)}(z_\nu,\bar{z}_\nu)
        \prod_{\lambda=1}^{m'} V_{(e_\lambda,n_\lambda)}(u_\lambda)\right\rangle
          = \delta({\textstyle \sum}_{\nu=1}^mn_\nu+{\textstyle \sum}_{\lambda=1}^{m'}
          n_\lambda+{\textstyle{\frac12}})
         \delta({\textstyle \sum}_{\nu=1}^me_\nu+{\textstyle \sum}_{\lambda=1}^{m'}e_\lambda)\\[3mm]
        & & \ \ \ \
         \times \ \prod_{\nu>\mu}|z_\nu-z_\mu|^{-2\alpha_{\nu\mu}}\prod_{\nu>\mu}
        |z_\nu-\bar{z}_\mu|^{-2\alpha_{\nu\mu}}
         \prod_{\nu,\lambda}|z_\nu-u_\lambda|^{-4\alpha_{\nu\lambda}}
         \prod_{\lambda>\kappa}|u_\lambda-u_\kappa|^{-4\alpha_{\kappa\lambda}}
            \\[5mm]
          &  & \quad  \text{where} \qquad \qquad \qquad \alpha_{\nu\mu}\ =\
        -n_\nu\frac{e_\mu}{k}-n_\mu\frac{e_\nu}{k}-\frac{e_\nu
        e_\mu}{k^2}
\nonumber
\end{eqnarray}
and $V_{(e_\nu,n_\nu)}= \exp(ieX+inY)$ are bosonic vertex
operators. As in the bulk theory it is easy to see that the all
expansions \eqref{corrdef} truncate after a finite number of
terms. In fact, the inserted bulk and boundary vertex operators on
the right hand side of eq.\ \eqref{corrdef} contain at most $2m +
m'$ fermionic fields $c_\pm$. Since each interaction term from
$S_{\text{int}}$ contributes at least one insertion of $b_\pm$, we
conclude that terms with $s \geq 2m + m'$ vanish.

\subsection{Solution of the boundary WZNW model}

A boundary conformal field theory is uniquely characterised by the
bulk-boundary and the boundary operator product expansions. We
shall now employ the perturbative calculational scheme we developed
in the previous section in order to determine these data. After
a short warm-up with the discussion of bulk 1-point functions, we
determine the bulk-boundary 2-point function in the second
subsection. The 3-point function of boundary fields is addressed
in subsection 4.3.

\subsubsection{Bulk 1-point function}

The bulk 1-point function is the simplest non-vanishing quantity
in a boundary conformal field theory. It contains the same
information as the boundary state. For volume filling branes, the
boundary state was determined in our previous work
\cite{Creutzig:2007jy}. Our first aim now is to reproduce our old
result through our new perturbative expansion.

The 1-point function of a typical bulk field $\Phi_{\<e,n\>}$ is
computed by inserting a single vertex operator \eqref{vertV} into
the expansion \eqref{corrdef}. Since bulk vertex operators contain
at most two fields $c$, the only non-zero terms can come from
$s=0,1$. The term with $s=0$ contains no insertion of the
interaction and it vanishes identically. So, let us see what
happens for $s=1$. In this case, only the insertion of the
boundary interaction can contribute. The results is
\begin{eqnarray*}
    & & \langle \Phi_{\langle e,n\rangle}(z,\bar{z})\rangle \ = \
    \frac{i}{2\pi} \int du\ \langle e^{-iY(u)/2}b_+(u)C(u)
    V_{\<e,n\>}(z,\bar z)\rangle\\[2mm]
   &  &  \ \ \ \ = \ E^1_1 \delta(e) \delta(n-1) \frac{1}{4\pi i}
    \int du\ \left(\frac{1}{u-\bar z} - \frac{1}{u-z}\right)
    \ = \  \int d\mu\ \varphi_{\langle e,n\rangle} \ .
\end{eqnarray*}
Here, $E^1_1$ is the elementary matrix which has zeroes everywhere
except in the lower right corner. Note that the only field with
non-vanishing 1-point function has conformal weight $\Delta =0$.
Hence, there is no dependence on the insertion point $(z,\bar z)$.
In the last line we have expressed the numerical result as an
integral of the matrix valued function \eqref{phi} over the
supergroup \GL. The integration is performed with the Haar
measure
\begin{equation} d\mu \ = \ 2^{-1} e^{-iy}dxdyd\etap d\etam\ \ .
\end{equation}
Since the Haar measure is \gl\ invariant, the integral of
$\varphi_{\<e,n\>}$ is an intertwiner from $\<e,n\> \otimes
\<e,n\>$ to the trivial representation. This proves that the
expectation value we computed has the desired transformation
behaviour.

\subsubsection{Bulk-boundary 2-point function}

Now we want to compute the full bulk-boundary 2-point function.
It is quite useful to determine the general form of this 2-point
function first before we enter the detailed calculations. Let us
suppose for a moment that our calculations were guaranteed to give
a \gl\ covariant answer. Then it is clear that the bulk-boundary
2-point function can be written as
\begin{eqnarray} \label{2pt}
\< \Psi_{\<2e',2n'\>}(0)\, \Phi_{\<-e,-n+1\>}(iy,-iy)\>
& = & \sum_{\nu=0,1} C_{\nu}(e)
  \frac{\<\psi_{\<2e',2n'\>} \, \varphi_{\<-e,-n+1\>}\>_\nu}
  {|y|^{2 \Delta_\nu} } \\[2mm]
  \text{ where }\quad   \Delta_0 \ = \
   \frac{2e}{k}\bigl(2n-1+\frac{e}{k}\bigr)
    & & \text{and}\qquad
    \Delta_1 \ = \ \frac{2e}{k}\bigl(2n-\frac{1}{2}+
   \frac{e}{k}\bigr)\ .
\end{eqnarray}
The structure constants $C_\nu(e)$ are not determined by the \gl\
symmetry. We will calculate them perturbatively below (see eqs.\
\eqref{C0} and \eqref{C1} below). The expressions in the numerator
on the right hand side are certain \gl\ intertwiners which are
defined by
\begin{eqnarray}  \label{int}
\<\psi_{\<2e',2n'\>} \, \varphi_{\<-e,-n+1\>}\>&\!\!\!=&
\hspace*{-3mm} \int d\mu\, \psi_{\<2e',2n'\>} \,
\phi_{\<-e,-n+1\>}
 =:\sum_{\nu=0,1} \<\psi_{\<2e',2n'\>} \,
  \varphi_{\<-e,-n+1\>}\>_\nu\quad   \\[2mm]
\text{where} \quad  \quad & & \hspace*{-5mm}
   \<\psi_{\<2e',2n'\>} \,
  \varphi_{\<-e,-n+1\>}\>_\nu \ = \
 \delta(e-e')\delta(n-n'-\nu/2)\  G_\nu
\label{intnu}
\end{eqnarray}
is the part of the full integral that contains the factor
$\delta(n-n'-\nu/2)$. Understanding the previous formulae requires
some input from the representation theory of \gl\ (see e.g.
\cite{Schomerus:2005bf} for all necessary details). Let us start
with the matrix $\varphi_{\<-e,-n+1\>}$. Under the  twisted
adjoint action of \gl\ this multiplet transforms in the tensor
product
$$\<-e,-n+1\> \otimes \<-e,-n+1\>  \ = \
  \< -2e,-2n + 2\> \oplus \< -2e,-2n+1\> \ \ . $$
Hence, there exist only two matrices $\psi_{\<2e',2n'\>}$ for
which the integral \eqref{int} does not vanish. These are the
matrices $\psi_{\<2e,2n\>}$ and $\psi_{\<2e,2n-1\>}$. The two
non-vanishing terms are used to define the the symbols
\eqref{intnu}. A similar analysis can now be repeated for the
fields in the WZNW model. We conclude immediately, that the
2-point function can only have two contributions. By \gl\
symmetry, these must be proportional to the intertwiners
\eqref{intnu}. The \gl\ symmetry, however, does not fix an overall
constant $C_\nu$ that can depend on the parameters of the fields.
Finally, the exponents $\Delta_\nu$ are simply determined by the
conformal dimensions of bulk and boundary fields. Let us point out
that the entire discussion leading to the expression \eqref{2pt}
is based on the global \gl\ symmetry. Since we have not yet shown
that our perturbative computations respect the action of \gl\ it
will be important to verify that the form of the 2-point function
comes out right.

In our perturbative computation, there are at most three fields
$c_\pm$ inserted and hence we only have to determine the expansion
terms for $s=0,1,2$. Contributions to the $\nu=0$ term in the
2-point function \eqref{2pt}, i.e. to the correlator with the
boundary field $\Psi_{\<2e,2n\>}$, can only come from $s=0$.
In fact, insertions of an interaction term - bulk or boundary -
would violate the conservation of $Y$-charge. Computation
without any insertion of an interaction are easily performed,
e.g.
\begin{equation}
\langle U^{11}_{\langle2e',2n'\rangle}(0)
  V^{00}_{\langle -e,-n+1\rangle}(iy,-iy)\rangle \ = \
   - \delta(n-n')\, \delta(e-e') |y|^{-4e/k(2n-1/2+e/k)}
\end{equation}
Here, we have introduced the notation $U^{\epsilon'\epsilon}$
and $V^{\epsilon'\epsilon}$ for matrix elements. The field
$U^{11}_{\<2e',2n'\>}$, for example, denotes the lower right
corner etc. The computation of the associated integral
\eqref{intnu} with $\nu=0$ is equally simple and allows us to
read off that
\begin{equation}  \label{C0}
C_0(e,n) \ = \ 1 \ \ .
\end{equation}
Let us note that there are other combinations of bulk and
boundary fields that can have a non-zero 2-point function
without any insertion of interactions. In all those cases
one may repeat the above calculation to find the same
coefficient $C_0 =1$, in agreement with \gl\ symmetry.

Next we would like to address the coefficient $C_1$ in the
expression \eqref{2pt}. $Y$-charge conservation implies that
its only contributions are associated with a single insertion
of the boundary interaction. This time, the computations are
slightly more involved. As an example we treat the
following 2-point function
\begin{equation}
    \begin{split}
             \langle U^{00}_{\langle2e',2n'\rangle}(0)
         V^{11}_{\langle-e,-n+1\rangle}&(iy,-iy)\, S^{\text{bdy}}_{\text{int}}   \rangle \, = \, \\[2mm]
         &= \, -\frac{\delta(n-n'-\frac12)\delta(e-e')}{|y|^{4\frac{e}{k}
   (2n-1+\frac{e}{k})}} \frac{y}{2\pi}\int du \ \frac{|u|^{2\alpha}}{|u^2+y^2|^{\alpha+1}} \\[2mm]
    &= \, -\frac{\delta(n-n'-\frac12)\delta(e-e')}{|y|^{4\frac{e}{k}
    (2n-1+\frac{e}{k})}} \frac{1}{2\pi}\int du \ |1+u^2|^{-\alpha-1} \\[2mm]
    &= \, -\frac{\delta(n-n'-\frac12)\delta(e-e')}{2|y|^{4\frac{e}{k}
    (2n-1+\frac{e}{k})}}2^{-2\alpha}
\frac{\Gamma(2\frac{e}{k}+1)}{\Gamma^2(\frac{e}{k}+1)} \\[2mm]
   &= \, -\frac{\delta(n-n'-\frac12)\delta(e-e')}{2|y|^{4\frac{e}{k}
  (2n-1+\frac{e}{k})}}
\frac{\Gamma(\frac{e}{k}+\frac12)}{\sqrt{\pi}\Gamma(\frac{e}{k}+1)}
\end{split}
\end{equation}
The second step is the substitution $u\rightarrow y/u$, then we can
apply \eqref{eq:1bdyint} which is a special case of the integral formula
in \cite{Fateev:2007wk}. The last step is the Euler doubling formula of
the Gamma function.
Comparison with the associated contribution to the minisuperspace
integral \eqref{int} gives
\begin{equation} \label{C1}
C_1(e) \ = \ \frac{\Gamma(e/k+1/2)}{\sqrt{\pi}\Gamma(e/k+1)}\ \ .
\end{equation}
Once more, one can perform similar computations with a single
insertion of a boundary interaction for other pairs of bulk
and boundary fields. All these calculations lead to the same
result for $C_1$, as predicted by \gl\ covariance.

At this point, we have computed all the data we were interested
in. But there are more contributions to the perturbative expansion
of the bulk-boundary 2-point function. As we stated above,
non-vanishing contributions arise from $s=0,s=1$ and $s=2$. We
have completely determined the $s=0$ term. At $s=1$, however, our
attention so far was restricted to the boundary interaction. The
other term with a single bulk insertion can also contribute since
it contains a product of only two $b_\pm$. Similarly, at $s=2$,
two insertions of the boundary interaction can lead to a
non-vanishing result. Products of bulk and boundary interactions
or two bulk interactions, on the other hand, involve too many
fields $b_\pm$ and vanish by simple zero mode counting. Hence, we
are left with two more terms to calculate, those arising from a
product of two boundary interactions $S^{\text{bdy}}_{\text{int}}$
and from a single bulk interaction $S^{\text{bulk}}_{\text{int}}$.
$Y$-charge conservation implies that the additional terms involve
a factor $\delta(n-n'-1)$. Such a term, if present,  would be
inconsistent with the global \gl\ symmetry. Our task therefore is
to show that the sum of the two aforementioned contributions
vanishes.

Let us begin with the computation of the term that arises from a
single insertion of the bulk interaction,
\begin{equation}\label{1bulk}
    \begin{split}
        \langle U^{11}_{\langle2e,2n-2\rangle}&(0)
    V^{11}_{\langle -e,-n+1\rangle}(iy,-iy)\ S^{\text{bulk}}_{\text{int}}
  \rangle \ \sim  \\[2mm]
        \sim \ & y^{-2\frac{e}{k}(4n-3+2\frac{e}{k})}\,
       \frac{y^3}{k\pi} \int_{UHP} d^2z\
      |z^2+y^2|^{-2(\frac{e}{k}+1)}|z^2|^{2\frac{e}{k}-1}(z-\bar{z}) \\[2mm]
        = \ & -y^{-2\frac{e}{k}(4n-3+2\frac{e}{k})} \frac{1}{e\sqrt{\pi}}
    \frac{\Gamma(2e/k+1/2)}{\Gamma(2\frac{e}{k}+1)}\\[2mm]
    \end{split}
\end{equation}
We have been a bit sloppy here by setting the parameters the
parameters $2e'=2e$ and $2n'-2$ to the values at which the
expectation value has a non-vanishing contribution. Strictly
speaking, this quantity is divergent, but the divergence is an
overall (volume) factor $\delta(0)$ which we suppressed
consistently. In the first equality we simply inserted the
relevant free field correlator. After the substitution
$z\rightarrow y/z$, the integral over the insertion point $u$ of
the boundary interaction can be evaluated using an integral
formula from \cite{Fateev:2007wk} (see also \eqref{eq:1bulkint}). Finally, the
answer is simplified by means of Euler's doubling formula for
Gamma functions.

Next we turn to the contributions coming from two boundary interactions.
Since the corresponding free field correlator is slightly more involved
in this case, we state an expression for the fermionic contribution
before going into the actual computation,
\begin{equation}
    \begin{split}
        \langle\bp(u_1)C(u_1)&\bp(u_2)C(u_2)(\cp-\cm)(0)C(0)
    \cp(-iy)\cm(iy)\rangle^{\text {F}} \ = \\[2mm]
        = \ &\frac{-4\pi y^3(u_2-u_1)}{u_1u_2(u_1^2+y^2)(u_2^2+y^2)}
        \Bigl[\sign(u_2-u_1)-\sign(u_2)+\sign(u_1)\Bigr]\ .\\
    \end{split}
\end{equation}
This result is inserted to compute
\begin{equation}\label{2bdy}
    \begin{split}
        \langle  U^{11}_{\langle2e,2n-2\rangle}&(0)
  V^{11}_{\langle -e,-n+1\rangle}(iy,-iy)\
 \left(S^{\text{bdy}}_{\text{int}}\right)^2 \rangle
   \ \sim \\[2mm]
        = \ & y^{-2\frac{e}{k}(4n-3+2\frac{e}{k})}
        \frac{y^3}{\pi k}\int du_1du_2\
      |u_1^2+y^2|^{-e/k-1}|u_2^2+y^2|^{-\frac{e}{k}-1}
      |u_1^2|^{\frac{e}{k}-1}\\[2mm]
        \ & |u_2^2|^{\frac{e}{k}-1}(u_2-u_1)
   \Bigl[\sign(u_2-u_1)-\sign(u_2)+
      \sign(u_1)\Bigr] \\[2mm]
        = \ &  y^{-2\frac{e}{k}(4n-3+2\frac{e}{k})}
  \frac{1}{\pi k}\int dx_1dx_2\
    |x_1^2+1|^{-\frac{e}{k}-1}|x_2^2+1|^{-\frac{e}{k}-1}|x_1-x_2| \\[2mm]
            = \ &  y^{-2\frac{e}{k}(4n-3+2\frac{e}{k})}
              \frac{2}{e\sqrt{\pi}}
  \frac{\Gamma(2\frac{e}{k}+\frac12)}{\Gamma(2\frac{e}{k}+1)}\\
    \end{split}
\end{equation}
The integral in the fourth line is again evaluated with a special
case of the integral formula of Fateev and Ribault
\eqref{eq:2bdyint}. Putting the results of eqs.\ \eqref{1bulk} and
\eqref{2bdy} together we arrive at
\begin{equation}
  \langle U^{11}_{\langle2e',2n'\rangle}(0)
   V^{11}_{\langle -e,-n+1\rangle}(iy,-iy) \left( S^{\text{bulk}}_{\text{int}}
    + \frac{1}{2!} \left( S^{\text{bdy}}_{\text{int}}\right)^2 \right)
     \  \rangle \ = \ 0 \ \ ,
\end{equation}
in agreement with \gl\ covariance of the 2-point function. Thereby,
we have now established the formula \eqref{2pt} through our
perturbative computations.

Before we leave the subject of bulk boundary 2-point functions, we
would like to make a few comments on the cases when $e/k$ is an
integer multiple of $1/2$. Consider inserting a bulk
vertex operator with $e$ momentum $e=-mk-k/2-k\varepsilon$ and
sending $\varepsilon$ to zero. In the limit, the second term of
eq.\ \eqref{2pt} develops a logarithmic singularity,
\begin{equation}
    \begin{split} \label{2ptlog}
        C_1(-mk-k/2-k\epsilon)|y|^{-\Delta_1} \ &= \
        \frac{(-1)^m}{m!\Gamma(-m+1/2) |y|^{2\Delta}} (\mathcal{Z} +
        \tilde{\Delta}\ln|y| + o(\epsilon)) \\[2mm]
        \text{where}\qquad \mathcal{Z} \ &= \
        \frac{1}{\epsilon}+\Psi(-m)-\Psi(-m+1/2) \ , \\[2mm]
        \Delta \ &= \ -(2m+1)(2n-m-1) \ . \\[1mm]
    \end{split}
\end{equation}
and $\tilde{\Delta} = 4n-4m-3$.  Here, $\Psi$ is the usual
Di-gamma function. The form of our bulk-boundary 2-point function
\eqref{2ptlog} resembles a similar expression in
\cite{Gaberdiel:2006pp}. A link between boundary
correlation functions of symplectic fermions and the
corresponding correlators in the \GL\ WZNW model may be
established following ideas in \cite{LeClair:2007aj}.

\subsubsection{Boundary 3-point functions}

The second object of interest for us is the boundary 3-point
function. Before we get there, we have to turn our attention to an
important detail that we glossed over in the previous subsection.
We recall that our $2\times 2$ matrices $\Psi_{\<e,n\>}, e \neq k
\mathbb{Z},$  of boundary fields contain two irreducible
multiplets $\< e,n\>$ under the unbroken global \gl\ symmetry.
These two multiplets have opposite fermion number, i.e.\ the state
with lower eigenvalue of $N$ is bosonic for one of them and
fermionic for the other. In general, the two multiplets are
allowed to have different couplings to the other fields in the
theory. When we studied bulk-boundary 2-point function, only one
of the two multiplets from each of the $2\times2$ matrices
$\Psi_{\<2e,2n\>}$ and $\Psi_{\<2e,2n-1\>}$ could have a
non-vanishing overlap with the bulk field $\Phi_{\<-e,-n+1\>}$,
simply because of fermion number conservation. Hence, the
bulk-boundary 2-point functions were parameterised by two
non-vanishing structure constants $C_\nu(e)$ rather than four. For
boundary 3-point functions, however, the distinction becomes
important. Consequently, we introduce the symbols
\begin{equation}
\begin{split} U^0_{\<-2e,-2n+1\>}(u)\ & =
  \ e^{ieX+inY}\, (\  1\  , \ \cp-\cm\ )  \\[2mm]
              U^1_{\<-2e,-2n+1\>}(u)\ & =
\ e^{ieX+inY}\, (\, C , (\cp-\cm)\, C\, )
\end{split}
\end{equation}
for the first and second row of the matrix \eqref{vertU}. The same
notation is used for the rows of the matrices $\psi$ of functions
and $\Psi$ of boundary fields.

Let us now begin with the 3-point function of three fields from
the first multiplet $\Psi^0$. These acquire contributions
exclusively from a single insertion of the boundary interaction. A
non-vanishing correlator requires that the parameters $e_i$ of the
three fields sum up to $\tilde e = e_1+e_2 + e_3 = 0$ and
similarly that $\tilde n = n_1 + n_2 + n_3 = 1$. Using the
integral formulae from Appendix A, the 3-point function of
fields $\Psi^0$ in the regime $0 < x < 1$ is found to be
\begin{equation}\label{3t000}
    \begin{split}
     &\langle \Psi^{0\epsilon_1}_{\langle -2e_1,-2n_1+1\rangle}(0)
   \Psi^{0\epsilon_2} _{\langle -2e_2,-2n_2+1\rangle}(1)
   \Psi^{0\epsilon_2}_{\langle -2e_3,-2n_3+1\rangle}(x)\rangle \ =
   \delta(\tilde e)\, \delta (\tilde n -1) \, \delta(\tilde \epsilon-2)\, \times \\[4mm]
        &\qquad\qquad  \times\  x^{2\Delta_{13}}(1-x)^{2\Delta_{23}}\
        \frac{\pi}{i} \frac{s(\alpha_1)+s(\alpha_2)+s(\alpha_3)}
   {s(\alpha_1)s(\alpha_2)s(\alpha_3)
 \Gamma(\alpha_1+\epsilon_1)\Gamma(\alpha_2+\epsilon_2)
 \Gamma(\alpha_3+\epsilon_3)}\\
    \end{split}
\end{equation}
where we defined the parameters $\alpha_i$ by $\alpha_i =2e_i/k$
and introduced the short-hands $s(z)$ and $\tilde \epsilon$ for
$s(z) = \sin(\pi z)$ and $\tilde \epsilon  = \sum \epsilon_i$. The
conformal weights are given by
$$ \Delta_{ij}\ = \ (n_i-1/2)\alpha_j+(n_j-1/2)
 \alpha_i+\alpha_i\alpha_j\ . $$
In the limit $k \rightarrow \infty$ the function $s(\alpha_i)$ can
be approximated by $s(\alpha) \sim 2 \pi e_i/k $ and the entire
3-point function is seen to vanish due to the conservation of $e$
momentum. This is consistent with the minisuperspace theory. In
fact, the corresponding  integral of functions on our brane is
easily seen to vanish,
$$ \langle \psi^{0\epsilon_1}_{\langle -2e_1,-2n_1+1\rangle}
   \psi^{0\epsilon_2} _{\langle -2e_2,-2n_2+1\rangle}
   \psi^{0\epsilon_2}_{\langle -2e_3,-2n_3+1\rangle}\rangle
    \ = \ 0\ .
    $$
This is so
because integration with the Haar measure needs a product of two
different fermionic zero modes in order to give a non-zero result.
Our functions $\psi^0$, however, only contain the zero mode
$\eta_+ -\eta_-$.

Let us now move on to discuss the 3-point in the case where a
single field from the second multiplet $\Psi^1$ is inserted.
Contributions to such correlators arise only from the leading term
$s=0$ of the perturbation series (see below). The result is
therefore straightforward to write down
\begin{equation}\label{3pt001}
    \begin{split}
    & \langle \Psi^{0\epsilon_1}_{\langle -2e_1,-2n_1+1\rangle}(0)
    \Psi^{0\epsilon_2}_{\langle -2e_2,-2n_2+1\rangle}(1)
    \Psi^{1\epsilon_3}_{\langle -2e_3,-2n_3+1\rangle}(x)\rangle
    \ = \ \qquad\qquad \qquad \\[3mm] & \qquad \qquad \qquad
     \qquad \qquad \qquad  \ = \ \delta(\tilde e)\,
       \delta (\tilde n - 1/2) \,
         \delta(\tilde \epsilon-1) \
        x^{2\Delta_{13}}(1-x)^{2\Delta_{23}}\ . \\
    \end{split}
\end{equation}
This coupling in independent of the level $k$ and it matches the
minisuperspace answer which is non-zero because the multiplet
$\psi^1$ contains both fermionic zero modes.

The most interesting 3-point coupling appears when we insert two
fields from the second multiplet $\Psi^1$. Once more,
non-vanishing terms can only arise from the insertion of a single
boundary interaction. They can be worked out with the help of
integral formulae in Appendix A,
\begin{equation}\label{3pt011}
    \begin{split}
        &\langle \Psi^{0\epsilon_1}_{\langle -2e_1,-2n_1+1\rangle}(0)
        \Psi^{1\epsilon_2}_{\langle -2e_2,-2n_2+1\rangle}(1)
        \Psi^{1\epsilon_3}_{\langle -2e_3,-2n_3+1\rangle}(x)\rangle
         \ = \ \delta(\tilde e)\, \delta (\tilde n - 1) \,
         \delta(\tilde \epsilon-2)\ \times \\[4mm]
        &\qquad  \times \ \frac{2\pi^2i}{k}\,
        x^{2\Delta_{13}}(1-x)^{2\Delta_{23}}\
       \frac{s(\alpha_1)-s(\alpha_2)-s(\alpha_3)}
{s(\alpha_1)s(\alpha_2)s(\alpha_3)
\Gamma(\alpha_1+\epsilon_1)\Gamma(\alpha_2+\epsilon_2)
 \Gamma(\alpha_3+\epsilon_3)}\ \ . \\
    \end{split}
\end{equation}
Note that the factor $\sim 1/k$ in the first term of the second
row is necessary in order for the whole expression to scale to a
finite value as we send the level $k$ to infinity. The expression
that arises in this limit can be checked easily in the
minisuperspace theory.

There remains one more case to consider, namely the 3-point
function for three fields from the second multiplet $\Psi^1$.  It
is given by
\begin{equation}\label{3pt111}
    \begin{split}
    & \langle \Psi^{1\epsilon_1}_{\langle -2e_1,-2n_1+1\rangle}(0)
    \Psi^{1\epsilon_2}_{\langle -2e_2,-2n_2+1\rangle}(1)
    \Psi^{1\epsilon_3}_{\langle -2e_3,-2n_3+1\rangle}(x)\rangle
    \ = \\[3mm]
     & \qquad \qquad\qquad \ = \ \delta(\tilde e)\, \delta (\tilde n - 1/2) \,
         \delta(\tilde \epsilon-1) \frac{2\pi}{k}\
        x^{2\Delta_{13}}(1-x)^{2\Delta_{23}}\ \ . \\
    \end{split}
\end{equation}
As in the previous formula  \eqref{3pt011}, the result contains a
factor $1/k$. Consequently, the 3-point coupling on the right hand
side of  eq.\ \eqref{3pt111} vanishes at $k \sim \infty$, in
agreement with the associated minisuperspace computation.

The last result \eqref{3pt111} was obtained without any insertion
of bulk or boundary interactions, though naively one might expect
to see contributions from one bulk or two boundary insertions. A
similar comment applies to the second case \eqref{3pt001} above.
It is indeed true that the insertion of $S^{\text
bulk}_{\text{int}}$ or $(S^{\text{bdy}}_{\text{int}})^2$ both lead
to non-vanishing expressions. But, as in the case of the bulk
boundary 2-point functions, their sum vanishes, i.e.\
$$\langle\,  U^{\epsilon'_1\epsilon_1}_{\langle e_1,n_1\rangle}(0)\,
 U^{\epsilon'_2\epsilon_2}_{\langle e_2,n_2\rangle}(1)\,
 U^{\epsilon'_3\epsilon'_3}_{\langle e_3,n_3\rangle}(u)\,
 \left( S^{\text{bulk}}_{\text{int}} + \frac{1}{2!} \left(
    S^{\text{bdy}}_{\text{int}}\right)^2\right)
    \rangle\ = \ 0 \ .
$$
The result is trivially fulfilled for $\tilde \epsilon' = 0,2$. It
requires rather elaborate computations when $\tilde \epsilon' =
1,3$. These can be performed with the help of the integral
formulae (\ref{eq:A3}-\ref{eq:A5}) we list in Appendix A.

Before closing this section we would like to add two more
comments. The first one concerns the logarithmic singularities
that appear in the 3-point functions whenever one of the
parameters $2e_i$ is an integer multiple of $k$. If we
consider joining two open strings with $e$ momentum $e_1 = e -
\varepsilon/2$ and $e_2= - e - \varepsilon/2$, for example, and
send $\varepsilon$ to zero, we obtain
\begin{equation}
    \begin{split} \label{eq:atypical}
        &\langle \Psi^{00}_{\langle -2e+\varepsilon,-2n_1+1\rangle}(0)
         \Psi^{11}_{\langle 2e+\varepsilon,-2n_2+1\rangle}(1)
         \Psi^{11}_{\langle -2\varepsilon,-2n_3+1\rangle}(u)\rangle \ \sim
   \\[2mm]
        &\qquad  \ \sim \ u^{2\Delta}(1-u)^{-2\Delta}
   \,  \delta (\tilde n-1) \
    \bigl(\mathcal{Z}+\mathcal{R}(\alpha)+A_{23}\ln|1-u|+A_{13}\ln|u|+ o(\varepsilon)\bigr)\\[4mm]
    &\qquad\text{where} \qquad \mathcal{Z} \ = \ \frac{1}{\varepsilon}+ \frac{4\varepsilon\gamma}{k}
     \qquad , \qquad \mathcal{R}(\alpha)\ = \ -2\pi\frac{1+c(\alpha)}{ks(\alpha)}  \\[2mm]
    & \qquad  A_{13} \ = \ \frac{1}{k}(2n_1-n_3-1/2+2\alpha)\qquad , \qquad
     A_{23}\ = \ \frac{1}{k}(2n_2-n_3-1/2-2\alpha)\\[2mm]
    \end{split}
\end{equation}
and $\Delta \ = \ \alpha(n_3-1/2)$. The function $c(\alpha)$
stands for $c(\alpha) = \cos( \pi \alpha)$ and $\gamma$ is the
Euler-Mascheroni constant. In the limit $\varepsilon \rightarrow 0$, the
constant $\mathcal{Z}$ diverges. This divergency can be regularised by
adding to $\Psi^{11}$ an appropriate field from the socle of the
involved atypical multiplet. In the following, we shall assume that
$\mathcal{Z}$ has been set to zero.

Our final comment deals with an interesting quantum symmetry of
the boundary 3-point functions. As in the bulk sector
\cite{Schomerus:2005bf}, the boundary 3-point function is periodic
under shifts of the $e$-momentum, in the following sense,
\begin{equation} \nonumber
    \begin{split}
          &\langle \Psi^{\epsilon_1\epsilon_1'}_{\langle 
                 -2e_1,-2n_1+1\rangle}(0)
          \Psi^{\epsilon_2\epsilon_2'}_{\langle 
                 -2e_2,-2n_2+1\rangle}(1)
          \Psi^{\epsilon_3\epsilon_3'}_{\langle 
                 -2e_3,-2n_3+1\rangle}(x)\rangle\ = \\[4mm]
          &\qquad(1-u)^{2n_3-1}u^{1-2n_3}
          \langle \Psi^{\epsilon_2\epsilon_2'}_{\langle 
                 -2e_1+k,-2n_1\rangle}(1)
          \Psi^{\epsilon_1\epsilon_1'}_{\langle 
                 -2e_2-k,-2n_2+2\rangle}(0)
          \Psi^{\epsilon_3\epsilon_3'}_{\langle 
                  -2e_3,-2n_3+1\rangle}(x)\rangle\ \ . \\
      \end{split}
\end{equation}
Further shifts by multiples of $\pm k$ can also be considered,
but necessarily involve inserting descendants of the tachyon
vertex operators. Our observation proves that the boundary \GL\
model for volume filling branes possesses spectral flow symmetry.
Shifts by integer multiples of the level $k$ are a symmetry of 
the affine representation theory. In principle, this symmetry 
could be broken by the boundary structure constants. The previous
formula asserts that, like in the bulk sector, the boundary 
operator product expansions preserve the spectral flow symmetry. 
The same is true for the bulk-boundary operator product 
expansions.

\subsection{Correlation functions involving atypical fields}
\label{sec:coratyp}

Throughout the last few sections we have learnt how to compute
correlation functions of bulk and boundary tachyon vertex
operators for a volume filling brane in the \GL\ WZNW model. We
now want to add a few comments on a particular set of correlation
functions that are essentially not effected by the interaction and
hence can be derived without cumbersome calculations. These will
include a non-vanishing annulus amplitude. We shall use the latter
to perform a highly non-trivial test on the proposed boundary
state of volume filling branes \cite{Creutzig:2007jy}.

\subsubsection{Correlators for special atypical fields}

In the previous sections we developed a first order formalism for
computations of correlation functions in the \GL\ WZNW model. Very
special correlators, however, can also be computed in the original
formulation. To begin with, let us explain the main idea at the
example of bulk correlators. We recall that the bulk action of the
\GL\ model is given by
\begin{equation}\label{eq:bulkSWZW}
    \begin{split}
        S_{\text{bulk}}\ =\ &-\frac{k}{4\pi i}\int_\Sigma d^2z\
        \left(
        \del X\bar{\del}Y+\del
        Y\bar{\del}X+2e^{iY}\del\cp\bar{\del}\cm\right)
    \end{split}
\end{equation}
The path integral is evaluated with the \gl\ invariant
measure \eqref{WZNWmeasure} on the space of fields.
A glance at the interaction term of the WZNW model and the measure
suggests to introduce the new coordinates $\chi_\pm=e^{iY/2}\cpm$.
After this substitution,  the path integral measure is the canonical
one,
\begin{equation}
    d\mu_{\text{WZW}} \ \sim \ \mathcal{D}X\mathcal{D}Y\mathcal{D}
     \chim\mathcal{D}\chip\ .
\end{equation}
Our bulk action $S_{\text{bulk}}= S_0 + Q$, on the other hand,
splits naturally into a free field theory $S_0$ and an interaction
term $Q$ where
\begin{equation}\label{eq:bulk}
    \begin{split}
    \qquad S_0\ =\ &-\frac{k}{4\pi i}\int_\Sigma d^2z\ \left( \del X\bar{\del}Y+
            \del Y\bar{\del}X+2\del\chip\bar{\del}\chim\right) \\[2mm]
    \qquad  \ \ Q\ = \ &\frac{k}{4\pi i}\int_\Sigma d^2z\ \left(i\chip\bar{\del}\chim\del Y+
     i\del\chip\chim\bar{\del}Y+\chip\chim\del Y\bar{\del}Y\right)  \ .\\
    \end{split}
\end{equation}
Due to the complicated form of $Q$, treating the WZNW model as a
perturbation by the interaction terms in $Q$ is not too useful for
most practical computations. Under very special circumstances,
however, the split into $S_0$ and $Q$ allows for a very
interesting conclusion. Observe that each term in the interaction
$Q$ contains at least one derivative $\partial Y$ or $\bar
\partial Y$. In our free field theory $S_0$, the only
non-vanishing contractions involving derivatives of $Y$ are those
with the field $X$. Hence, we can simply ignore the presence of
$Q$ for all correlation functions of tachyon vertex operators that
do not involve $X$. In other words, correlation functions of
fields without any $X$-dependence are given by their free field
theory expressions! This had already been observed in the results
of \cite{Schomerus:2005bf}. Our split of the action in $S_0$ and
$Q$ provides a rather simple and general explanation. Let us
stress again that this split is not helpful for any other
computation involving more generic typical fields.

It is clear that all this is not restricted to the bulk theory. In
fact, we can use the same substitution for the boundary terms of
the action \eqref{eq:bdySWZW},
\begin{equation}\label{eq:bdytwisted}
    \begin{split}
        S_{\partial 0}\ =\ &\frac{k}{8\pi i}\int_\Sigma du\ (\chip+\chim)
          \del_u(\chip+\chim)\ . \\
    \end{split}
\end{equation}
Since $S_{\partial 0}$ is quadratic in the fields $\chi_\pm$, it
gets added to the free bulk action $S_0$, i.e.\ we now work with a
free field theory on the upper half plane whose action is given by
$S_0+S_{\partial 0}$. There is no additional boundary contribution
to the bulk interaction $Q$. In the free theory, the fields
$\chi_\pm$ satisfy Neumann gluing conditions of the following
simple form,
\begin{equation}\label{eq:bdyfermions}
    \del \chipm(z,\bar{z}) \ = \ \mp \bar{\del} \chi_{\mp}(z,\bar{z})
    \qquad \text{for}\qquad z \ = \ \bar{z} \ .
\end{equation}
The gluing condition implies that fermions of the free boundary
theory are contracted as follows,
\begin{equation}
    \begin{split}
        \chim(z,\bar{z})\chip(w,\bar{w}) \ &\sim \ \frac{1}{k}\ln |z-w|^2 \ ,
        \\[2mm]
        \chipm(z,\bar{z})\chipm(w,\bar{w}) \ &\sim \
         \frac{1}{k}\ln (\bar{z}-w)-\frac{1}{k}\ln (\bar{w}-z) \ . \\
    \end{split}
\end{equation}
The bosonic fields $X,Y$ also obey simple Neumann boundary
conditions so that the evaluation of correlators in the free field
theory $S_0 + S_{\partial 0}$ is straightforward. Taking the
interaction $Q$ into account is a difficult task unless none of
the vertex operators in the correlation function contain the field
$X$. If all field are $X$ independent, then the correlator is
simply given by the free field theory formula, just as in the bulk
theory above.

One may apply the observation in the previous paragraph to the
evaluation of boundary 3-point functions of three atypical fields
for the volume filling brane. Note that we did not spell out a
formula for this particular correlator before. In principle, it
can be computed in the first order formalism, but the
corresponding calculation requires some care. Our new approach
allows to write down the result right away.
We shall discuss another interesting application of our new
approach to atypical correlation functions in the next subsection.
Let us mention in passing that we expect similar results to hold
for the completely atypical sectors in all $GL(N|N)$ and
$PSL(N|N)$ WZNW models. This will be discussed in more detail
elsewhere.

\subsubsection{Twisted boundary state and modular bootstrap}

We already discussed the twisted boundary state using the symplectic fermion correspondence. In this section, we give an alternative construction and relate the amplitude to a spectral density computed in the three-point correlation functions.

In order to construct a non-trivial quantity on the annulus, we
need to insert some fermionic zero modes, see e.g.\
\cite{Creutzig:2006wk} for similar tests in the simpler $bc$ ghost
system.  Let us anticipate that only atypical bulk fields couple to the volume
filling brane. Hence, if we insert fermionic zero modes through
some atypical bulk field, the entire amplitude is built from
atypical terms and should be computable through a simple free
field formalism, as explained in the previous subsection. Let us
see now how the details of this calculation work out.

To begin with, let us review the construction of the boundary
state $|\Omega\>$ for the volume filling brane. With the help of
our free field realisation, the formula becomes quite explicit. We
shall start from the boundary state $|\Omega\>_0$ of the free
theory. This state clearly factorises into a product of a bosonic
$|\Omega,B\rangle_0$ and a fermionic $|\Omega,F \rangle_0$
contribution. The latter two obey the following gluing conditions
\begin{equation} \label{glue01}
    (X_n + \bar X_{-n})\, |\, \Omega,B\rangle_0\ =\
    (Y_n + \bar Y_{-n})\, |\, \Omega,B\rangle_0 \ =\ 0
\end{equation}
and
\begin{equation}\label{glue02}
    ( \chi^\pm_n \mp \bar \chi^{\mp}_{-n})
    \, |\, \Omega,F\rangle_0\ =\  0\ \ .
\end{equation}
Here, $X_n$ and $\bar X_n$ are the modes of the currents $i
\sqrt{k} \del X$ and $i\sqrt{k}\bar \del X$ etc. Up to
normalisation, there exists a unique solution for these linear
constraints. For the bosonic and the fermionic sector, they are
given by the following coherent states,
\begin{eqnarray} \label{bound01}
|\, \Omega,B\rangle_0 & = & \exp \left( - \sum_{n=1}^\infty
\frac{1}{n} (Y_{-n} \bar X_{-n} + X_{-n} \bar Y_{-n}\right)
\,|0,0\rangle_B \\[2mm] \label{bound02}
|\, \Omega,F\rangle_0 & = & \exp \left( - \sum_{n=1}^\infty
\frac{1}{n} (\chi^+_{-n} \bar \chi^+_{-n} - \chi^-_{-n} \bar
\chi^-_{-n}\right)|0,0\rangle_F\ \ .
\end{eqnarray}
Here, $|0,0\rangle$ denote the vacua in the bosonic and the
fermionic theory. The product of the two components is the
boundary state of the free field theory, before the interaction is
taken into account. We now include the effects of the interaction
by multiplying the free boundary state with the exponential of the
interaction $Q$,
\begin{equation} \label{BS}
    |\, \Omega\rangle \ = \ \mathcal{N} e^{Q}\,  |\, \Omega\rangle_0 \ = \
     \mathcal{N}\ \left( \sum_{n=0}^\infty\
    \frac{Q^n}{n!}\right)  \ |\, \Omega,B\rangle_0 \times\,  |\, \Omega,F\rangle_0 \ ,
\end{equation}
where $\mathcal{N}=\sqrt{\pi/i}$ is a normalisation constant.
The operator $Q$ is defined as in eq.\
\eqref{eq:bulk}, but with the integration restricted to the
interior of the unit disc. It is possible to check that $\exp{Q}$
rotates the gluing conditions from the free field theory relations
\eqref{glue01} and \eqref{glue02} to their interacting
counterparts (see \eqref{eq:bdyeom}). The dual boundary state is
constructed analogously.

Our main aim now is to compute some non-vanishing overlap of the
twisted boundary state $|\Omega\rangle$. This requires the
insertion of the invariant bulk field $\Phi^{11}_{\<0,0\>}  =
\chim\chip $, i.e.\ we are going to study
\begin{equation}
    \begin{split}\label{eq:amplitudetwisted}
 Z_\Omega(q,z) \ := \ \langle\Omega\, |\, \tilde q^{L_0^c}(-1)^{F^c}\,
   \tilde z^{N_0^c}\, \Phi^{11}_{\<0,0\>}
         \, |\, \Omega\rangle\ ,
    \end{split}
\end{equation}
where $L_0^c = (L_0+\bar{L}_0)/2$ and $N_0^c = (N_0-\bar{N}_0)/2$
are obtained from the zero modes of the Virasoro field and the
current $N$. The corresponding expressions are standard, see e.g.
\cite{Schomerus:2005bf}. Our parameters $\tilde q$ and $\tilde z$
are defined in terms of $\mu,\tau$ through $\tilde q = \exp (-2\pi
i/\tau)$ and $\tilde z = \exp(2\pi i \mu/\tau)$.  We are now going
to argue that the computation of $Z_\Omega$ can be reduced to a
simple calculation in free field theory, i.e.\
\begin{equation}
    \begin{split}\label{eq:amplitudetilde}
        \langle\Omega\, |\, \tilde q^{L_0^c}(-1)^{F^c}\, \tilde z^{N_0^c}
       \,  \Phi^{11}_{\<0,0\>} \, |\, \Omega\rangle\ = \ \mathcal{N}^2\
    _0\langle\Omega\, |\, \tilde q^{{L}_0^c}(-1)^{F^c}\, \tilde
    z^{{N}_0^c}\, \Phi^{11}_{\<0,0\>} \, |\, \Omega\rangle_0\ .
    \end{split}
\end{equation}
The reasoning goes as follows. In a first step we write the
interacting boundary state as a product of the interaction term
$\exp Q$ and the free boundary state $|\Omega\>_0$. Next we
observe that all bosonic operators in between the two boundary
states involve derivatives  such as  $\del X$ etc. Hence, we can
use the gluing conditions \eqref{glue01} to express all these
terms through $Y_n$ and $X_n$. The modes $\bar Y_n$ and $\bar X_n$
of the anti-holomorphic derivatives only appear in the
construction \eqref{bound01} of the free bosonic boundary state
$|\Omega,B\>_0$. A non-vanishing term requires that the number of
$\bar X_n$ equals the number of $\bar Y_{-n}$. But since the $\bar
X_{-n}$ and $\bar Y_{-n}$ come paired with their holomorphic
partners $Y_{-n}$ and $X_{-n}$ in the boundary state, the operator
in between $_0 \< \Omega|$ and $|\Omega\>_0$ must have equal
numbers for $X_n$ and $Y_n$ modes in order for the corresponding
term not to vanish. In $Q$, all terms have an excess of $Y$ modes.
Since no term in $L_0^c$ or $N_0^c$ can compensate this through an
excess of $X$-modes, we can safely replace $\exp Q$ by its zeroth
order term, i.e.\ $\exp Q \sim 1$.

The computation of the overlap \eqref{eq:amplitudetilde} in free
field theory is straightforward. In a first step, the amplitude is
split into a product of bosonic and fermionic terms. The bosonic
contribution is the same as for extended branes in flat
2-dimensional space. The fermionic factor involves an insertion.
Its evaluation is reminiscent of a similar calculation in
\cite{Creutzig:2006wk}. We can express the result through a single
character of the affine \gl\ algebra,
\begin{equation}
 Z_\Omega(q,z) \  = \ \mathcal{N}^2\ \
  \hat{\chi}_{{\cal P}_0}(-1/\tau,\mu/\tau)\
  =\ \frac{2\pi}{k}\int de dn \ \frac{\hat{\chi}_{\langle e,n\rangle}
        (\tau,\mu)} {\sin(\pi e/k)} \ .
\end{equation}
The affine characters $\hat{\chi}$ along with their behaviour under
modular transformations can be found in the Appendix \ref{sc:App}. In order to achieve proper normalisation
(see below) we have set ${\cal N}^2=\pi/i$. Since the spectrum of
boundary operators on the volume filling brane is continuous, the
result involves some open string spectral density function. From
the result, this is read off as
\begin{equation} \label{SD}
 \rho(e,n) \ = \
\rho(e) \ = \ \frac{2\pi}{k\sin(\pi e/k)}\ \ .
\end{equation}
We would expect $\rho$ to be encoded in the boundary 3-point
function of $\Psi_{\langle e,n\rangle}$, $\Psi_{\langle
-e,-n\rangle}$ with the special boundary field
$\Psi^{11}_{\<0,0\>}$. One possible 3-point function that
contains the required information is a particular case
of our more general formula \eqref{eq:atypical}, i.e.
\begin{equation}
    \begin{split}
    &\langle \Psi^{00}_{\langle e,n\rangle}(0) \Psi^{11}_{\langle -e,-n\rangle}
      (1)\Psi^{11}_{\<0,0\>} \rangle \ \sim \ \\[3mm]
        &\qquad\qquad\qquad\ \sim \ u^{2\Delta}(1-u)^{-2\Delta}
    \bigl(\mathcal{Z} +\mathcal{R}(-\pi e/k)+A_{23}\ln|1-u|+
    A_{13}\ln|u|\bigr)\ \ . \\[2mm]
    \end{split}
\end{equation}
All quantities that appear on the right hand side were introduced
in equation \eqref{eq:atypical}. The additive constant ${\cal Z}$
is not universal. It is naively infinite, but can be made finite
by a proper regularisation prescription. We use the universal term
${\cal R}$ to determine the spectral density
\begin{equation}
    \frac{d}{de}\ln \mathcal{R}(-2 e/k) \ = \
    \frac{2\pi}{k} \frac{d}{d\alpha}\ln \frac{1+c(\alpha)}{s(-\alpha)}
     \ = \  \frac{2\pi}{k\sin(\pi e/k)} \ = \  \rho(e) \ .
\end{equation}
Here, we have used that $\alpha =  e/k$, as before. The result
agrees with the expression \eqref{SD} that was obtained through
modular transformation of the overlap \eqref{eq:amplitudetilde}.
Thereby, we have now been able to subject our formula \eqref{BS}
for the boundary state of the volume filling brane to a strong
consistency check.

There is another somewhat weaker but still non-trivial test for
the boundary state that arises from the minisuperspace limit of
the boundary WZNW model. In fact, in the particle limit we find
that
\begin{equation}
    \tr(z^{\ad^\Omega_N}(-1)^F \psi^{11}_{\langle 0, 0\rangle})
    \ = \ \int dedn \ \frac{\chi_{\langle e,n\rangle}(z)}{e} \ = \
    \lim_{k\rightarrow\infty} Z_\Omega(q,z)\ .
\end{equation}
In the first step we simply evaluated the trace directly in the
minisuperspace theory. We then observed in the second equality
that the result coincides with the modular transform of the
overlap \eqref{eq:amplitudetilde} in the appropriate limit $k
\rightarrow \infty$.

\subsection{Conclusions and open problems}

In this section we have solved the boundary theory for the volume
filling brane on \GL. We achieved this with the help of a
Kac-Wakimoto-like representation of the boundary theory. The first
order formalism we developed in section 2 is similar to the one
used in \cite{Fateev:2007wk} for $AdS_2$ branes in the Euclidean
$AdS_3$. The main difference is that we were forced to introduce
an additional fermion on the boundary. Such auxiliary boundary
fermions are quite common in fermionic theories (see e.g.\
\cite{Warner:1995ay,Hosomichi:2004ph} and references therein).
With the help of our first order formalism we were then able to
set up a perturbative calculational scheme for correlation
functions of bulk and boundary fields. The main features of this
scheme are similar to the pure bulk case \cite{Schomerus:2005bf}.
In particular, for any given correlator, only a finite number of
terms from the expansion can contribute. We computed the exact
bulk-boundary 2-point functions and the boundary 3-point
functions, thereby solving the boundary conformal field theory of
volume filling branes on \GL\ explicitly. Finally, we proposed a
second approach to correlation functions of atypical fields. It
singles out a particular subsector of the bulk and boundary \GL\
WZNW model that is not affected at all by the interaction. Hence,
within this subsector, all quantities agree with their free field
theory counterparts. The insight was then put to use for a
calculation of a particular non-vanishing annulus amplitude in
section 5.2. Together with our previous results on boundary
3-point functions, we obtained a strong test for the boundary
state of the volume filling brane in the \GL\ WZNW model.

There are several obvious extensions that should be worked out. To
begin with, it would be interesting to set up an equally efficient
framework to calculate correlation functions for the boundary
theories of point-like localised branes. Unfortunately, we have
not succeeded to calculate correlators from a finite number of
contributions, as in the case of the volume filling brane. It is
possible to develop a Kac-Wakimoto-like presentation for
point-like branes using the boundary conditions of
\cite{Creutzig:2006wk} for the $bc$ system. But since the gluing
conditions of \cite{Creutzig:2006wk} identify derivatives of $c$
with $\bar b$ etc., zero mode counting does not furnish simple
vanishing results. Therefore, an infinite number of terms can
contribute to any given correlation function. On the other hand,
the approach of section \ref{sec:coratyp} does generalise to
point-like branes. Since the boundary spectrum on a single
point-like brane is purely atypical, some interesting quantities
can be computed. This applies in particular to the boundary
3-point functions on a single point-like brane.
Correlation functions involving boundary condition changing fields
or typical bulk fields, however, are not accessible along these
lines. The symplectic fermion correspondence seems to be a better candidate to address these questions. 

It is certainly interesting to investigate how much of our program
extends to higher supergroups. Encouraged by the recent
developments on the bulk sector \cite{Quella:2007hr}, it seems
likely that most of our constructions may be generalised, at least
to supergroups of type I. This includes the superconformal
algebras psl(N$|$N) and many other interesting Lie superalgebras
(see e.g. \cite{Frappat:1996pb} for a complete list). We believe
that in all these cases there exists one class of branes which can
be solved through some appropriate square root of the bulk
formalism. Taking the proper square root will certainly involve a
larger number of fermionic boundary fields. Our second approach to
atypical correlation functions may also be extended to higher
supergroups and it provides interesting insights on the atypical
subsector of the WZNW models.

\chapter{N=2 superconformal field theories}

In this chapter, we introduce a new class of conformal field theories with even more symmetry. We investigate world-sheet and target space supersymmetric conformal field theories. The main result is that many Lie supergroup and supercoset $N=1$ world-sheet supersymmetric WZNW models possess an additional hidden $N=2$ superconformal symmetry.
In superstring theory, $N=2$ superconformal symmetry is a valueable ingredient for many reasons. The supersymmetry puts strong constraints on the dynamics of the string theory. One can twist the superconformal algebra to obtain a topological conformal field theory. The twisting promotes one of the two super-currents to a BRST-current which defines a cohomology theory. Correlation functions involving only fields in the cohomology are independent of their world-sheet positions. 

We start this chapter by introducing topological CFT and the gauged $N=1$ supersymmetric WZNW model of a Lie supergroup. Then we turn to the mathematically concepts, most importantly Manin triples. The main result is: If a Lie superalgebra $\g$ possesses a Manin decomposition then the gauged $N=1$ world-sheet supersymmetric WZNW model possesses a hidden $N=2$ superconformal symmetry. 
Our findings are in the spirit of the work of Kazama and Suzuki on cosets of compact Lie groups \cite{Kazama:1988qp}. It is remarkable that the generalisation to superspace not only includes many supercosets but also supergroups. This is due to the indefinite metric of the supergroup. 
This chapter is based on \cite{Creutzig:2009fh}.

\section{Topological conformal field theory}

In this section, we follow \cite{Dijkgraaf:1990qw}.
A topological conformal field theory is usually a subsector of a larger CFT.
Consider a conformal field theory with a BRST-operator $Q$ satisfying $Q^2=0$ and a BRST invariant action $S$.
Physical observables are fields that commute with $Q$, i.e.
\begin{equation}
 [Q,\phi]\ = \ 0\, .
\end{equation}
Further these states are defined up to a $Q$-exact term
\begin{equation}
 \phi \ \equiv \ \phi +[Q,\tilde\phi]\, .
\end{equation}
This means that the space of physical states is the cohomology of $Q$
\begin{equation}
 \mathcal{H}_{\text{phys}} \ = \ \frac{\text{kernel}(Q)}{\text{image}(Q)}\, .
\end{equation}
BRST invariance ensures that correlation functions of only physical fields are independent of the choice of representative for each $\phi$. The BRST-operator $Q$ defines a topological conformal field theory if there exists a holomorphic field $G(z)$ of dimension $(2,0)$ and also an anti-holomorphic field $\bar G(\bar z)$ of dimension $(0,2)$ satisfying 
\begin{equation}\label{eq:TQexact}
 T(z) \ = \ [Q,G(z)]\qquad\text{and}\qquad \bar T(\bar z) \ = \ [Q, \bar G(\bar z)]\, .
\end{equation}
This condition implies that physical correlation functions
\begin{equation}
 \langle \phi_1(z_1,\bar z_1)\dots\phi_n(z_n,\bar z_n)\rangle_\Sigma
\end{equation}
depend only on the inserted fields $\phi_i$ and the topology of the world-sheet $\Sigma$. But they are independent of the world-sheet positions $(z_i,\bar z_i)$ of the fields $\phi_i$. The topological CFT is then characterised by the cohomology ring $\mathcal{H}_{\text{phys}}$. The ring multiplication is given by the operator product expansion of physical fields
\begin{equation}
 \phi_i\phi_j \ \sim \ {c_{ij}}^k\phi_k\, .
\end{equation}
 \smallskip

A class of topological CFTs is obtained by twisting $N=2$ superconformal field theories. 
These are defined as follows. 
\begin{definition}\label{def:superconformalalgebra}
 The $N=2$ superconformal algebra of conformal central charge $c$ consists of the energy-stress tensor $T(z)$, a $U(1)$-current $U(z)$ of conformal dimension $1$ and two fermionic dimension $3/2$ super-currents $G^\pm$ subject to the operator product expansions
\begin{equation}
 \begin{split}
T(z)T(w)\ &\sim \ \frac{c/2}{(z-w)^4}+\frac{2T(w)}{(z-w)^2}+ \frac{\del T(w)}{(z-w)}\\
  G^+(z)G^-(w) \ &\sim \ \frac{c/3}{(z-w)^3} + \frac{U(w)}{(z-w)^2} + \frac{T(w)+\frac{1}{2}\del U(w)}{(z-w)} \\
G^\pm(z)G^\pm(w) \ &\sim \ 0 \\
U(z)G^\pm(w) \ &\sim \  \frac{\pm G^\pm(w)}{(z-w)}\\
U(z)U(w) \ &\sim \  \frac{c/3}{(z-w)^2}\, .\\
 \end{split}
\end{equation}
The anti-holomorphic partners $\bar T(\bar z), \bar G^\pm(\bar z)$ and $\bar U(\bar z)$ satisfy the analogous relations. 
\end{definition}
One distinguishes two kinds of twisted topological CFTs, the A-twists and the B-twists.
The twisted CFTs are obtained by defining the twisted energy-momentum tensor as follows
\begin{equation}
\begin{split}
 T^\pm_{\text{twisted}}(z)\ = \ T(z)\pm\frac{1}{2}\del U(z)\, ,\\ 
\bar T^\pm_{\text{twisted}}(\bar z)\ = \ \bar T(\bar z)\pm\frac{1}{2}\bar\del \bar U(\bar z)\, .\\ 
\end{split}
\end{equation}
Their OPEs 
\begin{equation}
\begin{split}
 T^\pm_{\text{twisted}}(z)T^\pm_{\text{twisted}}(w)\ &\sim \ 
\frac{2T^\pm_{\text{twisted}}(w)}{(z-w)^2}+ \frac{\del T^\pm_{\text{twisted}}(w)}{(z-w)}\ \ \ \text{and}\\
\bar T^\pm_{\text{twisted}}(\bar z)\bar T^\pm_{\text{twisted}}(\bar w)\ &\sim \ 
\frac{2\bar T^\pm_{\text{twisted}}(\bar w)}{(\bar z-\bar w)^2}+ \frac{\bar \del \bar T^\pm_{\text{twisted}}(\bar w)}{(\bar z-\bar w)}\\
\end{split}
\end{equation}
define a CFT of central charge $c=0$.

Consider the twisted theory given by $(T^-_{\text{twisted}},\bar T^-_{\text{twisted}})$.
Then the conformal dimension of $G^+(z)$ is $(2,0)$, the dimension of
$G^-(z)$ is $(1,0)$ and similarly those of $\bar G^+(\bar z)$ and $\bar G^-(\bar z)$ are $(0,2)$ and $(0,1)$. 
The operator $G^-_0+\bar G^-_0$ satisfies
\begin{equation}
\begin{split}
 (G^-_0+\bar G^-_0)^2 \ &= \ 0\ ,\\ 
[G^-_0+\bar G^-_0,G^+(z)]\ &= \ T^-_{\text{twisted}}(z)\ \text{and}\\ 
[G^-_0+\bar G^-_0,\bar G^+(\bar z)]\ &= \ \bar T^-_{\text{twisted}}(\bar z)\, .\\
\end{split}
\end{equation}
Hence, we have obtained a topological CFT with BRST-operator $G^-_0+\bar G^-_0$. 
This is an example of a B-twist. B-twists are those twists where $T$ and $\bar T$ are twisted in the same way, i.e.
\begin{equation}
  (T^\pm_{\text{twisted}}(z),\bar T^\pm_{\text{twisted}}(\bar z))\qquad Q_{\text{BRST}} \ = \ G^\pm_0+\bar G^\pm_0\,.
\end{equation}
If $T$ and $\bar T$ are twisted differently then one obtains an A-twisted topological CFT, i.e.
\begin{equation}
 (T^\pm_{\text{twisted}}(z),\bar T^\mp_{\text{twisted}}(\bar z))\qquad Q_{\text{BRST}}
 \ = \ G^\pm_0+\bar G^\mp_0\, .
\end{equation}
In the following, we want to look for $N=2$ superconformal field theories. We start with a presentation of the relevant models.

\section{The gauged $N=1$ WZNW model}

The $N=1$ world-sheet supersymmetric WZNW model of a compact Lie group is explained in \cite{DiVecchia:1984ep}. The generalisation to supergroups is straight forward and will be described in this section. Let $G$ be a Lie supergroup and $\g$ its Lie superalgebra. 
\smallskip

We start with some world-sheet supersymmetry considerations. The world-sheet is given by the usual bosonic world-sheet $\Sigma$, locally parameterised by coordinates $\tau, \sigma$. In addition there are two fermionic dimensions parameterised by $\theta$ and $\bar \theta$. Our notation follows \cite{Hori:2003ic}.

The covariant derivatives are
\begin{equation}
	D \ = \  -i\frac{\del}{\del \theta} - 2 \theta \del\qquad\text{and}\qquad 
	\bar D \ = \  -i\frac{\del}{\del \bar \theta} - 2 \bar \theta \bar \del\, 
\end{equation}
and the supercharges
\begin{equation}
	Q \ = \  -i\frac{\del}{\del \theta} + 2 \theta \del\qquad\text{and}\qquad 
	\bar Q \ = \  -i\frac{\del}{\del \bar \theta} + 2 \bar \theta \bar \del\, .
\end{equation}
The supercharges commute with the derivatives, and hence an action constructed just out of superfields and its derivatives is classical supersymmetric by construction.
The superfield has the following form
\begin{equation}
	\G\ = \ \exp (i\theta\chi)\ g \ \exp (-i\bar\theta\bar\chi)\, .
\end{equation}
Here $g$ is a Lie supergroup valued field and the fields $\chi=\chi^at_a$ and $\bar\chi=\bar\chi^at_a$ transform in the adjoint representation of $\g$ (the $\{t_a\}$ denote a basis of $\g$). The components of $\chi$ and $\bar \chi$ corresponding to the even directions of the Lie superalgebra are fermionic while those corresponding to the odd directions are bosonic.
Denote by $\str$ a nonzero metric of the Lie superalgebra $\g$. We included the level $k$ in the metric. 
The kinetic term of the action is\vspace{2mm}
\begin{equation}
	S_{\text{kin}}[\G] \ = \ \frac{1}{2\pi}\int d\tau d\sigma d^2\theta\  \str( \G^{-1}D \G\ \G^{-1}\bar{D}\G)  \vspace{2mm} 
\end{equation}
and the Wess-Zumino term is \cite{Witten:1983ar}\vspace{2mm}
\begin{equation}
	S_{\text{WZ}}[\tilde{\G}] \ = \ \frac{1}{6\pi}\int_B   d\tau d\sigma dtd^2\theta\
	\str( \tilde{\G}^{-1}\del_t \tilde{\G}\ \tilde{\G}^{-1}D \tilde{\G}\ \tilde{\G}^{-1}\bar D \tilde{\G}) 
	  \ .  \vspace{2mm}
\end{equation}
where $\tilde \G$ is an extension to $B$ as usual.
We compute the Polyakov-Wiegmann identity for the kinetic term, the WZ term and the WZNW action,\vspace{2mm}
\begin{equation}\nonumber
	\begin{split}\label{PolyakovWighman}
		S_{\text{kin}}[\G\HH] \, = \, &S_{\text{kin}}[\G] +S_{\text{kin}}[\HH] + 
		\frac{1}{2\pi}\int_\Sigma d\tau d\sigma d^2\theta\, \str( D \HH\HH^{-1}\ \G^{-1}\bar{D}\G)+\str( \G^{-1}D \G\ \bar{D}\HH\HH^{-1}) \\
		S_{\text{WZ}}[\tilde{\G}\tilde{\HH}] \, = \, &S_{\text{WZ}}[\tilde{\G}]+S_{\text{WZ}}[\tilde{\HH}] +
		\frac{1}{2\pi}\int_\Sigma d\tau d\sigma d^2\theta\, \str( D \HH\HH^{-1}\ \G^{-1}\bar{D}\G)-\str( \G^{-1}D \G\ \bar{D}\HH\HH^{-1}) \\
		S[\tilde{\G}\tilde{\HH}] \, = \,  &S[\tilde{\G}]+S[\tilde{\HH}]	+\frac{1}{\pi}\int_\Sigma d\tau d\sigma d^2\theta\, \str( D \HH\HH^{-1}\ \G^{-1}\bar{D}\G) \ .\vspace{2mm}
	\end{split}
\end{equation}
Further it is easy to see that
\begin{equation}
	S_{\text{WZ}}[\exp (i\theta\chi)] \ = \ S_{\text{WZ}}[\exp (-i\bar\theta\bar\chi)]\ = \ 0\, .
\end{equation}
The next computation is
\begin{equation}
	\begin{split}
		g^{-1}Dg \ &= \ -2\theta g^{-1}\del g\qquad, \qquad g^{-1}\bar Dg \ = \ -2\bar \theta g^{-1}\bar \del g \\
		\exp -i\theta\chi D \exp i\theta\chi \ &= \ \chi -i\theta\chi\chi \qquad, \qquad
	       \exp -i\theta\chi \bar D \exp i\theta\chi \ = \ 2i \theta\bar\theta\bar\del\chi \\
	       \exp -i\bar\theta\bar\chi D \exp i\bar\theta\bar\chi \ &= \ -2i \theta\bar\theta\del\bar\chi \qquad, \qquad
	       \exp -i\theta\chi \bar D \exp i\theta\chi \ = \ \bar\chi -i\bar\theta\bar\chi\bar\chi\, . \\
       \end{split}
\end{equation}
Then we read off
\begin{equation}
	\begin{split}
	S[\G]\ &= \ S[g]+S[\exp (i\theta\chi)]+S[\exp (-i\bar\theta\bar\chi)]\\
	&= \ S[g]+\frac{1}{2\pi}\int d\tau d\sigma d^2\theta\, \bigl(\str(\chi\bar\del\chi)+\str(\bar\chi\del\bar\chi)\bigr) \, ,\\
\end{split}
\end{equation}
where we have integrated out the world-sheet fermions with measure
\begin{equation}
	\int d^2\theta\ \theta\bar\theta \ = \ \frac{1}{4}\, .
\end{equation}
Thus the field content of the $N=1$ WZNW model is that of the ordinary WZNW model times free chiral and anti-chiral fermions and bosons transforming in the adjoint representation of the Lie superalgebra. 
Finally let us mention that in changing the path integral measure to the invariant path integral measure of the Lie supergroup times the free measure of the fermions and bosons the WZNW part of the action gets shifted by half the dual Coxeter number \cite{Tseytlin:1993my}, i.e. the final form of the action is
\begin{equation}\label{eq:N=1action}
        S[\G]\ = \ \Bigl(1+\frac{h^\vee}{k}\Bigr)S[g]+\frac{1}{2\pi}\int d\tau d\sigma\, \str(\chi\bar\del\chi)+\str(\bar\chi\del\bar\chi) \, \\
\end{equation}
with measure
\begin{equation}
 \mathcal{D}\mu(g)\prod_{a,b}\mathcal{D}\chi^a\mathcal{D}\bar\chi^b\, .
\end{equation}

The model is classically supersymmetric by construction, i.e. the action is invariant under the supersymmetry variation
\begin{equation}
	\begin{split}
		\delta \ &= \ \epsilon Q + \bar\epsilon\bar Q \\
		\delta g\ &=\ 2\epsilon\chi g -2\bar\epsilon g\bar\chi \\
		\delta \chi \ &= \ \epsilon(-i\del g g^{-1}-\{\chi,\chi\})\\
		\delta \bar\chi \ &= \ \bar\epsilon(ig^{-1}\bar\del g+\{\bar\chi,\bar\chi\})\, . \\
	\end{split}
\end{equation}
\smallskip

The gauged WZNW model of Lie groups has been described in e.g. \cite{Faddeev:1985iz, Bardakci:1987ee, Gawedzki:1988hq, Gawedzki:1988nj, Karabali:1989dk, Karabali:1988au, Tseytlin:1993my}. The formulation extends immediately to Lie supergroups. Let $\k$ be a Lie subsuperalgebra of the Lie  superalgebra $\g$,  $\{s_b\}$ a basis of $\k$ and $K$ the corresponding Lie subsupergroup. Further let $A(\tau,\sigma,\theta,\bar\theta)=A^bs_b$ and $\bar A(\tau,\sigma,\theta,\bar\theta)=\bar A^bs_b$ be two Lie subsuperalgebra valued gauge fields. Then the gauged $N=1$ WZNW action is 
\begin{equation}\nonumber
 \begin{split}
  S[\G,A,\bar A]\ &= \ S[\G]+\frac{1}{\pi}\int d\tau d\sigma d^2\theta\ \str\bigl(A\G^{-1}\bar D\G -D\G\G^{-1}+A\bar A-\Ad(\G)(A)\bar A\bigr)\, .
 \end{split}
\end{equation}
This action is invariant under the following gauge transformation
\begin{equation}
 \begin{split}
  \G\ &\rightarrow \ {\HH}\G\HH^{-1}\, , \\
   A\ &\rightarrow \ \Ad(\HH)A-\HH^{-1}D\HH\, ,\\
 \bar A\ &\rightarrow \ \Ad({\HH})\bar A-\HH^{-1}\bar D\HH\, 
 \end{split}
\end{equation}
for $\HH$ in $K$.
Thus the above action describes an $N=1$ world-sheet supersymmetric $G/K$ supercoset.
If we change fields according to
\begin{equation}\label{eq:supercosettrafo}
 \begin{split}
 A \ &= \ D\tilde{\HH}\tilde{\HH}^{-1}\qquad,\qquad \bar A \ = \ \bar D\tilde{\bar\HH}\tilde{\bar\HH}^{-1}\ ,\\
\tilde G \ &= \ \tilde{\bar\HH}^{-1}\G\tilde\HH\qquad\text{and}\qquad \HH\ = \ \tilde{\bar\HH}^{-1}\tilde\HH
 \end{split}
\end{equation}
 the gauged action becomes 
\begin{equation}
 S[\tilde\G]-S[\HH]\, .
\end{equation}
Further the Jacobian of the change of fields \eqref{eq:supercosettrafo} is field independent, i.e.
\begin{equation}
 \int \mathcal{D}\G\mathcal{D}A\mathcal{D}\bar A\ e^{-S[\G,A,\bar A]}\ = \ \mathcal{J}
\int \mathcal{D}\tilde\G\mathcal{D}\HH\ e^{-S[\tilde\G]+S[\HH]}\, 
\end{equation}
 for some constant $\mathcal{J}$ as explained in \cite{Tseytlin:1993my}.
The gauge fixing procedure requires to
introduce additional ghost fields. They come in four different
kinds. There are $\dim \k^{\bar 0}$ fermionic ghosts and $\dim
\k^{\bar 1}$ bosonic ones, each contributing a central charge $c =
-2$ and $c = +2$, respectively. These all have $N=1$
superpartners, i.e.\ there are $\dim \k^{\bar 0}$ bosonic ghosts
with central charge $c = -1$ and $\dim \k^{\bar 1}$ fermionic ones
with central charge $c=1$. Taking all these into account, the
ghost sector contributes  $c_{\text{ghosts}} = -3 \sdim \k$ so
that the total central charge is
\begin{eqnarray*}
 && \hspace*{-8mm} c(G/K)\ = \\[2mm]  & & \hspace*{-5mm}= \,
\left(\frac32-\frac{h^\vee_\s}{k}\right)\, \sdim \s
 + \left( \frac32 + \frac{h^\vee_\k}{k}\right) \sdim \k
 - 3\, \sdim \k \\[2mm] & & \hspace*{-5mm} =
 \ \left(\frac32-\frac{h^\vee_\s}{k}\right)\, \sdim \s
 - \left( \frac32 - \frac{h^\vee_\k}{k}\right) \sdim \k\ \ .
\end{eqnarray*}
The total Virasoro field $T_{\text{total}} = T_{\s \times \k} +
T_{\text{ghost}}$ possesses an $N=1$ superpartner
$G_{\text{total}} $. Both these fields descend to the state space
of the coset model. The latter is obtained by computing the
cohomology of the BRST operator $Q$. One may show that $T_{\text{total}}$
and $G_{\text{total}}$ are in the same cohomology class
as the Virasoro element $T_{G/K}$ and its superpartner $G_{G/K}$
in the coset conformal field theory. Details on how this works in
$N=1$ WZNW cosets $G/K$ of bosonic groups can be found in
\cite{Rhedin:1995um,FigueroaO'Farrill:1995pv}. The generalization
of these constructions to supergroups is entirely straightforward.
In the case of Lie groups, Kazama and Suzuki used the current
symmetry to show that some of the $N=1$ WZNW cosets admit an $N=2$
superconformal algebra \cite{Kazama:1988qp}. Their construction
may also be embedded into the product theory. In fact, it suffices
to show that the $N=1$ superconformal algebra of the WZNW model on
$S \times K$ admits an extension to $N=2$. The corresponding
fields of the $N=2$ superconformal algebra receive additional
contributions from the ghost sector to form a total $N=2$ algebra
whose basic $G^\pm_{\text{total}}$ and $U_{\text{total}}$ reside
in the same cohomology class as the associated fields in the coset
model. Our goal is to extend the analysis of Kazama and Suzuki to
the case in which $G$ and $K$ are Lie supergroups. According to
the remarks we have just made, all we need to do is to exhibit an
$N=2$ superconformal algebra in the $N=1$ WZNW model on the
product $G \times K$.

\section{Manin triples of Lie superalgebras}\label{section:manintriple}

We follow very closely the reasoning of \cite{Getzler:1993py}. Let us recall that article. The main statement is: Given a Lie algebra $\g$ which allows for a Manin triple, then the $N=1$ superconformal current algebra extends to an $N=2$ superconformal symmetry. These are exactly those models considered by Kazama and Suzuki \cite{Kazama:1988qp}. 

Inspired by these results, we will define a Manin triple for a Lie superalgebra, and derive the Kazama-Suzuki construction \`a la Getzler. 

\begin{definition}\label{def:manin}
A {\bf Manin triple} $(\g,\ap,\am)$ consists of a Lie superalgebra $\g$ possessing a consistent non-degenerate supersymmetric invariant bilinear form $(\,\cdot\, , \,\cdot\, )$ and isotropic Lie subalgebras  
$\apm$ such that 
\begin{equation}
	\g\ = \ \ap\oplus\am\, .
\end{equation}
Further denote the subspace of $\g$ orthogonal to the direct sum of the derived subalgebras of $\apm$ by $\azero$, i.e.
\begin{equation}
 \azero \ := \ \{\,x\,\in\,\g\,|\,(x,y)\,=\,0\ \forall\ y\,\in\,[\ap,\ap]\cup[\am,\am]\,\}\, .   
\end{equation}
\end{definition}
The nice property of Lie superalgebras is, that many of them already allow for a Manin triple.
\begin{example}\label{example:manindecomposition}
The most important Manin triples we
shall exploit arise from Lie superalgebras $\g = \s$, i.e. $K = \{ e\}$ .
Let us suppose that the even part $\g^{\bar 0}$ of $\g$
splits into two bosonic subalgebras $\g^{\bar 0} =\g^{\bar
0}_a\oplus\g^{\bar 0}_b$ of equal rank. This condition applies
to the Lie superalgebras $\g=gl(n|n),psl(n|n),sl(n|n\pm1)$ and
$\g=osp(2n+1|2n),osp(2n|2n)$. In all these examples, the bilinear form of
the Cartan subalgebra of one of these subalgebras is positive
definite while the other one is negative definite (with a proper choice of real form).
Consequently, we can perform an isotropic decomposition of the
Cartan subalgebra
\begin{equation}
    \h \ = \ \h_+\oplus \h_-\, .
\end{equation}
In order to extend the decomposition of $\h$ to an isotropic
decomposition of $\g$ we recall that any Lie superalgebra admits a
triangular decomposition  into the Cartan subalgebra $\h$, the
subalgebra of the positive root spaces $\nn_+$ and the subalgebra
of negative root spaces $\nn_-$:
\begin{equation}
 \g\ = \ \nn_-\oplus\h\oplus\nn_+\, .
\end{equation}
Hence the
triple $(\g,\ap=\h_+\oplus\nn_+,\am=\h_-\oplus\nn_-)$ is a
Manin triple. We also note that the derived
subalgebras $[\apm,\apm]$ of $\apm$ are contained in $\nn_\pm$ and
consequently,
\begin{equation}\label{eq:azerocartan}
 \azero \ \supseteq \ \h\, .
\end{equation}
There exist many other Manin triples, in particular when the
Lie superalgebra $\g$ is not simple.
\end{example}

Before we can turn to the $N=2$ superconformal algebra we need a variety of identities.
Denote by $x_i$ a basis of $\ap$ then this choice determines a dual basis $x^i$ of $\am$ with respect to the metric of the $N=1$ WZNW model. Recall that the metric already contains the level $k$ of the WZNW model.
The structure constants are defined as 
\begin{equation}
	\begin{split}
		[x_i,x_j]\ &= \ {c_{ij}}^kx_k\\
		[x^i,x^j]\ &= \ {f^{ij}}_kx^k\\
		[x_i,x^j]\ &= \ {c_{ki}}^jx^k+{f^{jk}}_ix_k\\
	\end{split}
\end{equation}
where the last equation follows from the first two. The Jacobi identity for $\apm$ in terms of structure constants is
\begin{equation}
	\begin{split}
		0\ &= \ (-1)^\parik {c_{il}}^m  {c_{jk}}^l +   (-1)^\parij {c_{jl}}^m  {c_{ki}}^l +  (-1)^\parjk {c_{kl}}^m  {c_{ij}}^l\qquad \text{and}\\
		0\ &= \ (-1)^\parik {f^{il}}_m  {f^{jk}}_l +   (-1)^\parij {f^{jl}}_m  {f^{ki}}_l +  (-1)^\parjk {f^{kl}}_m  {f^{ij}}_l\, .\\
	\end{split}
\end{equation}
Further the Jacobi identity of $\g$ implies the following cocycle formula
\begin{equation}\label{eq:cocycle}
	{c_{kl}}^m{f^{ij}}_m \ = \ -(-1)^\parij {f^{jm}}_l {c_{km}}^i-(-1)^\parij {f^{mi}}_k {c_{ml}}^j+{f^{im}}_l {c_{km}}^j+{f^{mj}}_k {c_{ml}}^i\, .\\
\end{equation}
We define
\begin{equation}\label{eq:rhotilde}
	\begin{split}
		\tilde\rho \ :&= \ -[x^i,x_i]\ = \ (-1)^i{c_{ki}}^ix^k+(-1)^i{f^{ik}}_ix_k \ ,\\
		 \tilde\rho_+ \ :&= \ (-1)^i{f^{ik}}_ix_k \qquad \text{and} \\
                 \tilde\rho_- \ :&= \ (-1)^i{c_{ki}}^ix^k\ . \\
	\end{split}
\end{equation}
The Jacobi identities of $\apm$ imply
\begin{equation}
	(-1)^m{f^{ml}}_m{f^{jk}}_l \ = \ (-1)^m{c_{ml}}^m{c_{jk}}^l \ = \ 0\, .
\end{equation}
Then it follows that
\begin{equation}
		\tilde\rho \ \in \ \azero \qquad \ \text{and}\qquad\
		[\tilde\rho_+,\tilde\rho_-]\ = \ 0 \, .
\end{equation}
Taking the supertrace over $i=l$ in the cocycle formula yields
\begin{equation}\label{eq:derivation}
	\begin{split}
		Dx_i \ := \ -[\tilde\rho,x_i]_+ \ &= \ (-1)^k({c_{jk}}^k{f^{jl}}_i+{c_{ji}}^l{f^{jk}}_k)x_l \\
		&= \ (-1)^{mn}{c_{mn}}^l{f^{mn}}_ix_l \qquad\text{and}\\
                \str(D) \ &= \ -(\tilde\rho,\tilde\rho)\, .\\
	\end{split}
\end{equation}
Further we need the Killing form in terms of the structure constants, in general that is
\begin{equation}
	\langle X^a,X^b\rangle \ = \ -(-1)^n {C^{na}}_m{C^{mb}}_n\, .
\end{equation}
More precisely, we need the special values of the Killing form
\begin{equation}\label{eq:killingofg}
	\begin{split}
		\langle x_i,x^j\rangle \ &= \ 2(-1)^{mn}{c_{ni}}^m{f^{nj}}_m+(-1)^{mn}{c_{mn}}^j{f^{mn}}_i  \ = \ 2A_i^j+D_i^j\qquad \text{where}\\
 A_i^j \ &= \  (-1)^{mn}{c_{ni}}^m{f^{nj}}_m \qquad \text{and} \\
D_i^j \ &= \ (-1)^{mn}{c_{mn}}^j{f^{mn}}_i\, .\\
	\end{split}
\end{equation}
This terminates our preparations.

\section{$N=2$ superconformal field theories}

We are prepared to show that a Manin triple $(\g,\ap,\am)$ of a Lie superalgebra $\g$
gives rise to an $N=2$ superconformal symmetry in the spirit of Kazama and Suzuki. 

Denote by $J_i(z)$ and $J^i(z)$ the chiral affine currents corresponding to the generators $x_i$ and $x^i$, then their OPEs are \cite{DiVecchia:1984ep}
\begin{equation}
	\begin{split}
		J_i(z)J_j(w)\ &\sim \ \frac{\frac{1}{2}\langle x_i,x_j\rangle}{(z-w)^2}+\frac{{c_{ij}}^kJ_k(w)}{(z-w)}\\
		J^i(z)J^j(w)\ &\sim \ \frac{\frac{1}{2}\langle x^i,x^j\rangle}{(z-w)^2}+\frac{{f^{ij}}_kJ_k(w)}{(z-w)}\\
		J_i(z)J^j(w)\ &\sim \ \frac{{\delta_i}^j+\frac{1}{2}\langle x_i,x^j\rangle}{(z-w)^2}+\frac{{f^{jk}}_iJ_k(w)+{{c_{ki}}^j}J^k(w)}{(z-w)}\\
	\end{split}
\end{equation}
where $\langle\ \ , \ \ \rangle$ is the Killing form and this shift by the Killing form in the metric is due to our parameterisation and its measure \eqref{eq:N=1action}. 
The fermions we denote by $a_i(z)$ and $a^i(z)$ and their OPE is 
\begin{equation}
	\begin{split}
		a_i(z)a^j(w)\ &\sim \ \frac{{\delta_i}^j}{(z-w)}\\
 		a_i(z)a_j(w)\ &\sim \ 0\\
		a^i(z)a^j(w)\ &\sim \ 0\ .
	\end{split}
\end{equation}
Their conformal dimension is $1/2$. We want to show that the following dimension $3/2$ currents generate an $N=2$ superconformal algebra
\begin{equation}
	\begin{split}
		G^+(z)\ &= \ J_i(z)a^i(z)-\frac{1}{2}(-1)^{ik}{c_{ij}}^k\,:a^i(z)a^j(z)a_k(z):\\
		G^-(z)\ &= \ J^i(z)a_i(z)-\frac{1}{2}(-1)^{jk}{f^{ij}}_k\,:a_i(z)a_j(z)a^k(z):\, .\\
	\end{split}
\end{equation}
We compute
\begin{equation}
	G^\pm(z)G^\pm(w)\ \sim\ 0\, .
\end{equation}
The next task is to compute the OPE of $G^+$ and $G^-$. We split that into several steps. First we introduce the notation 
\begin{equation}\label{eq:current}
	J \ = \ J_kx^k+(-1)^kJ^kx_k\, ,
\end{equation}
then
\begin{equation}
	\begin{split}
		J_i(z)a^i(z)&J^j(w)a_j(w) \ \sim \ \frac{A_3}{(z-w)^3}+\frac{A_2}{(z-w)^2}+\frac{A_1+B_1}{(z-w)}\\
		A_3 \ &= \ \frac{1}{2}\text{sdim}\g +\frac{1}{2}\langle x^i,x_i\rangle \\
		A_2 \ &= \ :a^i(w)a_i(w):+\frac{1}{2}\langle x_i,x^j\rangle :a^i(w)a_j(w):+(\tilde\rho,J(w))\\
		A_1 \ &= \ (-1)^iJ_i(w)J^i(w)+:\del a^i(w)a_i(w):+\frac{1}{2}\langle x_i,x^j\rangle:\del a^i(w)a_j(w):\\
		B_1 \ &= \ {f^{jk}}_i(-1)^\parjk J_k(w):a^i(w)a_j(w):+{c_{ki}}^j(-1)^\parjk J^k(w):a^i(w)a_j(w):\, . \\
	\end{split}
\end{equation}
The last term gets cancelled by
\begin{equation}
	\begin{split}
		(-\frac{1}{2}(-1)^{ik}{c_{il}}^k :a^i(z)a^l(z)a_k(z):)J^j(w)a_j(w)\ &\sim\ -\frac{{c_{ki}}^j(-1)^\parjk J^k(w):a^i(w)a_j(w):}{(z-w)}\\
		J_i(z)a^i(z)(-\frac{1}{2}(-1)^{jk}{f^{lj}}_k :a_l(w)a_j(w)a^k(w):)\ &\sim \ -\frac{{f^{jk}}_i(-1)^\parjk J_k(w):a^i(w)a_j(w):}{(z-w)}\, .\\
	\end{split}
\end{equation}
The next one is
\begin{equation}
	\begin{split}
		\frac{1}{4}(-1)^{ik+mn}&{c_{ij}}^k{f^{lm}}_n :a^i(z)a^j(z)a_k(z): :a_l(w)a_m(w)a^n(w): \ \sim \\ 
		&\qquad\qquad\qquad\qquad\qquad\qquad\qquad\sim \ \frac{C_3}{(z-w)^3}+\frac{C_2}{(z-w)^2}+\frac{C_1}{(z-w)} \\
		C_3 \ &= \ -\frac{1}{2}\text{str}(D) \\
		C_2 \ &= \ - A_j^m  :a^j(w)a_m(w):+\frac{1}{2}D^m_j:a^j(w)a_m(w): \\
		C_1 \ &= \ - A_j^m  :\del a^j(w)a_m(w):+\frac{1}{2}D^m_j:a^j(w)\del a_m(w):\, . \\
	\end{split}
\end{equation}
$A$ and $D$ where introduced in \eqref{eq:killingofg}.
Using \eqref{eq:current} we have
\begin{equation}
	(-1)^iJ_iJ^i\ = \ \frac{1}{2}((:J,J:)+(\tilde\rho,\del J))\, .
\end{equation}
The element $\tilde\rho$ was defined in \eqref{eq:rhotilde}.
Then putting all together and using \eqref{eq:killingofg} we arrive at
\begin{equation}
	\begin{split}
		G^+(z)G^-(w)\ \sim \ \frac{\frac{1}{2}\text{sdim}\,\g+\str D}{(z-w)^3}+\frac{U(w)}{(z-w)^2}+\frac{T(w)+\frac{1}{2}\del U(w)}{(z-w)}\, ,
	\end{split}
\end{equation}
where $T(z)$ is the Sugawara energy-stress tensor
\begin{equation}
	T(z)\ = \ \frac{1}{2}((:J,J:)+:\del a^ia_i:-:a^i\del a_i:)\, .
\end{equation}
And $U(z)$ has the form
\begin{equation}
	U(z)\ = \ :a^ia_i:+(\tilde\rho,J)+D^i_j:a^ja_i:\, .
\end{equation}
It remains to check that $G^\pm$ are correctly charged under $U$. 
This is rather tedious, we arrive at
\begin{equation}
	\begin{split}
		:a^i(z)a_i(z):G^+(w)\ &\sim \ \frac{G^+(w)}{(z-w)}-\frac{(\tilde\rho_-,a(w))}{(z-w)^2}\\
		((\tilde\rho,J(z))+D^i_j:a^j(z)a_i(z):)G^+(w) \ &\sim \ \frac{(\tilde\rho_-,a(w))}{(z-w)^2}\, .\\
	\end{split}
\end{equation}
For the computation of the second line we use the Jacobi identity as well as the cocycle formula to show that the first order term vanishes, while for the computation of the second order term we use \eqref{eq:derivation}. Analogously, we compute
\begin{equation}
	\begin{split}
		:a^i(z)a_i(z):G^-(w)\ &\sim \ -\frac{G^-(w)}{(z-w)}+\frac{(\tilde\rho_+,a(w))}{(z-w)^2}\\
		((\tilde\rho,J(z))+D^i_j:a^j(z)a_i(z):)G^-(w) \ &\sim \ -\frac{(\tilde\rho_+,a(w))}{(z-w)^2}\, .\\
	\end{split}
\end{equation}
In summary we have obtained the following result
\begin{prp}
Let $(\g,\ap,\am)$ be a Manin triple of a Lie superalgebra $\g$.
Then the $U(1)$-current
\begin{equation}
    U(z) \ = \ :a^ia_i:+(\tilde\rho,J)+D^i_j:a^ja_i:\, ,
\end{equation}
the energy-momentum tensor 
\begin{equation}
	T(z)\ = \ \frac{1}{2}((:J,J:)+:\del a^ia_i:-:a^i\del a_i:)\, 
\end{equation}
and the two super-currents
\begin{equation}
	\begin{split}
		G^+(z)\ &= \ J_i(z)a^i(z)-\frac{1}{2}(-1)^{ik}{c_{ij}}^k\,:a^i(z)a^j(z)a_k(z):\\
		G^-(z)\ &= \ J^i(z)a_i(z)-\frac{1}{2}(-1)^{jk}{f^{ij}}_k\,:a_i(z)a_j(z)a^k(z):\\
	\end{split}
\end{equation}
form an $N=2$ superconformal algebra of central charge 
\begin{equation}
 c \ = \ \frac{3}{2}\text{sdim}\,\g+3\str D\, .
\end{equation}
\end{prp}
Note that if we take a type I Lie superalgebra (i.e. $\g\,\in\{gl(n|n), psl(n|n),sl(n\pm1|n)\}$), then the central charge is $\frac{3}{2}\text{sdim}\,\g$, because $\str D=0$ in these cases ($D$ is introduced in \eqref{eq:derivation}). 
Let us also list some supercosets in table 6.1 to which above conditions apply.
\begin{table}[h]\label{table:supercosets}
\begin{center}
 \begin{tabular}{ | c | c | c |  }
    \hline
    $G$ & $H$ $\vphantom{\Bigl(\Bigr)}$ & $c\bigl(G/H\bigr)$ \\[1mm] \hline\hline
    GL($n|n$)  & GL($n-m|n-m$)$\vphantom{\Bigl(\Bigr)}$ $\qquad\quad\ n>m\geq0$ & 0\\  \hline
    GL($n|n$)  & SL($n-m|n-m\pm1$)$\vphantom{\Bigl(\Bigr)}$ $\qquad n>m>0$ & 0\\  \hline
    PSL($n|n$)  & PSL($n-m|n-m$)$\vphantom{\Bigl(\Bigr)}$ $\qquad\quad n>m\geq0$ & 0 \\  \hline    PSL($n|n$)  & SL($n-m|n-m\pm1$)$\vphantom{\Bigl(\Bigr)}$ $\qquad n>m>0$ & -3 \\  \hline
    SL($n|\tilde n$)$\ \, n\neq\tilde n$ &\qquad\ \ SL($n-m|\tilde n-m$)$\vphantom{\Bigl(\Bigr)}$ $\qquad\ \ \text{min}\{n,\tilde n\}\geq m\geq0$ & 0\\  \hline
\end{tabular}\caption{{\em Incomplete list of $N=2$ superconformal supercosets $G/H$ with central charge $c\bigl(G/H\bigr)$}}
\end{center}
\end{table}

\subsection{Deformations}

There exist more $N=2$ superconformal algebras, which are obtained from the previous ones by a deformation by an element $\alpha$ in $\azero$. Recall that $\azero$ is the subspace orthogonal to the direct sum of the derived subalgebras of $\ap$ and $\am$.

Consider an element $\alpha=p^ix_i+q_ix^i$ in $\azero$, this means that 
\begin{equation}\label{eq:alpha}
	{c_{ij}}^kq_k \ = \ {f^{ij}}_kp^k \ = \ 0\, .
\end{equation}
We deform the supercurrents as follows
\begin{equation}
	\begin{split}
		G^+_\alpha \ &= \ G^++q_i\del a^i \\
		G^-_\alpha \ &= \ G^-+p^i\del a_i \\
	\end{split}
\end{equation}
and since we want the supercurrents to be fermionic, we require $\alpha$ to be bosonic. 
Due to \eqref{eq:alpha} we get 
\begin{equation}
	G^\pm_\alpha(z)G^\pm_\alpha(w) \ \sim \ 0\, .
\end{equation}
We want to show that this deformation is still an $N=2$ superconformal algebra. 
First introduce
\begin{equation}
	\begin{split}
		I_i \ &= \ J_i-(-1)^{ik}{c_{ij}}^k:a^ia_k:-\frac{1}{2}(-1)^{ik}{f^{jk}}_ia_ja_k \\
		I^i \ &= \ J^i-(-1)^{jk}{f^{ij}}_k:a_ja^k:-\frac{1}{2}(-1)^{ij}{c_{jk}}^ia^ja^k\\
	\end{split}
\end{equation}
Then we compute
\begin{equation}
	\begin{split}
		G^+(z)p^i\del a_i(w) \ &\sim \ \frac{p^iI_i(w)}{(z-w)^2}+\frac{p^i\del I_i(w)}{(z-w)}\\
		q_i\del a^i(z)G^-(w) \ &\sim \ -\frac{q_iI^i(w)}{(z-w)^2}\\
		q_i\del a^i(z)p^i\del a_i(w)\ &\sim \ -\frac{2q_ip^i}{(z-w)^3}\\
	\end{split}
\end{equation}
This defines us our deformed U(1)-current and energy-stress tensor
\begin{equation}
	\begin{split}
		U_\alpha(z) \ &= \ U(z)+p^iI_i(z)-q_iI^i(z)\\
		T_\alpha(z) \ &= \ T(z)+\frac{1}{2}(p^i\del I_i(z)+q_i\del I^i(z))\\
	\end{split}
\end{equation}
It remains to check that $J_\alpha$ is indeed a U(1)-current, for this purpose we again use the cocycle formula as well as $\alpha$ being in $\azero$ and we 
get
\begin{equation}
	\begin{split}
		U_\alpha(z)G^\pm_\alpha(w)\ \sim \ \pm\frac{G^\pm_\alpha(w)}{(z-w)}\, .
	\end{split}
\end{equation}
In summary, we have shown the following.
\begin{prp} Let $\alpha=p^ix_i+q_ix^i$ in $\azero$, then the currents
\begin{equation}
\begin{split}
                U_\alpha(z) \ &= \ U(z)+p^iI_i(z)-q_iI^i(z)\\
		T_\alpha(z) \ &= \ T(z)+\frac{1}{2}(p^i\del I_i(z)+q_i\del I^i(z))\\G^+_\alpha \ &= \ G^++q_i\del a^i \\
		G^-_\alpha \ &= \ G^-+p^i\del a_i \\
\end{split}
\end{equation}
form an $N=2$ superconformal algebra of central charge 
\begin{equation}
 c \ = \ \frac{3}{2}\text{sdim}\,\g+3\str D\, -6q_ip^i.
\end{equation}
\end{prp}

\subsection{Spectral flow}

The physical state space of topological CFTs resulting from twisting the deformed $N=2$ superconformal field theories considered in the previous section in some cases coincides with the physical state space of topological CFTs obtained by twisting the undeformed $N=2$ superconformal field theories and acting by a spectral flow automorphism. 
The topological CFT is described by the cohomology of the BRST-operator $Q$. This operator is composed of the zero modes of the currents $G^\pm$ and $\bar G^\pm$. In this section, we show that the zero modes of some deformed super-currents $G^\pm_\alpha, \bar G^\pm_\alpha$ agree with the action of the spectral flow $\gamma_\alpha,\bar\gamma_\alpha$ on the zero modes of the undeformed super-currents, i.e.
\begin{equation}
 G^\pm_{\alpha,0} \ = \ \gamma_\alpha(G^\pm_0)\qquad\text{and}\qquad
\bar G^\pm_{\alpha,0} \ = \ \bar\gamma_\alpha(\bar G^\pm_0)\, .
\end{equation}
This implies that for every twist the two BRST-operators coincide and hence their cohomologies as well.
 
We restrict to type I Lie superalgebras $\g$ that allow for a Manin decomposition  as described in example (\ref{example:manindecomposition}). 
Recall that $\azero$ contains the Cartan subalgebra $\h$ in this example \eqref{eq:azerocartan}.
Denote by $x_i$ a basis of $\ap$ and by $x^i$ the dual basis of $\am$ with respect to the bilinear form $(\,\cdot\,,\,\cdot\,)$.
Then the generators of the affine Lie superalgebra $\hat\g$ are denoted by
$x_{i,n},x^i_n$, in addition the level is fixed to be $k$ and the derivation $d$ is identified with the Virasoro zero mode $L_0$. 
We restrict our attention to the holomorphic currents. The anti-holomorphic part is treated analogously. 

Recall the mode expansion of the affine currents
\begin{equation}
	\begin{split}
		J^i(z) \ &= \ \sum_{n\,\in\,\Z}x^i_{n}z^{-n-1} \\
		J_i(z) \ &= \ \sum_{n\,\in\,\Z}x_{i,n}z^{-n-1} \\
		a^i(z) \ &= \ \sum_{n\,\in\,\Z}a^i_{n}z^{-n-1/2} \\
		a_i(z) \ &= \ \sum_{n\,\in\,\Z}a_{i,n}z^{-n-1/2} \\
	\end{split}
\end{equation}
Then $G^\pm$ have the following mode expansions
\begin{equation}\nonumber
\begin{split}
	G^+(z)\, &= \, \sum_{n\,\in\,\Z}G^+_{n}z^{-n-3/2}\\
&= \, \sum_{n,m\,\in\,\Z} x_{i,n-m}a^i_{m}z^{-n-3/2}-\frac{1}{2}(-1)^i{c_{ij}}^k \sum_{n,m,r\,\in\,\Z}:a^i_{n-m-r}a^j_ma_{k,r}:z^{-n-3/2} \, ,\\
G^-(z)\, &= \, \sum_{n\,\in\,\Z}G^-_{n}z^{-n-3/2}\\
&= \, \sum_{n,m\,\in\,\Z} x^i_{n-m}a_{i,m}z^{-n-3/2}-\frac{1}{2}(-1)^j{f^{ij}}_k \sum_{n,m,r\,\in\,\Z}:a_{i,n-m-r}a_{j,m}a^k_r:z^{-n-3/2}\, .
\end{split} 
\end{equation} 
Here the normal ordering sign means that positive mode operators are to the right of negative mode operators. 
Further recall the action of spectral flow.
Let $(\,\cdot\,|\,\cdot\,)$ be the bilinear form on $\h^\star$ induced by $(\,\cdot\,,\,\cdot\,)$. Recall that the bilinear form contains the level $k$. 
Let $\beta$ be a coroot, further denote by $\alpha_i$ the root corresponding to $x_i$, then $-\alpha_i$ corresponds to $x^i$. Note that when $x_i$ is in the Cartan subalgebra this means that $\alpha_i=0$.
Then the action of a spectral flow automorphism is \eqref{eq:spectralflow} and \cite{Creutzig:2008ag},
\begin{equation}
	\begin{split}
		\tilde{T}_\beta:\ \ x_{i,n} \ &\mapsto\ x_{i,n-(\beta|\alpha_i)}+\beta(x_i)\delta_{\alpha_i,0}\delta_{n,0} \\[1mm]
	                            x^i_{n} \ &\mapsto\ x^i_{n+(\beta|\alpha_i)}+\beta(x^i)\delta_{\alpha_i,0}\delta_{n,0} \\[1mm]
				    a_{i,n} \ &\mapsto\ a_{i,n+(\beta|\alpha_i)} \\[1mm]
	                            a^i_{n} \ &\mapsto\ a^i_{n-(\beta|\alpha_i)}\, . \\[1mm]
	\end{split}
\end{equation}
This induces an action on $G^\pm$
\begin{equation}
	\begin{split}
		\tilde T_\beta:\ \,G^{+}(z)\ &\mapsto \ G^+(z)+\sum_{\substack{x_i\,\in\,\h_+\\ n\,\in\,\Z}}
\Bigl(\beta(x_i)+\sum_{\alpha_j\,\in\,\Delta_+}\alpha_j(x_i)(\beta|\alpha_j)\Bigr)a^i_nz^{-n-3/2}\, , \\
\tilde T_\beta:\ \,G^{-}(z)\ &\mapsto \ G^-(z)+\sum_{\substack{x^i\,\in\,\h_-\\ n\,\in\,\Z}}
\Bigl(\beta(x^i)+\sum_{\alpha^j\,\in\,\Delta_-}(-1)^j\alpha^j(x^i)(\beta|\alpha^j)\Bigr)a_{i,n}z^{-n-3/2}\, . \\
	\end{split}
\end{equation}
Hence the zero modes get shifted by
\begin{equation}
\begin{split}
 \tilde T_\beta:\ \,G^+_0\ &\mapsto G^+_0+\sum_{x_i\,\in\,\h_+}
\Bigl(\beta(x_i)+\sum_{\alpha_j\,\in\,\Delta_+}\alpha_j(x_i)(\beta|\alpha_j)\Bigr)a^i_0\\
\tilde T_\beta:\ \,G^-_0\ &\mapsto G^-_0+\sum_{x^i\,\in\,\h_-}\Bigl(\beta(x^i)+\sum_{\alpha^j\,\in\,\Delta_-}(-1)^j\alpha^j(x^i)(\beta|\alpha^j)\Bigr)a_{i,0}\, .\\
\end{split}
\end{equation}
But this is the same shift as the one induced by a deformation with an appropriate Cartan subalgebra element $\gamma$ 
\begin{equation}
 \tilde T_\beta(G^\pm_0)\ = \ G^\pm_{\alpha_\beta,0}\, ,
\end{equation}
where 
\begin{equation}
\begin{split}
         \alpha_\beta \ = \ &-2\sum_{x_i\,\in\,\h_+}
\Bigl(\beta(x_i)+\sum_{\alpha_j\,\in\,\Delta_+}\alpha_j(x_i)(\beta|\alpha_j)\Bigr)x_i\ +\\
&-2\sum_{x^i\,\in\,\h_-}\Bigl(\beta(x^i)+\sum_{\alpha^j\,\in\,\Delta_-}(-1)^j\alpha^j(x^i)(\beta|\alpha^j)\Bigr)x^i\ \, .
\end{split}
\end{equation}
In summary, we have shown that the zero modes of the supercurrents of the deformed $N=2$ superconformal field theory (deformed by $\alpha_\beta$) coincide with the image of the spectral flow automorphism $\tilde T_\beta$ on the zero modes of the undeformed supercurrents. 
It follows that the BRST operators of the twisted topological CFTs also coincide and hence their cohomology groups.

Note that one could also consider spectral flow induced by elements in $\h^\star$ that are not coroots. Then the affine Lie superalgebra is mapped to a twisted affine Lie superalgebra. The correspondence to deformation still holds.

\subsection{Example \GL}

In this section we want to compute the space of physical observables in a B-twist of the $N=1$ \GL\ WZNW model. 
Let $\psi^\pm,E,N$ be the generators of \GL, and introduce $\tilde{N}=N+\frac{E}{2k}$.
Then the $N=2$ superconformal currents are given by
\begin{equation}
\begin{split}
	G^+\ &=\ \frac{1}{\sqrt{k}}(J_E\xi+J_+\gamma) \\
	G^-\ &=\ \frac{1}{\sqrt{k}}(J_{\tilde{N}}\eta+J_-\beta+\eta:\gamma\beta:) \\
	U \ &= \ :\xi\eta:+:\beta\gamma:-\frac{J_E}{k}\\
\end{split}
\end{equation}
and their anti-holomorphic counterparts read
\begin{equation}
\begin{split}
	\bar G^+\ &=\ \frac{1}{\sqrt{k}}(\bar J_E\bar \xi+\bar J_+\bar\gamma) \\
	\bar G^-\ &=\ \frac{1}{\sqrt{k}}(\bar J_{\tilde{N}}\bar \eta+\bar J_-\bar\beta+\bar \eta:\bar \gamma\bar\beta:) \\
	\bar U \ &= \ :\bar\xi\bar\eta:+:\bar\beta\bar\gamma:-\frac{\bar J_E}{k}\, .\\
\end{split}
\end{equation}
Here $\eta(z),\xi(z),\bar\eta(\bar z),\bar\xi(\bar z)$ are fermionic chiral fields and $\gamma(z), \beta(z), \bar\gamma(\bar z),\bar\beta(\bar z)$ are bosonic chiral fields with OPEs
 \begin{equation}
 \begin{split}
  \eta(z)\xi(w) \ &\sim \ \frac{1}{(z-w)}\qquad,\qquad
\beta(z)\gamma(w) \ \sim \ \frac{1}{(z-w)}\, ,\\
 \bar\eta(\bar z)\bar\xi(\bar w) \ &\sim \ \frac{1}{(\bar z-\bar w)}\qquad,\qquad
\bar\beta(\bar z)\bar\gamma(\bar w) \ \sim \ \frac{1}{(\bar z-\bar w)}\, .\\
 \end{split}
\end{equation}
Consider the B-twisted topological CFT defined by
\begin{equation}\label{eq:GLtwist}
	T^+_{\text{twisted}}(z)\ = \ T(z)+\frac{1}{2}\del U(z)\qquad\text{and}\qquad
	\bar T^+_{\text{twisted}}(\bar z)\ = \ \bar T(\bar z)+\frac{1}{2}\bar\del\bar U(\bar z)\, .
\end{equation}
Then the conformal dimensions in the twisted theory are as follows
\begin{equation}
 \begin{split} 
  \Delta(\xi) \ &= \ \Delta(\gamma) \ = \ (0,0)\, , \\
\Delta(\eta) \ &= \ \Delta(\beta) \ = \ (1,0)\, , \\
 \Delta(\bar\xi) \ &= \ \Delta(\bar\gamma) \ = \ (0,0)\, , \\
\Delta(\bar\eta) \ &= \ \Delta(\bar\beta) \ = \ (0,1)\, . \\
 \end{split}
\end{equation}
 Further the BRST-operator is
\begin{equation}\label{eq:BRSToperator}
	\begin{split}
	Q_{\text{BRST}}\ &= \ G^+_0+\bar G^+_0\\
	                 &= \ \frac{1}{\sqrt{k}}(J_Ec+J_+\gamma)_0+\frac{1}{\sqrt{k}}(\bar J_E\bar c+\bar J_+\bar\gamma)_0\, .
	 \end{split}
 \end{equation}
 A representative of a physical observable can always be chosen to have conformal dimension $(0,0)$ due to \eqref{eq:TQexact}.
 Thus we can restrict our attention to fields of zero conformal dimension. Consider a family of automorphisms $\tau_{\alpha,\bar\alpha}$ of \agl\,$\times$\,\agl\ induced by \eqref{eq:transautomorphismgl11}. They are defined as follows
\begin{equation}\label{eq:autogl}
	\tau_{\alpha,\bar\alpha}(N_0) \ = \ N_0+\alpha E_0 \qquad,\qquad 
\tau_{\alpha,\bar\alpha}(\bar N_0) \ =\ \bar N_0+\bar\alpha\bar E_0
\end{equation}
and leaving all other operators invariant. 
These automorphisms leave the BRST-operator invariant. Recall the bulk fields of \GL\ \eqref{eq:polvertexope1} 
\begin{equation}
    V_{\<-e,-n+1\>}\ =\ \normord{e^{eX+nY}}\begin{pmatrix}
                           1 & c_-\\
                           c_+ & c_-c_+ \\
                         \end{pmatrix},
\end{equation}
and their conformal dimension in the twisted theory is (compare with \eqref{eq:polconfdim})
\begin{equation}
    \De({V_{\<-e,-n+1\>}})\ =\ \Bigl(\frac{e}{2k}(2n-2+\frac{e}{k}),\frac{e}{2k}(2n+\frac{e}{k})\Bigr)\, .
\end{equation}
Thus, when $e\neq0$ the above automorphisms \eqref{eq:autogl} ensure that every primary and every descendant of $V_{\<-e,-n+1\>}$ is isomorphic to a field of non-integer conformal dimension and hence cannot contribute to the physical observables.

Hence we restrict our attention to fields with $e=0$. The conformal dimension zero fields are
\begin{equation}
	\Phi(n,m,\bar m, \lambda,\bar \lambda) \ := \ V_{\<0,-n+1\>}\gamma^m\bar\gamma^{\bar m}\xi^\lambda\bar \xi^{\bar\lambda}
\end{equation}
for non-negative integers $m,\bar m$ and $\lambda,\bar \lambda\in\{0,1\}$. 
In \cite{Schomerus:2005bf} it is shown that the vertex operator $V_{\<0,-n+1\>}$
transforms as follows
\begin{equation}
	\begin{split}
	J_+(z)V^{1a}_{\<0,-n+1\>}(w,\bar w) \ &\sim \ k_{a}\frac{V^{0a}_{\<0,-n+1\>}(w,\bar w)}{(z-w)} \\
        J_+(z)V^{0a}_{\<0,-n+1\>}(w,\bar w) \ &\sim \ 0 \\
        \bar J_+(\bar z)V^{1a}_{\<0,-n+1\>}(w,\bar w) \ &\sim \ \bar k_{a}\frac{V^{0a}_{\<0,-n\>}(w,\bar w)}{(\bar z-\bar w)} \\
        \bar J_+(z)V^{0a}_{\<0,-n+1\>}(w,\bar w) \ &\sim \ 0 \\
\end{split}
\end{equation}
for some non-zero constants $k_{a}, \bar k_{a}$. 	
Thus the image of $Q_{\text{BRST}}$ for these fields is
\begin{equation}
	\begin{split}
	\bigl[Q_{\text{BRST}},\Phi^{1,a}(n,m,\bar m, \lambda,\bar \lambda)\bigr]\ = \ 
	&k_{a}\Phi^{0,a}(n,m+1,\bar m, \lambda,\bar \lambda)+\\
	&\bar k_{1a}\Phi^{0,a}(n-1,m,\bar m+1, \lambda,\bar \lambda)\, .
\end{split}
\end{equation}
On the other hand the kernel is 
\begin{equation}
	\bigl[Q_{\text{BRST}},\Phi^{0,a}(n,m,\bar m, \lambda,\bar \lambda)\bigr]\ = \ 0\, .
\end{equation}
Thus a basis of representatives of the space of physical observables is
\begin{equation}
	\{ \Phi^{0,a}(n,m,0, \lambda,\bar \lambda)\, |\, m \geq0 \ \text{and}\ a,\lambda, \bar\lambda \in \{0,1\}\}\, .
\end{equation}
Their OPEs are 
\begin{equation}
	\begin{split}
	\Phi^{0,a}(n_1,m_1,0, \lambda_1,\bar \lambda_1)\Phi^{0,b}(n_2,&m_2,0,\lambda_2,\bar \lambda_2)\ \sim \ (1-ab)(1-\alpha_1\alpha_2)(1-\bar\alpha_1\bar\alpha_2)\ \times\\[2mm]
	&\times \Phi^{0,a+b}(n_1+n_2,m_1+m_2,0, \lambda_1+\lambda_2,\bar \lambda_1+\bar\lambda_2)\, .
\end{split}
 \end{equation}
 We conclude that the space of physical observables of the $N=1$ \GL\ WZNW model for the twist \eqref{eq:GLtwist} is
 \begin{equation}
	 \mathcal{H}_{\text{phys}} \ = \ \R\times\Z_{\geq0}\times\Z_2\times\Z_2\times\Z_2\, .
 \end{equation}
In the \GL\ WZNW model the $V^{0a}_{\<0,-n+1\>}(z,\bar z)$ span a maximal set of fields whose OPEs are independent of their world-sheet positions.

\section{Branes}

From now on, we restrict our attention to type I Lie supergroups and the Manin decomposition of example \ref{example:manindecomposition}. For branes in $N=2$ superconformal cosets of Lie groups see e.g. \cite{Lindstrom:2002vp}. 

We want to investigate branes that preserve the $N=2$ superconformal symmetry but also the affine Lie superalgebra symmetry. 
The $N=2$ superconformal algebra is preserved if we require the following gluing conditions
\begin{equation}\label{eq:Agluing}
	\begin{split}
		G^\pm(z)\ &= \ \eta\bar G^\mp(\bar z)\, ,\\
		U(z)\ &= \ -\bar U(\bar z)\, ,\\
		T(z)\ &= \ \bar T(\bar z)\, \qquad\qquad\text{for}\, z\ = \ \bar z\, .\\
\end{split}
\end{equation}
Here, $\eta=\pm1$. These gluing conditions preserve the A-twist
\begin{equation}
  T(z)\pm\frac{1}{2} \del U(z) \ =\ \bar T(\bar z)\mp\frac{1}{2}\bar\del \bar U(\bar z)\,  
\qquad\text{for}\ z\ = \ \bar z\, .	
\end{equation}
Another choice for gluing conditions is
\begin{equation}
	\begin{split}
		G^\pm(z)\ &= \ \eta\bar G^\pm(\bar z)\, ,\\
		U(z)\ &= \ \bar U(\bar z)\, ,\\
		T(z)\ &= \ \bar T(\bar z)\, \qquad\qquad\text{for}\, z\ = \ \bar z\, .\\
\end{split}
\end{equation}
In this case the B-twist is preserved
\begin{equation}
  T(z)\pm\frac{1}{2} \del U(z) \ =\ \bar T(\bar z)\pm\frac{1}{2}\bar\del \bar U(\bar z)\,
\qquad\text{for}\ z\ = \ \bar z\, .	
\end{equation}
Preserving affine Lie superalgebra symmetry means that the gluing conditions of the currents are given by a metric preserving automorphism $\Omega$
\begin{equation}
 J(z)\ = \ \Omega(\bar J(\bar z))\qquad\text{for}\ z\ = \ \bar z\, .
\end{equation}
Let $\g$ be the Lie superalgebra of type I allowing a Manin triple, i.e.
$\g\,\in\{gl(n|n), psl(n|n)$, $sl(n\pm1|n)\}$, and $\g=\ap\oplus\am$ the Manin decomposition of example \ref{example:manindecomposition}. Then in order to preserve the B-twist $\Omega$ has to be a Lie superalgebra automorphism of $\ap$ and also of $\am$. A natural candidate is the identity automorphism.

\subsection{B-branes}

We start with the gluing conditions for the currents $J$ and $a$, we take
\begin{equation}
	J(z) \ = \ \bar J(\bar z)\qquad , \qquad a(z) \ = \ \eta\bar a(\bar z)\qquad\text{for}\, z\ = \ \bar z\,
\end{equation}
where $\eta=\pm1$. Inserting these conditions in the $N=2$ currents gives
\begin{equation}
	\begin{split}
		G^\pm(z)\ &= \ \eta\bar G^\pm(\bar z)\, ,\\
		G_b^\pm(z)\ &= \ \eta\bar G_b^\pm(\bar z)\, ,\\
		U(z)\ &= \ \bar U(\bar z)\, ,\\
		T(z)\ &= \ \bar T(\bar z)\, \qquad\qquad\text{for}\, z\ = \ \bar z\, .\\
	\end{split}
\end{equation}
Thus, these conditions preserve the B-twist
\begin{equation}
	T(z)\pm \frac{1}{2}\del U(z)\ = \ \bar T(\bar z)\pm\frac{1}{2}\bar\del\bar U(\bar z))\, \qquad\text{for}\ z \ = \ \bar z\, .
\end{equation}

\subsection{A-branes}

The case of A-branes is more subtle and the gluing conditions on the supercurrents will differ from \eqref{eq:Agluing}.
We employ the automorphism $\Omega=(-st)$ \eqref{eq:minussupertranspose}. 
Its action on the generators of the affine currents is as follows
\begin{equation}
 \begin{split}
\Omega(J_i(z)) \ &= \
  \Bigl\{ \begin{array}{cc}
	       \ -J^i(z)&\qquad \text{if} \ x_i \ \text{in}\ \nn_+    \\
	       \ -J_i(z)&\qquad \text{if} \ x_i \ \text{in}\ \h_+ \\
                        \end{array} \\
\Omega(J^i) \ &= \
  \Bigl\{ \begin{array}{cc}
	       \ -(-1)^iJ_i(z)&\qquad \text{if} \ x^i \ \text{in}\ \nn_-    \\
	       \ -J^i(z)&\qquad \text{if} \ x^i \ \text{in}\ \h_- \\
                        \end{array}\, .
 \end{split}
\end{equation}
Further for the fields $a(z)$ we choose the following automorphism
\begin{equation}
 \omega(a^i(z)) \ = \ a_i(z)\qquad\text{and}\qquad
\omega(a_i(z)) \ = \ (-1)^ia^i(z)\, .
\end{equation}
These two automorphisms induce an isomorphism $\tilde\Omega$ on $N=2$ superconformal algebras
\begin{equation}
 \begin{split}
  \tilde\Omega(G^+) \ &= \ \tilde G^+ \ = \ -J^ia_i-\frac{1}{2}(-1)^j{c_{ij}}^k:a_ia_ja^k:\\
  \tilde\Omega(G^-) \ &= \ \tilde G^- \ = \ -J_ia^i-\frac{1}{2}(-1)^i{f^{ij}}_k:a^ia^ja_k:\\
\tilde\Omega(U) \ &= \ -U\\
 \tilde\Omega(T) \ &= \ T\, .\\
 \end{split}
\end{equation}
Thus we obtained a second copy of an $N=2$ superconformal algebra. Note that the $U(1)$-currents only differ by a sign.
Analgously we define anti-holomorphic supercurrents 
\begin{equation}
 \begin{split}
  \tilde\Omega(G^+) \ &= \ \tilde{\bar G}^+ \ = \ -\bar J^i\bar a_i-\frac{1}{2}(-1)^j{c_{ij}}^k:\bar a_i\bar a_j\bar a^k:\\
  \tilde\Omega(G^-) \ &= \ \tilde{\bar G}^- \ = \ -\bar J_i\bar a^i-\frac{1}{2}(-1)^i{f^{ij}}_k:\bar a^i\bar a^j\bar a_k:\\
\tilde\Omega(U) \ &= \ -\bar U\\
 \tilde\Omega(T) \ &= \ \bar T\, .\\
 \end{split}
\end{equation}

After this preparation, we use the automorphisms considered above as gluing automorphisms, i.e. we demand the following boundary conditions
\begin{equation}
 J(z)\ = \ \Omega(\bar J(\bar z))\quad\text{and}\quad
a(z)\ = \ \omega(\bar a(\bar z))\quad\text{for}\ z \ = \ \bar z\, .
\end{equation}
These imply the boundary conditions for the $N=2$ superconformal algebra
\begin{equation}
 \begin{split}
  G^+(z) \ &= \ \tilde{\bar G}^+(\bar z)\\
  G^-(z) \ &= \ \tilde{\bar G}^-(\bar z)\\
U(z) \ &= \ -\bar U(\bar z)\\
T(z) \ &= \ \bar T(\bar z)\qquad\text{for}\ z \ = \ \bar z\, .\\
 \end{split}
\end{equation}
Thus the A-twist is preserved
\begin{equation}
	T(z)\pm \frac{1}{2}\del U(z)\ = \ \bar T(\bar z)\mp\frac{1}{2}\bar\del\bar U(\bar z))\, \qquad\text{for}\ z \ = \ \bar z\, .
\end{equation}

\section{Conclusion}

In this chapter, we have introduced a new family of $N=2$ superconformal field theories. 
The construction is an extension of the findings of Kazama and Suzuki, who considered the Lie group case. While only cosets of compact Lie groups allow for an $N=2$ superconformal algebra, Lie supergroups provide a richer variety of realizations of superconformal algebras. E.g. even the Lie supergroups $gl(n|n), psl(n|n), sl(n\pm1|n), osp(2n|2n)$ and $osp(2n+1|2n)$ allow for the construction of an $N=2$ superconformal algebra. Also solvable Lie (super)groups like Heisenberg supergroups and even Heisenberg groups allow for the construction. These cases are very interesting as they are the Penrose limit of models with AdS target space. 

Moreover, we considered deformations of $N=2$ superconformal algebras. Then we explained that the physical state space of topological CFTs resulting from twisting the deformed $N=2$ superconformal field theories in some cases coincides with the physical state space of topological CFTs obtained by twisting the undeformed $N=2$ superconformal field theories and acting by a spectral flow automorphism.

In the example of a B-twist of the \GL\ model we computed the cohomology ring. Finally, we considered branes in the $N=2$ superconformal Lie supergroup models. We found gluing conditions that preserve the A-twist and we also found gluing conditions that preserve the B-twist.

\chapter{Outlook}

\section{Results}

The aim of this thesis was to initiate a systematic study of Lie supergroup boundary WZNW models. We started with symplectic fermions. 
We showed that they possess an $SL(2)$ family of boundary conditions.
We constructed their boundary states in the twisted and also in the untwisted sectors. Furthermore, amplitudes were computed. In order to obtain a non-trivial amplitude it was sometimes necessary to insert additional fermionic fields.

As a simple prototypical example of a boundary supergroup WZNW model we chose the \GL\ model. This model possesses two families of current algebra preserving gluing conditions. For both of them, we constructed the boundary states and computed overlaps. The results agree with fusion, similar to the case of WZNW models on compact Lie groups. We also found typical features of logarithmic CFT as the appearance of indecomposable but reducible spectra, overlaps of Ishibashi states with log q dependence and Ishibashi states with zero norm. In order to get a non-vanishing amplitude we again sometimes had to insert additional fields.

For the twisted boundary conditions, we set up a first order formulation to solve the model. The novel feature in this set-up was the introduction of an additional fermionic boundary degree of freedom. We then solved the model, that is we computed bulk one-point functions, bulk-boundary two-point functions and boundary three-point functions. Logarithmic singularities appeared in certain bulk-boundary two-point functions and boundary three-point functions. 

Previously, the bulk \GL\ WZNW model was solved using the first order formulation. We showed that this model is equivalent to a pair of scalars plus symplectic fermions. The non-triviality of this model lies in the twisted symplectic fermion sectors. Thus, we gave a different approach to the bulk model. 

For general Lie supergroups, we showed that geometrically a branes' worldvolume is a twisted superconjugacy classes and we constructed their actions. Further, we identified superconjugacy classes with representations of the affine Lie superalgebra. Whenever the superconjugacy class is localised in some fermionic direction the associated representation is atypical. 
Moreover there are regions in the supergroup, which are not covered by any superconjugacy class. We suspect that there exist also branes covering these regions. In the case of \GL\ appropriate gluing conditions and actions for these new atypical branes exist \cite{Creutzig:2008ag}. This again involved the additional introduction of extra boundary degrees of freedom. 

Finally, we turned to world-sheet and target space supersymmetric theories. The celebrated Kazama-Suzuki cosets are $N=1$ world-sheet supersymmetric cosets of compact Lie groups and they possess a hidden $N=2$ superconformal symmetry. We show that this result does not only extend to many cosets of Lie supergroups, but also to some Lie supergroups as $GL(n|n), PSL(n|n),SL(n\pm1|n), OSP(2n|2n), OSP(2n+1|2n)$ and Heisenberg (super)groups. Moreover, there exist deformations of these models and we show that in some cases these deformations coincide with spectral flow on the level of the twisted topological field theory. Finally, we explain that each supergroup possesses two families of branes, one that preserves the A-twist and one that preserves the B-twist.

\section{Open problems}

Our findings leave a variety of interesting directions for future research. 

In the \GL\ WZNW model it would be interesting to solve the boundary theory of point-like branes. For this purpose the computation of bulk-boundary two-point functions is missing. This is difficult, because a first order formulation gives a perturbative description which does not terminate after a finite number of steps. On the other hand we can employ the symplectic fermion correspondence to this problem. We believe that this is doable. 

The methods we developed in this thesis should be applied to more sophisticated supergroups. The apparent open problem is the extension of the first order formalism to any type I Lie supergroup and appropriate gluing automorphism, i.e. $\Omega=(-st)$. This will require the additional introduction of fermionic boundary degrees of freedom and a boundary screening charge which looks like a square root of the bulk screening charge.

This problem resembles matrix factorisation in open string Landau Ginzburg models. 
Landau Ginzburg models possess an $N=2$ superconformal symmetry. Warner showed that it is necessary to introduce additional fermionic boundary degrees of freedom in order to preserve the superconformal symmetry at the boundary \cite{Warner:1995ay}.
Further the bulk super potential factorises into the boundary super potentials. 

One question is to understand the connection between world-sheet and target space supersymmetric theories. 

An important goal is to understand the newly introduced $N=2$ superconformal field theories. Let us list some questions.
\begin{itemize}
 \item What is the chiral ring of such a model? \footnote{Very recently an example has been investigated \cite{Giribet:2009eb}.} 
\item Can we use mirror symmetry to understand correspondences and dualities?
\item Are there models with even more supersymmetry like $N=4$?
\item Are there deformations of the $N=1$ PSL(n$|$n) WZNW model that preserve the $N=2$ (or $N=4$) superconformal symmetry?
\end{itemize}
Some of these questions are already under investigation.

\section{Applications beyond WZNW models}

Let us conclude with two problems that go beyond WZNW models and affine Lie superalgebras. 

Due to its underlying affine Lie superalgebra symmetry the WZNW model on a supergroup is well treatable. But for some Lie supergroups there exist many more CFTs with less symmetry. If the Killing form of a supergroup vanishes then there exists an additional one-parameter family of conformally invariant sigma models on this supergroup \cite{Bershadsky:1999hk}. 
These additional CFTs can be described as an exactly marginal perturbation of the WZNW model. In view of the AdS/CFT correspondence the PSU(1,1|2) sigma models play an important role.
Computations in these sigma models are not easy. The strategy is to restrict attention to some quantities that are protected by symmetry. In section \ref{section:gl11symplecticfermions} and also in section \ref{sec:coratyp} we found that in the \GL\ WZNW model correlators consisting only of atypical fields are protected, i.e. they could be computed in free field theory.
Also in the PSU(1,1|2) sigma models we succeeded to compute boundary spectra of branes that are localised in the bosonic directions while extending completely into the fermionic ones \cite{Quella:2007sg}. The perturbative computation of these spectra could be performed because of many cancellations due to the symmetry of the model.

The goal is to extend this analysis and to find other quantities that are also protected by symmetry. The idea is to use a fermionic symmetry $Q$ of the supergroup sigma model that squares to zero, $Q^2=0$, and thus defines a cohomology. We then want to employ such a symmetry to argue that the cohomology is protected, i.e. it is not or only partially influenced by a perturbation \cite{Creutzig:2009b}.
\smallskip

The second interesting area we would like to mention is logarithmic CFT in the context of three-dimensional gravity. One believes that three-dimensional pure gravity has a dual conformal field theory description \cite{Witten:2007kt}. The relevant CFTs are extremal of central charge an integer multiple of 24. Moreover it was recently observed that the CFT should be logarithmic \cite{Grumiller:2008es}. 
An important question in conjunction with gravity is to find logarithmic extensions of extremal CFTs. More general one would like to understand how to extend a chiral conformal field theory to a logarithmic CFT. The search might profit from Lie supergroup sigma models where the origin of the logarithmic singularities is understood. Especially a Lie supergroup WZNW model can be treated as an exactly marginal perturbation of a Lie group WZNW model plus some fermionic ghost systems \cite{Quella:2007hr}. The unperturbed model is not logarithmic and the perturbation generates the logarithmic behaviour.
Moreover, in appendix \ref{section:ghosts} we consider a CFT that is non-logarithmic in the bulk, but the boundary theory possesses logarithmic singularities.

Furthermore, the extremal CFT of central charge 24 is the monster CFT \cite{Frenkel:1988xz}. It is famous for its relevance in the proof of moonshine by Richard Borcherds \cite{Borcherds}.
Another key ingredient in the proof is that the monster CFT has an underlying infinite dimensional Lie algebra whose denominator identity is an automorphic product.
It turns out that only ten Lie algebras of a similar kind exist \cite{Scheithauer}. Especially the denominator identities of these infinite dimensional Lie algebras are also automorphic products. Four out of these ten Lie algebras can be constructed from a conformal field theory \cite{Hoehn,Creutzig:2008gm}. For the remaining ones it is conjectured. Further some Lie superalgebras are also known \cite{Scheithauer2}. Thus there exists another class of conformal field theories, besides WZNW models, with an underlying infinite dimensional Lie (super)algebra. 
Moreover, not only the monster CFT has appeared in relation to gravity, but also infinite dimensional Lie algebras whose denominator identity is an automorphic product describe Dyon spectra in CHL compactifications and degeneracies of corresponding black holes \cite{Dijkgraaf:1996it}.

\vspace{2cm}
\smallskip
\qquad\qquad\qquad\qquad
\qquad\qquad{\bf Acknowledgements}\smallskip\newline
Most of all I would like to express my gratitude to Volker Schomerus for sharing his ideas and supporting me over the last years.
Further I am gratefull to Thomas Quella and Peter R\o nne for their collaboration and discussions.
Many thanks to Peter R\o nne and Volker Schomerus for numerous remarks on the draft of this thesis.
I also would like to thank our string theory group and its former members for 
exchanging ideas and many helpfull discussions.
 
For scientific support and interesting conversations, I would especially like to thank Nils Scheithauer and Vincent Bouchard.
Many thanks to Kai Keller and Thomas Hack for their help on computer problems and an enjoyable working atmosphere.

Besonders dankbar bin ich meinen Eltern, meiner Schwester Eva und meinem Bruder Felix, dass sie mich immer unterst\"utzt  haben und f\"ur mich da waren.
Finalement, un grand merci $\grave{\text{a}}$ Maryse pour l'amour et le bonheur que tu as apport$\acute{\text{e}}$s dans ma vie. 
\appendix

\chapter{The $bc$-ghost system}\label{section:ghosts}

We shortly present the bulk model.
The $bc$-ghost system involves two
sets of chiral bulk fields $c,\bar c$ and $b,\bar b$ of conformal
dimension $h_c = 0$ and $h_b = 1$, respectively. The action of the bulk model is 
\begin{equation} 
	S \ = \ \frac{1}{2\pi} \int d^2z \left[ b\,\bar \partial c +
           \bar b\,\partial\bar c \right]\, .
\end{equation}
The energy-momentum tensor is 
\begin{equation}
	T(z)\ = \ -b\del c\qquad ,\qquad \bar T(\bar z)\ = \ -\bar b\bar\del\bar c
\end{equation}
and the operator product expansion is
\begin{equation}\label{eq:bcope}
	b(z)c(w)\ \sim \ \frac{1}{(z-w)}\, \qquad\text{and}\qquad\bar b(\bar z)\bar c(\bar w)\ \sim \ \frac{1}{(\bar z-\bar w)}\, .
\end{equation}
The world-sheet is again the complex plane. On the complex plane, correlation functions to be non-zero require the insertion of a zero-mode of the fields $c$ and $\bar c$. We normalise them as follows
\begin{equation}
	\langle c(z)\bar c(\bar w)\rangle \ = \ 1\, .
\end{equation}
Arbitrary correlation functions are computed using the above contraction \eqref{eq:bcope}.
We now turn to the description of the boundary theory. 

\section{Twisted boundary conditions in the $bc$ $c=-2$ ghost system}

This section is the content of \cite{Creutzig:2006wk}.
In this section we study a new boundary condition for the $bc$
system with central charge $c=-2$. In the
conventional setup, we would glue $c$ to $\bar c$ and $b$
to $\bar b$ along the boundary \cite{Callan:1987px}. But there exists 
another possibility: namely, to glue $b$ to a derivative of $\bar c$
and  vice versa. More precisely, we can demand that
\begin{equation} \label{bc}
b(z) \, = \, \mu \bar \partial \bar c(\bar z) \  \ , \  \
\bar b(\bar z) \, = \, - \mu \partial c(z) \  \  \text{ for }
  \  z  = \bar z\ \ .
\end{equation}
These relations guarantee trivial gluing conditions for the
energy momentum tensor $T=-b\partial c$.
It is not difficult to check that the action of the $bc$ system
is invariant under variations respecting (\ref{bc}) provided
we add an appropriate boundary boundary term,
\begin{equation} \label{eq:actbcbdy}
S \ = \ \frac{1}{2\pi} \int d^2z \left[ b\,\bar \partial c +
           \bar b\,\partial\bar c \right]
  - \frac{i\mu}{4\pi} \int du \ c\,\partial_u \bar c \ \ .
  \end{equation}
Our aim here is to solve the theory that is defined by
the action (\ref{eq:actbcbdy}) and the boundary condition
(\ref{bc}). We shall set $\mu =1$ throughout our
discussion. Formulae for the general case are easily
obtained from the ones we display below.

In order to construct the state space and the fields
explicitly, we introduce an algebra that is generated
by the modes $c_n, b_n$ and two additional zero
modes $\xib_0, \xic_0$ subject to the conditions
\begin{eqnarray} \label{comm}
\{ c_n,b_m\} & = & n\,\delta_{n,-m} \ \ ,\\[2mm]
 \{ \xic_0,b_0 \} \ = \ 1 \ \ \ \ & ,  & \ \ \ \ %
 \{\xib_0 , c_0 \} \ = \ 1 \ \ .
\end{eqnarray}
All other anti-commutators in the theory are assumed to
vanish. The state space of our boundary theory is
generated from a ground state with the properties
\begin{equation}\label{vac}
c_n | 0\rangle \ = \ b_n |0\rangle\ = \ 0
\ \ \ \text{ for } \ \ \ n \ \geq \ 0 
\end{equation}
by application of `raising operators', including
the zero modes $\xib_0$ and $\xic_0$. On this
space we can introduce the local fields $c,\bar c,
b,\bar b$ through the prescription
\begin{eqnarray} \label{b6} 
b(z) &= & \ \sum_{n\in\QZ} b_n z^{-n-1} \\[2mm]
c(z) & = & \sum_{n\neq0} \frac{c_n}{n} z^{-n} +
 c_0 \ln z + \xic_0\\[2mm]
\bar b(\bar z) & = & \sum_{n\neq 0} c_n \bar z^{-n-1} - c_0 \bar
z^{-1}
\\[2mm]
\bar c(\bar z)  & = & - \sum_{n\neq 0} \frac{b_n}{n} \bar z^{-n} +
b_0 \ln \bar z - \xib_0 \label{bc9} 
\end{eqnarray}
It is not difficult to check with the help of
eqs.\ (\ref{comm}) that these fields satisfy
the correct local anti-commutation relations
$$
\bigl\{ b(z), c(w) \bigr\} \ = \ \delta(z-w) \ \ , \ \
\bigl\{ \bar b(\bar z),\bar c(\bar w) \bigr\} \ = \
\delta(\bar z- \bar w)\ \
$$
in the interior of the upper half plane.
Needless to stress that they also fulfil our
boundary conditions (\ref{bc}) with $\mu = 1$.

For later use let us also spell out the construction
of the Virasoro generators in terms of fermionic modes,
$$ L_n \ = \ \sum_{m \neq 0} \,  -:b_{n-m} c_m : - b_n
    c_0\ \ . $$
It is important to stress that - due to the term $c_0 b_0$
--  the element $L_0$ satisfies $L_0 \xic_0\xib_0 |0\rangle = |0\rangle$. 
Since $L_0$ vanishes on all other ground states, it is
non-diagonalisable. In other words, our boundary theory is an
example of a logarithmic conformal field theory. The logarithms in
this model, however, are restricted to the boundary sector since
the Hamiltonian of the bulk theory is diagonalisable (see below).

Before we can calculate correlation functions
in our boundary theory, we need to introduce
a dual vacuum with the properties
\begin{eqnarray}\label{cav}
\langle 0| c_n \ = \ \langle 0| b_n  = 0
\ \ \ & \text{ for } & \ \ \ n \ \leq \ 0 \\[2mm]
\langle 0 | \xic_0 \xib_0 |0\rangle 
 =  (2\pi)^{-1} \ \ \ & \text{and} & \ \ \ \langle 0 | 0 \rangle  =  0 \ \ .
\end{eqnarray}
Our particular normalisation of the dual vacuum $\langle 0|$ 
will turn out to be convenient below. With the help of our 
formulae (\ref{b6})-(\ref{bc9}), we can compute arbitrary 
correlators. As long as there are no insertions of $\bar c$, 
correlators take the following simple form
\begin{equation}
 \langle0| \prod^{n_c}_{\mu} c(w_\mu) \prod^{n_{\bar b}}_{\bar \nu} 
\bar b(\bar z_{\bar \nu})
\prod^{n_b}_{\nu} b(z_\nu) |0\rangle \ = \ \prod_{\bar \nu} (-\partial_{\bar \nu})
\frac{\prod_{\nu<\nu'} x_{\nu\nu'}\prod_{\mu < \mu'}x_{\mu\mu'}
\prod_{\bar \nu< \bar\nu'}x_{\bar \nu\bar \nu'}}
 { \prod_{\bar\nu,\mu} x^{-1}_{\bar \nu \mu }\prod_{\nu,\mu} x_{\nu\mu}
 \prod_{\nu,\bar \nu}x_{\nu\bar\nu}}\, ,
\end{equation}
where $x_{\nu\mu} = z_\nu -z_\mu, x_{\nu\bar \mu} = z_\nu-
\bar z_{\bar \mu}$ etc and $\partial_{\bar \nu}$ denote derivatives
with respect to $\bar z_{\bar \nu}$. Insertions of the field $\bar c$ 
may be removed one after the other using the following rules for 
contractions
$$
\bar c(\bar z)  c(w) \ \sim\  \ln (\bar z-w) \ \ \ , \ \ \
\bar c(\bar z) \bar b(\bar w) \ \sim \ (\bar z-\bar w)^{-1}
$$
that can be derived from our explicit operator realisation
of the basic fields. The other two types of contractions
with fields $\bar c$ or $b$ vanish identically.

Next we would like to display the boundary state $|N\rangle$ 
for our new boundary condition. Before we provide explicit 
formulae let us briefly recall that the bulk fields are 
obtained as 
$$ 
c(z) \ =  \ \xic_0 + \sum_{n \neq 0} \, \frac{c_n}{n} \, z^{-n}  
\ \ \ , \ \ \ b(z) \ =  \ \sum_{n\in Z} \, b_n \, z^{-n-1} 
$$
and similarly for their anti-holomorphic counterparts. Note that 
there are no modes $c_0,\bar c_0$ and $\xib_0,\bar\xib_0$ in the
bulk of our $bc$ ghost system. This feature distinguishes the 
$c=-2$ ghosts from the closely related symplectic fermions. 
According to the standard rules, the boundary state for our 
boundary theory must satisfy the following Ishibashi conditions 
\cite{Ishibashi:1988kg}
\begin{equation} \label{Ish1} 
(b_n -  \bar c_{-n}) |N\rangle \ = \ 0 \ \ , \ \ (c_n + \bar
b_{-n}) |N\rangle \ = \ 0 \ \
\end{equation}
for $n\neq 0$ and $b_0 |N\rangle = \bar b_0 |N\rangle = 0$. As
one may easily check, the unique solution to these conditions 
is given by
\begin{equation}\label{eq:BS}
{|N\rangle} \, = \, 
\exp\left(-\sum_{m=1}^\infty\,
 ( \frac{c_{-m} \bar c_{-m}}{m} +
  \frac{b_{-m} \bar b_{-m}}{m}) \right) |0\rangle
\end{equation}
where $|0\rangle$ is a state in the bulk theory that satisfies
conditions of the form (\ref{vac}) for both chiral and anti-chiral
modes. There also exists a dual boundary state $\langle N|$,
satisfying the conditions
\begin{equation}
\langle N| (b_n + \bar c_{-n}) \ = \ 0 \ \ , \ \ \langle N | (c_n
- \bar b_{-n}) \ = \ 0 \ \
\end{equation}
for $n\neq 0$ and $\langle N| b_0  = \langle N| \bar b_0 = 0$.
These linear relations are related to eqs. (\ref{Ish1}) by 
conjugation using that $c_n^* = - c_{-n}$ and $b^*_n = b_{-n}$ 
etc. The dual boundary state is given by the following explicit 
formula
\begin{equation}\label{dBS}
{\langle N|} \, = \, \langle 0|\, \exp\left(\sum_{m=1}^\infty \, 
  \frac{1}{m}\, ( c_{m} \bar c_{m} +
    b_{m} \bar b_{m}) \right)
\end{equation}
involving a dual closed string ground state $\langle 0|$ that
obeys conditions of the form (\ref{cav}) for modes of chiral and
anti-chiral fields and that is normalised by $\langle 0 | 
\xic_0 \bar \xic_0 |0\rangle=1$ 
\footnote{In order to have $SL(2,C)$ invariant vacua $|0\rangle$ and 
$\langle 0 |$, they have to be annihilated by the zero modes $b_0,\bar b_0$ 
(resp. by $b_0,c_0$ for our boundary theory). This implies 
$\langle 0|0\rangle = \langle 0|\{b_0,\xic_0\}|0\rangle =0$. 
The first non-vanishing expressions are $\langle 0|\xic_0\xib_0|0\rangle$ for our boundary theory,
and $\langle 0|\xic_0\bar \xic_0 |0\rangle$ in the bulk. 
This is described in detail in \cite{Flohr:2001zs}.}

As a first non-trivial test for our theory, we would like
to verify that it satisfies world sheet duality.
Let us stress that in this note
we consider a theory in which bulk and boundary
theory consist of Ramond sectors only, a choice that
we shall comment in more detail below. In such a
model, world-sheet duality relates quantities
that are periodic in both world-sheet space and
time. The simplest such quantity in our boundary
theory would be $\tr[q^{L_0 + 1/12} (-1)^F]$ which
vanishes since bosonic and fermionic states come
in pairs on each level of the state space. The
same is certainly true for $\langle N| q^{L_0  + 
1/12} (-1)^{F} |N\rangle$, in agreement 
with world-sheet duality. In order to probe finer 
details of the theory, we need to 
consider quantities with additional insertions 
of fields or zero modes. Here, we shall establish 
the relation
\begin{equation} \label{test1}
 \tr \left(q^{H^o} (-1)^F c(z) \bar c(\bar z)  \right)  \ = \ 
 \langle N|
 \tilde q^{\frac12 H^c} (-1)^{\frac12 F^c} c(\xi) \bar 
  c(\bar \xi)  |N\rangle\, ,
\end{equation}
where $H^o = L_0 + 1/12, q = \exp(2\pi i \tau), \xi = 
\exp(-\frac{1}{\tau}\ln z)$ and $F^c = F + \bar F$, as usual. 
The closed string Hamiltonian is given by
$$
H^c \ = \ \sum_{m\in\QZ}-\Bigl[:b_{-m} c_m: + :\bar b_{-m} 
\bar c_m:\Bigr]   + 1/6 \ \ .
$$
Validity of eq.\ (\ref{test1}) is required by the definition 
of boundary states (see e.g.\ \cite{Recknagel:1997sb}).   
Starting with the left hand side, it is rather easy to 
see that
\begin{eqnarray} 
 \tr \left(q^{L_0 + 1/12} (-1)^F c(z) \bar c(\bar z)
 \right) & = & \ \tr \left(q^{L_0 + 1/12} (-1)^F \xic_0 \xib_0
 \right) \nonumber \\[2mm] & & \hspace*{-3cm} \ = \  
  - i \tau \eta(q)^2 \ = \ \eta(\tilde q)^2\ \ . 
\end{eqnarray} 
In the computation we split off the
term $c_0 b_0$ from $H^o$ and use it to saturate the
fermionic zero modes. The rest is then straightforward. We can
reproduce the same result if we insert our explicit formulae for
the boundary states $|N\rangle$ and $\langle N|$ into the right
hand side of eq.\ (\ref{test1}).

It is possible to perform another similar test of our
boundary theory using the usual trivial boundary conditions
of the ghost system. In this case, the field $c(z)$ is
identified with its own anti-holomorphic partner $\bar
c (\bar z)$ along the boundary and likewise for
the pair $b$ and $\bar b$. Let us recall that the boundary
state $|\id\rangle $ and its dual $\langle \id |$ take the
form \cite{Callan:1987px}
\begin{eqnarray} \nonumber
    |\id \rangle & = &
 \text{exp}\Biggl(\sum_{m=1}^\infty \Bigl(\frac{c_{-m}\bar{b}_{-m}}{m}+
  \frac{\bar{c}_{-m} b_{-m}}{m}\Bigr)\Biggr)(\xic_0-\bar{\xic}_0)|0\rangle
  \\[2mm] \langle \id |  & = & i\langle0|(\xic_0-\bar{\xic}_0)\ \text{exp}
 \Biggl(\sum_{m=1}^\infty\, \frac{1}{m}\,  \Bigl(\bar{b}_{m}c_{m}+b_{m}
 \bar{c}_{m}\Bigr)\Biggr)
\end{eqnarray}
where we use the same notations as before. For the exchange 
of closed string modes between $|N\rangle$ and $\langle \id |$ 
the above formulae imply
\begin{eqnarray}
        \langle\id |\,  \tilde q^{\frac12 H^c }\,  (-1)^{\frac12 F^c}\,
c(\xi) \, |N\rangle & = &  \langle\id |\,  
  \tilde q^{\frac12 H^c }\,  (-1)^{\frac12 F^c}\, \xic_0 \, |N\rangle
\nonumber \\[2mm]
 & & \hspace*{-3cm} \ = \  \tilde{q}^{\frac{1}{12}}\prod_{n=1}^\infty
\bigl(1+\tilde{q}^{2n}\bigr)
\ = \
\sqrt{\frac{\theta_2(2\tilde{\tau})}{2\eta(2\tilde{\tau})}}\ \ .
\label{res1}
\end{eqnarray}
Once more we had to insert the field $c(z)$ in order to get
a non-vanishing result. For comparison with a world-sheet dual,
we need to quantise the ghost system on a strip or, equivalently,
on the upper half plane with trivial boundary conditions on the
positive real axis and our non-trivial ones on the other half. 
A moment of reflection reveals that the following
combinations $\chi^+(z)=2^{-1/2}(b(z)+i\partial c(z))$ and
$\chi^-(z)=2^{-1/2} (ib(z)+\partial c(z))$ obey the simple
periodicity relations $\chi^\pm(e^{2\pi i}
z) = \pm i \chi^\pm(z)$. Hence, they may be constructed through
fermionic $h=1$ twist fields \cite{Kausch:2000fu}
\begin{eqnarray*}
\chi^\pm(z) &= &\sum_{r\in\QZ\mp\frac{1}{4}}\chi_r^\pm\,z^{-r-1}\ \ .
\end{eqnarray*}
The modes $\chi^\pm_r$ obey the same canonical commutation
relations, $\{\chi^+_r,\chi^-_s\}=r \delta_{r,-s}$, as before.
Formulae for the Virasoro generators can easily be worked out. 
For us, it suffices to display the zero mode $\tilde L_0$,
\begin{equation}
 \tilde L_0 \ = \ -\sum_{r\in\QZ-\frac{1}{4}}:\chi^+_r\chi^-_{-r}:- 
\frac{3}{32}\ \ .
\end{equation}
The constant shift by $3/32$ is needed in order to obtain
standard Virasoro relations with the other generators (see 
also \cite{Saleur:1991hk} for a closely related analysis 
of twisted sectors in the bulk theory). The state space of 
our boundary theory contains two ground states $|\Omega_\pm
\rangle$ which are related to each other by the action of a
zero mode $\xic$. On this space we can introduce the 
field $c$ through 
$$ c(z) \ = \ \xic + \frac{i}{\sqrt2} 
\sum _{r\in\QZ -\frac{1}{4}}\frac{\chi_r^+}{r} \,z^{-r}
- \frac{1}{\sqrt2} \sum _{r\in\QZ+\frac{1}{4}}\frac{\chi_r^-}{r} 
\,z^{-r}\ \ . 
$$
>>>From the construction of the state space and our formula for
$\tilde H^o = \tilde L_0 + 1/12$ we infer the following
expression for the mixed open string amplitude,
\begin{equation}
    \tr\Bigl(q^{\tilde H^o} (-1)^F c(z) \Bigr) \ = \ q^{-\frac{1}{96}}
\prod_{n=0}^\infty\Bigl(1-q^{\frac{1}{2}(n+1/2)}\Bigr)\ =\
\sqrt{\frac{\theta_4(\tau/2)}{\eta(\tau/2)}}\, , 
\end{equation}
which reproduces exactly the previous result (\ref{res1})
upon modular transformation and concludes our investigation 
of the new boundary theory.

The choice of our new gluing condition for the $bc$ system 
was motivated by the interest in branes on supergroups. As we 
shall discuss in the next chapter, maximally symmetric branes 
in a WZNW model on a supergroup turn generically out to satisfy Neumann-type 
boundary conditions in the fermionic coordinates. This implies 
that all fermionic zero modes must act non-trivially on the 
space of open string states. In our toy model, the role of the 
fermionic coordinates is played by $c$ and $\bar c$. Hence, we 
needed to find boundary conditions with a four-fold degeneracy
of ground states. For the standard boundary conditions of the 
$bc$ system, $c = \bar c$ along the boundary and hence only one 
fermionic zero mode survives, giving rise to a 2-dimensional 
space of ground states. In this sense, the usual boundary 
conditions of the $bc$ systems are localised in one of the 
fermionic directions. Our boundary conditions come with two 
non-vanishing zero modes $\xib_0$ and $\xic_0$ (and their dual
momenta $c_0$ and $b_0$). This property makes them a  good 
model for maximally symmetric branes on supergroups.

There exist various extensions of our theory that we want to 
briefly comment about. In our analysis we focused on the RR 
sector of the $bc$ ghost system in the bulk. It is certainly 
straightforward to include an NSNS sector in case this is 
required by the application. Furthermore, we can also replace 
the bulk theory by its logarithmic cousin, the symplectic 
fermion model, we will do that in the next section.  

In the case of the $bc$ ghost system, the boundary state 
$|N\rangle$ has a rather novel feature: it describes a logarithmic 
boundary theory in a non-logarithmic bulk. Put differently, the 
$bc$ ghost system possesses a diagonalisable bulk Hamiltonian 
$H^c$. Nevertheless, the Hamiltonian $H^o$ of our new boundary 
theory is non-diagonalisable. Hence, logarithmic singularities 
can appear, but {\em only} when two boundary fields approach 
each other. To the best of our knowledge, such a behaviour has 
never been encountered before.

\chapter{\label{sc:App} The Representation Theory of $\mathbf{\widehat{gl}(1|1)}$}

\section{\label{sc:SpecFlow}Spectral flow automorphisms}

   A useful tool for the investigation of the current algebra
   $\widehat{gl}(1|1)$ and its representations are spectral flow
   automorphisms. The first one, $\gamma_m$, leaves the modes $N_n$
   invariant and acts on the remaining ones as
\begin{equation}
   \label{eq:SFE}
   \gamma_m(E_n)
   \ =\ E_n+km\delta_{n0} \ , \ \ \
\gamma_m(\Psi^\pm_n)\ =\ \Psi^\pm_{n\pm m}\ \ .
\end{equation}
   The previous transformation also induces a modification
   of the energy momentum tensor which is determined by
\begin{equation}
   \gamma_m(L_n)\ =\ L_n+mN_n\ \ .
\end{equation}
   Since the rank of \GL\ is two, there is a second one
   parameter family of spectral flow automorphisms
   $\tilde{\gamma}_\zeta$ which is parameterised by a continuous
   number $\zeta$. It is rather trivial in the sense that its
   action does not act on the mode numbers,
\begin{equation}
   \tilde{\gamma}_\zeta(N_n) \ =\ N_n+k\,\zeta\,\delta_{n0} \ \ \text{and}
   \ \ \tilde{\gamma}_\zeta(L_n) \ =\ L_n+\zeta\,E_n\ \ .
\end{equation}
   All other modes of the currents are left invariant.
\smallskip

   The two spectral flow symmetries above induce a map on
   the set of representations of \agl. Given any
   representation $\rho$ we obtain two new ones by defining
   $\rho_m=\rho\circ\gamma_m$ and
   $\tilde{\rho}_\zeta=\rho\circ\tilde{\gamma}_\zeta$. The latter
   is not very exciting but the former will play a crucial
   role below. Let us thus state in passing that the
   super-characters of these representations are related by
\begin{equation}
   \chi_{\rho_m}(\mu,\tau)\ =\ \chi_\rho(\mu+m\tau,\tau)\ \ .
\end{equation}
   This formula gives severe restrictions on the nature
   of the representations $\rho_m$.

\section{\label{sc:Theta}Some formulae concerning Theta functions}

   Let us recall some facts about the theta function in one variable,
   the reference is Mumford's first book \cite{Mumford}.
   $\theta(\mu,\tau)$ is the unique holomorphic function on
   $\mathbb{C}\times\mathbb{H}$, such that
\begin{equation}
   \begin{split}
     \theta(\mu+1,\tau)
     &\ =\ \theta(\mu,\tau),\\[2mm]
     \theta(\mu+\tau,\tau)
     &\ =\ e^{-\pi i\tau}e^{-2\pi i \mu}\theta(\mu,\tau),\\[2mm]
     \theta(\mu+\frac{1}{2},\tau+1)
     &\ =\ \theta(\mu,\tau),\\[2mm]
     \theta(\mu/\tau,-1/\tau)
     &\ =\ \sqrt{-i\tau}e^{\pi i\mu^2/\tau}\theta(\mu,\tau)\\[2mm]
     \lim_{\text{Im}(\tau)\rightarrow \infty}\theta(\mu,\tau)
     &\ =\ 1\ \ .
   \end{split}
\end{equation}
   The theta functions has a simple expansion as an infinite product,
\begin{equation}
   \theta(\mu,\tau)
   \ =\ \prod_{m=0}^\infty\bigl(1-q^m\bigr)
        \prod_{n=0}^\infty\bigl(1+u^{-1}q^{n+1/2}\bigr)\bigl(1+uq^{n+1/2}\bigr)\
\ ,
\end{equation}
   where $q=e^{2\pi i\tau}$ and $u=e^{2\pi i \mu}$. The \agl\ characters
   in the RR sector we shall present in the next section have a simple
   expression in terms of the variant
\begin{equation}
   \theta\Bigl(\mu-\frac{1}{2}(\tau+1),\tau\Bigr)
   \ =\
(1-u)\prod_{n=1}^\infty\bigl(1-q^n\bigr)\bigl(1-uq^n\bigr)\bigl(1-u^{-1}q^n\bigr)\
\ .
\end{equation}
   Its behaviour under modular $S$ transformations which send the
   arguments of the theta function to $\tilde{\tau}=-1/\tau$ and
   $\tilde{\mu}=\mu/\tau$ can be deduced from the properties above. One
   simply finds
\begin{equation}
   \theta\Bigl(\tilde{\mu}-\frac{1}{2}(\tilde{\tau}+1),\tilde{\tau}\Bigr)
   \ =\ i\sqrt{-i\tilde{\tau}}\,e^{\pi i\tilde{\mu}^2/\tilde{\tau}}\,
        u^{1/2}\tilde{u}^{-1/2}\,q^{-1/8}\tilde{q}^{1/8}\,
        \theta\Bigl(\mu-\frac{1}{2}(\tau+1),\tau\Bigr)\ \ .
\end{equation}

\section{\label{sc:Reps}Representations and their characters}

   In this appendix we review the representations of the current
   superalgebra \agl\ that are relevant for our discussion in the main
   text. We shall slightly deviate from the presentation in
   \cite{Schomerus:2005bf} in putting even more emphasis on the role of
   the spectral flow automorphism \eqref{eq:SFE}. The latter is the
   only constituent which leads to a substantial difference between the
   representation theory of the finite dimensional subalgebra \gl\ and
   that of its affinization \agl.
\smallskip

   All irreducible representations of \agl\ are quotients of Kac
   modules. Just as for \gl, we distinguish between Kac modules
   $\langle e,n\rangle$ and anti Kac modules $\overline{\langle
     e,n\rangle}$. These symbols have been chosen since the ground
   states transform in the corresponding representations of the
   horizontal subalgebra \gl.\footnote{We would like to stress that the
   representations $\langle mk,n\rangle$ and $\overline{\langle
     mk,n\rangle}$ are inequivalent for $m\in\mathbb{Z}$ even though
   their ground states transform identically as long as $m\neq0$. The
   reason becomes clear below.} For $e\not\in k\mathbb{Z}$ both types
   of representations will be called typical, otherwise
   atypical. Typical representations are irreducible and one has the
   equivalence $\langle e,n\rangle\cong\overline{\langle
     e,n\rangle}$. The super-character of (anti) Kac modules can easily
   be found to be
\begin{equation}
   \label{eq:CharTyp}
   \hat{\chi}_{\langle e,n\rangle}(\mu,\tau)
   \ = \ \hat{\chi}_{\overline{\langle e,n\rangle}}(\mu,\tau)
   \ = \ u^{n-1}q^{\frac{e}{2k}(2n-1+e/k)+1/8}
   \theta\Bigl(\mu-\frac{1}{2}(\tau+1),\tau\Bigr)\bigr/\eta(\tau)^3\ \ .
\end{equation}
   When writing down this expression we assumed the ground state with
   quantum numbers $(E_0,N_0)=(e,n)$ to be fermionic. The spectral flow
   $\gamma_m$ transforms the characters of Kac modules according to
\begin{equation}
   \label{eq:SFKac}
   \gamma_m: \quad \chi_{\langle e,n\rangle }(\mu,\tau)
             \ \mapsto\ (-1)^m\chi_{\langle e+mk,n-m\rangle }(\mu,\tau)\ \ .
\end{equation}
   This equation should be interpreted as defining a map between
   representations. We recognise that $\langle e,n\rangle$ is
   transformed into $\langle e+mk,n-m\rangle$ under $\gamma_m$ and that
   the parity of the module is changed if $m$ is odd. A change of
   parity occurs if the interpretation of what are bosonic and what
   are fermionic states is altered compared to the standard choice.
\smallskip

   The equivalence between Kac modules and anti Kac modules is
   destroyed for $e\in k\mathbb{Z}$. For these values the
   representations $\langle mk,n\rangle$ and $\overline{\langle
     mk,n\rangle}$ degenerate and exhibit a single singular vector
   which can be found on energy level $|m|$, see
   \cite{Schomerus:2005bf} for details.\footnote{In order to avoid
   confusion we would like to emphasise that the construction in
   \cite{Schomerus:2005bf} gives rise to Kac modules for $m<0$
   and anti Kac modules for $m>0$. The remaining modules cannot
   be obtained through Verma modules of the sort considered
   there.} This statement is particularly clear for $m=0$ when
   the singular vector is a ground state. In view of
   eq.~\eqref{eq:SFKac} the attentive reader will have
   anticipated that the residual cases $e=mk$ simply arise by
   applying the spectral flow automorphism $\gamma_m$.
\smallskip

   The structure of the Kac modules may be inferred from
   their composition series. According to our previous statements
   the Kac module $\langle mk,n\rangle$ contains precisely one
   irreducible submodule denoted by $\langle n-1\rangle^{(m)}$. The
   quotient of $\langle mk,n\rangle$ by the submodule
   $\langle n-1\rangle^{(m)}$ turns out to be the irreducible
   representation $\bigl(\langle n\rangle^{(m)}\bigr)'$. Hence,
   one can describe the representation using the composition series
\begin{equation}
   \label{eq:CompS}
   \begin{split}
     \langle mk,n\rangle:&\quad \bigl(\langle n\rangle^{(m)}\bigr)'
     \ \longrightarrow \ \langle n-1\rangle^{(m)}\ \ .
   \end{split}
\end{equation}
   Again, all this can be understood best for $m=0$ where the statement
   reduces to well-known facts about Kac modules of the finite dimensional
   subalgebra \gl. This remark especially implies that the atypical irreducible
   representations $\langle n\rangle^{(0)}$ are built over the
   one-dimensional \gl-module $\langle n\rangle$. They are transformed into
   the remaining representations $\langle n\rangle^{(m)}$ under the
   spectral flow automorphism $\gamma_m$. For $m\neq0$, the ground states
   of $\langle n\rangle^{(m)}$ can easily be seen to form the \gl-module
   $\langle mk,n-m\rangle$.
   The information contained in the composition series \eqref{eq:CompS} may
   be used to calculate the super-characters of the atypical irreducible
   representations $\langle n\rangle^{(m)}$. Following the ideas of
   \cite{Rozansky:1992td} one simply finds
\begin{equation}
   \label{eq:CharAtyp}
   \begin{split}
     \hat{\chi}_{\langle n\rangle}^{(m)}(\mu,\tau)
     &\ =\ \sum_{l=0}^\infty\hat{\chi}_{\langle mk,n+l+1\rangle}(\mu,\tau)\\[2mm]
     &\ =\ \frac{u^{n}}{1-uq^m}\,\frac{q^{\frac{m}{2}(2n+m+1)+1/8}
         \theta\Bigl(\mu-\frac{1}{2}(\tau+1),\tau\Bigr)}{\eta(\tau)^3}\ \ .
   \end{split}
\end{equation}
   Analogous results hold for anti Kac modules.
\smallskip

   Finally we need to discuss the projective covers of irreducible
   representations. The typical representations $\langle e,n\rangle$ with
   $e\not\in k\mathbb{Z}$ are projective themselves. But the atypical
   representations $\langle n\rangle^{(m)}$ have more complicated
   projective covers whose composition series reads
\begin{equation}
   \mathcal{P}_n^{(m)}:\quad
   \bigl(\langle n\rangle^{(m)}\bigr)' \ \longrightarrow \ %
   \langle n+1\rangle^{(m)} \oplus \langle n-1\rangle^{(m)}
   \ \longrightarrow \ \bigl(\langle n\rangle^{(m)}\bigr)'\ \ .
\end{equation}
   An alternative description of the projective covers is in terms
   of their Kac composition series
   $\mathcal{P}_n^{(m)}:\langle mk,n\rangle\to\langle mk,n+1\rangle'$.
   Consequently, the characters of projective covers are given by
\begin{equation}
   \hat{\chi}_{\mathcal{P}_n^{(m)}}(\mu,\tau)
   \ = \ \hat{\chi}_{\langle
mk,n\rangle}(\mu,\tau)-\hat{\chi}_{\langle
mk,n+1\rangle}(\mu,\tau) \ .
\end{equation}
   These statements can once again be checked explicitly for $m=0$
   and then generalised to arbitrary values of $m$ by means of the
   spectral flow transformation. For future convenience we shall
   silently omit the superscript $^{(m)}$ in the case that $m=0$.

\section{\label{sc:Mods}Some modular transformations}

   In this section we list the modular
transformations of all the affine characters
   appearing in the previous section. Since all these representations may be
   expressed in terms of Kac modules it is
sufficient to know the transformation
\begin{equation}
   \hat{\chi}_{\langle e',n'\rangle }(\mu,\tau)\ =\
   -\frac{1}{k}\int dedn\ \exp\frac{2\pi
i}{k}\Bigl[e'(n-1/2)+e(n'-1/2)+e'e/k\Bigr]\,
   \hat{\chi}_{\langle e,n\rangle }(\tilde{\mu},\tilde{\tau}) \ \ .
\end{equation}
   to derive the remaining ones. Using the series representation
   \eqref{eq:CharAtyp} one, e.g., obtains the following behaviour
   for characters of atypical representations,
\begin{equation}
   \hat{\chi}_{\langle n'\rangle}^{(m) }(\mu,\tau)\ =\
   \frac{1}{2ki}\int dedn\ \frac{\exp2\pi
i\bigl[e/k(n'+m)+m(n-1/2)\bigr]}{\sin(\pi e/k)}\,
   \hat{\chi}_{\langle e,n\rangle }(\tilde{\mu},\tilde{\tau})\ \ .
\end{equation}
   Similarly, using the Kac composition series for projective
   covers we deduce
\begin{equation}
   \begin{split}
     \hat{\chi}_{\mathcal{P}_{n'}^{(m)}}(\mu,\tau)
     &\ =\ \hat{\chi}_{\langle mk,n'\rangle
}(\mu,\tau)-\hat{\chi}_{\langle mk,n'+1\rangle }(\mu,\tau)\\
     &\ =\ \frac{2i(-1)^m}{k}\int dedn\ \exp2\pi i\Bigl[e/k(n'+mk)+mn\bigr]\,
          \sin(\pi e/k)\,\hat{\chi}_{\langle
e,n\rangle }(\tilde{\mu},\tilde{\tau})\ \ .
   \end{split}
   \raisetag{48pt}
\end{equation}
   The alternating signs in these formulae arise since the spectral flow
   changes the parity of representations for odd values of $m$.

\section{\label{sc:AffFus}Fusion rules of the \agl\ current algebra}

   Up to the need to incorporate the spectral flow automorphism and
   the additional atypical representations induced from it, the fusion
   rules of \agl\ agree precisely with the tensor product decomposition
   of \gl-modules, see e.g.\ \cite{Gotz:2005jz}. Given any two integers,
   $m_1, m_2 \in \mathbb{Z}$, we thus find
\begin{align}
     \langle e_1,n_1 \rangle &\otimes \langle e_2,n_2 \rangle \!\!\!\!\!&\cong\ &%
     \begin{cases}
       \langle e_1+e_2,n_1+n_2\rangle' \oplus \langle e_1+e_2,n_1+n_2-1\rangle
         &,\ e_1{+}e_2\not\in k\mathbb{Z}\\[2mm]
       \mathcal{P}^{(m)}_{n_1+n_2-1}
         &,\ e_1{+}e_2=mk
     \end{cases}\nonumber\\[2mm]
     \langle n_1 \rangle^{(m_1)}&\otimes \langle
n_2 \rangle^{(m_2)} \!\!\!\!\!&\cong\ &\langle n_1+n_2 \rangle^{(m_1+m_2)} \nonumber \\[2mm]
     \langle n_1 \rangle^{(m_1)}&\otimes \langle
e_2,n_2 \rangle \!\!\!\!\!&\!\!\!\!\cong\ &\langle m_1k+e_2, n_1+n_2 \rangle\ \ .
   \raisetag{16pt}
\end{align}
   The prime $'$ in the first line indicates that the representation
   has the opposite parity compared to our standard choice.

\chapter{Some integral formulae}

In this section, we provide a complete list of integral formulae
needed for the computation of the correlation functions. As
reference we use \cite{abramowitz}.

We start with the formulae needed in the computation of boundary
three-point functions. First recall the integral representations
of the hypergeometric function $F(\alpha,\beta;\gamma|x)$
\begin{equation}
    \begin{split}
        &\int_1^\infty du \ |u|^{-\alpha}|u-1|^{-\beta}|u-x|^{-\gamma} \ = \\[2mm]
        &\qquad\qquad\qquad\qquad
        \frac{\Gamma(\alpha+\beta+\gamma-1)\Gamma(1-\beta)}{\Gamma(\alpha+\gamma)}F(\gamma,\alpha+\beta+\gamma-1;\alpha+\gamma\ | \ x) \\[2mm]
        &\int_0^x du \ |u|^{-\alpha}|u-1|^{-\beta}|u-x|^{-\gamma} \ = \\[2mm]
        &\qquad\qquad\qquad\qquad\qquad x^{1-\alpha-\gamma}\frac{\Gamma(1-\alpha)\Gamma(1-\gamma)}{\Gamma(2-\alpha-\gamma)}
        F(\beta,1-\alpha;2-\alpha-\gamma\ | \ x) \\[2mm]
        &\int_{-\infty}^0 du \ |u|^{-\alpha}|u-1|^{-\beta}|u-x|^{-\gamma} \ = \\[2mm]
        &\qquad\qquad\qquad\qquad\frac{\Gamma(\alpha+\beta+\gamma-1)\Gamma(1-\alpha)}{\Gamma(\beta+\gamma)}
        F(\gamma,\alpha+\beta+\gamma-1;\beta+\gamma\ | \ 1-x) \\[2mm]
        &\int_x^1 du \ |u|^{-\alpha}|u-1|^{-\beta}|u-x|^{-\gamma} \ = \\[2mm]
        &\qquad\qquad\qquad\qquad(1-x)^{1-\beta-\gamma}\frac{\Gamma(1-\beta)\Gamma(1-\gamma)}{\Gamma(2-\beta-\gamma)}
        F(\alpha,1-\beta;2-\beta-\gamma\ | \ 1-x) \\[2mm]
    \end{split}
\end{equation}
these integrals converge for $|x|<1$.

If only the first order boundary interaction contributes, we need
the special case $\alpha+\beta+\gamma=2$ of the above integrals
which can be expressed as
\begin{equation}
    \begin{split}
        \int_{[-\infty,0]\ \cup\ [1,\infty]} du \ &|u|^{-\alpha}|u-1|^{-\beta}|u-x|^{-\gamma}  \ = \
          (1-x)^{\alpha-1}x^{\beta-1}\frac{\Gamma(1-\alpha)\Gamma(1-\beta)}{\Gamma(\gamma)}\\[2mm]
\int_{[0,x]}\qquad\qquad  du \
&|u|^{-\alpha}|u-1|^{-\beta}|u-x|^{-\gamma} \ = \
          (1-x)^{\alpha-1} x^{\beta-1}\frac{\Gamma(1-\alpha)\Gamma(1-\gamma)}{\Gamma(\beta)}\\[2mm]
\int_{[x,1]}\qquad\qquad  du \
&|u|^{-\alpha}|u-1|^{-\beta}|u-x|^{-\gamma} \ = \
        (1-x)^{\alpha-1}x^{\beta-1}\frac{\Gamma(1-\beta)\Gamma(1-\gamma)}{\Gamma(\alpha)}\ .\\[2mm]
    \end{split}
\end{equation}

If the bulk interaction term contributes, we have to evaluate the
following integral for $\alpha+\beta+\gamma=0$
\begin{equation}
    \begin{split} \label{eq:A3}
        &\int d^2z \ \frac{(z-\bar{z})}{|z|^{2\alpha+2}|z-1|^{2\beta+2}|z-x|^{2\gamma+2}} \ = \\[2mm]
        &\ \ \ = \ \ \
    \frac{1}{\gamma x+\beta}\int d^2z\ \bar{\del}\ \Bigl(\frac{\bar{z}(\bar{z}-1)(\bar{z}-x)}{|z|^{2\alpha+2}|z-1|^{2\beta+2}|z-x|^{2\gamma+2}}\Bigr)\ +\\[2mm]
    &\ \ \ \ \ \ \ - \frac{1}{\gamma x+\beta}\int d^2z \ \del\ \Bigl(\frac{z(z-1)(z-x)}{|z|^{2\alpha+2}|z-1|^{2\beta+2}|z-x|^{2\gamma+2}}\Bigr) \\[2mm]
    &\ \ \ = \ -\frac{2}{\gamma x+\beta}\int du \ \frac{u(u-1)(u-x)}{|u|^{2\alpha+2}|u-1|^{2\beta+2}|u-x|^{2\gamma+2}} \\[2mm]
    &\ \ \ = \ -\frac{1}{\gamma(\gamma x+\beta)}\frac{d}{dx}\ \Bigl(\ \int_{[-\infty,0]\ \cup\ [1,\infty]} du \ \frac{1}{|u|^{2\alpha+1}|u-1|^{2\beta+1}|u-x|^{2\gamma}}\ +\\[2mm]
    &\ \ \ \ \ \ \ -\int_0^1 du \ \ \frac{1}{|u|^{2\alpha+1}|u-1|^{2\beta+1}|u-x|^{2\gamma}}\ \Bigr)\\[2mm]
        & \ \ \ = \ -4(1-x)^{2\alpha-1}x^{2\beta-1}\Bigl(\frac{\Gamma(-2\alpha)\Gamma(-2\beta)}{\Gamma(2\gamma+1)}+
        \frac{\Gamma(-2\alpha)\Gamma(-2\gamma)}{\Gamma(2\beta+1)}+\frac{\Gamma(-2\beta)\Gamma(-2\gamma)}{\Gamma(2\alpha+1)}\Bigr)\\[2mm]
    \end{split}
\end{equation}
and if two boundary interactions contribute, we need (again
$\alpha+\beta+\gamma=0$)
\begin{equation}
    \begin{split} \label{eq:A4}
      &\int_{a_1}^{b_1} du_1\ \int_{a_2}^{b_2}du_2 \ \frac{|u_1-u_2|}{|u_1u_2|^{\alpha+1}|(u_1-1)(u_2-1)|^{\beta+1}|(u_1-x)(u_2-x)|^{\gamma+1}} \ = \\[2mm]
      &\ \ \qquad\qquad = \ x^{2\beta-1}(1-x)^{2\alpha-1}\int_{c_1}^{d_1} du_1\
      \int_{c_2}^{d_2}du_2 \ \frac{|u_1-u_2|}{|(u_1-1)(u_2-1)|^{\beta+1}|u_1u_2|^{\gamma+1}} \ ,\\[2mm]
     \end{split}
 \end{equation}
 where $c_i=\frac{b_i^{-1}-x^{-1}}{1-x^{-1}}$ and $d_i=\frac{a_i^{-1}-x^{-1}}{1-x^{-1}}$. For these integrals one has to evaluate
 \begin{equation}
     \begin{split} \label{eq:A5}
         &\int_{1}^{\infty} du_1\
     \int_{1}^{u_1}du_2 \ \frac{(u_1-u_2)}{|(u_1-1)(u_2-1)|^{\beta+1}|u_1u_2|^{\gamma+1}} \ = \
     4\frac{\Gamma(-2\alpha)\Gamma(-2\beta)}{\Gamma(2\gamma+1)} \\[2mm]
         &\int_{0}^{1} du_1\
     \int_{0}^{u_1}du_2 \ \frac{(u_1-u_2)}{|(u_1-1)(u_2-1)|^{\beta+1}|u_1u_2|^{\gamma+1}} \ = \
     4\frac{\Gamma(-2\gamma)\Gamma(-2\beta)}{\Gamma(2\alpha+1)} \\[2mm]
         &\int_{-\infty}^{0} du_1\
     \int_{-\infty}^{u_1}du_2 \ \frac{(u_1-u_2)}{|(u_1-1)(u_2-1)|^{\beta+1}|u_1u_2|^{\gamma+1}} \ = \
     4\frac{\Gamma(-2\gamma)\Gamma(-2\alpha)}{\Gamma(2\beta+1)}  \\[2mm]
     \end{split}
\end{equation}
where we used the following special form of the Gamma doubling
formula
\begin{equation}
             \frac{\Gamma(1/2-\alpha)\Gamma(-\alpha)\Gamma(1/2-\beta)\Gamma(-\beta)}{\Gamma(1/2)\Gamma(\gamma+1/2)\Gamma(\gamma+1)} \ = \
         4\frac{\Gamma(-2\alpha)\Gamma(-2\beta)}{\Gamma(2\gamma+1)} \ .
\end{equation}

For the computation of bulk-boundary 2-point functions we use some
special cases of an integral formula that can be found in the recent
work of Fateev and Ribault \cite{Fateev:2007wk}. In case of a single
insertion of the bulk interaction we need
\begin{equation}
    \begin{split}\label{eq:1bulkint}
        \int d^2z\ \frac{|z-\bar{z}|}{|1+z^2|^{2(\alpha+1)}} \ = \ & -2i\pi^{3/2} 2^{-4\alpha}
        \frac{\Gamma(2\alpha+1/2)\Gamma(2\alpha)}{\Gamma^2(\alpha+1)\Gamma^2(\alpha+1/2)} \ .
    \end{split}
\end{equation}
To treat the insertion of one boundary interaction we employ
\begin{equation}
    \begin{split}\label{eq:1bdyint}
        \int du\ |1+u^2|^{-(\alpha+1)} \ = \ & \pi  2^{-2\alpha}\frac{\Gamma(2\alpha+1)}{\Gamma^2(\alpha+1)} \ .
    \end{split}
\end{equation}
The insertion of boundary interactions may be evaluated by means of the
following formula
\begin{equation}
    \begin{split}\label{eq:2bdyint}
        \int du_1du_2\ \frac{|u_1-u_2|}{|1+u_1^2|^{1+\alpha}|1+u_2^2|^{1+\alpha}} \ = \ & 4\pi^{3/2} 2^{-4\alpha}
        \frac{\Gamma(2\alpha+1/2)\Gamma(2\alpha)}{\Gamma^2(\alpha+1)\Gamma^2(\alpha+1/2)} \ .
    \end{split}
\end{equation}


\addcontentsline{toc}{chapter}{Bibliography}


\providecommand{\href}[2]{#2}\begingroup\raggedright\endgroup

\end{document}